\documentclass[12pt]{report}	

\usepackage{utdiss2}  		

\usepackage{amsmath,amsthm,amsfonts,amscd, bm, mathrsfs} 
\usepackage[mathscr]{euscript}

\usepackage{verbatim}      	
\usepackage{makeidx}       	
\usepackage{psfig}         	
\usepackage{epsfig}         	
\usepackage{epstopdf}
\usepackage{citesort}         	%
\usepackage{url}		
\usepackage{algorithm}
\usepackage{algpseudocode}
\usepackage[authoryear, comma]{natbib}
\usepackage[makeroom]{cancel}
\usepackage[labelformat=simple]{subcaption}

\RequirePackage[colorlinks,allcolors=black]{hyperref}
\usepackage{slashbox}
\usepackage{booktabs}

\usepackage{rotating}
\makeatletter
\renewcommand\paragraph{%
  \@startsection{paragraph}
    {4}
    {\z@}
    {3.25ex \@plus1ex \@minus.2ex}
    {-1em}
    {\normalfont\normalsize\bfseries\maybe@addperiod}%
}
\newcommand{\maybe@addperiod}[1]{%
  #1\@addpunct{.}%
}
\makeatother

\usepackage[dvipsnames,usenames]{color}
\definecolor{brown}{rgb}{0.8, 0.33, 0.1}

\DeclareMathOperator*{\argmax}{arg\,max}
\DeclareMathOperator*{\argmin}{arg\,min}

\newcommand{\iidsim}{\stackrel{iid}{\sim}}
\newcommand{\eqdis}{\stackrel{d}{=}}

\newcommand{\Mn}{\mbox{Mn}}

\newcommand{\Dir}{\mbox{Dir}}
\newcommand{\unif}{\mbox{Unif}}
\newcommand{\ti}[1]{\tilde{#1}}
\newcommand{\tpi}{\tilde\pi}
\newcommand{\tC}{\tilde{C}}
\newcommand{\tN}{\tilde{N}}
\newcommand{\tn}{\tilde{n}}

\newcommand{\tp}{\tilde{p}}
\newcommand{\tbx}{\tilde{\bm{x}}}
\newcommand{\phat}{\hat{p}}
\newcommand{\pbar}{\bar{p}}
\newcommand{\Be}{\mbox{Be}}
\newcommand{\Choose}{\mbox{Choose}}
\newcommand{\Categorical}{\mbox{Categorical}}
\newcommand{\du}{\mbox{Discrete-Unif}}

\def\bh {\bm h}
\def\bL {\bm L}

\def\bpi {\bm \pi}
\def\bs {\bm s}

\def\bz {\bm z}
\def\bn {\bm n}
\def\brho {\bm \rho}
\def\btheta {\bm \theta}
\def\by {\bm y}
\def\bx {\bm x}
\def\bZ {\bm Z}
\def\bw {\bm w}
\def\bu {\bm u}
\def\tbu {\tilde{\bm{u}}}
\newcommand{\Cmin}{C_{\mbox{min}}}
\newcommand{\Cmax}{C_{\mbox{max}}}

\def\bY {\bar{Y}}
\def\tY {\tilde{Y}}

\def\bX {\bm X}
\def\bomega {\bm \omega}
\def\bomegae {\bm \omega_{E}}
\def\bomegao {\bm \omega_{O}}
\def\tby {\tilde{\bm y}}
\def\bv {\bm v}
\def\ts {\tilde{s}}
\def\tj {\tilde{j}}
\def\tkappa {\tilde{\kappa}}
\def\btheta {\bm \theta}

\newcommand{\Chat}{\hat{C}}
\newcommand{\Tauhat}{\hat{\Tau}}
\newcommand{\Zhat}{\hat{\bZ}}
\newcommand{\what}{\hat{\bw}}
\newcommand{\ptkghat}{\hat{p}_{tkg}}

\newcommand{\umle}{\bm u^*}
\newcommand{\wts}{w_{t\star}}

\newcommand{\MM}{\mathcal{M}}
\newcommand{\ZZ}{\mathcal{Z}}
\newcommand{\LL}{\mathcal{L}}
\newcommand{\DD}{\mathcal{D}}
\newcommand{\HH}{\mathcal{H}}
\newcommand{\true}{^{\text{TRUE}}}

\newcommand{\BC}{\text{BC}}

\newcommand{\Tau}{\mathcal{T}}
\newcommand{\tTau}{\tilde{\Tau}}
\newcommand{\tpois}{\text{Trunc-Pois}}
\newcommand{\binomial}{\text{Binom}}

\def\bh {\bm h}
\def\tf {\tilde{f}}
\def\tbx {\tilde{\bm x}}
\def\bc {\bm c}
\def\bd {\bm d}
\def\bD {\bm D}
\def\bz {\bm z}
\def\bZ {\bm Z}
\def\bw {\bm w}
\def\bv {\bm v}
\def\bV {\bm V}
\def\bN {\bm N}
\def\bxi {\bm \xi}
\def\bmu {\bm \mu}

\def\bbeta {\bm \beta}
\def\tbbeta {\tilde{\bm \beta}}
\def\bone {\bm 1}
\def\bphi {\bm \phi}
\def\tphi {\tilde{\phi}}
\def\bPhi {\bm \Phi}
\def\btheta {\bm \theta}
\def\by {\bm y}
\def\ba {\bm a}
\def\bb {\bm b}
\def\tb {\tilde{b}}
\def\bs {\bm s}
\def\bby {\bar{\bm y}}
\def\bY {\bar{Y}}

\def\br {\bm r}
\def\bomega {\bm \omega}
\def\det {\text{det}}
\def\logit {\text{logit}}
\def\GP {\mathcal{GP}}
\def\IG {\text{IG}}

\def\Unif {\text{Unif}}
\def\bernoulli {\text{Bernoulli}}
\def\E {\text{E}}
\def\Var {\text{Var}}
\def\Cov {\text{Cov}}
\def\vec {\text{vec}}
\def\diag {\text{diag}}

\def\bpi {\bm \pi}
\def\bC {\bm C}
\def\bD {\bm D}

\makeatletter
\newcommand{\pushright}[1]{\ifmeasuring@#1\else\omit\hfill$\displaystyle#1$\fi\ignorespaces}
\newcommand{\pushleft}[1]{\ifmeasuring@#1\else\omit$\displaystyle#1$\hfill\fi\ignorespaces}
\makeatother

\def\DP {\mathcal{DP}}
\newcommand\eqind{\stackrel{\text{d}}{=}}

\author{Tianjian Zhou}  	

\address{9905 Chukar Circle\\ Austin, Texas 78758}  

\title{Bayesian Nonparametric Models for Biomedical Data Analysis}

\supervisor
	{Peter Mueller}

\committeemembers
	[Michael Daniels]
	[Yuan Ji]
	{Sinead Williamson}

\graduationmonth{August}      
\oneandonehalfspacequote

\theoremstyle{definition}

\theoremstyle{remark}

\newcommand{\latexe}{{\LaTeX\kern.125em2%
                      \lower.5ex\hbox{$\varepsilon$}}}

\chardef\bslash=`\\	

\makeatletter		
\def\square{\RIfM@\bgroup\else$\bgroup\aftergroup$\fi
  \vcenter{\hrule\hbox{\vrule\@height.6em\kern.6em\vrule}%
                                              \hrule}\egroup}
\makeatother		

\makeindex    

\begin{document}

\copyrightpage          

%
%
%
\commcertpage           

\titlepage              

%
\begin{dedication}
\index{Dedication@\emph{Dedication}}%
Dedicated to my family.
\end{dedication}

\begin{acknowledgments}		
\index{Acknowledgments@\emph{Acknowledgments}}%
After completing my four-years life as a Ph.D. student, I would like to take this opportunity to thank many people who helped me. 

First of all, I would like to thank my advisor and mentor, Professor Peter Mueller, without whom I would not be able to write this dissertation. His wonderful lectures have introduced me to Bayesian statistics and MCMC algorithms, and his guidance has prepared me for doing statistics research. He is experienced and is good at explaining complex ideas, so that he can always help me out when I get in trouble in research. He is kind, patient and considerate, so that I always feel supported in my Ph.D. life. I would also like to thank Peter's wife, Gautami Shah, who is also very kind and considerate. I will miss the great memories our group had together (at UT, restaurants, Peter's house, and so on).

Second, I would like to thank Professor Michael Daniels, who has mentored me in part of my dissertation (the missing data part) and has financially supported me in the last two years of my Ph.D. life through research assistantship (NIH CA 183854). 
I would also like to thank Mike for his comprehensive Statistical Modeling 1 class and Reading in Statistics class.

Third, I would like to thank Professor Yuan Ji, who has mentored me in part of my dissertation (the tumor heterogeneity part), has financially supported me for two summers through internship positions at NorthShore University HealthSystem (NIH CA 132897) and has kindly provided me a postdoc position which I have accepted. 

I would also like to thank several other professors in my department. I would like to thank Professor Sinead Williamson, who has served as my dissertation committee member and has provided insightful comments for my dissertation. I would like to thank Professor James Scott for his wonderful Statistical Modeling 2 class and Reading in Statistics class. I would like to thank Professor Carlos Carvalho for his wonderful Time Series and Dynamic Models class and Reading in Statistics class.

I would like to thank the Department of Statistics and Data Sciences for financially supporting me through teaching assistantships over the years. I would like to thank the staffs in our department, in particular Vicki Keller, for always being helpful and supportive. 

I would like to thank my girlfriend, Lan Liang, who has supported me throughout the days and has made the last year of my Ph.D. life an unforgettable memory.  I would like to thank Haoyu Zhang, my friend from college, for helping me throughout the four years and supporting my decisions. 
I would like to thank my collaborator and friend, Subhajit Sengupta, for helping me out when I was in trouble in the tumor heterogeneity project and supporting me when I did my internship at NorthShore.
I would also like to thank my friends, Yanyan Dai, Yu Ding, Chao Ji, Yan Jin, Anastasiya Travina, Li Wang, Mengjie Wang, Yisi Wang, Carlos Pagani Zanini, Anao Zhang, and a lot more, for the fantastic memories we had together.

Finally, I would like to thank my family members, who always trust me and support me in every decision I made. Without them, I am not able to become a grown man.

\end{acknowledgments}

%
\utabstract
\index{Abstract}%
\indent
In this dissertation, we develop nonparametric Bayesian models for biomedical data analysis. In particular, we focus on inference for tumor heterogeneity and inference for missing data. 
First, we present a Bayesian feature allocation model for tumor subclone reconstruction using mutation pairs.
The key innovation lies in the use of short reads mapped to pairs of proximal single nucleotide variants (SNVs). In contrast, most existing methods use only marginal reads for unpaired SNVs.
In the same context of using mutation pairs, in order to recover the phylogenetic relationship of subclones, we then develop a Bayesian treed feature allocation model. In contrast to commonly used feature allocation models, we allow the latent features to be dependent, using a tree structure to introduce dependence. Finally, we propose a nonparametric Bayesian approach to monotone missing data in longitudinal studies with non-ignorable missingness. In contrast to most existing methods, our method allows for incorporating information from auxiliary covariates and is able to capture complex structures among the response, missingness and auxiliary covariates.  Our models are validated through simulation studies and are applied to real-world biomedical datasets.

\tableofcontents   

\listoftables      
\listoffigures     

%
%

\chapter{Introduction}
\label{chap:intro}
\index{Introduction@\emph{Introduction}}%

\section{Overview}

This dissertation develops nonparametric Bayesian
models, corresponding Markov chain Monte Carlo (MCMC) algorithms,
and applications for biomedical data analysis. 
Chapters \ref{chap:PairClone} and \ref{chap:PairCloneTree} are about applications to genomic data analysis, and Chapter \ref{chap:bnpmis} discusses applications to longitudinal missing data analysis.
Nonparametric Bayesian methods provide flexible and highly adaptable
approaches for statistical inference.  In applications with
biostatistics data such methods can often better address biological
research problems than more restrictive parametric methods.

In Chapter \ref{chap:PairClone}, we talk about inference on tumor heterogeneity.  
During tumor growth, tumor cells acquire somatic mutations that 
allow them to gain advantages compared to normal cells.  
As a result, tumor cell
populations are typically heterogeneous consisting of multiple
subpopulations with unique genomes, characterized by different subsets of mutations. This is known as tumor
heterogeneity. The homogeneous subpopulations are known as subclones
and are an important target in precision medicine. We propose a
Bayesian feature allocation model to reconstruct tumor subclones using
next-generation sequencing (NGS) data. The key innovation is the use
of (phased) pairs of proximal single nucleotide variants (SNVs) for the
subclone reconstruction. We utilize parallel tempering to achieve a
better mixing Markov chain with highly multi-modal posterior
distributions.
We also develop trans-dimensional MCMC algorithms with transition
probabilities that are based on splitting the data into training and
test data sets to efficiently implement trans-dimensional MCMC
sampling. Through simulation studies we show that inference under our model
outperforms models using only marginal SNVs by recovering the number of
subclones as well as their structures more accurately. 
This is the case despite significantly smaller number of phased pairs than the number of marginal SNVs.
Estimating our
model for four lung cancer tissue samples, we successfully
infer their subclone structures.  
For this work, I collaborate with Peter Mueller (The University of Texas at Austin), Subhajit Sengupta (NorthShore University HealthSystem) and Yuan Ji (NorthShore University HealthSystem and The University of Chicago). 

In Chapter \ref{chap:PairCloneTree}, we address another important aspect of
statistical inference for tumor heterogeneity, aiming to recover the
phylogenetic relationship of subclones.  
Such inference can significantly
enrich our understanding of subclone evolution and
cancer development. We develop a tree-based feature allocation
model which explicitly models dependence structure among
subclones. 
That is, in contrast to commonly used feature allocation models, we allow the latent features to be dependent, using a tree structure to introduce dependence.
In the application to inference for tumor heterogeneity this tree structure is interpreted as a phylogenetic tree of tumor cell subpopulations.
We adapt our MCMC sampling techniques to efficiently
search the tree space. We analyze a lung cancer data set and infer
the underlying evolutionary process. 
For this work, I collaborate with Subhajit Sengupta, Peter Mueller and Yuan Ji. 

In Chapter \ref{chap:bnpmis}, we model missing data in longitudinal studies.
In longitudinal clinical studies, the research objective is often to make inference on a subject's full data response conditional on covariates that are of primary interest; for example, to calculate the treatment effect of a test drug at the end of a study.
The vector of responses for a research subject is often incomplete due to dropout.
Dropout is typically non-ignorable and in such cases the joint distribution of the full data response and missingness needs to be modeled. In addition to the covariates that are of primary interest, we would often have access to some auxiliary covariates (often collected at baseline) that are not desired in the model
for the primary research question. Such variables can often provide information about the missing responses and missing data mechanism. In this setting, auxiliary covariates should be incorporated in the joint model as well, and we should proceed with inference unconditional on these auxiliary covariates.
As a result, we consider a joint model for the full data response, missingness and auxiliary covariates. 
In particular, we specify a nonparametric Bayesian model for the observed data via Gaussian process priors and Bayesian additive regression trees. These model specifications allow us to capture non-linear and non-additive effects, in contrast to existing parametric methods. We then separately specify the conditional distribution of the missing data response given the observed data response, missingness and auxiliary covariates (i.e. the extrapolation distribution) using identifying restrictions. We introduce meaningful sensitivity parameters that allow for a simple sensitivity analysis. Informative priors on those sensitivity parameters can be elicited from subject-matter experts. We use Monte Carlo integration to compute the full data estimands.
Our methodology is motivated by, and applied to, data from a clinical trial on treatments for schizophrenia.
For this work, I collaborate with Michael Daniels (The University of Texas at Austin).

The remainder of this chapter is organized as follows, Section \ref{sec:bnp} contains basics of Bayesian inference and Bayesian nonparametrics. Sections \ref{sec:lcm}, \ref{sec:lfm} and \ref{sec:reg} present three classes of statistical models, discuss how nonparametric Bayesian methods can be used, and demonstrate applications related to succeeding chapters. These models include latent class models, latent feature models and regression.

\section[Bayesian Nonparametrics]{Bayesian Nonparametrics}
\label{sec:bnp}
\paragraph{Bayesian Inference}
By way of introducing notation, we briefly review the setup of \emph{Bayesian inference}.
Bayesian inference, named for Thomas Bayes, is a particular approach to statistical inference. Let $y$ denote the observed \textit{data}, $\theta$ denote the unobserved \textit{parameters} of interest, and $\tilde{y}$ denote unknown but potentially observable quantities (such as a data point that is not yet observed) of interest. In the Bayesian framework, we update our belief on the unobserved parameters according to evidences in the observed data based on Bayes' rule:
\begin{align}
p(\theta \mid y) = \frac{p(y \mid \theta) p(\theta)}{p(y)}.
\label{eq:bayesrule}
\end{align}
In equation \eqref{eq:bayesrule}, $p(\theta)$ is called the \textit{prior distribution}, $p(y \mid \theta)$ is called the \textit{sampling distribution} (when regarded as a function of $y$ with fixed $\theta$) or the \textit{likelihood} (when regarded as a function of $\theta$ with fixed $y$).
The denominator $p(y)$ is the marginal distribution of $y$, which is calculated by $p(y) = \int p(y \mid \theta) p(\theta) d\theta$.
Inference on $\theta$ is given by the \textit{posterior distribution} $p(\theta \mid y)$. 
We can then make inference on  $\tilde{y}$ based on the posterior \textit{predictive distribution}
\begin{align*}
p(\tilde{y} \mid y) = \int p(\tilde{y} \mid \theta) p(\theta \mid y) d\theta.
\end{align*}
Bayesian statistical inference is stated in terms of probability statements conditional on the observed values of $y$. For a review of Bayesian statistics, see, for example, \cite{gelman2014bayesian} or \cite{hoff2009first}.

\paragraph{Exchangeability}
\textit{Exchangeability} plays an important role in statistics. Suppose we have $N$ random variables (which can be data points or parameters) with a joint distribution $p(y_1, \ldots, y_N)$.  The random variables are called exchangeable if their joint distribution is invariant to permutation. Let $[N] = \{1, \ldots, N \}$ and denote by $\sigma: [N] \rightarrow [N]$ a permutation of $[N]$. (Finite) exchangeability states that
\begin{align*}
y_1, \ldots, y_N \eqdis y_{\sigma(1)}, \ldots, y_{\sigma(N)}
\end{align*}
for any $\sigma$, where $\eqdis$ means equal in distribution. 
Furthermore, an infinite sequence of random variables $y_1, y_2, \ldots$ is called infinitely exchangeable if 
\begin{align*}
y_1, y_2, \ldots \eqdis y_{\sigma(1)}, y_{\sigma(2)}, \ldots,
\end{align*}
where $\sigma : \mathbb{N} \rightarrow \mathbb{N}$ is a finite permutation. That is, for some finite value $N_\sigma$, $\sigma(n) = n$ for all $N > N_\sigma$.

The importance of exchangeability is due to de Finetti's theorem \citep{de1931funzione, hewitt1955symmetric, de1974theory}, which states that
\footnote{This is a rephrased simpler version from \cite{gelman2014bayesian}. The original version is a statement about probability measure.  De Finetti's original paper \cite{de1931funzione} is for the case of binary random variables, and \cite{hewitt1955symmetric} extended it to any real valued random variables.} 
if $y_1, y_2, \ldots$ are infinitely exchangeable random variables, their joint distribution can be expressed as a mixture of independent and identical distributions
\begin{align}
p(y_1, \ldots, y_N) = \int \left( \prod_{i=1}^N p(y_i \mid \theta) \right) p(\theta) d\theta.
\label{eq:definetti1}
\end{align}
The theorem can be rephrased from a more general perspective 
\footnote{Another simpler version from \cite{teh2011dirichlet}.}.
If $y_1, y_2, \ldots$ are infinitely exchangeable, there exists a random distribution $F$ such that the sequence is composed of i.i.d. draws from it,
\begin{align}
p(y_1, \ldots, y_N) = \int \prod_{i=1}^N F(y_i)  dp(F).
\label{eq:definetti}
\end{align}
That is, $\theta$ in Equation \eqref{eq:definetti1} can be interpreted as indexing a probability measure $F$, or $\theta$ can even be the probability measure $F$ itself.

\paragraph{Bayesian Nonparametrics}
A model is called \textit{parametric} if it only has a finite (and usually small) number of parameters, i.e. $\theta$ lives in a finite dimensional space. In contrast, a \textit{nonparametric} model has a potentially infinite number of parameters, i.e. $\theta$ or $F$ are in an infinite dimensional space. Thus, nonparametric Bayesian inference requires constructing probability distributions on an infinite dimensional parameter space. Such probability distributions are called \textit{stochastic processes} with sample paths in the parameter space. 
Nonparametric Bayesian methods avoid the often restrictive assumptions of parametric models and provide flexible and highly adaptable approaches for statistical modeling. Reviews of Bayesian nonparametrics include
\cite{hjort2010bayesian, ghosh2003bayesian, muller2015bayesian, walker1999bayesian, muller2004nonparametric, orbanz2011bayesian, gershman2012tutorial} and \cite{orbanz2014notes} \footnote{\cite{hjort2010bayesian} includes a set of introductory and overview papers, \cite{ghosh2003bayesian} focuses on posterior convergence, and \cite{muller2015bayesian} has more discussion on data analysis problems.}.

Nonparametric Bayesian approaches have been widely used in many statistical inference problems, including density estimation, clustering, feature allocation, regression, classification, and graphical models. In the next sections, we give examples to show how Bayesian nonparametric approaches can be applied to address those important problems.

\section{Latent Class Models}
\label{sec:lcm}
Suppose we have $N$ objects $y_1, \ldots, y_N$. In a \emph{latent class model} (for a review, see \citealp{griffiths2011indian}), each object $y_n$ belongs to a latent class $c_n = k$, $k = 1, \ldots, K$, where $K$ is the number of possible classes, and $K = \infty$ is allowed. 
When $K = \infty$, the model has an infinite number of classes and thus has an infinite number of parameters. 
In this case, the model is nonparametric.
We are interested in how the classes are related to the objects, $p(\by \mid \bc)$, and the distribution over class assignments, $p(\bc)$. For $p(\by \mid \bc)$, we assume conditional independence,
\begin{align}
p(\by \mid \bc) &= \prod_{n = 1}^N p(y_n \mid c_n), \nonumber\\
p(y_n \mid c_n = k, \mu_k^*) &= G(\cdot \mid \mu_k^*). \label{eq:lcm1}
\end{align}
That is, for an object belonging to class $k$, we assume it has a distribution $G$ with parameter $\mu_k^*$. We then put some prior distribution $F_0$ on the $\mu_k^*$'s,
\begin{align}
\mu_1^*, \ldots, \mu_K^* \iidsim F_0. 
\label{eq:lcm2}
\end{align}

Next, we specify $p(\bc)$. Specifying $p(\bc)$ is equivalent to defining a distribution on a \emph{random partition} of the index set $[N] := \{1, 2, \ldots, N \}$. The formal definition of random partition follows in Section \ref{sec:random_partition}.
The equivalence can be seen by noticing that the unique values of $\bc$ correspond to a partition of $[N]$, $f_N = \{ A_1, \ldots, A_K\}$, with $n \in A_k$ if $c_n = k$, and vice versa.

\subsection{Random Partition}
\label{sec:random_partition}
We briefly summarize the definition of random partitions as given in \cite*{broderick2013cluster}. See there for more details and discussion.
Let $[N] := \{1, 2, \ldots, N \}$ denote the index set of $N$ objects.  A \textit{partition} $f_N$ of $[N]$ is a collection of mutually exclusive, exhaustive, nonempty subsets $A_1, \ldots, A_K$ of $[N]$ called \textit{blocks}. 
Let $f_N = \{ A_1, \ldots, A_K\}$, where $K$ is the number of blocks. 
Here $N = \infty$ is allowed, in which case the index set becomes $\mathbb{N} = \{1, 2, 3, \ldots \}$, and $K = \infty$ is also allowed. 

Let $\mathcal{F}_N$ be the space of all partitions of $[N]$. A \emph{random partition} $F_N$ of $[N]$ is a random element of  $\mathcal{F}_N$. The probability $p(F_N = f_N)$ is called the partition probability function of $F_N$.

Exchangeability of a random partition can be defined as follows. Let $\sigma : \mathbb{N} \rightarrow \mathbb{N}$ be a finite permutation. That is, for some finite value $N_\sigma$, $\sigma(n) = n$ for all $N > N_\sigma$. Furthermore, for any block $A \subset \mathbb{N}$, denote the permutation applied to the block as $\sigma(A) := \{ \sigma(n): n \in A\}$. For any partition $\Pi_N$, denote the permutation applied to the partition as  $\sigma(\Pi_N) := \{ \sigma(A): A \in \Pi_N\}$.
A random partition $\Pi_N$ is called \emph{exchangeable} if $\Pi_N \stackrel{d}{=} \sigma(\Pi_N)$ for every permutation of $[N]$. The importance of exchangeability has been stated in Section \ref{sec:bnp}.

\subsection{Chinese Restaurant Process}
The Chinese restaurant process (CRP) defines an exchangeable random partition. The CRP can be derived in multiple ways, such as from the Dirichlet process \citep{blackwell1973ferguson}, or by taking limit of a finite mixture model \citep{green2001modelling, neal1992bayesian, neal2000markov}. We take the latter approach, following \cite{griffiths2011indian}.

\paragraph{A Finite Mixture Model}
We assume object $i$ belongs to class $k$ with probability $\pi_k$, 
\begin{align*}
p(c_n = k \mid \pi_k) = \pi_k.
\end{align*}
Note that the distribution of $\bc$ always induces a distribution on a random partition of $[N]$. In the above case, the assignment of an object to a class is independent of the assignments of the other objects conditional on $\bpi$, and the latent class model is called a \emph{mixture model}. To complete the model, we put a symmetric Dirichlet distribution prior on $(\pi_1, \ldots, \pi_K)$,
\begin{align*}
(\pi_1, \ldots, \pi_K) \sim  \Dir(\alpha / K, \ldots, \alpha / K),
\end{align*}
$\pi_k \geq 0$ and $\sum_{k=1}^K \pi_k = 1$.

Integrating out the $\pi_k$'s, the marginal distribution of $\bc$ is
\begin{align}
p(\bc) = \frac{\prod_{k=1}^K \Gamma(m_k + \frac{\alpha}{K})}{\Gamma \left( \frac{\alpha}{K} \right)^K} \frac{\Gamma(\alpha)}{\Gamma(N + \alpha)},
\label{eq:crp1}
\end{align}
where $m_k = \sum_{n=1}^N I(c_n = k)$ is the number of objects assigned to class $k$. This distribution is exchangeable, since it only depends on the counts and does not depend on the ordering of objects.

\paragraph{Equivalence Classes}
A partition $\bc$ includes an ordering of the $K$ blocks by increasing labels $k = 1, \ldots, K$. In many applications, we are only interested in the division of objects, and the ordering of the blocks does not matter. For example, $f_3 = \{ \{ 1\}, \{ 2, 3\} \}$ and $f_3' = \{ \{ 2, 3\}, \{ 1 \} \}$ correspond to the same division of objects, where the only difference is the choice of labels of the blocks. If the order of the blocks is not identifiable, it is helpful to define an equivalence class of assignment vectors, denoted by $[\bc]$, with two assignment vectors $\bc$ and $\bc'$ belonging to the same equivalence class if they imply the same division of objects.

We therefore focus on the equivalence classes $[\bc]$. Let $K_+$ be the number of classes for which $m_k > 0$, and $K_0$ be the number of classes for which $m_k = 0$, so $K = K_0 + K_+$. 
The cardinality of $[\bc]$ is $\left( K! / K_0! \right)$. 
Taking the summation over all assignment vectors that belong to the same equivalence class, and expanding \eqref{eq:crp1}, we obtain
\begin{align}
p([\bc]) &= \sum_{\bc \in [\bc]} p(\bc) \nonumber\\
&= \frac{K!}{K_0!} \left(\frac{\alpha}{K} \right)^{K_+} \left( \prod_{k=1}^{K_+}  \prod_{j=1}^{m_k-1} \left( j + \frac{\alpha}{K} \right) \right) \frac{\Gamma(\alpha)}{\Gamma(N+\alpha)}.
\label{eq:crp2}
\end{align}

\paragraph{Taking the Infinite Limit}
When the number of classes $K \rightarrow \infty$, taking the limit in Equation \eqref{eq:crp2}, we get
\begin{align}
\lim_{K \rightarrow \infty} p([\bc]) = \alpha^{K_+} \left( \prod_{k=1}^{K_+} (m_k - 1)! \right) \frac{\Gamma(\alpha)}{\Gamma(N+\alpha)}.
\label{eq:crp3}
\end{align}
See details in \cite{griffiths2011indian}. Note that this distribution is still exchangeable, just as in the finite case.

\paragraph{Chinese Restaurant Analogy}
The Chinese restaurant analogy \citep{aldous1985exchangeability} 
\footnote{Aldous credits this analogy to Jim Pitman and Lester Dubins.}
comes from the fact that the distribution in Equation \eqref{eq:crp3} can be described by a Chinese restaurant metaphor.
Let the $N$ objects be customers in a restaurant, and the $K$ classes be tables at which they sit, $K = \infty$. The customers enter the restaurant one by one, and each chooses a table at random.
At time 1, the first customer comes in and chooses the first table to sit. At time $n$, the $n$-th customer comes in, chooses an occupied table with probability proportional to the number of customers sitting at that table, or the first unoccupied table with probability proportional to $\alpha$, $n = 2, \ldots, N$. The customers and tables form a partition of $\bc$, if we treat the tables as partition blocks. 
Denote by $c_n = k$ the event that customer $n$ sits at table $k$, 
$m_k$ the number of customers sitting at table $k$ after time $n-1$, and
$K_+ =  \max \{ c_1, \ldots, c_{n-1} \}$ .
Mathematically, the CRP can be written as
\begin{align}
c_{n} \mid c_1, \ldots, c_{n-1} = 
\begin{cases}
k, \quad &\text{w. pr. $\frac{m_k}{\alpha + n - 1}$, for $k \leq K_+$;}\\
K_+ + 1, \quad &\text{w. pr. $\frac{\alpha}{\alpha + n - 1}$,}
\end{cases}
\label{eq:crp4}
\end{align}
for $n = 1, \ldots, N$. The probability of a partition of $\bc$ given by Equation \eqref{eq:crp4} is identical to what given in Equation \eqref{eq:crp3}.

\paragraph{Dirichlet Process}
The CRP defines an exchangeable random partition of $[N]$. We can further extend he model by assigning each table $k$ a value $\mu_k^*$, with $\mu_k^*$ generated from some fixed distribution $F_0$. We then assign customer $n$ a value $\mu_n = \mu_k^*$ if the customer sits at table $k$. 
The predictive distribution of $\mu_{n}$ given the values $\mu_1, \ldots, \mu_{n-1}$ of the first $n-1$ customers is
\begin{align}
\mu_{n} \mid \mu_1, \ldots, \mu_{n-1} = 
\begin{cases}
\mu_k^*, \quad &\text{w. pr. $\frac{m_k}{\alpha+n-1}$, for $k \leq K_+$;}\\
\mu_{K_+ + 1}^* \sim F_0, \quad &\text{w. pr. $\frac{\alpha}{\alpha+n-1}$.}
\end{cases}
\label{eq:polya_urn}
\end{align}
It can be shown that the random sequence $\mu_1, \mu_2, \ldots$ is infinitely exchangeable. By Equation \eqref{eq:definetti}, there exists a random distribution $F$ such that $\mu_n \mid F \iidsim F$ and $F \sim \upsilon$.
Here $\upsilon$ is a prior over the random distribution $F$, which is known as the Dirichlet process (DP) \citep{ferguson1973bayesian}. We denote by $\DP(\alpha, F_0)$ a DP with concentration parameter $\alpha$ and base distribution $F_0$. Equation \eqref{eq:polya_urn} is also called the P\'{o}lya urn representation \citep{blackwell1973ferguson} of the DP. Using the notion of DP, we can re-parameterize Equations \eqref{eq:lcm1}, \eqref{eq:lcm2} and \eqref{eq:crp4} with a hierarchical model
\begin{align}
\begin{split}
y_n \mid \mu_n &\sim G( \cdot \mid \mu_n) \\
\mu_n \mid F &\sim F \\
F \mid \alpha, F_0 &\sim \DP(\alpha, F_0)
\end{split}
\label{eq:dpm}
\end{align}
The model \eqref{eq:dpm} is called a Dirichlet process mixture (DPM) model.

The DP is probably the most popular nonparametric Bayesian model. Discussions and extensions of the DP include \cite{blackwell1973discreteness, blackwell1973ferguson, antoniak1974mixtures, lo1984class, sethuraman1994constructive, pitman1997two, maceachern2000dependent, muller2004method, teh2006hierarchical, rodriguez2008nested, lijoi2010models, adams2010tree, teh2011dirichlet, de2015gibbs}, where \cite{sethuraman1994constructive} proposes the stick-breaking construction of the DP, \cite{pitman1997two} extend the DP to the Pitman-Yor process, \cite{maceachern2000dependent} proposes the dependent DP, \cite{teh2006hierarchical} develop the hierarchical DP, and \cite{rodriguez2008nested} develop the nested DP. Literature about posterior inference methods for the DP includes \cite{west1994hierarchical, escobar1995bayesian, maceachern1998estimating, neal2000markov, rasmussen2000infinite, ishwaran2001gibbs, jain2004split}, and \cite{blei2006variational}.

\subsection{Related Applications}
\label{sec:lcm_app}
Latent class models, in particular, the DPM model and its variations, have been extensively used in many data analysis problems. We highlight their applications to inference for tumor heterogeneity and inference for missing data because of the relevance to the following chapters.

\paragraph{Inference for Tumor Heterogeneity}
Tumor cell populations are typically heterogeneous consisting of multiple homogeneous subpopulations with unique genomes. Such subpopulations are known as subclones. 
Our goal is reconstructing such subclones from next-generation sequencing (NGS) data \citep{mardis2008next}.  See more details in Chapters \ref{chap:PairClone} and \ref{chap:PairCloneTree}. One approach to this problem is to model the observed read count data using a latent class model. This approach is taken by PyClone \citep{roth2014pyclone} and PhyloWGS \citep{jiao2014inferring, deshwar2015phylowgs}. Tumor evolution is a complex process involving many biological details, such as tumor purity, copy number variations and tumor phylogeny. Also, NGS data are subject to sequencing error and are often overdispersed. For simplicity of illustration we consider only one pure tumor tissue sample, ignore all the complexities mentioned above and also ignore the zygosity of the mutation sites. For detailed discussions see, for example, \cite{roth2014pyclone, jiao2014inferring, deshwar2015phylowgs} and Chapters \ref{chap:PairClone} and \ref{chap:PairCloneTree}.

Consider $S$ single nucleotide variants (SNVs). 
Here SNVs refer to the loci of the nucleotides (base pairs) for which we record variants. Variants are defined relative to some reference genome. 
The SNVs are the objects in the latent class model.
In an NGS experiment, DNA fragments are first produced by extracting
the DNA molecules from the cells in a tumor sample. The fragments are
then sequenced using short reads. The short reads are mapped to the reference genome, and counts are recorded for each locus (i.e. base pair).
In the end, for each SNV locus $s$ ($s = 1, \ldots, S$), denote by $N_s$ and $n_{s}$ the total number of reads and number of variant reads covering the locus, respectively. The total number of reads $N_s$ is usually treated as a fixed number. PyClone uses a DPM model for $n_{s}$,
\begin{align*}
n_s \mid p_s &\sim \binomial (N_s; p_s), \\
p_s \mid F &\sim F, \\
F \mid  \alpha, F_0 &\sim \DP(\alpha, F_0),
\end{align*}
where the base distribution $F_0$ is chosen to be $\unif(0,1)$. Here $p_s$ is known as the cellular prevalence of mutation $s$, i.e. the fraction of cancer cells harbouring a mutation. 
The DP prior for $F$ allows multiple mutations to share the same cellular prevalence. 
The critical step towards subclone reconstruction is the following.
Mutations having the same cellular prevalence are thought of having occurred at the same point in the clonal phylogeny. 
Thus, latent classes of mutations can be used as markers of subclone populations \citep{roth2014pyclone}. 
We note that this is essentially an application of latent class models to \emph{clustering}. PhyloWGS, on the other hand, uses
the tree-structured stick breaking process (TSSB)
\citep{adams2010tree} as the prior for $F$,
\begin{align*}
F \mid  \alpha, \gamma, F_0 &\sim \text{TSSB}(\alpha, \gamma, F_0),
\end{align*}
which allows it to infer tumor phylogeny.

One restriction of the latent class model is that each object can only belong to one class. In the tumor heterogeneity application, this restriction implies that each mutation can only occur once in the clonal phylogeny.
Therefore, subclone reconstruction methods based on latent class models usually rely on the infinite site assumption (ISA) \citep{kimura1969number}, which can be summarized as \citep{roth2014pyclone}
\begin{enumerate}
\item Subclone populations follow a \emph{perfect} phylogeny. That is, no SNV site mutates more than once in its evolutionary history;
\item Subclone populations follow a \emph{persistent} phylogeny. That is, mutations do not disappear or revert.
\end{enumerate}
However, ISA is not necessarily valid, in which case we should model the observed read count data using latent feature models. See Section \ref{sec:lfmapp}.

\paragraph{Inference for Missing Data}
Missing data are very common in real studies.
Missingness is typically non-ignorable \citep{rubin1976inference, little2014statistical}, and in such cases the joint
distribution of the full data response and missingness needs to be modeled. 
We focus on the missing outcome case. See \cite{linero2017general} for a general review of Bayesian nonparametric approach to missing outcome data. 
Let $Y_{ij}$ denote the outcome that was planned to
be collected for subject $i$ at time $j$, and $R_{ij}$ be the missingness indicator with $R_{ij} = 1$ or $0$ accordingly as $Y_{ij}$ is observed or not, $i = 1, \ldots, N$, $j = 1, \ldots, J$. 
Let $\bm X_i$ denote the covariates that are of primary interest to the study. In longitudinal clinical trial setting, $\bm X_i$ is usually an indicator of treatment.
We often treat $\bm X_i$ as fixed and do not proceed with inference on it.
The full data for subject $i$ are $(\bm Y_{i}, \bm R_i, \bm X_i)$.
The observed data for subject $i$ are $(\bm Y_{i, \text{obs}}, \bm R_i, \bm X_i)$, where $\bm Y_{i, \text{obs}} = (Y_{ij} \mid j: R_{ij} = 1)$.
We assume $(\bm Y_i, \bm R_i \mid \bm X_i) \iidsim p(\by, \br \mid \bx)$. We stratify the model by $\bx$ and suppress the conditional on $\bx$ hereafter to simplify notation. The extrapolation factorization \citep{daniels2008missing} factorizes
\begin{align*}
p(\by, \br) = p(\by_{\text{mis}} \mid \by_{\text{obs}}, \br) p(\by_{\text{obs}}, \br).
\end{align*}
The observed data distribution $p(\by_{\text{obs}}, \br)$ is identified by the observed data, while the extrapolation distribution $p(\by_{\text{mis}} \mid \by_{\text{obs}}, \br)$ is not. Identifying the extrapolation distribution relies on untestable assumptions such as parametric models for the full data distribution or identifying restrictions. See Chapter \ref{chap:bnpmis} or \cite{linero2017general}.

For current discussion, we focus on specifying $p(\by_{\text{obs}}, \br)$. One way of specifying $p(\by_{\text{obs}}, \br)$ is specifying $p(\by, \br)$, and set
\begin{align*}
p(\by_{\text{obs}}, \br) = \int p(\by, \br) d\by_{\text{mis}}.
\end{align*}
This is known as the working model idea \citep{linero2015flexible, linero2017nonparametric, linero2017general}.
\cite{linero2017general} discuss a Bayesian nonparametric framework for modeling a complex joint distribution of outcome and missingness, which sets
\begin{align}
p(\by, \br) = \sum_{k=1}^K \pi_k \, G( \cdot \mid \bmu_k).
\label{eq:intro_bnpmis}
\end{align}
Equation \eqref{eq:intro_bnpmis} is another way of writing a latent class model
\begin{align*}
\bm Y_i, \bm R_i \mid c_i = k, \bmu_k &\sim G(\cdot \mid \bmu_k), \\
p(c_i = k \mid \pi_k) &= \pi_k.
\end{align*}
When $K = \infty$, mixture models of the form \eqref{eq:intro_bnpmis} can approximate any joint distribution for $(\bm Y_i, \bm R_i)$ (subject to technical constraints).
We note that this is essentially an application of latent class models to \emph{density estimation}.

\cite{linero2017general} use a model of the form \eqref{eq:intro_bnpmis} to analyze data from the Breast Cancer Prevention Trial. In this trial, $Y_{ij} = 1$ or $0$ represent subject $i$ is depressed or not at time $j$. They model
\begin{align*}
p(\by, \br) = \sum_{k=1}^{\infty} \pi_k \left\{ \prod_{j=1}^J \gamma_{kj}^{r_j} 
(1 - \gamma_{kj})^{1 - r_j} \right\} \left\{ \prod_{j=1}^J \beta_{kj}^{y_j} 
(1 - \beta_{kj})^{1 - r_j} \right\}.
\end{align*}

\cite{linero2015flexible} discussion is another example of applying latent class models to inference for missing data. They analyze data from an acute schizophrenia clinical trial, where the outcome $Y_{ij}$ is a continuous variable called the positive and negative syndrome scale (PANSS) score. Missingness is monotone in this application. That means, if $Y_{ij}$ was unobserved then $Y_{i,j+1}$ was also unobserved. Let $S_i$ denote the dropout time, i.e. if $S_i = j$ then $Y_{ij}$ was observed but $Y_{i,j+1}$ was not. For monotone missingness, $S$ captures all the information about missingness. Denote by $\bm \bY_{ij} = (Y_{i1}, \ldots, Y_{ij})$ the history of outcomes through the first $j$ times. They model
\begin{align*}
p(\by, s) = \sum_{k=1}^{\infty} \pi_k f(\by \mid \bmu_k, \Sigma_k) g(s \mid \by, \bm \zeta_k, \bm \gamma_k),
\end{align*}
with 
\begin{align*}
&f(\by \mid \bmu, \Sigma) = N(\by; \bmu, \Sigma), \\
&\logit [g(S = j \mid S \geq j, \by, \bm \zeta, \bm \gamma)] =  \zeta_{j} + \bm \gamma_{j}^T \bby_j.
\end{align*}

\section{Latent Feature Models}
\label{sec:lfm}
Suppose we have $N$ objects, $y_1, \ldots, y_N$. In a \emph{latent feature model} (for a review, see \citealp{griffiths2011indian}), each object $y_n$ is represented by a vector of latent feature values $\bd_n = (d_{n1}, \ldots, d_{nK})$, where $K$ is the number of features.
Similar to the latent class model case, $K = \infty$ is allowed, in which case the model is nonparametric.
Examples of latent feature models include probabilistic principle component analysis \citep{tipping1999probabilistic} and factor analysis \citep{roweis1999unifying}. We focus on the case that there is no upper bound on the number of features.

We can break the vector $\bd_n$ into two components: a binary vector $\bz_n$ with $z_{nk} = 1$ if object $n$ has feature $k$ and $0$ otherwise, and a second vector $\bv = (v_1, \ldots, v_K)$ indicating the value of each feature. 
The vector $\bd_n$ can be expressed as the elementwise product of $\bz_n$ and $\bv$, i.e.
\begin{align}
d_{nk} = z_{nk} v_k, \quad k = 1, \ldots, K.
\label{eq:lfm0}
\end{align}
Let $\bD = (\bd_1, \ldots,  \bd_N)^T$ and $\bZ = (\bz_1, \ldots,  \bz_N)^T$ denote $N \times K$ matrices with columns $\bd_n$ and $\bz_n$, respectively.
We are interested in how the feature values are related to the data, $p(\by \mid \bD)$, and the distribution over feature values, $p(\bD)$. For $p(\by \mid \bD)$, we generally assume conditional independence
\begin{align}
p(\by \mid \bD) &= \prod_{n=1}^N p(y_n \mid \bd_n), \nonumber\\
p(y_n \mid \bd_n) &\sim G( \cdot \mid \bd_n). \label{eq:lfm1}
\end{align}
The distribution $p(\bD)$ can be broken into two components, with $p(\bD)$ being determined by $p(\bZ)$ and $p(\bv)$. We assume an independent prior on $\bv$,
\begin{align*}
v_{1}, \ldots, v_K \iidsim F_0.
\end{align*}

We then focus on defining a prior on $\bZ$. Specifying $p(\bZ)$ is equivalent to defining a distribution on a \emph{random feature allocation} of the index set $[N]$. The formal definition of random feature allocation follows in Section \ref{sec:random_fa}.
The equivalence can be seen by noticing that the values of $\bZ$ correspond to a feature allocation $f_N = \{ A_1, \ldots, A_K\}$, with $n \in A_k$ if $z_{nk} = 1$ and $n \notin A_k$ if $z_{nk} = 0$, and vice versa.

\subsection{Random Feature Allocation}
\label{sec:random_fa}
We briefly summarize the definition of random feature allocations from \cite*{broderick2013cluster}. See there for more details and discussion.
Feature allocations could be seen as a generalization of partitions which relaxes the restriction to mutually exclusive and exhaustive subsets. Consider an index set $[N]$. A \emph{feature allocation} $f_N$ of $[N]$ is a multiset of non-empty subsets of $[N]$ called \emph{features}, such that no index $n$ belongs to infinitely many features. Denote by $f_N = \{ A_1, \ldots, A_K\}$, where $K$ is the number of features. 
Here $N = \infty$ and $K = \infty$ are allowed.
For example, a feature allocation of $[6]$ is $f_6 = \{ \{1, 4\}, \{1, 3, 6\}, \{4\}, \{4\}, \{4\}\}$.

Let $\mathcal{F}_N$ be the space of all feature allocations of $[N]$. A \emph{random feature allocation} $F_N$ of $[N]$ is a random element of $\mathcal{F}_N$. 

Let $\sigma : \mathbb{N} \rightarrow \mathbb{N}$ be a finite permutation. That is, for some finite value $N_\sigma$, $\sigma(n) = n$ for all $N > N_\sigma$. Furthermore, for any feature $A \subset \mathbb{N}$, denote the permutation applied to the feature as $\sigma(A) := \{ \sigma(n): n \in A\}$. For any feature allocation $F_N$, denote the permutation applied to the feature allocation as  $\sigma(F_N) := \{ \sigma(A): A \in F_N\}$. Let $F_N$ be a random feature allocation of $[N]$.
A random feature allocation $F_N$ is called \emph{exchangeable} if $F_N \stackrel{d}{=} \sigma(F_N)$ for every permutation of $[N]$.

\begin{figure}[h!]
\begin{center}
\begin{subfigure}[t]{.48\textwidth}
\centering
\includegraphics[width=\textwidth]{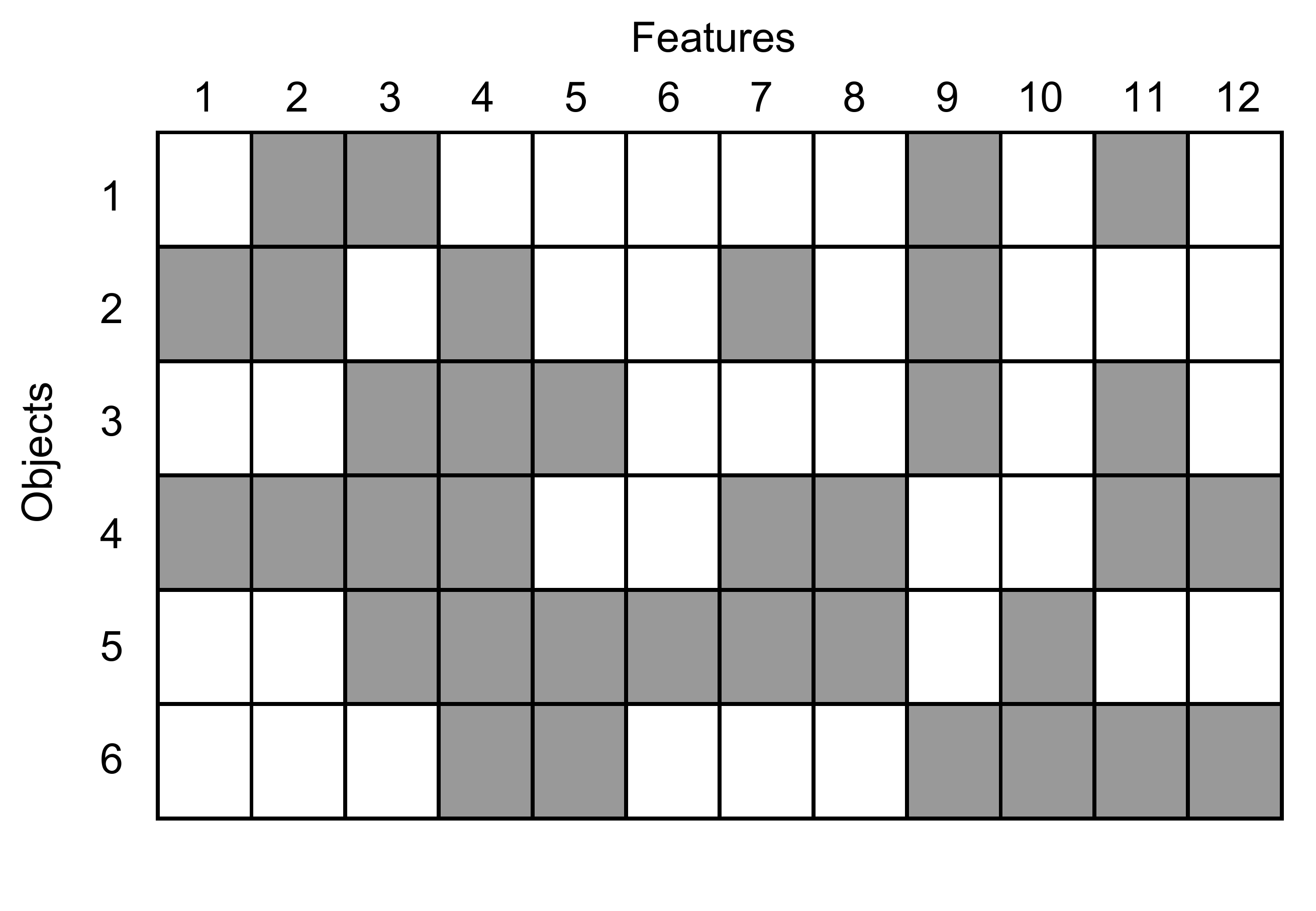}
\caption{}		
\end{subfigure}
\begin{subfigure}[t]{.48\textwidth}
\centering
\includegraphics[width=\textwidth]{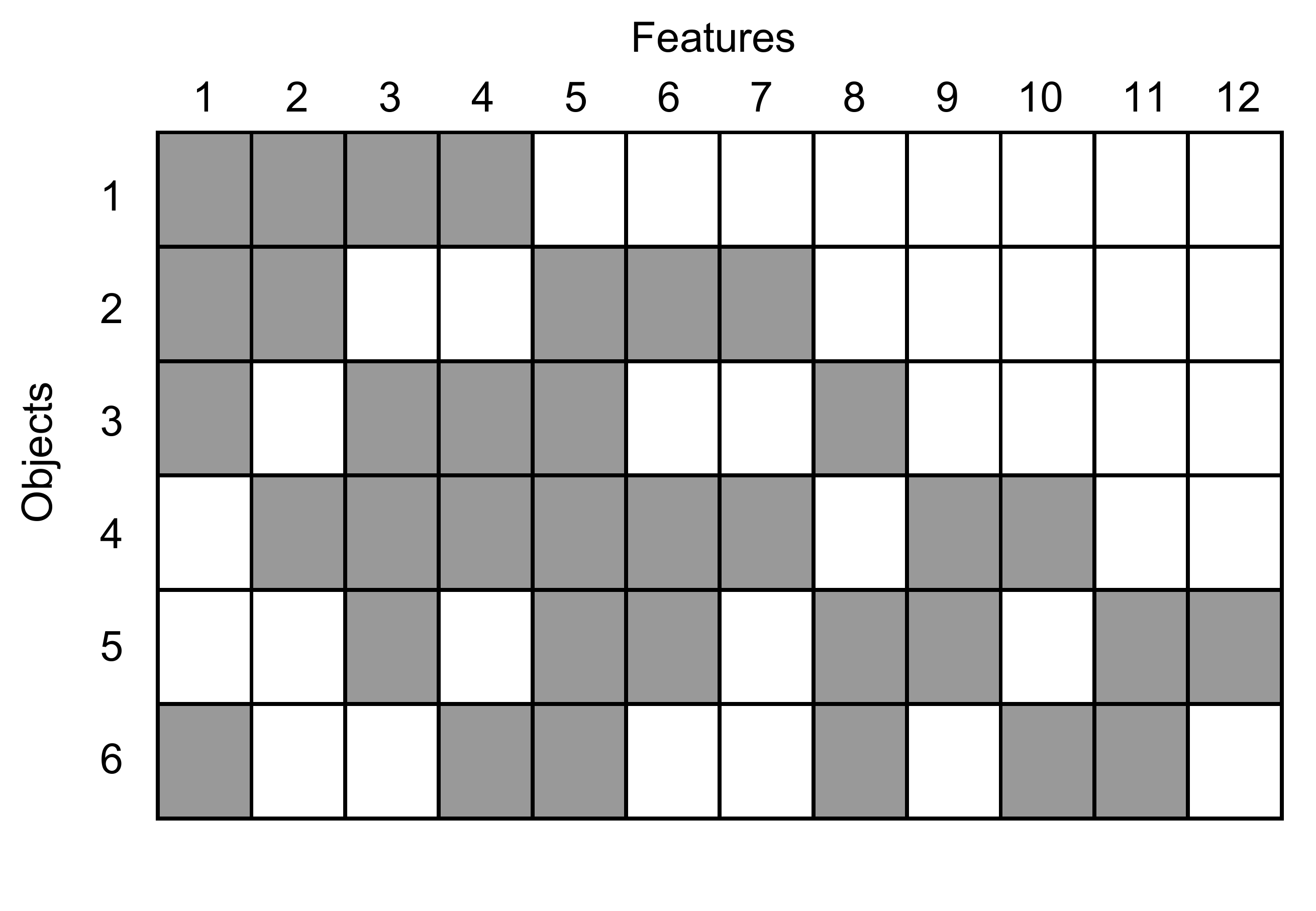}
\caption{}		
\end{subfigure}
\end{center}
\caption{An example of binary matrix representation of feature allocation. A shaded rectangle indicates the corresponding matrix element $z_{nk} = 1$. The binary matrix on (a) is transformed into the left-ordered binary matrix on (b) by the function $lof(\cdot)$.}
\label{fig:fa_example1}
\end{figure}

\paragraph{Matrix Representation}
Suppose $N$ objects and $K$ features are present. A feature allocation can be represented by a $N \times K$ binary matrix, denoted by $\bm Z$.
Rows of $\bm Z$ correspond to the index set $[N]$, and columns of $\bm Z$ correspond to features.
Each element $z_{nk}$, $n = 1, \ldots, N$, $k = 1, \ldots, K$, is a binary indicator, where $z_{nk} = 0$ or $1$ indicates index $n$ belongs or does not belong to feature $k$, i.e. $n \notin A_k$ or $n \in A_k$, respectively.
For example, Figure \ref{fig:fa_example1}(a) shows a feature allocation of $[6]$ with $12$ features, $f_6 = \{\{2, 4\}$, $\{1, 2, 4\}$, $\{1, 3, 4, 5\}$, $\{2, 3, 4, 5, 6\}$, $\{3, 5, 6\}$, $\{5\}$, $\{2, 4, 5\}$, $\{4, 5\}$, $\{1, 2, 3, 6\},$ $\{5, 6\}$, $\{1, 3, 4, 6\}$, $\{4, 6\} \}$.

\subsection{Indian Buffet Process}
The Indian buffet process (IBP) \citep{griffiths06, griffiths2011indian}) is a popular example of an exchangeable random feature allocation. 
Using the matrix representation of feature allocation, the IBP defines a distribution on binary matrices (with an unbounded random number of columns).

\paragraph{A Finite Feature Model}
The IBP can be defined as the limit of a finite feature model.
Suppose we have $N$ objects and $K$ features. We use a binary variable $z_{nk}$ to indicate object $n$ has feature $k$, thus $z_{nk}$ form a binary $N \times K$ matrix $\bm Z$. We assume that each object possesses feature $k$ with probability $\pi_k$, and the features are independent. Furthermore, beta distribution priors are put on $\pi_k$'s. That is,
\begin{align*}
z_{nk} \mid \pi_k &\sim \text{Bernulli}(\pi_k); \\
\pi_k &\sim \text{Beta}(\alpha / K, 1).
\end{align*}
Integrating out the $\pi_k$'s, the marginal distribution of $\bm Z$ is
\begin{align*}
p(\bm Z) = \prod_{k = 1}^K \frac{\frac{\alpha}{K} \Gamma(m_k + \frac{\alpha}{K} )\Gamma(N - m_k + 1)}{\Gamma(N + 1 + \frac{\alpha}{K})},
\end{align*}
where $m_k = \sum_{n = 1}^N z_{nk}$ is the number of objects possessing feature $k$. This distribution is exchangeable, since it only depends on the counts and does not depend on the ordering of the objects.

\paragraph{Left-ordered Constraint and Equivalence Classes}
A feature allocation indicates an ordering of the $K$ features. In many applications, the ordering of the features is not identifiable. When the labels of the features are arbitrary, it is helpful to define an equivalence class of binary matrices, denoted by $[\bm Z]$.
We first introduce an order constraint on binary matrices called the \emph{left-ordered constraint}. For a binary matrix $\bm Z$, its corresponding left-ordered binary matrix, denoted by $lof(\bm Z)$, is obtained by ordering the columns of $\bm Z$ from left to right by the magnitude of the binary number expressed by that column, taking the first row as the most significant bit. For example, Figure \ref{fig:fa_example1}(b) shows the corresponding left-ordered binary matrix of Figure \ref{fig:fa_example1}(a).  In the first row of the left-ordered matrix, the columns for which $z_{1k} = 1$ are grouped at the left. In the second row, the columns for which $z_{2k} = 1$ are grouped at the left of the sets for which $z_{1k} = 1$. This grouping structure persists throughout the matrix.

We can then define equivalence classes with respect to the function $lof(\cdot)$. 
This function maps binary matrices to left-ordered binary matrices, as described before. The function $lof(\cdot)$ is many-to-one: many binary matrices reduce to the same left-ordered form, and there is a unique left-ordered form for every binary matrix. Any two binary matrices $\bm Y$ and $\bm Z$ are $lof(\cdot)$ equivalent if $lof(\bm Y) = lof(\bm Z)$. In models where feature order is not identifiable, performing inference at the level of $lof$-equivalence classes is appropriate.
The probability of a particular $lof$-equivalence class of binary matrices $[\bm Z]$ is $p([\bm Z]) = \sum_{\bm Z \in [\bm Z]} p(\bm Z)$.

The matrix left-ordered form motivates the following definition. The \emph{history} of feature $k$ at object $n$ is defined to be $(z_{1k}, \ldots, z_{(n-1)k})$. When $n$ is not specified, history refers to the full history of feature $k$, $(z_{1k},...,z_{Nk})$. The histories of features are individuated using the decimal equivalent of the binary numbers corresponding to the column entries. For example, at object 3, features can have one of four histories: 0, corresponding to a feature with no previous assignments, 1, being a feature for which $z_{2k} = 1$ but $z_{1k} = 0$, 2, being a feature for which $z_{1k} = 1$ but $z_{2k} = 0$, and 3, being a feature possessed by both previous objects were assigned. The number of features possessing the history $h$ is denoted by $K_h$, with $K_0$ being the number of features for which $m_k = 0$ and $K_+ = \sum_{h = 1}^{2^N - 1} K_h$ being the number of features for which $m_k >0$, so $K = K_0 + K_+$. The function $lof$ thus places the columns of a matrix in ascending order of their histories.

Using the notion above, the cardinality of $[\bZ]$ is $\left( K! /  \prod_{h = 0}^{2^N - 1} K_h ! \right)$. Thus,
\begin{align}
p([\bm Z]) = \frac{K!}{\prod_{h = 0}^{2^N - 1} K_h !} \cdot \prod_{k = 1}^K \frac{\frac{\alpha}{K} \Gamma(m_k + \frac{\alpha}{K} )\Gamma(N - m_k + 1)}{\Gamma(N + 1 + \frac{\alpha}{K})}.
\label{eq:ibp2}
\end{align}

\paragraph{Taking the Infinite Limit}
Taking the limit $K \rightarrow \infty$ in Equation \eqref{eq:ibp2},
\begin{align}
\lim_{K \rightarrow \infty} p([\bm Z]) = \frac{\alpha^{K_+}}{\prod_{h = 1}^{2^N - 1} K_h !} \cdot \exp \{- \alpha H_N \} \cdot \prod_{k = 1}^{K_+} \frac{(N - m_k)! (m_k - 1)!}{N !},
\label{eq:limit_Z}
\end{align}
where $H_N$ is the $N$-th harmonic number, $H_N = \sum_{j = 1}^N 1/j$. See \cite{griffiths2011indian} for details.
This distribution is still exchangeable.  In practice, we usually drop all columns with all zeros, since they corresponds to the features that no object possesses, and it should not be included in the feature allocation as features are non-empty sets. It can be proved that we can obtain a matrix with finite columns with probability 1 by deleting the columns with all zeros.

\paragraph{Indian Buffet Analogy}
The probability distribution defined in Equation \eqref{eq:limit_Z} can be derived from a simple stochastic process, which is referred to as the IBP. Think about an Indian buffet where customers (objects) choose dishes (features). In the buffet, $N$ customers enter one after another, and each customer encounters infinitely many dishes arranged in a line. The first customer starts at the left of the buffet and takes a serving from each dish, stopping after a $\text{Poisson}(\alpha)$ number of dishes. The $n$-th customer moves along the buffet, sampling dishes in proportion to their popularity, serving himself with probability $m_k / n$, where $m_k$ is the number of previous customers who have sampled a dish. Having reached the end of all previously sampled dishes, the $n$-th customer then tries a $\text{Poisson}(\alpha / n)$ number of new dishes. We use a binary matrix $\bm Z$ with $N$ rows and infinitely many columns to indicate which customers chose which dishes, where $z_{nk} = 1$ if the $n$-th customer sampled the $k$-th dish. The matrices produced by this process are generally not in left-ordered form, and customers are not exchangeable under this distribution. However, if we only record the $lof$-equivalence classes of the matrices generated by this process, one obtains the exchangeable distribution $p([\bm Z])$ given by Equation \eqref{eq:limit_Z}.

\paragraph{Beta Process}
Similar to the relationship between the DP and the CRP, the de Finetti's measure underlying the exchangeable distribution produced by the IBP is the beta process (BP) \citep{hjort1990nonparametric}. See full details in \cite{thibaux2007hierarchical}. 

Other discussions of the IBP and the BP include \cite{teh2007stick, teh2009indian, doshi2009variational, paisley2010stick, williamson:2010, knowles2011nonparametric}, and \cite{miller2012phylogenetic}.

\subsection{Latent Categorical Feature Models}

In a latent feature model of the form \eqref{eq:lfm0} and \eqref{eq:lfm1}, a feature can either have the same effect on the objects possessing it, or have no effect on the objects not possessing it. This is sometimes too restrictive.
One relaxation is to assume different effects of each feature on different objects, as seen in \cite{griffiths2011indian}. That is, in Equation \eqref{eq:lfm0}, $v_k$ can depend on $n$, and $d_{nk} = z_{nk} v_{nk}$. We can then calibrate prior on $v_{nk}$ according to the specific application.

In some applications, for example, the tumor heterogeneity application in Chapters \ref{chap:PairClone} and \ref{chap:PairCloneTree}, it is natural to categorize the features. To elaborate, let $z_{nk} = 0, 1, \ldots, Q$ indicating object $n$ does not possess feature $k$, if $z_{nk} = 0$, or possesses category $q$ of feature $k$, if $z_{nk} = q$, $q = 1, \ldots, Q$. A feature has the same effect on the objects possessing the same category of it, while the feature has different effects on the objects possessing different categories of it. That is, in Equation \eqref{eq:lfm0},
\begin{align}
d_{nk} = g(z_{nk}) \cdot v_k, \quad k = 1, \ldots, K,
\label{eq:clfm}
\end{align}
where $g: \{0, \ldots, Q \} \mapsto \mathbb{R}$ represents those $(Q+1)$ types of effects. We do not restrict $g(0) \equiv 0$. That is, a feature can have effect on the objects that do not possess it. We will see this is the case in the application in Chapters \ref{chap:PairClone} and \ref{chap:PairCloneTree}. 
We hereafter refer to this type of latent feature models (Equations \eqref{eq:lfm1} and \eqref{eq:clfm}) as \emph{latent categorical feature models}.
To complete the model, we need to define a distribution on categorical valued matrices.

\subsection{Categorical Indian Buffet Process}
The categorical Indian buffet process (cIBP) \citep{Sengupta2013cIBP} is a categorical extension of the IBP, which defines a distribution on categorical valued matrices (with a random and unbounded number of columns).
Each entry of the matrix $\bm Z$ can take values from a set of integers $\{ 0, 1, \ldots, Q\}$, where $Q$ is fixed.
Here $z_{nk} = 0$ indicates object $n$ does not possess feature $k$, and $z_{nk} = q$, $q = 1, \ldots, Q$ represents object $n$ possesses category $q$ of feature $k$.

\paragraph{A Finite Feature Model}
Similar to the construction of the IBP, the cIBP can be derived as a limit of finite feature allocation models. Assume that each object possesses category $q$ of feature $k$ with probability $\pi_{kq}$, i.e. $Pr(z_{nk} = q) = \pi_{kq}$, and the features are independent. Furthermore, beta-Dirichlet distribution \citep{kim2012bayesian} priors are put on $\bm \pi_k$'s. That is, $\pi_{k0} = Pr(z_{nk} = 0)$ follows a beta distribution with parameters 1 and $\alpha / K$, i.e. $(1 - \pi_{k0}) = Pr(z_{nk} \neq 0) \sim \text{Beta} (\alpha / K, 1)$. 
Let $\tilde{\pi}_{kq} = \pi_{kq} / (1 - \pi_{k0})$, $q = 1, \cdots, Q$. Then
$(\tilde{\pi}_{k1}, \ldots, \tilde{\pi}_{kQ})$ follows a Dirichlet distribution with parameters $(\beta_1, \ldots, \beta_Q)$. In summary,
\begin{align*}
z_{nk} \mid \bm \pi_k &\sim \Categorical( \bm \pi_k),\\
\bm \pi_k &\sim \text{Beta-Dirichlet}(\alpha/K, 1, \beta_1, \ldots, \beta_Q).
\end{align*}
For simplicity we only consider a symmetric Dirichlet distribution $\beta_1 = \cdots = \beta_Q = \beta$, which is sufficient in most cases.
Integrating out the $\bm \pi_k$'s, the marginal distribution of $\bZ$ is
\begin{multline}
p(\bZ) = \prod_{k=1}^K \bigg \{ \frac{\Gamma(Q \beta) \Gamma(\frac{\alpha}{K} + 1)}{\Gamma(\frac{\alpha}{K}) (\Gamma(\beta))^Q} \cdot \\
\frac{\left[ \prod_{q=1}^Q \Gamma(\beta + m_{kq}) \right] \Gamma(N - m_{k\cdot} + 1) \Gamma(\frac{\alpha}{K} + m_{k\cdot}) }{\Gamma(\frac{\alpha}{K} + N + 1) \Gamma(Q\beta + m_{k\cdot})} \bigg\},
\label{eq:cibp1}
\end{multline}
where $m_{kq} = \sum_{n = 1}^N I(z_{nk} = q)$ denotes the number of objects from total $N$ objects possessing
category $q$ of feature $k$, and $m_{k \cdot} = \sum_{q = 1}^Q m_{kq}$ denotes the number of objects possessing feature $k$. The distribution is exchangeable since it depends on the counts $m_{kq}$ only.

\begin{figure}[h!]
\begin{center}
\begin{subfigure}[t]{.48\textwidth}
\centering
\includegraphics[width=\textwidth]{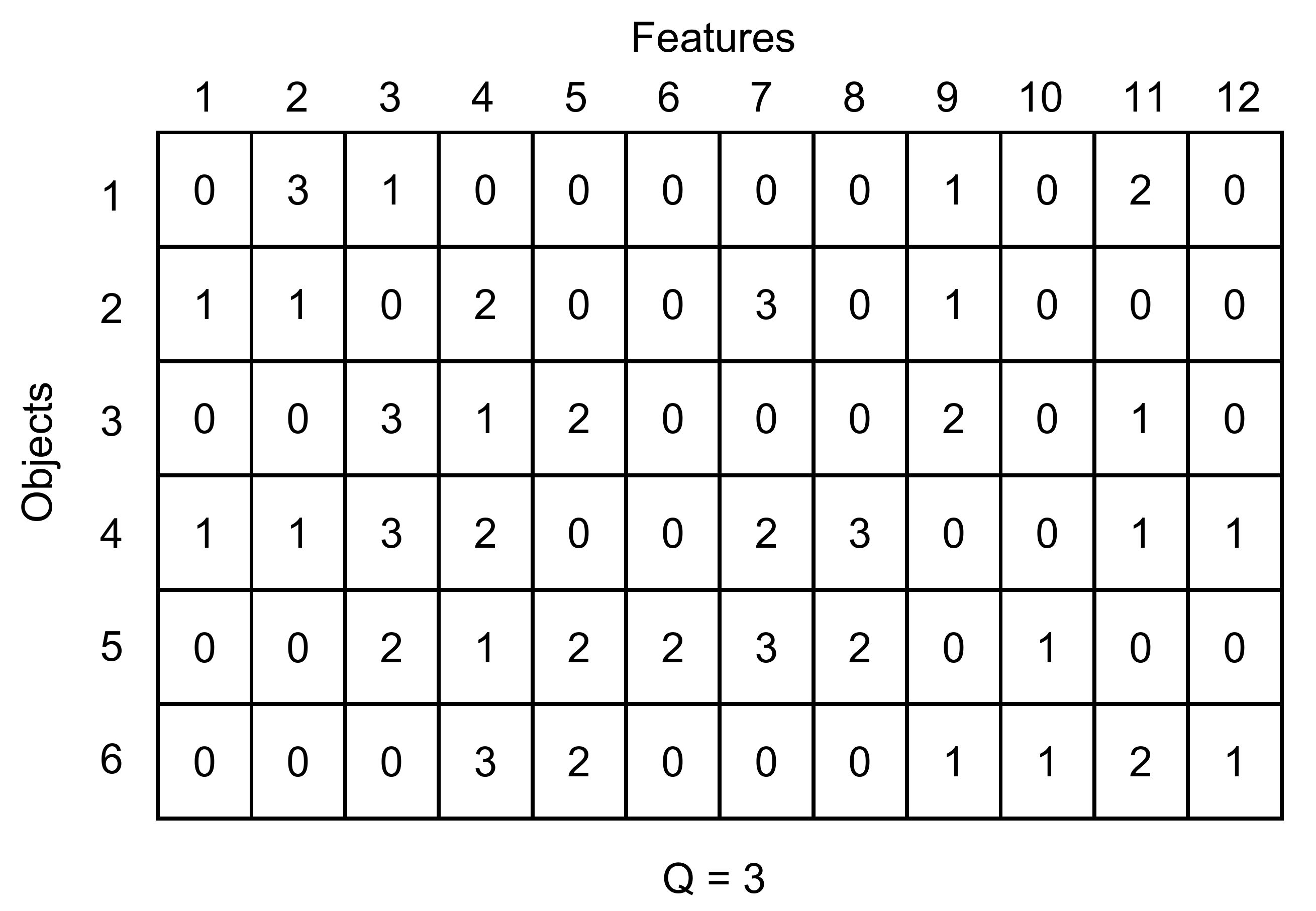}
\caption{}		
\end{subfigure}
\begin{subfigure}[t]{.48\textwidth}
\centering
\includegraphics[width=\textwidth]{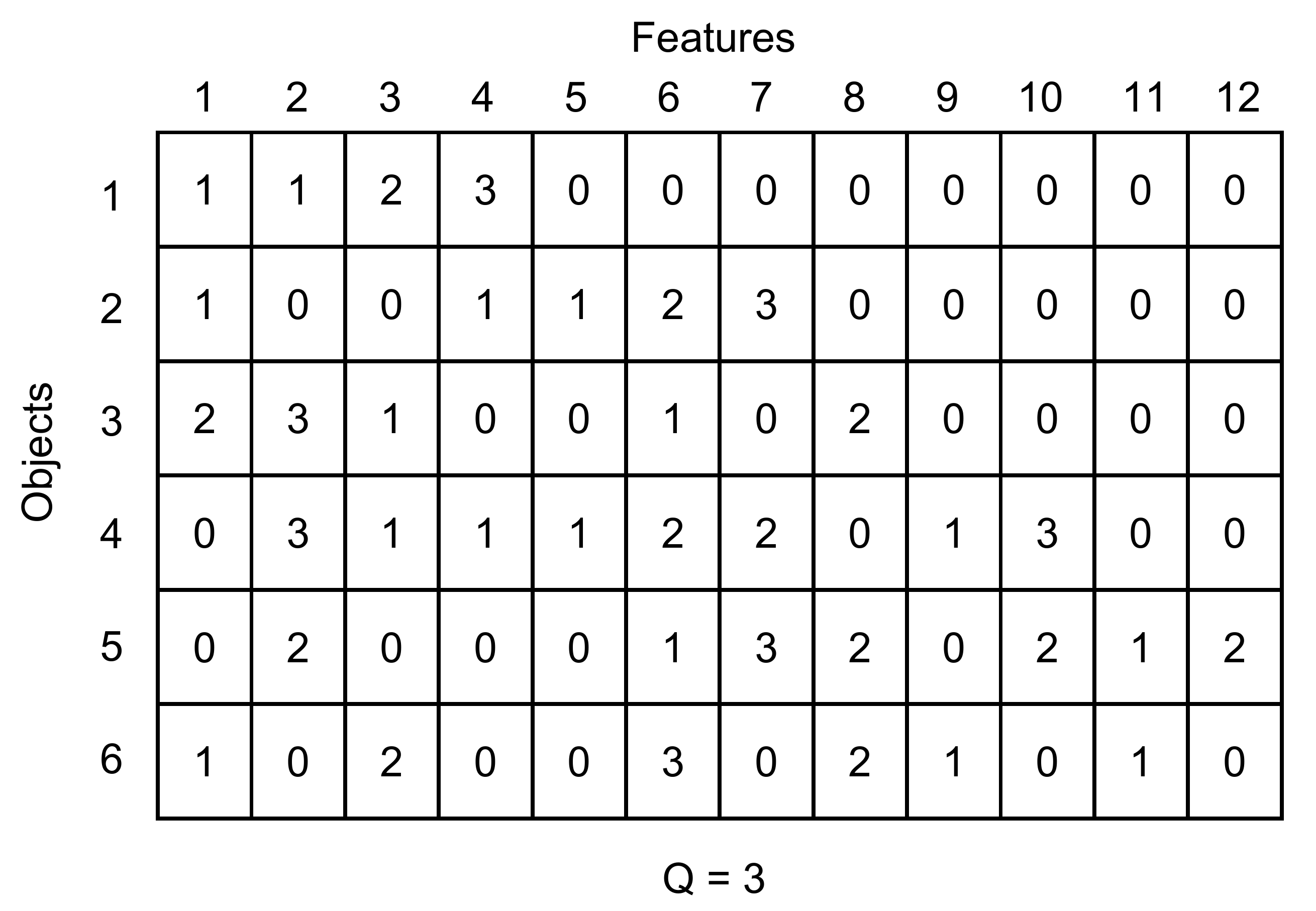}
\caption{}		
\end{subfigure}
\end{center}
\caption{An example of $(Q + 1)$-nary matrix with $Q = 3$. The matrix on (a) is transformed into the left-ordered matrix on (b) by the function $lof(\cdot)$.}
\label{fig:cibp_example1}
\end{figure}

\paragraph{Left-ordered Constraint and Equivalent Classes}
Similar to what was defined for binary matrices, we can define  \emph{left-ordered form} and \emph{history} on $(Q+1)$-nary matrices.
A left-ordered $(Q+1)$-nary matrix or $lof(\bm Z)$ is obtained by ordering the columns of $\bm Z$ from left to right by the magnitude of the $(Q+1)$-nary number (i.e represented in base $(Q+1)$) expressed by that column taking first row as the most significant bit. Figure \ref{fig:cibp_example1}(a) shows an example of $(Q+1)$-nary matrix, where $Q = 3$, and Figure \ref{fig:cibp_example1}(b) shows the corresponding left-ordered matrix. The $lof$-equivalence class of matrix $\bm Z$ is still denoted by $[\bm Z]$.
The \emph{history} of feature $k$ at object $n$ is defined to be the decimal equivalent of the $(Q+1)$-nary number represented by the vector $(z_{1k}, \ldots, z_{(n-1)k})$, and the full history of feature $k$ refers to the decimal equivalent of the $(Q+1)$-nary number of $(z_{1k}, \ldots, z_{Nk})$. The number of features having history $h$ is denoted by $K_h$, with $K_0$ being the number of features for which $m_{k \cdot} = 0$ and $K_+ = \sum_{h = 1}^{(Q+1)^N - 1} K_h$ being the number of features for which $m_{k \cdot} > 0$, and $K = K_0 + K_+$.

Using the notions above, the cardinality of $[\bZ]$ is $\left( K! /  \prod_{h = 0}^{(Q+1)^N - 1} K_h ! \right)$, and 
\begin{align}
p([\bZ]) = \sum_{\bZ \in [\bZ]} p(\bZ) = \frac{K!}{\prod_{h = 0}^{(Q+1)^N - 1} K_h !} \cdot p(\bZ).
\label{eq:cibp2}
\end{align}

\paragraph{Taking the Infinite Limit}
Taking the limit $K \rightarrow \infty$ in Equations \eqref{eq:cibp2} and \eqref{eq:cibp1}, we obtain
\begin{multline}
\lim_{K \rightarrow \infty} p([\bm Z]) = \frac{(\alpha / Q)^{K_+}}{\prod_{h = 1}^{(Q+1)^N - 1} K_h !} \cdot \exp \{ - \alpha H_N \} \cdot  \\
\prod_{k = 1}^{K_+} \Bigg\{ \frac{(N - m_{k \cdot})! (m_{k \cdot} - 1)! }{N!} \cdot
\frac{1}{\prod_{j = 1}^{m_{k \cdot} - 1}(j + Q \beta)} \frac{1}{\beta} \prod_{q = 1}^Q \frac{\Gamma(\beta + m_{kq})}{\Gamma(\beta)} \Bigg\}.
\label{eq:limit_Z_cibp}
\end{multline}
Details in \cite{Sengupta2013cIBP}. This distribution is exchangeable as in the finite case. The last step is dropping the columns with all zeros.

\paragraph{Indian Buffet Analogy}
The probability distribution defined in Equation \eqref{eq:limit_Z_cibp} can also be derived from a stochastic process similar to the IBP, which is referred to as the cIBP. The customers are the objects and the dishes are the features. The $Q$ categories of a feature can be seen as different spice levels of a dish. Start with $N$ customers and an infinite number of dishes. As the $i$-th customer walks in, he/she chooses the $k$-th dish with a particular spice level $q$ with probability $\left[ m_{k \cdot} \cdot (\beta_{q} + m_{kq}) \right] / \left[ i \cdot (\beta_* + m_{k \cdot})\right]$, where $m_{kq}$ denotes the number of customers who have tasted the dish $k$ with spice level $q$, $m_{k \cdot} = \sum_{q = 1}^Q m_{kq}$ denotes the number of customers who have tasted the
dish $k$ in total, and $\beta_* = \sum_{q = 1}^Q \beta_q$. Then the $i$-th customer tastes new dishes with spice level $q$ determined by a draw from $\text{Poisson}[(\beta_q \cdot \alpha) / (\beta_* \cdot i)]$. Using a $(Q+1)$-nary matrix $\bm Z$ with $N$ rows and infinitely many columns to indicate which customers chose which
dishes, where $z_{nk} = 0$ if the $n$-th customer did not choose the $k$-th dish, and $z_{nk} = q$ if the $n$-th customer chose the $k$-th dish with spice level $q$. If one only pays attention to the $lof$-equivalence classes of the matrices generated by this process, one obtains the exchangeable distribution $p([ \bm Z])$ given by Equation \eqref{eq:limit_Z_cibp}.

\subsection{Inference for Tumor Heterogeneity Using Feature Allocation Models}
\label{sec:lfmapp}
We continue the discussion of inference for tumor heterogeneity in Section \ref{sec:lcm_app}. 
Recall that one approach to this problem is to model the read counts using a latent class model, where the SNVs are the objects, and the subclones are the classes. 
We have discussed in Section \ref{sec:lcm_app} that subclone reconstruction methods based on latent class models usually rely on the ISA. 
However, ISA is not necessarily valid because multiple tumor subclones might acquire the same mutation in convergent evolution. See more discussions in \cite{marass2017phylogenetic}. 
Such concerns inspire another approach to this problem: modeling the read counts using a latent feature model. Clomial \citep{zare2014inferring}, \cite{lee2015bayesian}, BayClone \citep{sengupta2015bayclone}, Cloe \citep{marass2017phylogenetic} and PairClone (Chapters \ref{chap:PairClone}, \ref{chap:PairCloneTree}) take this approach. We briefly summarize BayClone and Cloe here.

Still, ignoring many biological details, consider $S$ SNVs. Let $N_s$ and $n_s$ denote the total and variant read counts covering locus $s$, respectively, $s = 1, \ldots, S$. BayClone models the variant read counts using a latent (categorical) feature model
\begin{align*}
&n_s \mid p_s \sim \binomial (N_s; p_s), \\
&p_s \mid \bw, \bZ =  \sum_{c = 1}^C w_{c} \, g(z_{sc}), \\
&p(z_{sc} = q \mid \pi_{cq}) = \pi_{cq}, \\
&(\pi_{c0}, \ldots, \pi_{cQ}) \sim \text{Beta-Dirichlet}(\alpha/C, 1, \beta, \ldots, \beta)
\end{align*}
where $c = 1, \ldots, C$ represent the $C$ subclones (i.e. $C$ features), $w_c$ is the cellular proportion of subclone $c$ (i.e. value of feature $c$), and $z_{sc}$ is the genotype, with $z_{sc} = 0$, $1$ or $2$ corresponding to a homozygous wild-type, heterozygous variant or homozygous variant at site $s$ of subclone $c$. The mapping $g$ maps $g(0) = 0$, $g(1) = 0.5$ and $g(2) = 1$, which indicates the contributions of three different genotypes to the expected variant allele fraction $p_s$.

Cloe, on the other hand, uses a column-dependent phylogenetic prior for $\bZ$, which allows it to infer tumor phylogeny.

\section{Regression}
\label{sec:reg}
Given two or more observables, \emph{regression analysis} is concerned with the relationship between a dependent variable (or response), denote by $y$, and one or more independent variables (or predictors), denote by $\bx$. We usually treat $\bx$ as fixed quantities and treat $y$ as random variables. We focus on the case that $y \in \mathbb{R}$ is continuous and is normally distributed. For more general cases (e.g. generalized linear models) see, for example, \cite{dobson2008introduction} or \cite{dey2000generalized}.
Suppose we have $N$ observations $\{ (y_n, \bx_n) : n = 1, \ldots, N \}$. For observation $n$, we assume
\begin{align}
y_n = f(\bx_n) + \varepsilon_n,
\label{eq:regr1}
\end{align}
where
\begin{align*}
\varepsilon_n \iidsim N(0, \sigma^2) 
\end{align*}
is normally distributed random error.

In the linear regression setting, the function $f$ in Equation \eqref{eq:regr1} is specified as a linear function
\begin{align*}
f(\bx) = \bx^T \bbeta,
\end{align*}
where $\bbeta$ is a parameter vector, and the elements of $\bbeta$ are called regression coefficients. The model is completed with a prior on $\bbeta$, which is usually a normal conjugate prior. Linear regression model is a traditional statistical model and has been studied extensively. For a review, see, for example, \cite{gelman2014bayesian} or \cite{christensen2011plane}.

In many applications, the linearity assumption is too restrictive. 
One possible generalization using nonparametric Bayesian models is to replace the linear model with a less restrictive flexible prior on $f$. 
A popular prior specification for $f$ is the Gaussian process, which we will describe in detail in Section \ref{sec:GP}. Another popular specification for $f$ is based on a function basis $\mathcal{G} = \{g_1, g_2, \ldots \}$ and a representation of $f$ as $f(\bx) = \sum_j \beta_j g_j(\bx)$. A prior model for $\bbeta = (\beta_1, \beta_2, \ldots)$ induces a prior model for $f$. See \cite{muller2004nonparametric} for a review.

\subsection{Gaussian Process}
\label{sec:GP}
The Gaussian process (GP) \citep{o1978curve, neal1998regression, rasmussen2006gaussian} defines a distribution over random functions (stochastic processes). A GP is a collection of random variables $\{ f(\bx): \bx \in \mathcal{X} \}$ such that, for any finite number of indices $\bx_1, \ldots, \bx_N \in \mathcal{X}$, the joint distribution of $[f(\bx_1), \ldots, f(\bx_N)]^T$ is multivariate normal.

A GP is completely characterized by its mean function 
$m(\bx): \mathcal{X} \rightarrow \mathbb{R}$ and covariance function 
$C(\bx, \bx'): \mathcal{X} \times \mathcal{X} \rightarrow \mathbb{R}^+$, where
\begin{align*}
m(\bx) &= \E[f(\bx)], \\
C(\bx, \bx') &= \Cov[f(\bx), f(\bx')].
\end{align*}
We denote by $\GP(m(\bx), C(\bx, \bx'))$ a GP with mean function $m$ and covariance function $C$, and write $f(\bx) \sim \GP(m(\bx), C(\bx, \bx'))$ if $f$ has a GP prior.

A common choice of the mean function is the linear function,
\begin{align*}
m(\bx) = \bx^T \bbeta,
\end{align*}
which means we center the GP on a linear model.
Suppose $\bx = (x_1, \ldots, x_p)$. A common choice of the covariance function is the squared exponential covariance function,
\begin{align*}
C(\bx, \bx') = \tau^2 \exp \left[  - \sum_{j=1}^p \frac{(x_j - x_j')^2}{2l_j^2}  \right] + \tau_0^2 \delta(\bx, \bx'),
\end{align*}
where $\tau^2, \tau_0^2, l_1^2, \ldots, l_p^2$ are hyperparameters, and $\delta(\bx, \bx')$ is the Kronecker delta function that takes the value 1
if $\bx = \bx'$ and 0 otherwise. Here $\tau^2$ controls the magnitude of $C$,  and $l_1^2, \ldots, l_p^2$ (called length scales) control the smoothness of $C$. The function $\delta(\bx, \bx')$ is used to introduce small nugget for the diagonal covariances, which overcomes near-singularity of the covariance matrices and improves numerical stability. The nugget term $\tau_0^2$ is usually chosen small, e.g. $\tau_0^2 = 0.01$. For simplicity, sometimes we set $l_1^2 = \ldots = l_p^2 = l^2$, in which case $C$ is called isotropic. For a detailed discussions of different covariance functions, see \cite{rasmussen2006gaussian} (Chapter 4).

\paragraph{Basis Expansions}
An alternative way to derive the GP is using basis expansions. Consider $H$ basis functions $\phi_1(\bx), \ldots, \phi_H(\bx)$ and let $\bphi(\bx) = [\phi_1(\bx), \ldots, \phi_H(\bx)]^T$. Let
\begin{align}
f(\bx) = \bphi(\bx)^T \bbeta, \quad \bbeta \sim N(\bbeta_0, \Sigma_{\beta}).
\label{eq:gp2}
\end{align}
Integrating out the $\bbeta$, for any finite number of indices $\bx_1, \ldots, \bx_N$, we have
\begin{align*}
[f(\bx_1), \ldots, f(\bx_N)]^T \sim N[(m(\bx_1), \ldots, m(\bx_N))^T,  S],
\end{align*}
where
\begin{align*}
m(\bx) &= \bphi(\bx)^T \bbeta_0, \\
C(\bx, \bx') &= \bphi(\bx)^T \Sigma_{\beta} \bphi(\bx'),
\end{align*}
and $S$ is a covariance matrix with $(i,j)$-th element $S_{ij} = C(\bx_i, \bx_j)$.
Therefore, $f$ is a GP. When $\bphi(\bx) = \bx$, Equation \eqref{eq:gp2} reduces to a linear model, and we can see linear regression is a special case of GP with covariance function $C(\bx, \bx') = \bx^T \Sigma_{\beta} \bx'$.
The number of basis functions $H$ needs not to be finite. For example, when 
\begin{align*}
\phi_h(x) = \exp \left[ - \frac{(x - h)^2}{2l^2} \right],
\end{align*}
and $H \rightarrow \infty$ (consider scalar $x$ for simplicity), Equation \eqref{eq:gp2} leads to a GP with squared exponential covariance function (details in \citealp{rasmussen2006gaussian}). 

\paragraph{Inference}
We are usually interested in predicting the value of $f$ at some location $\tbx$ given the observed data $\{ (y_n, \bx_n) : n = 1, \ldots, N \}$. Denote by $\by = (y_1, \ldots, y_N)^T$, $\bm m = [m(\bx_1), \ldots, m(\bx_N) ]^T$,  $\tf = f(\tbx)$ and $\tilde{m} = m(\tbx)$. The joint distribution of $\by$ and $\tf$ is
\begin{align*}
\left(\begin{array}{c}
        \by  \\
        \tf  \\
\end{array}\right) \sim N \left[  
\left(\begin{array}{c}
        \bm m  \\
        \tilde{m}  \\
\end{array}\right), 
\left(\begin{array}{cc}
        C(X, X) + \sigma^2 I & C(X, \tbx) \\
        C(\tbx, X) & C(\tbx, \tbx)  \\
\end{array}\right)
\right].
\end{align*}
The posterior predictive distribution of $\tf$ is thus
\begin{multline*}
\tf \mid \by \sim N \bigg[ \tilde{m}  + C(\tbx, X) [C(X, X) + \sigma^2 I]^{-1} (\by - \bm m), \\ 
C(\tbx, \tbx) - C(\tbx, X) [C(X, X) + \sigma^2 I]^{-1} C(X, \tbx) \bigg].
\end{multline*}
We can then add one more step to predict the response value $\tilde{y}$ at $\tbx$,
\begin{align*}
\tilde{y} \mid \tf   \sim N(\tf, \sigma^2).
\end{align*}

Recent literature on results and generalizations of GP priors includes the following.
\cite{neal1995bayesian} 
reveals the connection between neural networks (with one hidden layer and an infinite number of units) and Gaussian processes, \cite{ghosal2006posterior} and \cite{choi2007posterior} discuss posterior consistency, 
\cite{gramacy2008bayesian} develop treed Gaussian processes, and \cite{banerjee2008gaussian, banerjee2013efficient, hensman2013gaussian} and \cite{datta2016hierarchical} develop efficient computational algorithms.

\subsection{Inference for Missing Data Using Nonparametric Regression Models}

We continue the discussion of inference for monotone missing data in Section \ref{sec:lcm_app}. 
Recall that one approach to this problem is to model the joint distribution of the full data response $\bm Y_i$ and dropout $S_i$ (conditional on the covariates of primary interest $\bm X_i$) using a latent class model. In many cases, in addition to $(\bm Y_i, S_i, \bm X_i)$, we would have access to a set of \emph{auxiliary covariates}, denoted by $\bV_i$. Such covariates, although are not of direct interest, can often provide information about the missing responses and missing data mechanism. See \cite{daniels2008missing} and \cite{daniels2014fully} for more discussion. In this setting, we should incorporate $\bV_i$ and consider a joint model for $(\bm Y_i, S_i, \bm V_i \mid \bm X_i)$, denoted by $p(\by, s, \bv \mid \bx)$. 
Here we proceed with inference unconditional on $\bv$, because the primary interest is in $p(\by, s \mid \bx)$, and
\begin{align*}
p(\by, s \mid \bx) = \int p(\by, s, \bv \mid \bx) d \bv.
\end{align*}

Still, we stratify the model by $\bx$ and suppress the conditional on $\bx$. Under the extrapolation factorization,
\begin{align*}
p(\by, s, \bv) = p(\by_{\text{mis}} \mid \by_{\text{obs}}, s, \bv) p(\by_{\text{obs}}, s, \bv).
\end{align*}
In Chapter \ref{chap:bnpmis}, we specify $p(\by_{\text{obs}}, s, \bv)$  based on pattern-mixture modeling \citep{little1993pattern},
\begin{align*}
p(\by_{\text{obs}}, s, \bv) = p(\by_{\text{obs}} \mid s, \bv) p(s \mid \bv) p(\bv).
\end{align*}
The models $p(\by_{\text{obs}} \mid s, \bv)$ and $p(s \mid \bv)$ are regression models. We then specify 
\begin{align*}
[Y_j \mid \bm \bar{Y}_{j-1} = \bby_{j-1}, S = s, \bV = \bv ] &= a(\bv, j, s) + \bby_{j-1}' \bPhi_{js} + \varepsilon_{js}, \; \; (j = 1, \ldots, s);\\
p(S = k \mid S \geq k, \bv, \bm f) &= F_N(f_k(\bv)),
\end{align*}
where
\begin{align*}
a(\bv, j, s) \sim \GP(\mu, C),
\end{align*}
$F_N$ denotes the standard normal cdf (probit link), and $f_k(\bv)$ is the Bayesian additive regression trees (BART) model \citep{chipman2010bart}. BART is also a popular Bayesian nonparametric model for regression. See Chapter \ref{chap:bnpmis} for further details.

\section{Contributions}
This dissertation makes the following contributions in methodology and applications. 

In Chapter \ref{chap:PairClone}, we propose a Bayesian feature allocation model for tumor subclone reconstruction using mutation pairs. With respect to methodology, we develop a feature allocation model with categorical matrix-valued features. We also develop a trans-dimensional MCMC algorithm based on splitting the data into training and test data sets, which is specially tailored to the feature allocation model. 
In terms of application, we model subclones characterized by phased pairs of (diploid) variant alleles. 
Our approach is a substantial improvement over current methods which all work with marginal counts only. 
We make inference for tumor heterogeneity on the basis of the proposed model and show that the model with (few) phased mutation pairs provides more accurate inference than current models with (far more) marginal SNVs. We also develop an open source software package \texttt{PairClone} which is available at \url{http://www.compgenome.org/pairclone}.

In Chapter \ref{chap:PairCloneTree}, we propose a Bayesian treed feature allocation model for tumor subclone reconstruction using mutation pairs. Regarding methodology, we develop a feature allocation model with \textit{a priori} dependent features, where the dependence is modeled with a tree prior. We develop a computationally efficient posterior simulation method on the tree.  In terms of application, our model allows for inference for phylogenetic trees of tumor cell subpopulations. We also develop an open source software package \texttt{PairCloneTree} which is available at \url{http://www.compgenome.org/pairclonetree}. This project uses the same data as the first project. However, the tree-based prior on dependent features is entirely different from the IBP model.

In Chapter \ref{chap:bnpmis}, we propose a nonparametric Bayesian approach to monotone missing data in longitudinal studies. With regard to methodology, we develop nonparametric Bayesian regression models and shrinkage priors for observed data responses and dropout across dropout times and patterns. Such models can effectively capture non-linear and non-additive relationships and allow for borrow of information across times and patterns. In particular, the model is built on a GP prior for a nonparametric regression of observed data responses (conditional on dropout pattern and auxiliary covariates), combined with a BART model for dropout as a function of auxiliary covariates.
In terms of application, our model allows for utilizing information from auxiliary covariates that are not desired in the primary research question. The inclusion of such auxiliary covariates can ideally reduce the extent of sensitivity analysis that is needed for drawing accurate inferences.

\chapter{A Bayesian Feature Allocation Model for Tumor Subclone Reconstruction Using Mutation Pairs}
\label{chap:PairClone}
Tumor cell populations can be thought of as being composed of
heterogeneous cell subpopulations, with each subpopulation being
characterized
by overlapping sets of single nucleotide variants (SNVs).  
Such
subpopulations are known as subclones and are an important target for
precision medicine.  Reconstructing such subclones from
next-generation sequencing (NGS) data is one of the major challenges
in precision medicine.  We present PairClone as a new tool to
implement this reconstruction.
The main idea of PairClone is to model short reads mapped to
pairs of proximal
SNVs. In contrast, most existing methods use only marginal 
reads for
unpaired SNVs.  Using Bayesian nonparametric models, we estimate
posterior probabilities of the number, genotypes and population
frequencies of subclones in one or more tumor sample.  We use the
categorical Indian buffet process (cIBP)
as a prior probability model for subclones that are represented
as vectors of categorical matrices that record the corresponding
sets of mutation  pairs.  Performance of PairClone is assessed
using simulated and real 
datasets. 
An open source software package can be obtained at \url{http://www.compgenome.org/pairclone}.

\section{Introduction}
We explain intra-tumor heterogeneity by representing tumor cell
populations as a mixture of 
subclones. We reconstruct unobserved  subclones by
utilizing information from pairs of proximal mutations
that are obtained from next-generation sequencing (NGS) data. 
We exploit the fact that some short reads
in NGS data cover pairs of phased mutations that reside on
two sufficiently proximal loci. Therefore haplotypes of the
mutation pairs can be observed and used for subclonal inference. 

We develop a suitable sampling model that represents
the paired nature of the data, and construct a nonparametric Bayesian
feature allocation model as a prior for the hypothetical
subclones. Both models together allow us to develop a fully probabilistic
description of the composition of the tumor as a mixture of
homogeneous underlying subclones, including the genotypes and number of
such subclones.

\subsection{Background}
\begin{figure}[h!]
\begin{center}
\begin{subfigure}[t]{.26\textwidth}
\centering
\includegraphics[width=\textwidth]{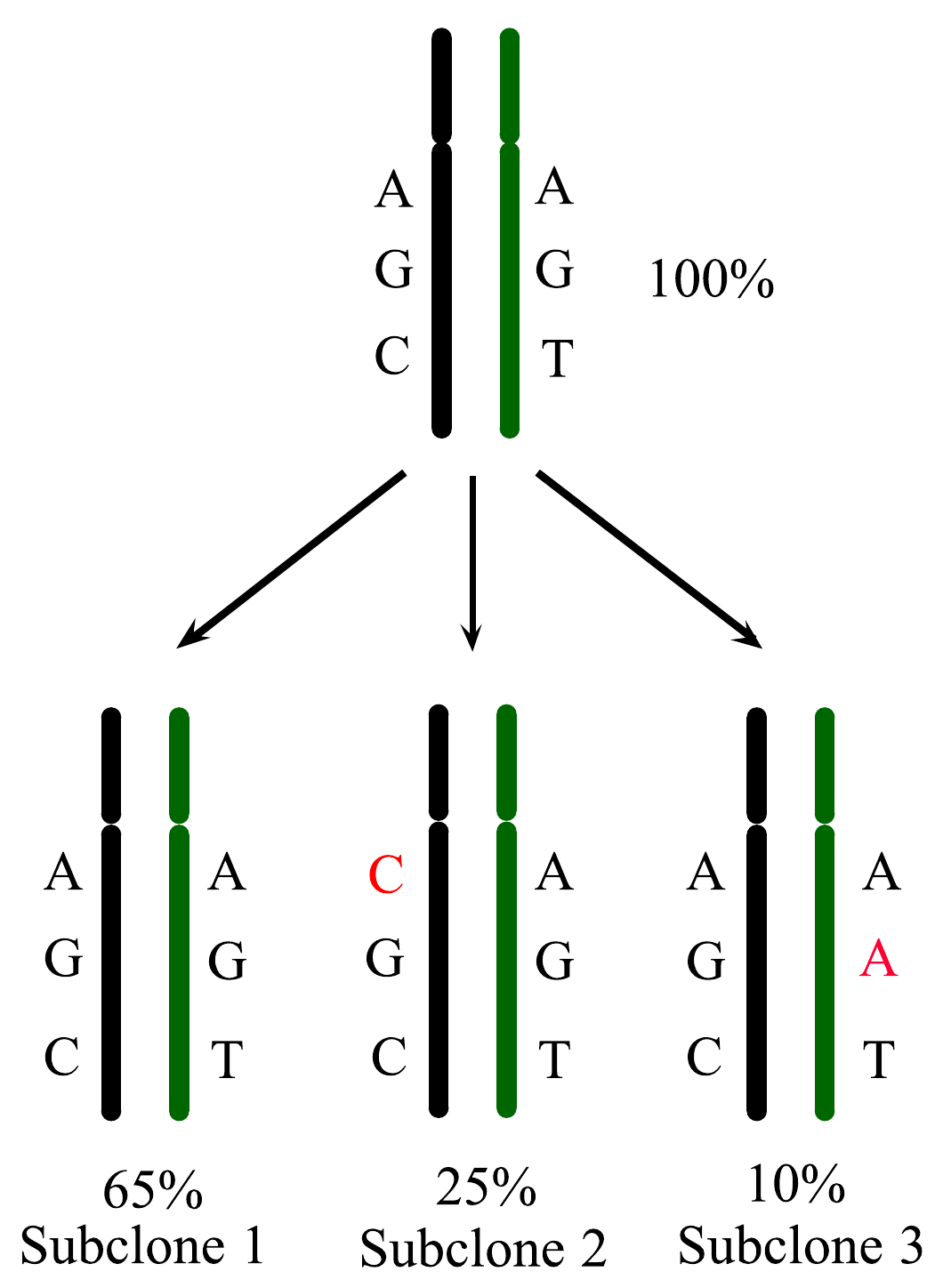}
\caption{Tumor evolution}
\label{fig:tm_evo}	
\end{subfigure}
\hspace{5mm}
\begin{subfigure}[t]{.65\textwidth}
\centering
\includegraphics[width=\textwidth]{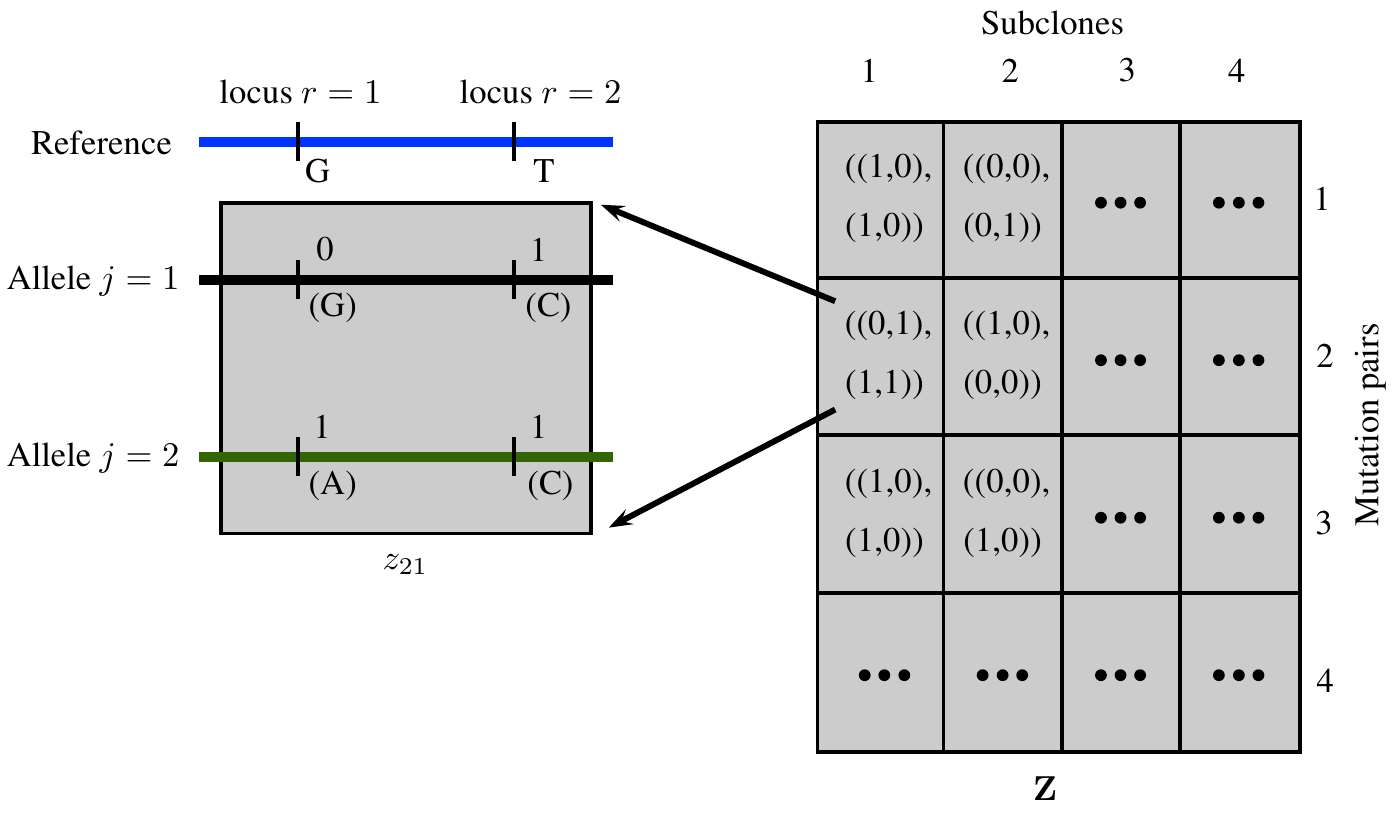}
\caption{Subclone structure matrix $\bZ$}
\label{fig:Z_example}	
\end{subfigure}
\begin{subfigure}[t]{.85\textwidth}
\centering
\vspace{7mm}
\includegraphics[width=\textwidth]{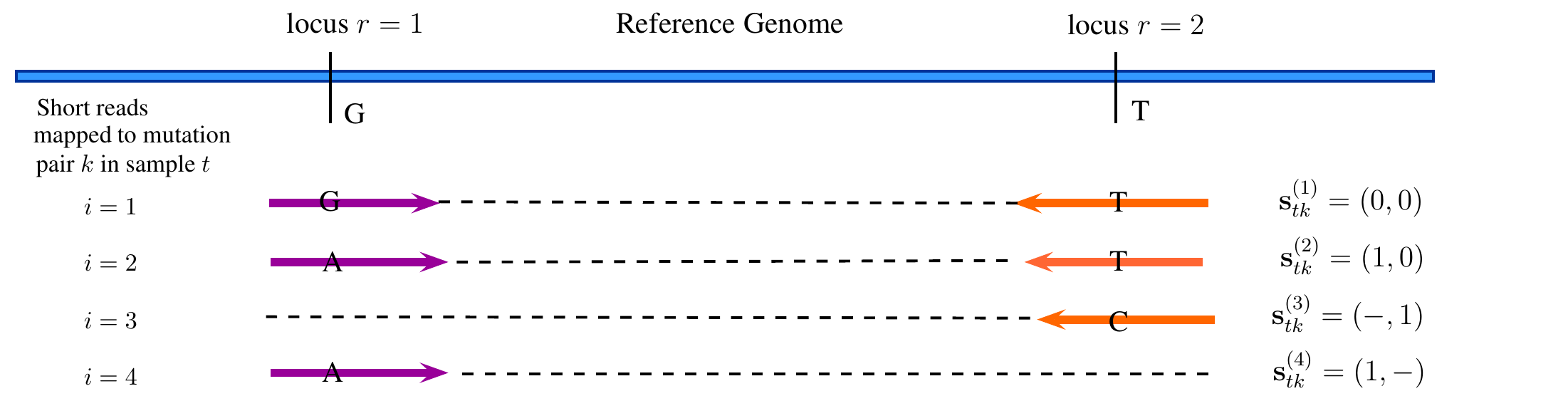}
\caption{Paired-end reads data}
\label{fig:mut_example}
\end{subfigure}
\end{center}
\caption{(a) Illustration of tumor evolution, emergence of subclones and
  their population frequencies. 
  (b) Illustration of the subclone structure matrix $\bZ$.  
  Right panel: A subclone is represented by one column of $\bZ$. Each
  element of a column represents the subclonal genotypes for a
  mutation pair. For example, the genotypes for mutation  pair
  2 in subclone 1 is ((0, 1), (1, 1)), which is shown in detail on the
  left panel. 
  Left panel: The reference genome for mutation pair $2$ is (G, T) and
  the corresponding genotype of subclone $1$ is ((G, C), (A, C)),
  which gives rise to $\bz_{21} = ((0, 1), (1, 1))$.  
  (c) Illustration of  paired-end reads   data for a mutation pair.  Shown are four short reads mapped to mutation pair $k$ in sample $t$. Some reads are mapped to both loci of the mutation pair, and others are mapped to only one of the two loci. The two ends of the same read are marked with opposing arrows in purple and orange.  }
\end{figure}

NGS technology~\citep{mardis2008next}
has enabled researchers to develop bioinformatics tools that are
being used to understand the landscape of tumors within and across
different samples.  An important related task is to reconstruct cellular
subpopulations in one or more tumor samples, known as subclones.
Mixtures of such subclones with varying population frequencies 
across spatial
locations in the same tumor, across tumors from different
time points,  or across tumors from the primary and metastatic
sites can provide information about the
mechanisms of tumor evolution and metastasis. 
 Heterogeneity of cell populations is seen, for example, in varying
frequencies of distinct somatic mutations.
The hypothetical tumor subclones are homogeneous. That is, a subclone
is characterized by unique genomic variants in its genome~
\citep{marjanovic2013cell,almendro2013cellular,polyak2011heterogeneity,stingl2007molecular,shackleton2009heterogeneity,dexter1978heterogeneity}.
Such subclones arise as the result of cellular evolution,
which can be described by a phylogenetic tree that records how a
sequence of somatic mutations gives rise to different cell subpopulations.
Figure~\ref{fig:tm_evo} provides a stylized and
simple illustration in which a homogeneous sample with one original normal 
clone evolves into a heterogeneous sample with three
subclones. Subclone 1 is the original parent cell population, and
subclones 2 and 3 are descendant subclones of subclone 1, each
possessing somatic mutations marked by the red letters. Each
subclone  possesses  two homologous chromosomes (in black and green), and
each chromosome in Figure~\ref{fig:tm_evo} is marked by a triplet of letters
representing the nucleotide on  the three genomic loci. Together, the
three subclones include four different haplotypes,
(A, G, C), (A, G, T), (C, G, C), and (A, A, T), at these three genomic
loci.  
In addition, each subclone has a different population frequency shown
as the percentage values in
Figure~\ref{fig:tm_evo}.

We use NGS data to infer such tumor heterogeneity.
In an NGS experiment, DNA fragments are first produced by extracting
the DNA molecules from the cells in a tumor sample. The fragments are
then sequenced using short reads.  For the three subclones in
Figure~\ref{fig:tm_evo}, there are the four aforementioned haplotypes at
the three loci. Consequently, short reads that cover some of these
three loci may manifest different alleles. 
For example, if a large
number of 
reads cover the first two loci, we might observe the four alleles (A, G), (C, G), (A, T) and
(C, T) 
for the mutation pair. Observing four alleles 
is direct evidence for the presence of subclones~\citep{sengupta2016ultra}.
This is because in the absence of copy number variations there can be
only two haploid genomes at any locus for a homogeneous human sample.
Therefore, one can use mutation pairs in copy
neutral regions to develop statistical inference for the presence and
frequency of subclones. This is the goal of the paper.

Almost all mutation-based subclone-calling methods in the literature
use only single nucleotide variants (SNVs)~\citep{oesper2013theta, strino2013trap, jiao2014inferring, miller2014sciclone, roth2014pyclone, zare2014inferring, deshwar2015phylowgs, sengupta2015bayclone,lee2015bayesian,lee2016bayesianjrssc}.
Instead of examining mutation pairs, SNV-based methods
use marginal counts for each recorded locus only.
Consider, for example,  the first locus in
Figure~\ref{fig:tm_evo}. At this locus, 
the reference genome 
has an ``A''
nucleotide while subclones 2 and 3 have a ``C'' nucleotide. In the entire
sample, the ``C'' nucleotide is roughly present in 17.5\% of the DNA
molecules based on the population frequencies illustrated in
Figure~\ref{fig:tm_evo}. 
The percentage of a mutated allele is called variant allele fraction (VAF).
If a sample is 
homogeneous and assuming no copy number variations at the locus, 
the population frequency for the ``C''
nucleotide should be close to  0, 50\%, or 100\%, depending on the
heterozygosity of the locus.  
Therefore,
if the
population frequency of ``C'' deviates from 0\%, 50\%, or 100\%,
the sample is likely to be heterogeneous.
Based on this argument,
SNV-based subclone callers search for SNVs with
VAFs
that are different from these frequencies
(0, 50\%, 100\%), which are evidence for the presence of different (homogeneous)
subpopulations.
In the event of copy
number variations, a similar but slightly more sophisticated reasoning 
can be applied, see for example, Lee et al. (2016).

\subsection{Using mutation pairs} 
NGS data usually contain 
substantially fewer mutation pairs than marginal SNVs.
However
mutation pairs carry important 
phasing information that improves the
accuracy of subclone reconstruction. 
We find that the phased data for (far fewer) pairs provide more information than the (far more) marginal VAFs.
For example, imagine a tumor
sample that is a mixture of subclones 2 and 3 in Figure
\ref{fig:tm_evo}. Suppose a sufficient amount of short reads cover the
first two loci, we should observe relatively large reads counts for
four alleles (C, G), (A, T), (C, T) and (A, G). One can then reliably
infer that there are heterogeneous cell subpopulations in the tumor
sample.  
In contrast, if we ignore the phasing information and only consider
the (marginal) VAFs for each SNV, then the observed VAFs for both SNVs
are 50\%, which could be heterogeneous mutations from a single cell
population. See Simulation 1 for an illustration. In summary, we
leverage the power of using mutation pairs over marginal SNVs by
incorporating partial phasing information in our model. 
Besides the simulation study we will later also empirically
confirm these considerations in actual data analysis.

The relative advantage of using mutation pairs over marginal
SNV's can also be understood as a special case of a more general
theme. In biomedical data it is often important to avoid
overinterpretation of noisy data and to distill a relatively weak signal. A
typical example is the probability of expression (POE) model of
\cite{POE:2002}.
Similarly, the 
modeling of mutation pairs is a way to extract the pertinent information
from the massive noisy data.  Due to noise and artifacts in NGS data,
such as base-calling or mapping error, many called SNVs might record
unusual population frequencies, for reasons unrelated to the presence
of subclones~\citep{li2014toward}.  Direct modeling of all marginal
read counts one ends up with noise swamping the desired signal
\citep{nik2012mutational, jiao2014inferring}. See our analysis of a
real data set in Section \ref{sec:realdata} for an example. To
mitigate this challenge, most methods use clustering of the VAFs,
including, for
example, \cite{roth2014pyclone}. One would then use the resulting
cluster centers to infer subclones, which is one way of extracting
more concise information. In addition, the vast majority of the methods in the literature show that even though a tumor sample could possess thousands to millions of SNVs, the number of inferred subclones usually is in the low single digit, no more than 10. To this end, we propose instead an alternative approach
to extract useful information by modeling (fewer) mutation pairs, as
mutation pairs contain more information and are of higher quality. We show in our numerical examples later that with a few dozens of these mutation pairs, the inference on the subclones is strikingly similar to cluster-based subclone callers using much more SNVs.


Finally,
using mutation pairs does not exclude the possibility of making use of marginal SNVs. 
In Section \ref{sec:mrc}, we show it is straightforward to jointly model mutation pairs and SNVs. Other biological complexities, such as tumor purity and copy number variations, can also be incorporated in our model.  See Sections \ref{sec:purity} and \ref{sec:cnv} for more details. 
Incorporating
CNVs greatly increases the complexity in modeling and is not addressed by
most existing methods. Thus, we mainly concern ourselves with mutation pairs in copy
number neutral regions and leave incorporation of CNVs to future work.

\subsection{Representation of subclones}
\label{sec:repsubclone}
We construct a $K \times C$ categorical valued matrix $\bZ$ (Figure
\ref{fig:Z_example}) to represent the subclone structure. Rows of
$\bZ$ are indexed by $k$ and  represent mutation pairs, and columns of $\bZ$, denoted by $\bm
z_c = (\bz_{1c}, \cdots, \bz_{Kc})$, record the  phased mutation
pairs on the two homologous chromosomes    of subclone $c$, $c = 1,
\ldots, C$.
As in Figure \ref{fig:Z_example},
let  $j = 1, 2$ index the two homologous
chromosomes,
 $r=1, 2$ index the two mutation loci,
$\bz_{kc} = (\bz_{kcj}, j=1,2)$ be  the genotype
consisting of   two alleles for mutation pair $k$ in subclone $c$,
and  $\bz_{kcj}=(z_{kcjr}, r=1,2)$ denote the allele of the $j$-th
homologous chromosome.
 Therefore, each entry $\bz_{kc}$ of the matrix $\bZ$ is a $2 \times
2$ binary submatrix itself. 
For example,  in Figure \ref{fig:Z_example} the
entry $z_{21}$ is a pair of 2-dimension binary row vectors, $(0, 1)$
and $(1, 1)$, representing the genotypes for both alleles at mutation
pair $k=2$ of subclone $c=1$; each vector indicates the allele for the mutation pair on
a homologous chromosome. The first vector $(0, 1)$ indicates that locus
$r = 1$ harbors no mutation (0) and locus $r=2$ harbors a mutation (1).
Similarly, the second vector
$(1, 1)$ marks two mutations on both loci.  

In summary, each entry of $\bZ$,
$$
\bz_{kc} = \left( \bz_{kc1}, \bz_{kc2} \right) = \left( (z_{kc11}, z_{kc12}), (z_{kc21}, z_{kc22}) \right)
$$
is a $2 \times 2$ matrix (with the two row vectors horizontally
displayed for convenience).
Each $z_{kcjr}$ is a binary indicator and $z_{kcjr} = 1$ (or $0$)
indicates a mutation (or reference).  Thus, $\bz_{kc}$
can take $Q = 16$ possible values. That is,
$\bz_{kc} \in 
\{\bz^{(1)}, \ldots, \bz^{(16)}\} =$
$\{(00, 00),$ $(00, 01),$ $\ldots,$ $(11,11) \}$,
where we write $00$ short for $(0,0)$ etc., and
$z^{(1)} = (00, 00)$ refers to  the genotype on the
reference genome. Formally, $\bz_{kc}$ is a $2 \times 2$ binary matrix,
and $\bZ$ is a matrix of such binary matrices.
Moreover, we can collapse some $\bz^{(q)}$ values as we do not have
phasing across mutation pairs.
For example, $\bz_{kc}=(01,10)$ and $\bz_{kc}=(10,01)$, etc. have
mirrored rows and are indistinguishable in defining a subclone (a
column of $\bZ$). (More details in Section \ref{sec:prior}).
Typically distinct mutation pairs are distant from each
other, and in NGS data they are almost never phased. Therefore, we can
reduce the number of possible outcomes of $z_{kc}$ to $Q = 10$, due to
the mirrored outcomes. We list them below for later reference:
$z^{(1)} = (00, 00), z^{(2)} = (00, 01), z^{(3)} = (00, 10), z^{(4)} =
(00, 11), z^{(5)} = (01, 01), z^{(6)} = (01, 10), z^{(7)} = (01, 11),
z^{(8)} = (10, 10), z^{(9)} = (10, 11)$ and $z^{(10)} = (11, 11)$. In
summary, the entire matrix $\bZ$ fully specifies the genomes of each
subclone at all the mutation pairs.

Suppose $T$ tumor samples are available from the same patient,
obtained either at different time points (such as initial diagnosis
and relapses), at the same time but from different spatial
locations   within the  same   tumor, or from tumors at different
metastatic sites. We assume those $T$ samples
share the same subclones, while the  subclonal population frequencies
may   vary across samples.
For clinical decisions it can be important to know the population frequencies
of the subclones. 
To facilitate such inference, we introduce a $T
\times (C+1)$ matrix $\bm w$ to represent the population frequencies
of subclones. The element $w_{tc}$ refers to the proportion of
subclone $c$ in sample $t$, where $0 < w_{tc} < 1$ for all $t$ and
$c$, 
and $\sum_{c = 0}^C w_{tc} = 1$. A background subclone, which has no
biological meaning and is indexed by $c = 0$, is included to account
for artifacts and experimental noise. We will discuss more about
this later.

The remainder of this article is organized as follows. In
Sections~\ref{sec:probmodel} and  \ref{sec:posterior}, we propose a
Bayesian feature allocation model and the corresponding posterior
inference scheme to  estimate   the latent subclone structure. In
Section~\ref{sec:simulation}, we  evaluate   the model with three
simulation studies. 
Section~\ref{sec:pipeline} extends the models to accommodate other
biological complexities and present additional simulation results.
Section~\ref{sec:realdata} reports the analysis results for a lung
cancer patient with multiple tumor biopsies.  We conclude with a final
discussion in
Section~\ref{sec:conclusion}.

\section{The PairClone Model}
\label{sec:probmodel}

\subsection{Sampling Model}
\label{sec:splmodel}
Suppose paired-end short reads data are obtained by deep DNA
sequencing of multiple tumor samples. In such data, a short read is
obtained by sequencing two ends of the same DNA fragment. Usually a
DNA fragment is much longer than a short read, and the two ends
do not overlap and must be mapped separately. 
However, since the
paired-end reads are from the same DNA fragment, they are naturally
phased and can be used for inference of alleles and
subclones.
We use \texttt{LocHap}~\citep{sengupta2016ultra} 
to find pairs of mutations that are no more than a fixed number, say 500,
base pairs apart. 
 Such mutation pairs can be mapped by 
paired-end reads,
making them eligible for PairClone analysis. 
See Figure \ref{fig:mut_example} for an example.
For each mutation pair, a number of short reads are mapped to at least
one of the two loci.  Denote the two sequences on short read $i$
mapped to mutation pair $k$ in tissue sample $t$ by $\bs_{tk}^{(i)}
= \left(s_{tkr}^{(i)}, r = 1, 2\right) = \left(s_{tk1}^{(i)},
s_{tk2}^{(i)}\right)$, where $r=1,2$ index the two loci,
$s_{tkr}^{(i)} = 0 \mbox{ or } 1$ indicates that the short read
sequence is a reference or mutation.  Theoretically, each
$s_{tkr}^{(i)}$ can take four values, A, C, G, T, the four nucleotide
sequences. However, at a single locus, the probability of observing
more than two sequences across short reads is negligible since it
would require the same locus to be mutated twice throughout the life
span of the person or tumor, which is unlikely.
We therefore code $s_{tkr}^{(i)}$ as a binary value.
Also, sometimes a short read may cover only one of the two loci in a pair, and we use
$s_{tkr}^{(i)} = -$ to represent a missing base when there is no overlap between a short read and the corresponding SNV. 
Therefore,
$s_{tkr}^{(i)} \in \{ 0, 1, -\}$.  For example, in Figure
\ref{fig:mut_example} locus $r = 1$, $s_{tk1}^{(1)} = 0$ for read $i =
1$, $s_{tk1}^{(2)} = 1$ for read $i = 2$, and $s_{tk1}^{(3)} = -$ for
read $i = 3$. Reads that are not mapped to either locus are excluded
from analysis since they do not provide any information for
subclones. Altogether, $\bs_{tk}^{(i)}$ can take $G = 8$ possible
values, and its sample space is denoted by $\HH = \{\bh_1,
\ldots, \bh_G \} = \{ 00, 01, 10, 11, -0, -1, 0-, 1-\}$.
Each value corresponds to an allele of two loci,
with $-$ being a special ``missing'' coverage.  For mutation pair $k$
in sample $t$, the number of short reads bearing allele $\bh_g$ is
denoted by $n_{tkg} = \sum_i I \left(\bs_{tk}^{(i)} = \bh_g
\right)$, where $I(\cdot)$ is the indicator function,
 and the total number of reads mapped to the mutation pair is then $N_{tk} = \sum_g n_{tkg}$.
Finally, depending upon whether a read covers both loci or only one
locus we distinguish three cases:
(i) a read maps to both loci (complete), taking values
$\bs_{tk}^{(i)} \in \{\bh_1,\ldots,\bh_4\}$;
(ii) a read maps to the second locus only (left missing), $\bs_{tk}^{(i)} \in \{\bh_5,\bh_6\}$;
and (iii) a read maps to the first locus only (right missing), $\bs_{tk}^{(i)} \in \{\bh_7,\bh_8\}$.
We assume a multinomial sampling model for the observed read counts
\begin{align}
(n_{tk1}, \ldots, n_{tk8}) \mid N_{tk} \sim
\Mn(N_{tk};\; p_{tk1}, \ldots, p_{tk8}).
\label{eq:multi}
\end{align}
Here $\bm p = \{ p_{tkg}, g = 1, \ldots, 8\}$ are the probabilities for the 8 possible values of $\bs_{tk}^{(i)}$. For the upcoming
discussion, we separate out the probabilities for the three missingness cases.
Let $v_{tk1}, v_{tk2}, v_{tk3}$ denote the probabilities of observing a short read satisfying cases (i), (ii) and (iii),
respectively. We write $p_{tkg} = v_{tk1} \, \tp_{tkg}, g = 1, \ldots,
4$, $p_{tkg} = v_{tk2} \, \tp_{tkg}, g = 5, 6$, and $p_{tkg} = v_{tk3}
\, \tp_{tkg}, g = 7, 8$. 
Here $\tp_{tkg}$ are the probabilities conditional on case (i),
(ii) or (iii).
That is, $\sum_{g=1}^4 \tp_{tkg} = \sum_{g=5,6} \tp_{tkg} =
\sum_{g=7,8} \tp_{tkg}=1$. We still use a single running index, $g=1,\ldots,8$,
to match the notation in $p_{tkg}$. 
Below we link the multinomial sampling model with the underlying subclone structure by expressing $\tp_{tkg}$ in terms of $\bZ$ and $\bm w$.
Regarding $v_{tk1}, v_{tk2}, v_{tk3}$ we assume non-informative missingness and therefore do not proceed with inference on them (and $v$'s remain constant factors in the likelihood).

\subsection{Prior Model}
\label{sec:prior}

\paragraph{Construction of $\tp_{tkg}$}
The construction of a prior model for $\tp_{tkg}$ is based on
the following generative model.
To generate a short read, we first select a subclone $c$ from which the read
arises, using the population frequencies $w_{tc}$ for sample
$t$. Next we select with probability $0.5$ one of the two 
DNA strands, 
$j= 1 , 2$.
Finally, we record the read $\bh_g$, $g=1,2,3$ or $4$, corresponding to the chosen allele
$\bz_{kcj}=(z_{kcj1},z_{kcj2})$.
In the case of left (or right) missing locus we observe $\bh_g$, $g=5$
or $6$ (or $g=7$ or $8$), corresponding to the observed locus of the chosen allele.
Reflecting these three generative steps, we denote the probability of
observing a short read $\bh_g$ that bears sequence $\bz_{kcj}$ by 
\begin{equation}
 A(\bh_g, \bz_{kc}) =
   \sum_{j = 1}^2 0.5\,\times\,I(h_{g1} = z_{kcj1})\, I(h_{g2}=z_{kcj2}),
 \label{eq:A}
\end{equation}
with the understanding that $I(- = z_{kcjr}) \equiv 1$ for missing reads.
Implicit in \eqref{eq:A} is the restriction
$A(\bh_g, \bz_{kc}) \in \{0, 0.5, 1\}$, depending on the arguments.

Finally, using the definition of $A(\cdot)$ we model the probability
of observing a short read $\bh_g$ as 
\begin{equation}
   \tp_{tkg} = \sum_{c = 1}^C w_{tc}\,A(\bh_g, \bz_{kc}) +
   w_{t0} \, \rho_g. \label{pprior1}
\end{equation}
In \eqref{pprior1} we include $w_{t0} \rho_g$ to model a background subclone
denoted by $c = 0$ with  population frequency   $w_{t0}$. The
background subclone does not exist and has no biological
interpretation. It is only used as a mathematical
device to account for noise and artifacts in the NGS data (sequencing
errors, mapping errors, etc.). The weights $\rho_g$ are the
conditional probabilities of observing a short read $\bs_{tk}^{(i)}$ harboring
allele  $\bh_g$ if the recorded read were due to experimental noise.
Note that $\rho_1+\ldots + \rho_4= \rho_5+\rho_6 = \rho_7+\rho_8
= 1$.

\paragraph{Prior for $C$}
We assume a geometric prior distribution on $C$, $C \sim
\text{Geom}(r)$, to describe the random number of subclones (columns
of $\bZ$), $p(C) = (1 - r)^{C} r, C \in \{1, 2, 3,
\ldots\}$. {\it A priori} $E(C) = 1/r$.  

\paragraph{Prior for $\bZ$}
We use the finite version of the categorical Indian buffet process
(cIBP)~\citep{Sengupta2013cIBP} as the prior for the latent
categorical matrix $\bZ$. The cIBP is a categorical extension of the
Indian buffet process~\citep{griffiths2011indian} and defines feature
allocation~\citep{broderick2013feature} for categorical matrices.
In our application, the mutation pairs are the objects, and the
subclones are the latent features chosen by the objects. The number of
subclones $C$ is random, with the geometric prior $p(C)$. 
Conditional on $C$, we now introduce for
each column of $\bZ$ a vector $\bm{\pi}_c = (\pi_{c1},
\pi_{c2}, \ldots, \pi_{cQ})$, where $p(\bz_{kc} = \bz^{(q)}) =
\pi_{cq}$, and $\sum_{q=1}^Q \pi_{cq} = 1$.  Recall that $\bz^{(q)}$ are
the possible genotypes for the mutation pairs defined in Section \ref{sec:repsubclone},
$q=1,\ldots,Q$, for $Q=10$ possible genotypes.

As prior model for $\bm \pi_c$, we use a Beta-Dirichlet distribution
\citep{kim2012bayesian}. Let $\tpi_{cq}= \pi_{cq}/(1-\pi_{c1})$, $q=2,\ldots,Q$.
Conditional on $C$, $\pi_{c1} \sim \Be(1,\alpha / C)$ follows a beta distribution, and
$(\tpi_{c2}, \ldots, \tpi_{cQ}) \sim \Dir(\gamma_2, \ldots,
\gamma_Q)$ follows a Dirichlet distribution. Here $\bz_{kc} = \bz^{(1)}$ corresponds to the
situation that subclone $c$ is not chosen by mutation pair $k$,
because $\bz^{(1)}$ refers to the reference genome. We write
\begin{align*}
\bm \pi_c \mid C \sim \text{Beta-Dirichlet}(\alpha/C, 1, \gamma_2, \ldots, \gamma_Q).
\end{align*}
This construction includes a positive probability for all-zero columns $\bz_c = \bm 0$. In our
application, $\bz_c = \bm 0$ refers to normal cells
with no somatic mutations, which could be included in the
cell subpopulations.

In the definition of the cIBP prior, we would have one more step of dropping all zero columns. This
leaves a categorical matrix $\bZ$ with at most $C$ columns. As shown in \cite{Sengupta2013cIBP}, the marginal limiting
distribution of $\bZ$ follows the cIBP as $C \rightarrow \infty$.

\paragraph{Prior for $\bw$}
We assume $\bm w_t$ follows a Dirichlet prior, 
\begin{align*}
\bm w_t \mid C \stackrel{iid}{\sim} \text{Dirichlet}(d_0, d, \cdots, d),
\end{align*}
for $t = 1, \cdots, T$.
We set $d_0 < d$ to reflect the nature of $c=0$ as a background
noise and model mis-specification term.

\paragraph{Prior for $\bm \rho$}
We complete the model with a prior for $\brho = \{\rho_g\}$.
Recall $\rho_g$ is the
conditional probability of observing a short read with  allele
$\bh_g$ due to experimental noise. We consider complete read, left
missing read and right missing read separately, and assume
\begin{align*}
\rho_{g_1} \sim \text{Dirichlet}(d_1, \ldots, d_1); \quad
\rho_{g_2} \sim  \text{Dirichlet}(2d_1, 2d_1); \quad
\rho_{g_3} \sim  \text{Dirichlet}(2d_1, 2d_1),
\end{align*}
where $g_1=\{1,2,3,4\}$, $g_2=\{5,6\}$ and $g_3 = \{7,8\}$.

\section{Posterior Inference}
\label{sec:posterior}
Let $\bx = (\bZ, \bm \pi, \bm w, \bm \rho )$ denote the unknown
parameters except $C$, where $\bZ = \{z_{kc}\}$, $\bm \pi =
\{\pi_{cq}\}$, $\bm w = \{w_{tc}\}$, and $\bm \rho = \{\rho_{g}\}$.
We use Markov chain Monte Carlo (MCMC) simulations
to generate samples from
the posterior $\bx^{(l)} \stackrel{iid}{\sim}  p(\bx \mid \bn,C)$,
$l = 1, \ldots, L$.  
With fixed $C$ such MCMC simulation is straightforward. 
See, for example, \cite{brooks2011} for a review of MCMC.
Gibbs sampling transition probabilities are used to
update $\bZ$ and $\bm \pi$, and Metropolis-Hastings transition
probabilities are used to update $\bm w$ and $\bm \rho$.
Since $ p(\bx \mid \bn, C)$ is 
expected to be highly multi-modal, we use additional
parallel tempering to improve mixing of the Markov chain.
Details of MCMC simulation and parallel
tempering are described in Appendix \ref{app:sec:mcmc}. 

\paragraph{Updating $C$}
Updating the value of $C$ is more difficult as it involves
trans-dimensional MCMC \citep{green1995reversible}.
At each iteration, we propose a new value $\tilde{C}$ by
generating from a proposal distribution $q(\ti C \mid C)$.
In the later examples we assume that $C$ is {\em a priori} restricted to
$\Cmin \leq C \leq \Cmax$, and use
a uniform proposal
$q(\ti C \mid C) \sim \unif\{\Cmin, \ldots, \Cmax \}$.

Next, we split the data into a training set $\bn'$ and a test
set $\bn''$ with $n_{tkg}' = b n_{tkg}$ and $n_{tkg}'' = (1 - b)
n_{tkg}$, respectively,  for $b \in (0,1)$. 
Denote by $p_b(\bx \mid C) = p(\bx \mid
\bn', C)$ the posterior of $\bx$ conditional on $C$ evaluated on
the training set only. We use $p_b$ in two instances. First, we
replace the original prior $p(\bx \mid C)$ by $p_b(\bx \mid C)$,
and second, we use $p_b$ as a proposal distribution for $\tbx$, 
as $q( \tbx \mid \tC ) = p_b(\tbx \mid
\tC)$. Finally, we evaluate the acceptance probability of
$(\tC,\tbx)$ on the test data by
\begin{align}
  p_{\text{acc}}(C,\bx, \tC,\tbx) = 1 \wedge
  \frac{p(\bn'' \mid \tbx, \tC)}
       {p(\bn'' \mid \bx, C)} \cdot
  \frac{p(\tC) \cancel{p_b(\tbx \mid \tC)}}
       {p(C)   \cancel{p_b(\bx  \mid C  )}} \cdot
  \frac{q(C \mid \tC) \cancel{q(\bx \mid C)}}
  {q(\tC \mid C) \cancel{q(\tbx \mid \tC)}}.
  \label{eq:fbf}
\end{align}
The use of the prior $p_b(\tbx \mid \tC)$ is
similar to the construction of the fractional Bayes factor (FBF)
\citep{ohagan1995} which uses a fraction of the data to define an
informative prior that allows the evaluation of Bayes factors.
In contrast, here $p_b$ is used as an informative proposal
distribution for $\tbx$. Without the use of a training sample it would
be difficult to generate proposals $\tbx$ with reasonable acceptance
rate. In other words, we use $p_b$ to achieve a better mixing Markov
chain Monte Carlo simulation.
The use of the same $p_b$ to replace the original prior avoids the otherwise
prohibitive evaluation of $p_b$ in the acceptance probability
\eqref{eq:fbf}. See more details in Appendix
\ref{app:sec:updatec} and \ref{app:sec:calib}.

\paragraph{Point estimates for parameters}
We use the posterior mode $\Chat$ as a point estimate of
$C$. Conditional on $\Chat$, we follow \cite{lee2015bayesian} to find a
point estimate of $\bZ$. For any two $K \times \Chat$ matrices $\bZ$
and $\bZ'$, a distance between the $c$-th column of $\bZ$ and the
$c'$-th column of $\bZ'$ is defined by $\DD_{cc'}(\bZ, \bm
Z') = \sum_{k = 1}^{K} \| \bz_{kc} - \bz'_{kc'}\|_1$, where $1 \leq c,
c' \leq \Chat$, and we take the vectorized form of $\bz_{kc}$ and $
\bz'_{kc'}$ to compute $L^1$ distance between them.  Then, we define
the distance between $\bZ$ and $\bZ'$ as $d(\bZ, \bZ') =
\min_{\bm \sigma} \sum_{c = 1}^{\Chat} \DD_{c, \bm \sigma_c}(\bm
Z, \bZ')$, where $\bm \sigma = (\sigma_1, \ldots, \sigma_{\Chat})$ is
a permutation of $\{1, \ldots, \Chat\}$, and the minimum is taken over
all possible permutations.
This addresses the potential label-switching
issue across the columns of $\bZ$.  Let $\{\bZ^{(l)}, l = 1, \ldots,
L\}$ be a set of posterior Monte Carlo samples of $\bZ$. A posterior
point estimate for $\bZ$, denoted by $\hat{\bZ}$, is reported as
$\hat{\bZ} = \bZ^{(\hat{l})}$, where
\begin{align*}
   \hat{l} = \argmin_{l \in \{1, \ldots, L\}} \sum_{l' = 1}^L d(\bZ^{(l)}, \bZ^{(l')}).
\end{align*} 
Based on $\hat{l}$, we report posterior point estimates of
$\bm w$ and $\bm \rho$, given by $ \hat{\bm w} = \bm w^{(\hat{l})}$ and $
\hat{\bm \rho} = \bm \rho^{(\hat{l})}$, respectively.

\section{Simulation Studies}
\label{sec:simulation}
We evaluate the proposed model with three simulation studies.
In the first simulation we use single sample data ($T = 1$),
since in 
most current applications only a single sample is available for
analysis. Inferring subclonal structure accurately under only one
sample is a major challenge, and not completely resolved in the 
current literature. The single sample does not rule out meaningful
inference, as the relevant sample size is the number of SNVs or
mutation pairs, or the (even larger) number of reads. In the second
and third simulations we consider multi-sample data,  
similar to the lung cancer data that we analyze later. 
In all simulations, we assume the missing probabilities
$v_{tk2}$ and $v_{tk3}$ to be 30\% or 35\%.  Recall that these
probabilities represent the probabilities that a short read will only
cover one of the two loci in the mutation pair. 

\subsection{Simulation 1}
\label{app:sim1}
\begin{figure}[h!]
\begin{center}
\begin{subfigure}[t]{.325\textwidth}
\centering
\includegraphics[width=\textwidth]{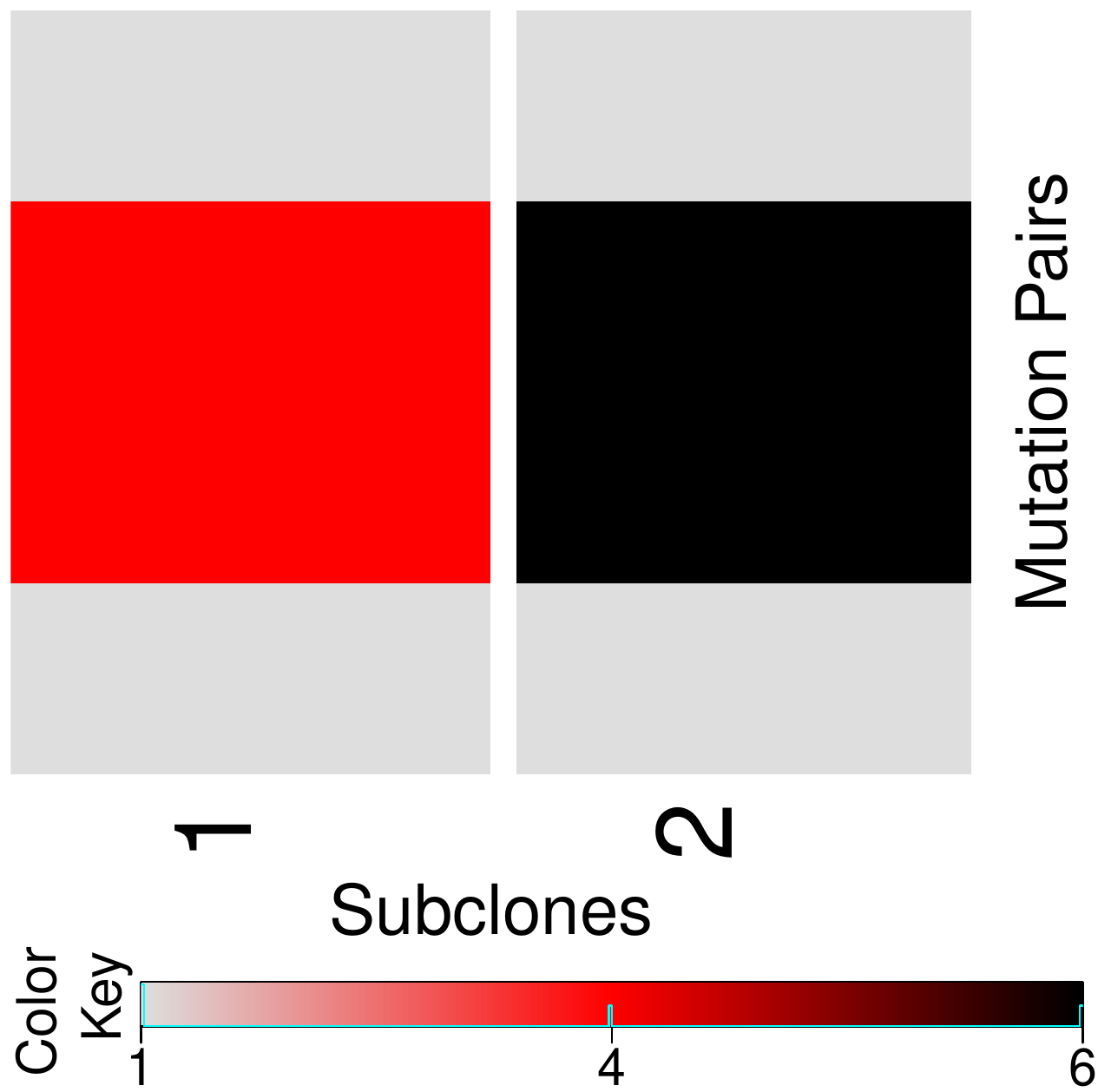}
\caption{$\bZ\true$}
\end{subfigure}
\begin{subfigure}[t]{.325\textwidth}
\centering
\hspace{-2mm}\includegraphics[width=\textwidth]{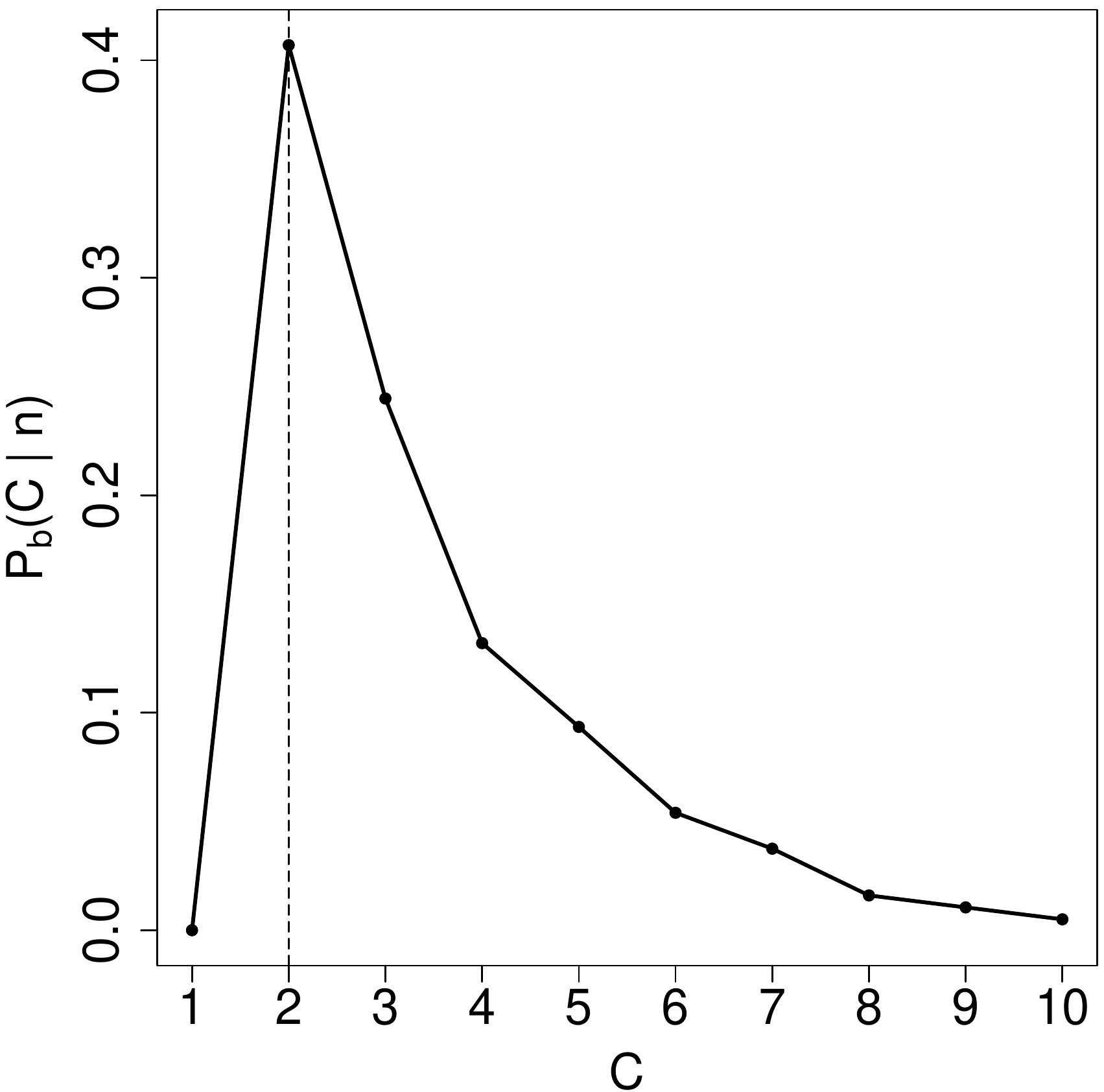}
\caption{$p_b(C \mid \bn'')$}		
\end{subfigure}
\begin{subfigure}[t]{.325\textwidth}
\centering
\includegraphics[width=\textwidth]{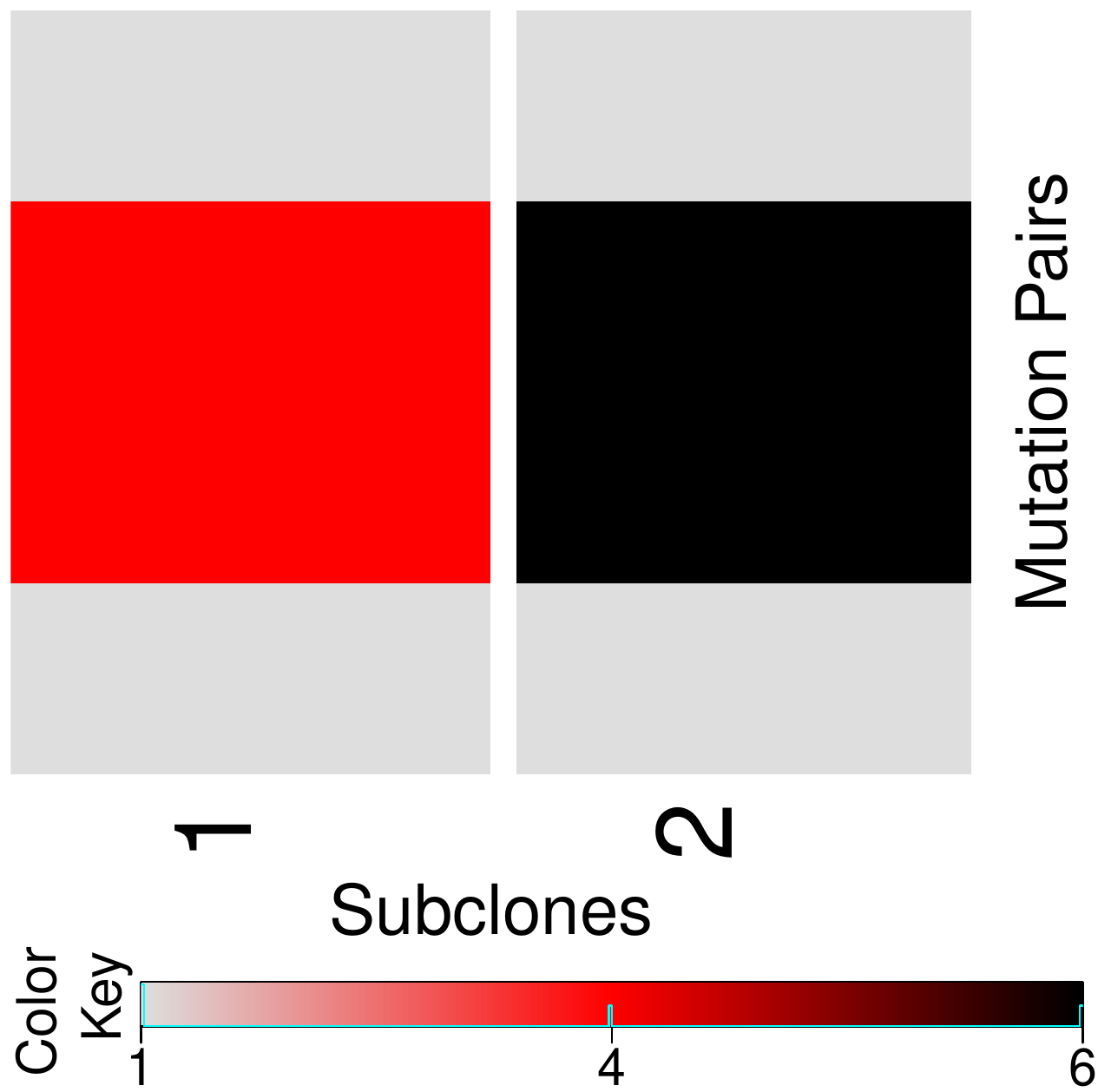}
\caption{$\Zhat$}		\vspace{5mm}
\end{subfigure}
\begin{subfigure}[t]{.325\textwidth}
\centering
\includegraphics[width=\textwidth]{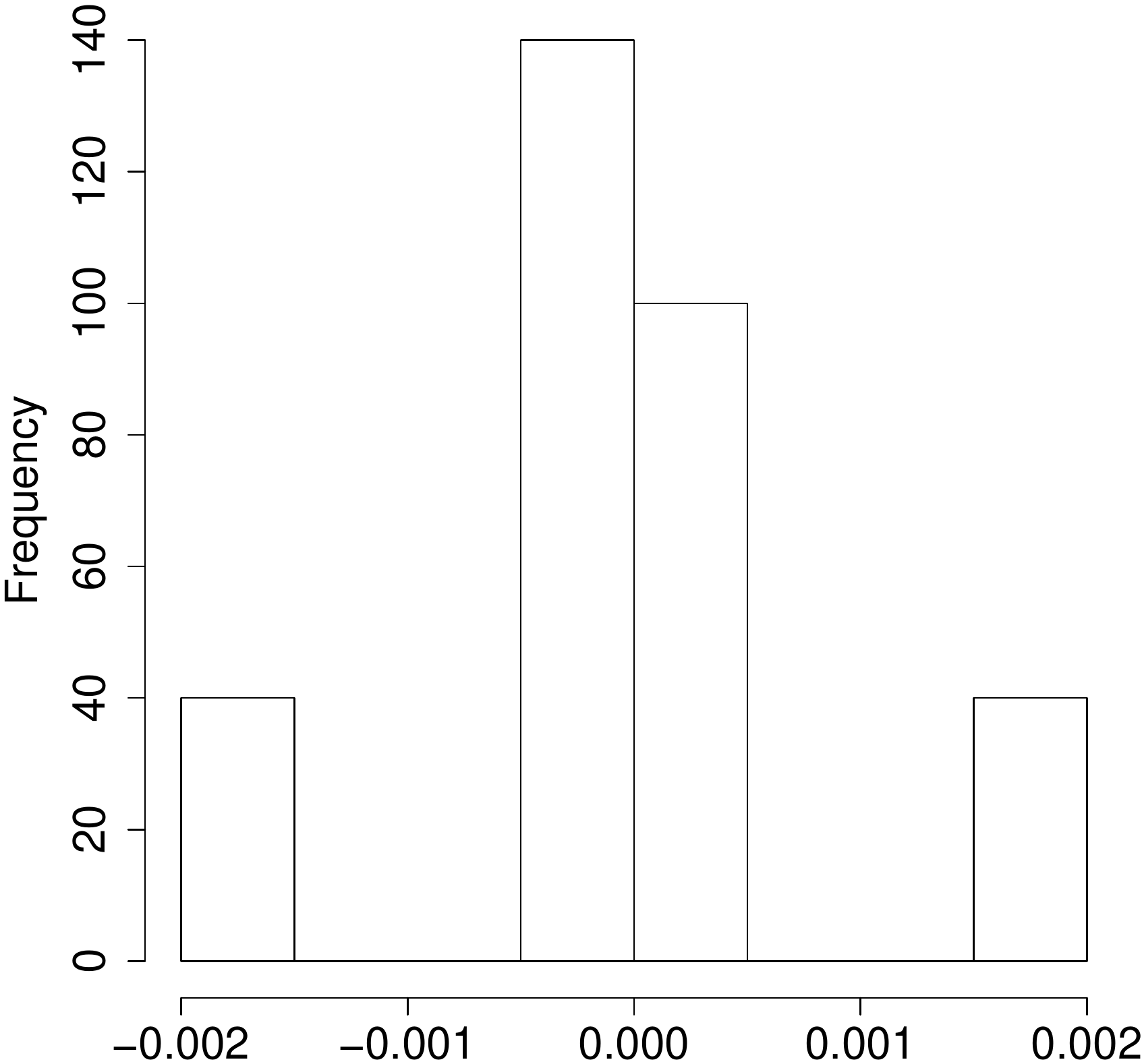}
\caption{ Histogram of $(\ptkghat - p_{tkg}\true)$}
\end{subfigure}
\begin{subfigure}[t]{.325\textwidth}
\centering
\includegraphics[width=\textwidth]{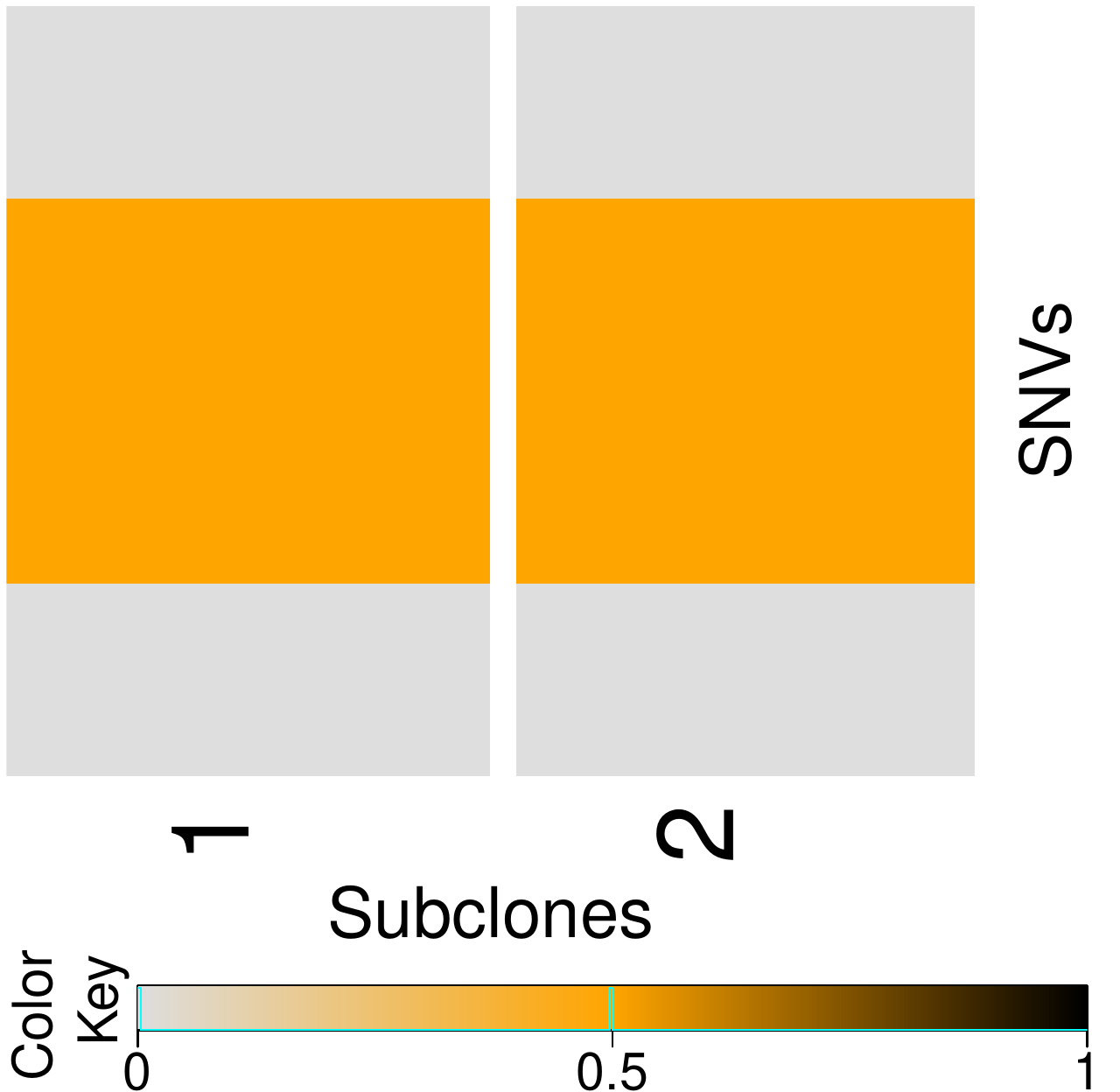}
\caption{$\bZ_{\text{BC}}\true$}		
\end{subfigure}
\begin{subfigure}[t]{.325\textwidth}
\centering
\includegraphics[width=\textwidth]{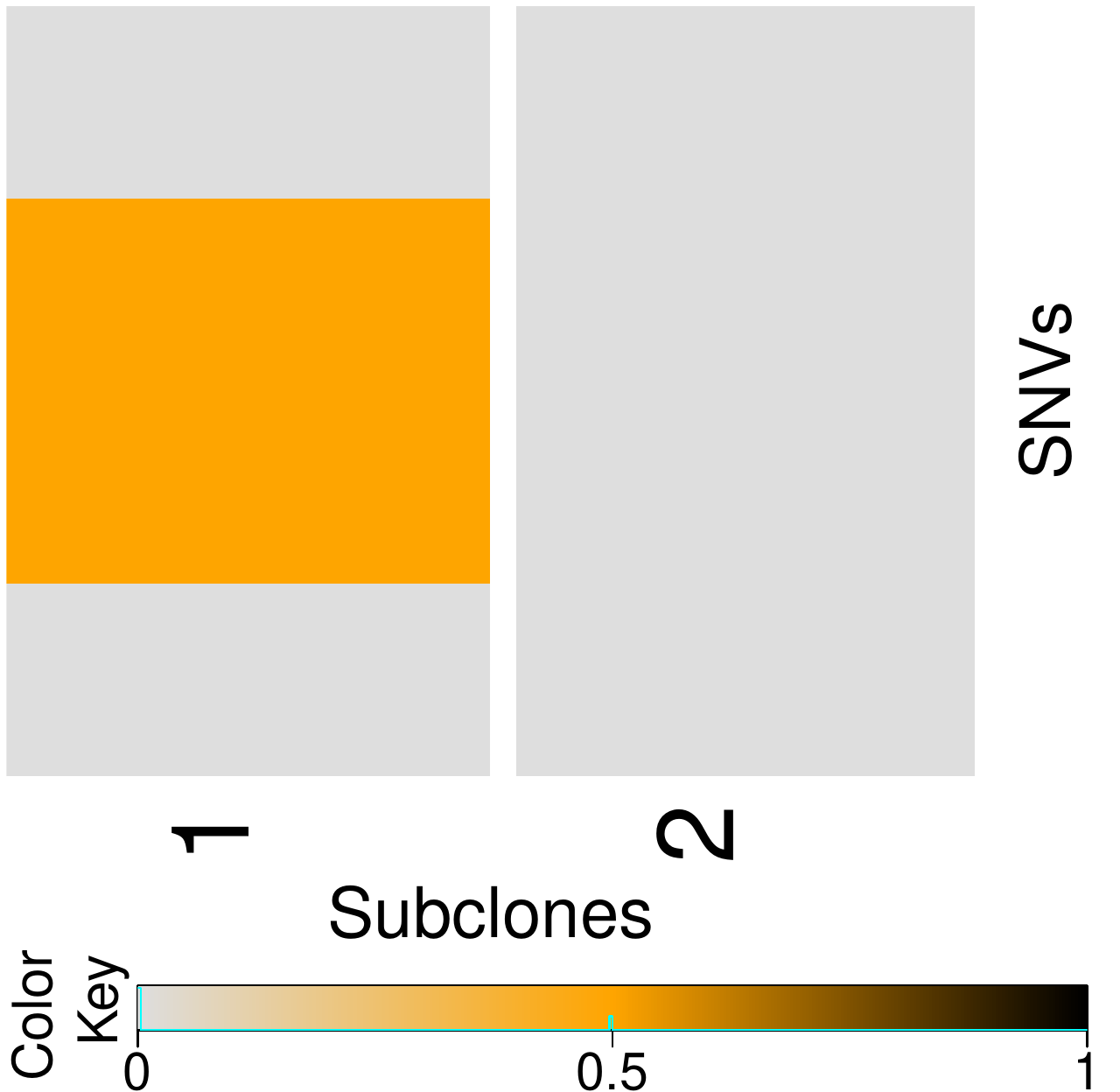}
\caption{$\Zhat_{\text{BC}}$}		
\end{subfigure}
\end{center}
\caption{Simulation 1. Simulation truth $\bZ\true$ (a, e), and posterior
  inference under PairClone (b, c, d) and under BayClone (f).}
\label{app:figsim1}
\end{figure}

\paragraph{Setup. }
In the first simulation, we illustrate the advantage of using mutation
pair data over marginal SNV counts.
We generate hypothetical short reads data for $T = 1$ sample and $K = 40$
mutation pairs.  Based on our own experiences, for a whole-exome
sequencing data set, we usually obtain dozens of mutation pairs with
decent coverage. See \cite{sengupta2016ultra} for a discussion.  We
assume there are $C\true = 2$ latent subclones, and set their
population frequencies as $\bw\true = (1.0 \times 10^{-7}, 0.8,
0.2)$, where $1.0 \times 10^{-7}$ refers to the proportion of the
hypothetical background subclone $c=0$.
The subclone matrix $\bZ\true$ is shown in
Figure~\ref{app:figsim1}(a) (as a heat map). Light grey, red and black
colors are used to represent genotypes $\bz^{(1)}$, $\bz^{(4)}$ and
$\bz^{(6)}$. For example, subclone 1 
has genotype $\bz^{(1)}$ (wild type) for mutation pairs 1--10 and 31
-- 40, and $\bz^{(4)}$ for mutation pairs 11--30.  We generate $\brho\true$ from its prior with hyperparameter $d_1 = 1$. Next we set the probabilities of observing left and right missing reads as
 $v_{tk2} = v_{tk3} = 0.3$ for all $k$ and $t$, to mimic a typical missing rate observed in the real data.  We calculate multinomial
probabilities $\{p_{tkg}\true\}$ shown in equations
\eqref{pprior1} and \eqref{eq:A} from the simulated $\bm
Z\true$, $\bm w\true$ and $\bm
\rho\true$. Total read counts $N_{tk}$ are generated as
random numbers ranging from $400$ to $600$, and finally we generate
read counts $n_{tkg}$ from the multinomial distribution given $N_{tk}$
as shown in equation \eqref{eq:multi}.

We fit the model with hyperparameters fixed as follows: $\alpha = 4$,
$\gamma_2 = \cdots = \gamma_Q = 2$, $d = 0.5$, $d_0 = 0.03$, $d_1 =
1$, and $r = 0.4$. We set $C_{\text{min}} = 1$ and $C_{\text{max}} =
10$ as the range of $C$. The fraction $b$ needs to be calibrated. We
choose $b$ such that the test sample size $(1 - b) \sum_{t = 1}^{T}
\sum_{k = 1}^{K} N_{tk}$ is approximately equal to
$160/T$. See Section \ref{app:sec:calib} for a discussion of
this choice.

We run MCMC simulation for $30,000$
iterations, discarding the first $10,000$ iterations as initial
burn-in, and keep one sample every $10$ iterations. The initial values
are randomly generated from the priors.

\paragraph{Results. }
Figure~\ref{app:figsim1}(b) shows $p_b(C \mid \bn'')$, where the
vertical dashed line marks the simulation truth. The posterior
mode $\Chat = 2$ recovers the truth. Figure~\ref{app:figsim1}(c) shows the
point estimate of $\bZ\true$, given by $\Zhat$. The true subclone
structure is perfectly recovered. The estimated subclone weights are
$\what = (2.27 \times 10^{-116}, 0.8099, 0.1901)$, which is also
very close to the truth. We use $\Zhat$ and $\what$ to calculate
estimated multinomial probabilities, denoted by $\{\ptkghat\}$.
Figure~\ref{app:figsim1}(d) shows a histogram of the differences
$(\ptkghat - p_{tkg}\true)$  as a residual plot to assess model
fitting. The histogram is centered at zero with little variation,
  indicating   a reasonably good model fit.
In summary, this simulation shows that the proposed inference
can almost perfectly recover the truth in a simple scenario with a single sample.

\subsection{Comparison with BayClone and PyClone}
We compare the proposed inference under PairClone versus
inference under SNV-based subclone callers ,i.e., based on marginal (un-paired) counts of point mutations, 
including BayClone\citep{sengupta2015bayclone} and 
PyClone~\citep{roth2014pyclone}. 

\begin{figure}[h!]
\begin{center}
\begin{subfigure}[t]{.45\textwidth}
\centering
\includegraphics[width=\textwidth]{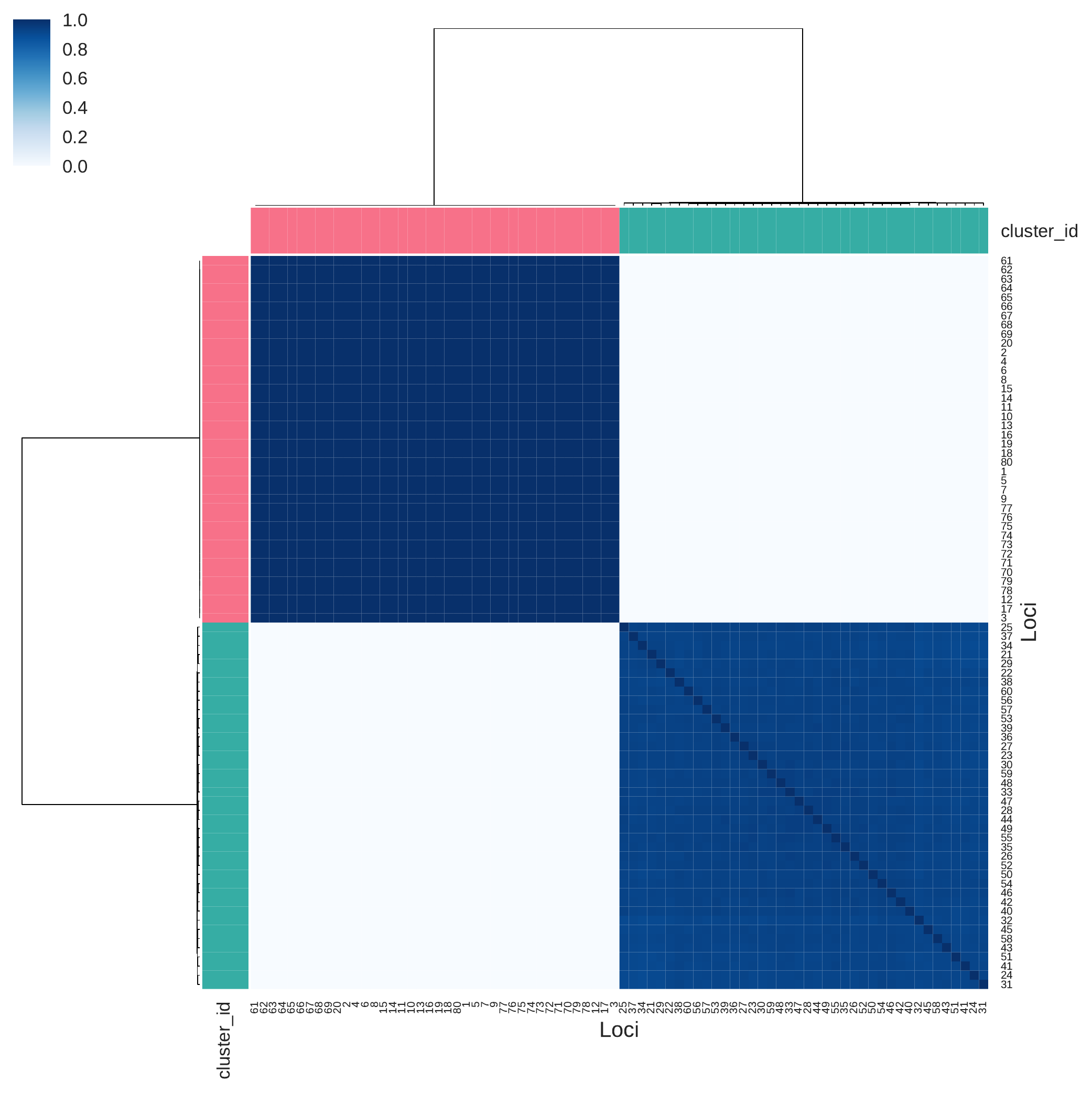}
\caption{Posterior similarity matrix}
\end{subfigure}
\begin{subfigure}[t]{.35\textwidth}
\centering
\includegraphics[width=\textwidth]{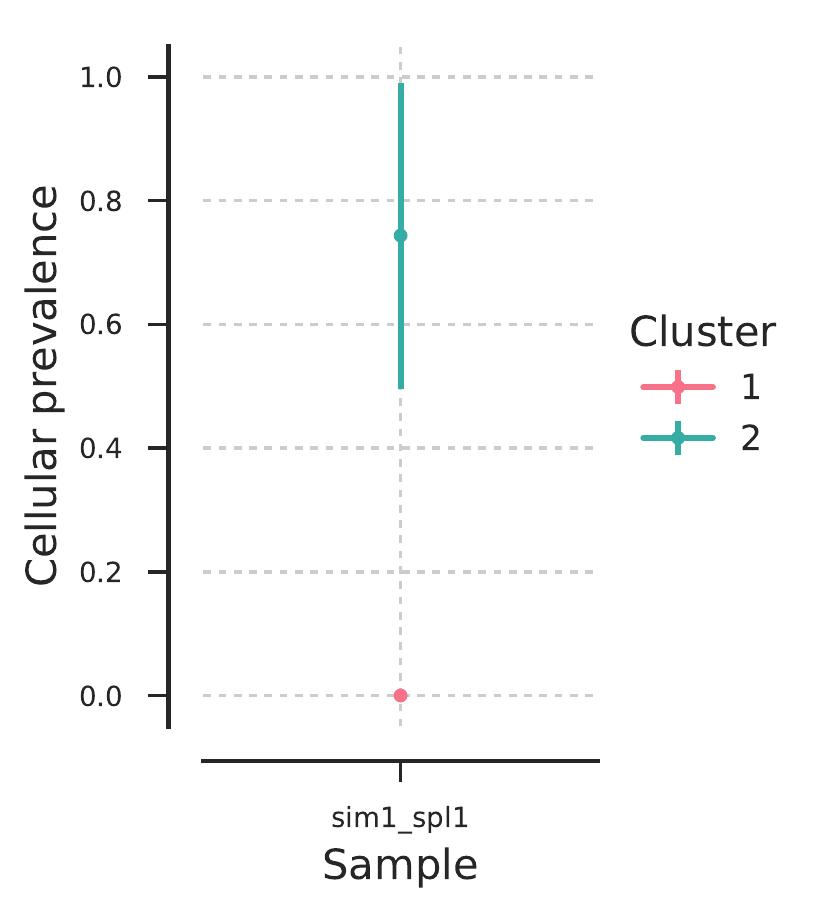}
\caption{Cellular prevalence}		
\end{subfigure}
\end{center}
\caption{Simulation 1. Posterior inference under PyClone.}
\label{app:figsim1_pyclone}
\end{figure}

\underline{BayClone} infers the subclone structure
based on marginal allele frequencies of the recorded SNVs, and chooses the
number of subclones based on log pseudo marginal likelihood (LPML)
model comparison.
Under the LPML criterion, the estimated number of subclones reported
by BayClone is $\Chat = 2$, which also recovers the truth.
Figure~\ref{app:figsim1}(e) displays
the true genotypes of the unpaired SNVs, denoted by
$\bZ_{\text{BC}}\true$, based on the true genotypes in
Figure~\ref{app:figsim1} for the mutation pairs. That is, we derive the
corresponding marginal genotype for each SNV in the mutation pair
based on the truth $\bZ\true$.  Figure~\ref{app:figsim1}(f) shows
the heat map of estimated matrix $\Zhat_{\text{BC}}$, where
$z_{sc} = 0$ (light grey), $0.5$ (orange) and $1$
(black) refer to homozygous wild-type, heterozygous variant and
homozygous variant at SNV locus $s$, respectively. The estimated
subclone proportions are $\what_{\text{BC}} = (0.008, 0.988, 
0.004)$.

\underline{PyClone}, on the other hand, clusters mutations based on
allele frequencies of the recorded SNVs using
the implied clustering under a Dirichlet process mixture model.
PyClone does not report subclonal genotypes and thus is not
directly comparable with PairClone. 
Posterior inference is summarized in 
Figure
\ref{app:figsim1_pyclone}.
Panel (a) indicates that the 80 SNV loci form two
clusters, with one cluster corresponding to loci 1--20 and
61--80, and the other cluster corresponding to loci 21--60, which
agrees with the truth. Panel (b) shows the cellular prevalence of the two
clusters across samples, where the middle point represents the
posterior mean, and the error bar indicates posterior standard
deviation. 
The cellular prevalence is defined as fraction of clonal population
harbouring a mutation.  In the PyClone MCMC samples, the estimated
cellular prevalence of cluster 2 fluctuates between 0.5 and 1 and thus
includes high posterior uncertainty, 
while the true cellular prevalence of cluster 2 is 1.

The estimates under SNV-based subclone callers do not fully recover
the simulation truth. The main reason is probably that the phasing
information of paired SNVs is 
lost in the marginal counts that are used in BayClone and PyClone,
making the subclone estimation less accurate than under PairClone.
For example, the two subclones with genotypes
$\bz^{(4)} = (00, 11)$ and $\bz^{(6)} = (01, 10)$ lead
to exactly the same allele frequency (50\%) for both loci.
BayClone can not distinguish between these two different subclones
based on the 50\% allele frequency for each locus.
Although BayClone correctly reports the number of subclones, inference
mistakenly includes a normal subclone with negligible weight, and thus
fails to recover the true population frequencies.  
On the other hand,  PyClone can not identify if cluster 2 contains homozygous (corresponding to cellular prevalence of 0.5) or heterozygous (corresponding to cellular prevalence of 1) variants. In contrast, using the phasing information, PairClone is able to infer two subclones having genotypes $(00, 11)$ and $(01, 10)$ for mutation pairs 11--30, and we know cluster 2 contains only heterozygous variants for sure.

\subsection{Simulation 2}
\label{app:sim2}

\begin{figure}[h!]
\begin{center}
\begin{subfigure}[t]{.325\textwidth}
\centering
\includegraphics[width=\textwidth]{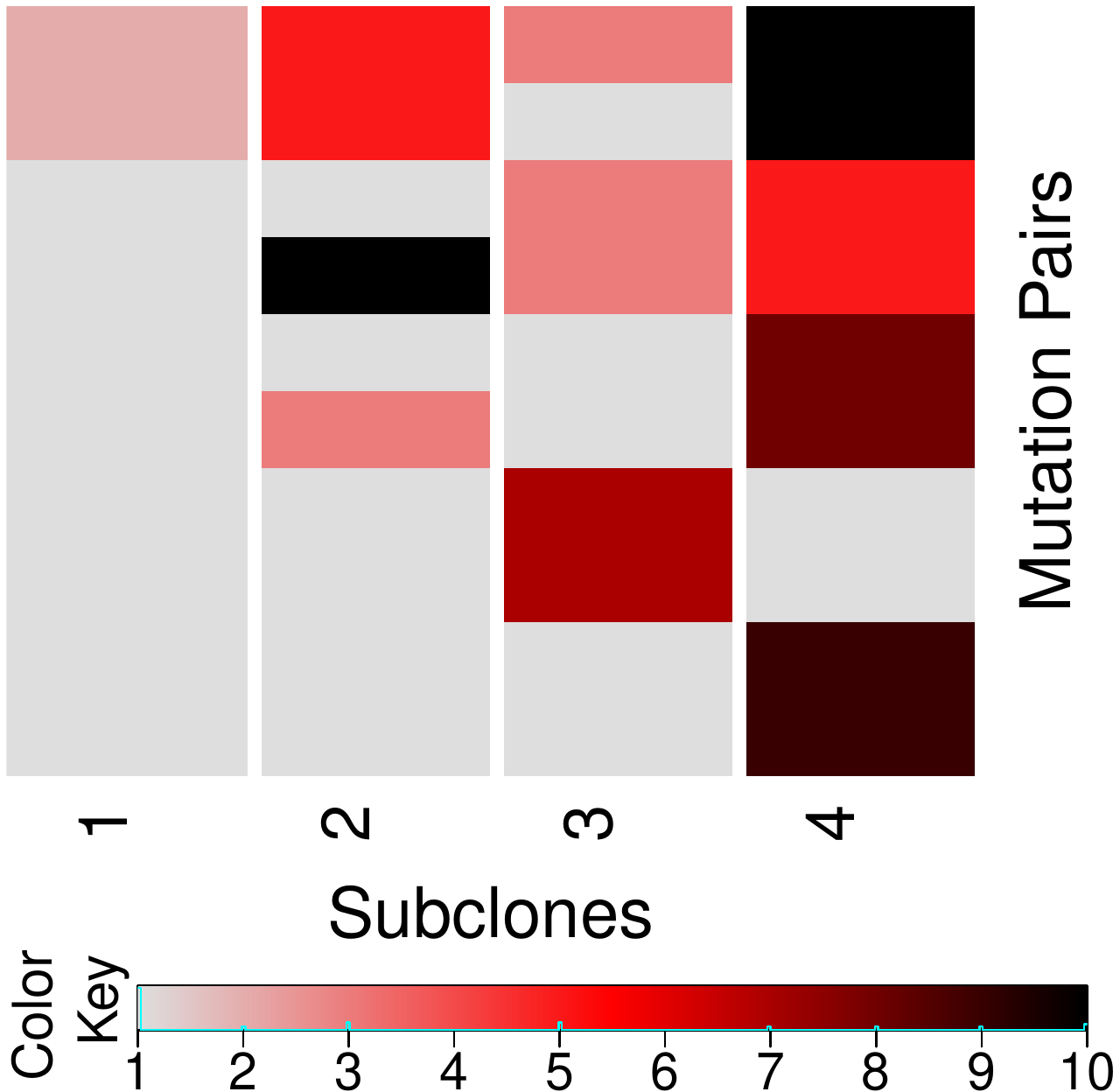}
\caption{$\bZ\true$}
\end{subfigure}
\begin{subfigure}[t]{.325\textwidth}
\centering
\includegraphics[width=\textwidth]{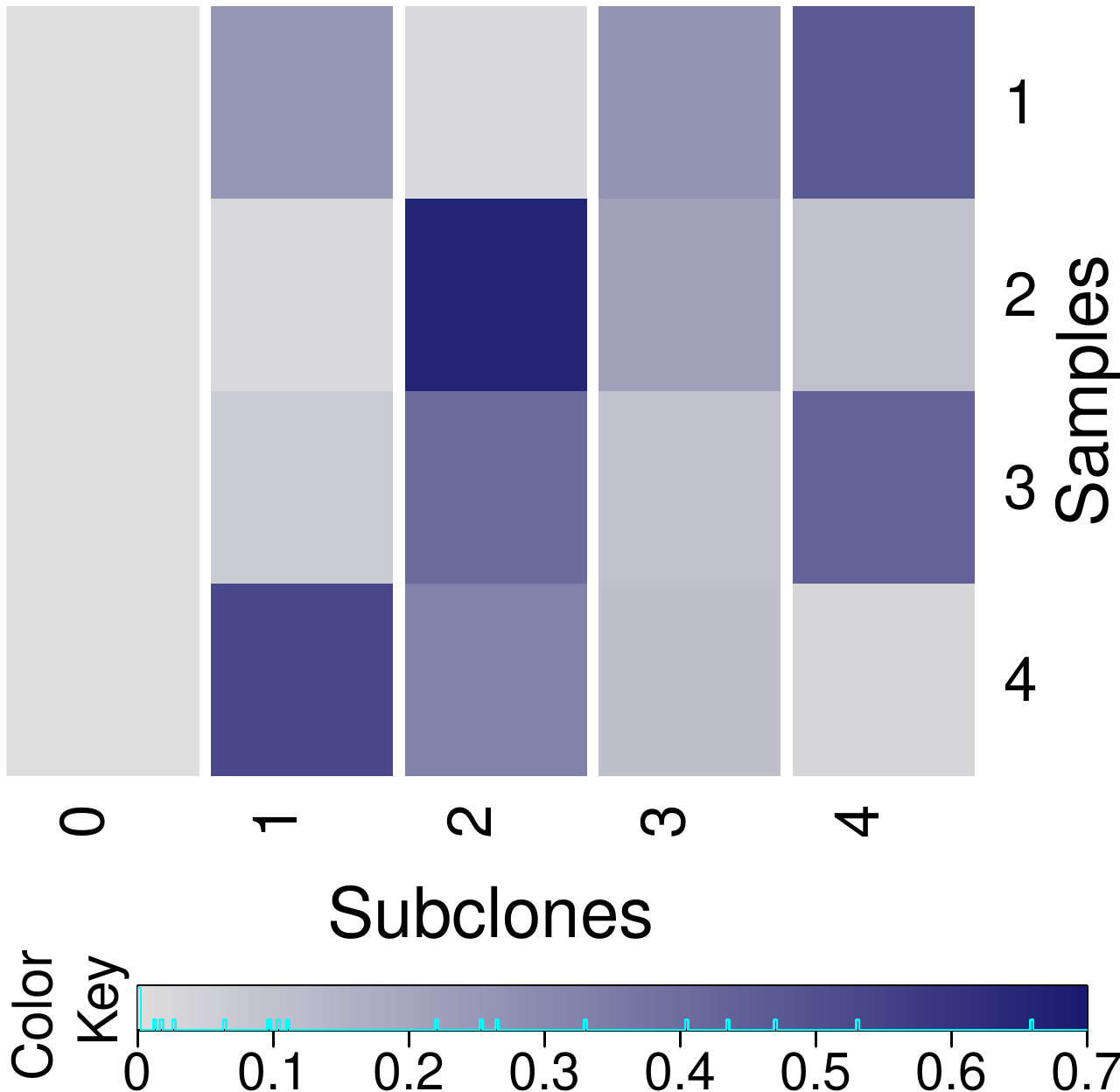}
\caption{$\bw\true$}
\end{subfigure}
\begin{subfigure}[t]{.325\textwidth}
\centering
\includegraphics[width=\textwidth]{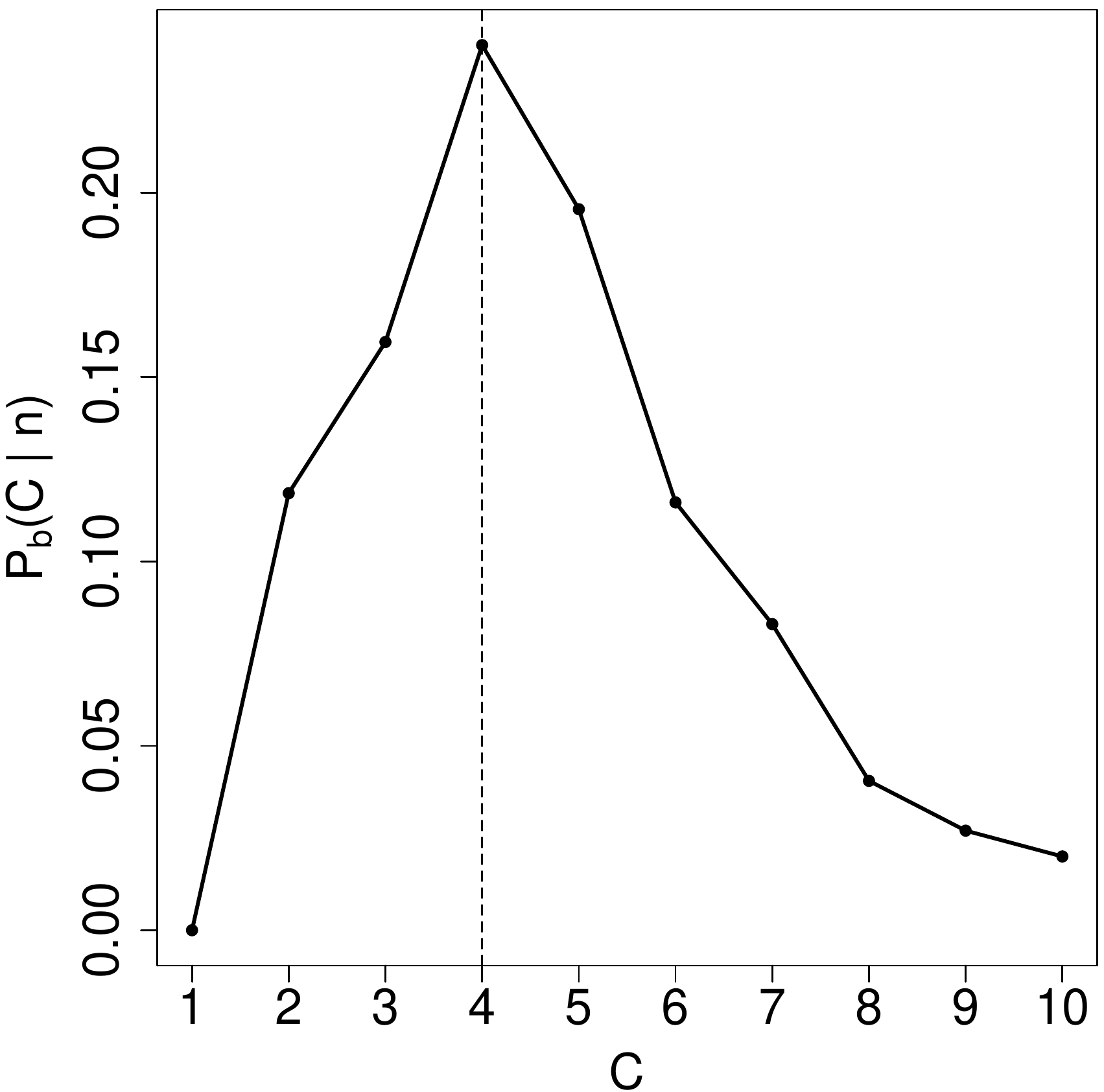}
\caption{$p_b(C \mid \bn'')$}
\end{subfigure}
\begin{subfigure}[t]{.325\textwidth}
\centering
\includegraphics[width=\textwidth]{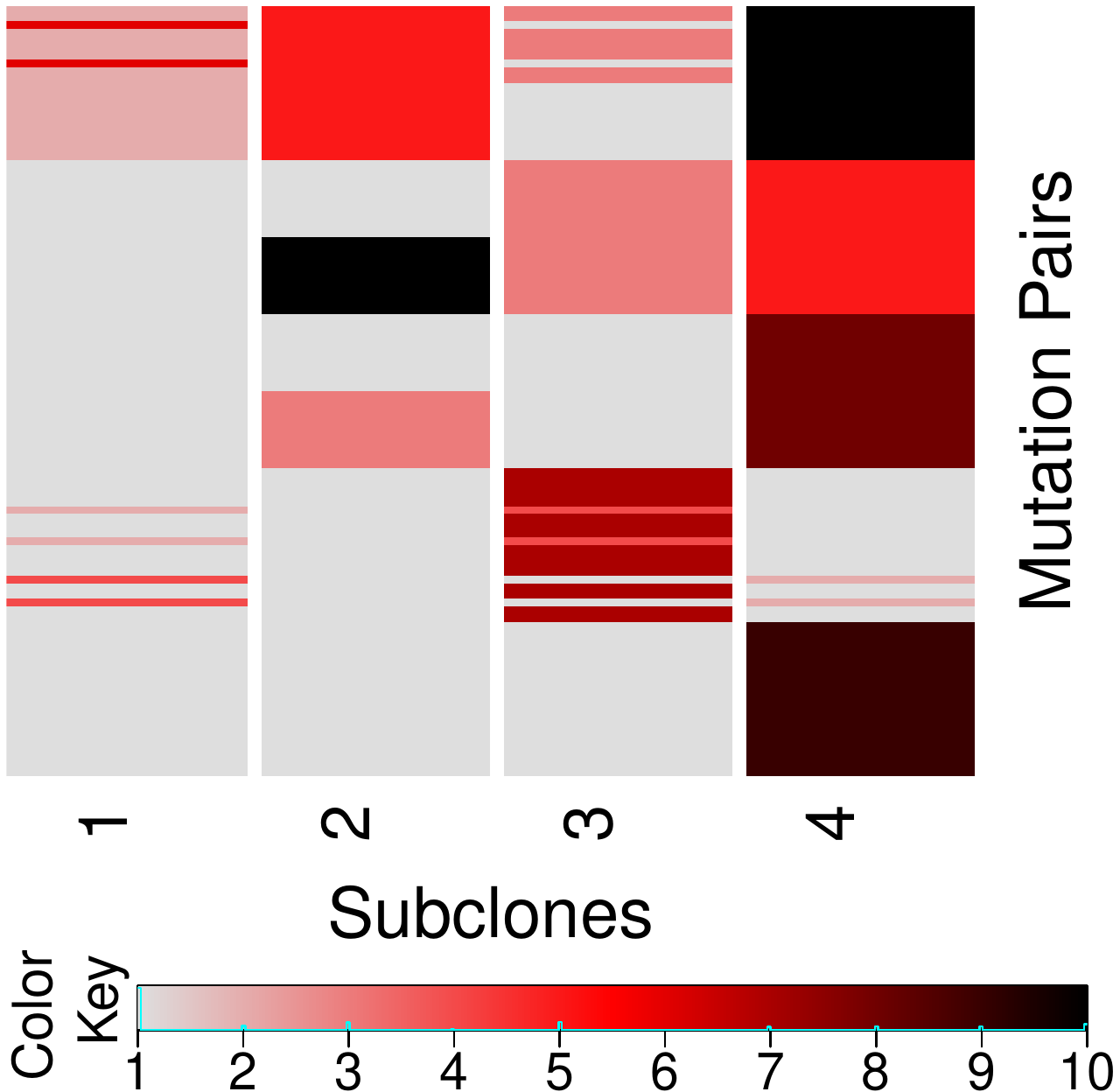}
\caption{$\Zhat$}
\end{subfigure}
\begin{subfigure}[t]{.325\textwidth}
\centering
\includegraphics[width=\textwidth]{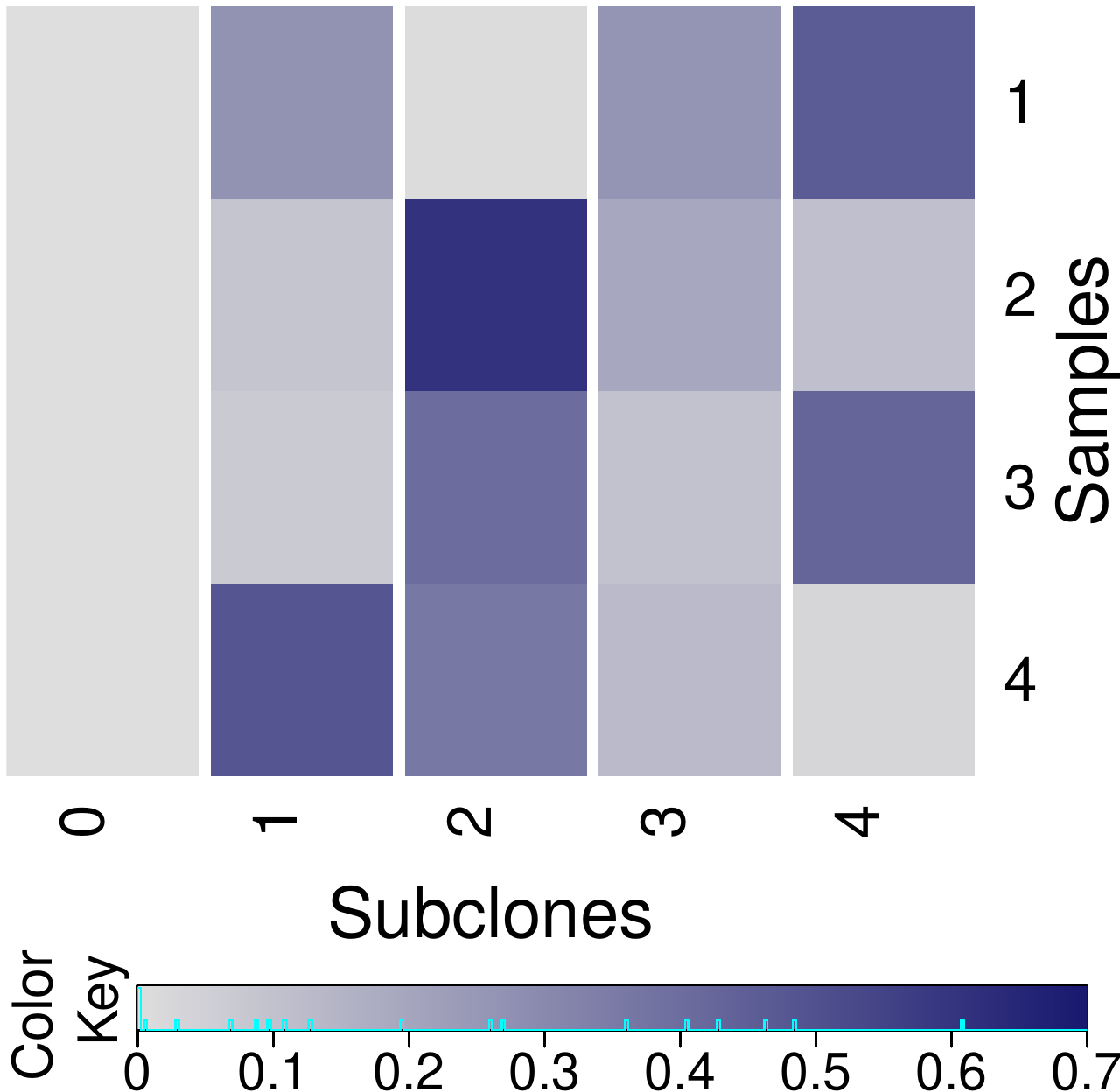}
\caption{$\what$}
\end{subfigure}
\begin{subfigure}[t]{.325\textwidth}
\centering
\includegraphics[width=\textwidth]{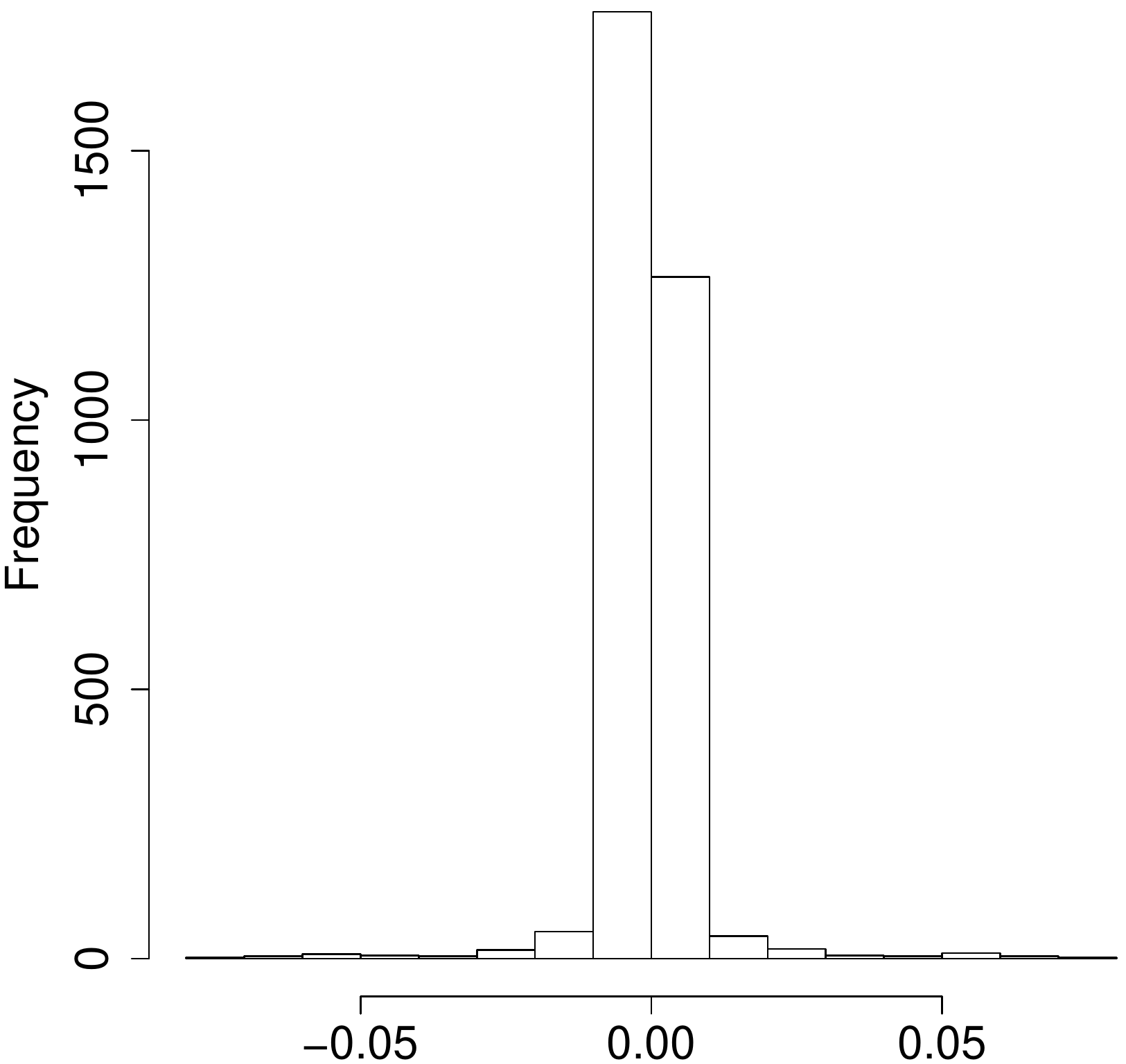}
\caption{Histogram of $(\ptkghat - p_{tkg}\true)$}
\end{subfigure}
\end{center}
\caption{Simulation 2. Simulation truth $\bZ\true$ and $\bw\true$ (a, b), and posterior inference under PairClone (c, d, e, f).}
\label{app:figsim2}
\end{figure}

In the second simulation, we  consider   data
with $K = 100$ mutation pairs    and a more complicated subclonal
structure with $C\true = 4$ latent 
subclones. 
We generate hypothetical data for $T = 4$ samples. 
The subclone matrix $\bZ\true$ is shown in
Figure~\ref{app:figsim2}(a). Colors on a scale from light grey to red, to
black (see the scale in the figure) are used to represent genotype
$\bz^{(q)}$ with $q = 1, \ldots, 10$. For example, subclone 4 has
genotype $\bz^{(10)}$  for mutation pairs 1--20, $\bz^{(5)}$  for
mutation pairs 21--40, $\bz^{(8)}$  for mutation pairs 41--60,
$\bz^{(1)}$  for mutation pairs 61--80, and $\bz^{(9)}$ for mutation
pairs 81--100. 
For each sample $t$, we generate the subclone proportions from a
Dirichlet distribution, $\bw_t\true \sim \Dir(0.01, \sigma(20, 10, 5,
2))$, where $\sigma(20, 10, 5, 2)$ is a random permutation of $(20,
10, 5, 2)$.  
The subclone proportion matrix $\bw\true$ is shown in Figure
\ref{app:figsim2}(b), where darker blue color indicates higher abundance
of a subclone in a sample, and light grey color represents low
abundance.  
The parameters $\bm \rho\true$ and $N_{tk}$ are generated using the
same approach as before, and we use
$v_{tk2} = v_{tk3} = 0.3$ for $k = 1, \ldots, 50$ and all $t$, and
$v_{tk2} = v_{tk3} = 0.35$ for $k = 51, \ldots, 100$ and all $t$.
Finally, we calculate $\{p_{tkg}\true\}$ and generate read counts $n_{tkg}$ from equation (\ref{eq:multi}) similar to previous simulation.

We fit the model with the same set of hyperparameters and MCMC
parameters as in simulation 1. Figure~\ref{app:figsim2}(c) shows
$p_b(C \mid \bn'')$. Again, the posterior mode $\Chat = 4$ recovers
the truth. Figure~\ref{app:figsim2}(d) shows the estimate $\Zhat$; the
truth is nicely approximated. 
Some mismatches are
expected under this more complex subclone structure.
The estimated subclone proportions $\what$ are shown in Figure~\ref{app:figsim2}(e), again close to the
truth. Figure~\ref{app:figsim2}(f) shows the histogram of ($\ptkghat -
p_{tkg}\true$) which indicates a good model fit.

\begin{figure}[h!]
\begin{center}
\begin{subfigure}[t]{.325\textwidth}
\centering
\includegraphics[width=\textwidth]{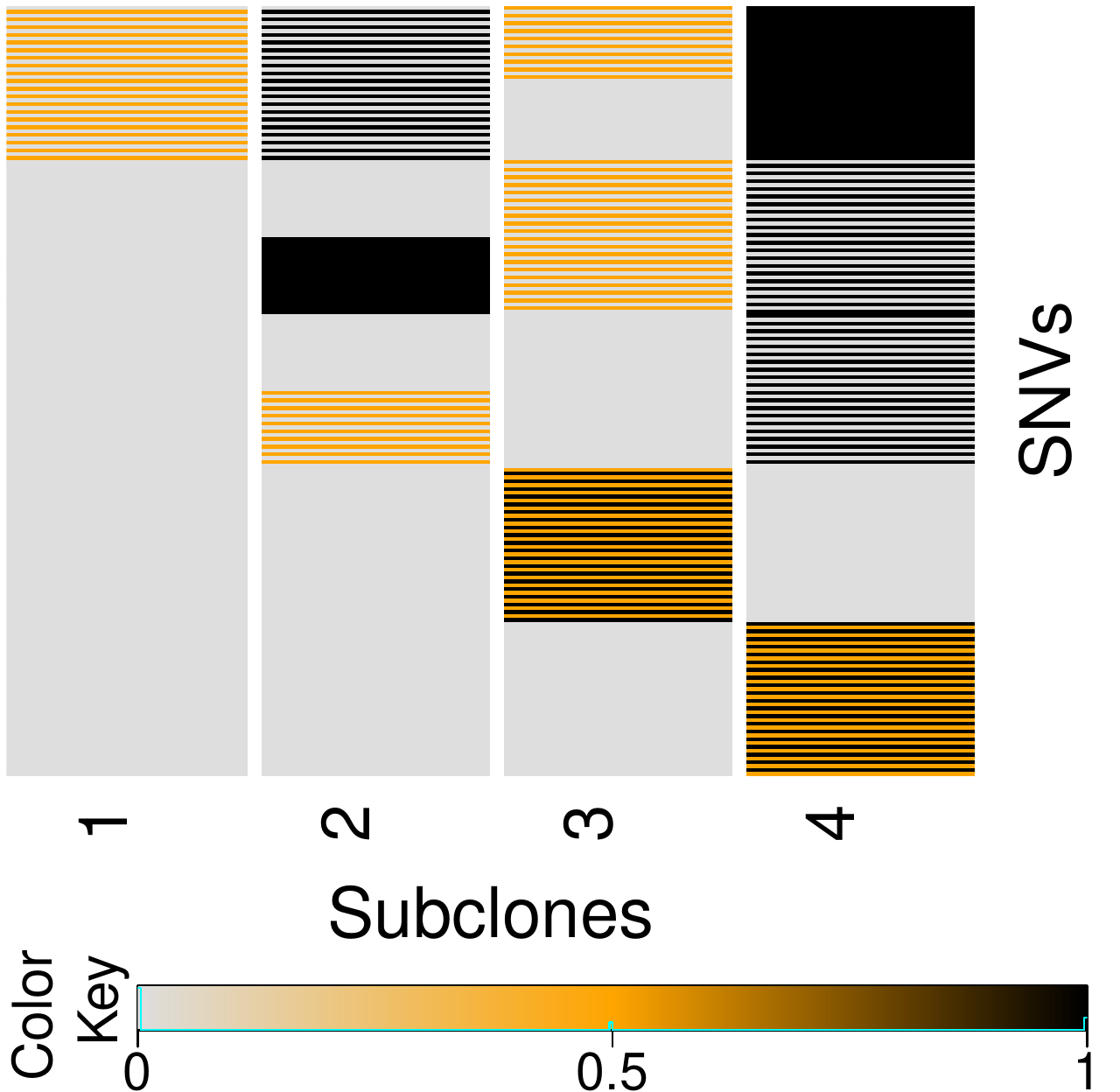}
\caption{$\bZ_{\BC}\true$}
\end{subfigure}
\begin{subfigure}[t]{.325\textwidth}
\centering
\includegraphics[width=\textwidth]{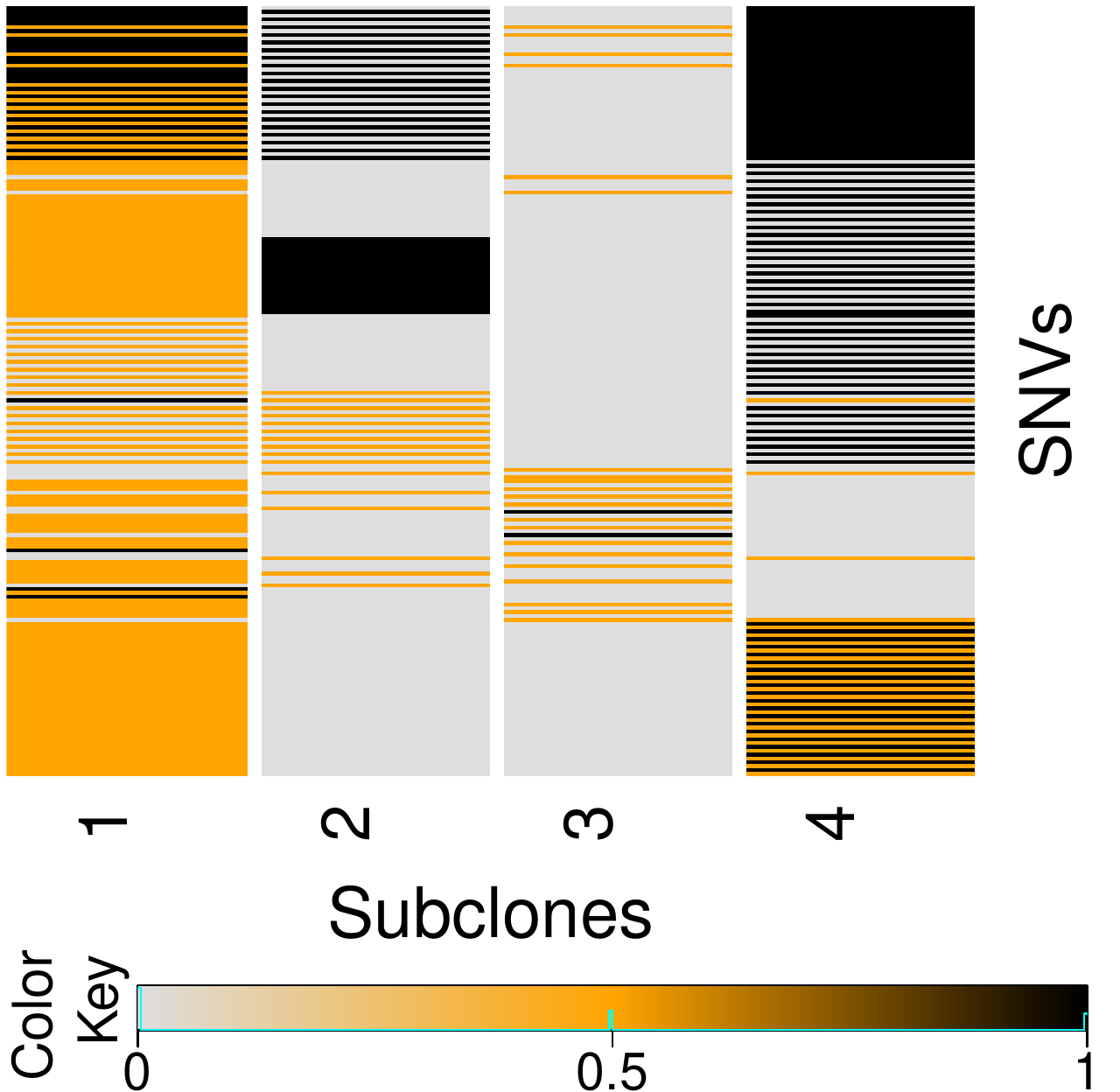}
\caption{$\Zhat_{\BC}$}
\end{subfigure}
\begin{subfigure}[t]{.325\textwidth}
\centering
\includegraphics[width=\textwidth]{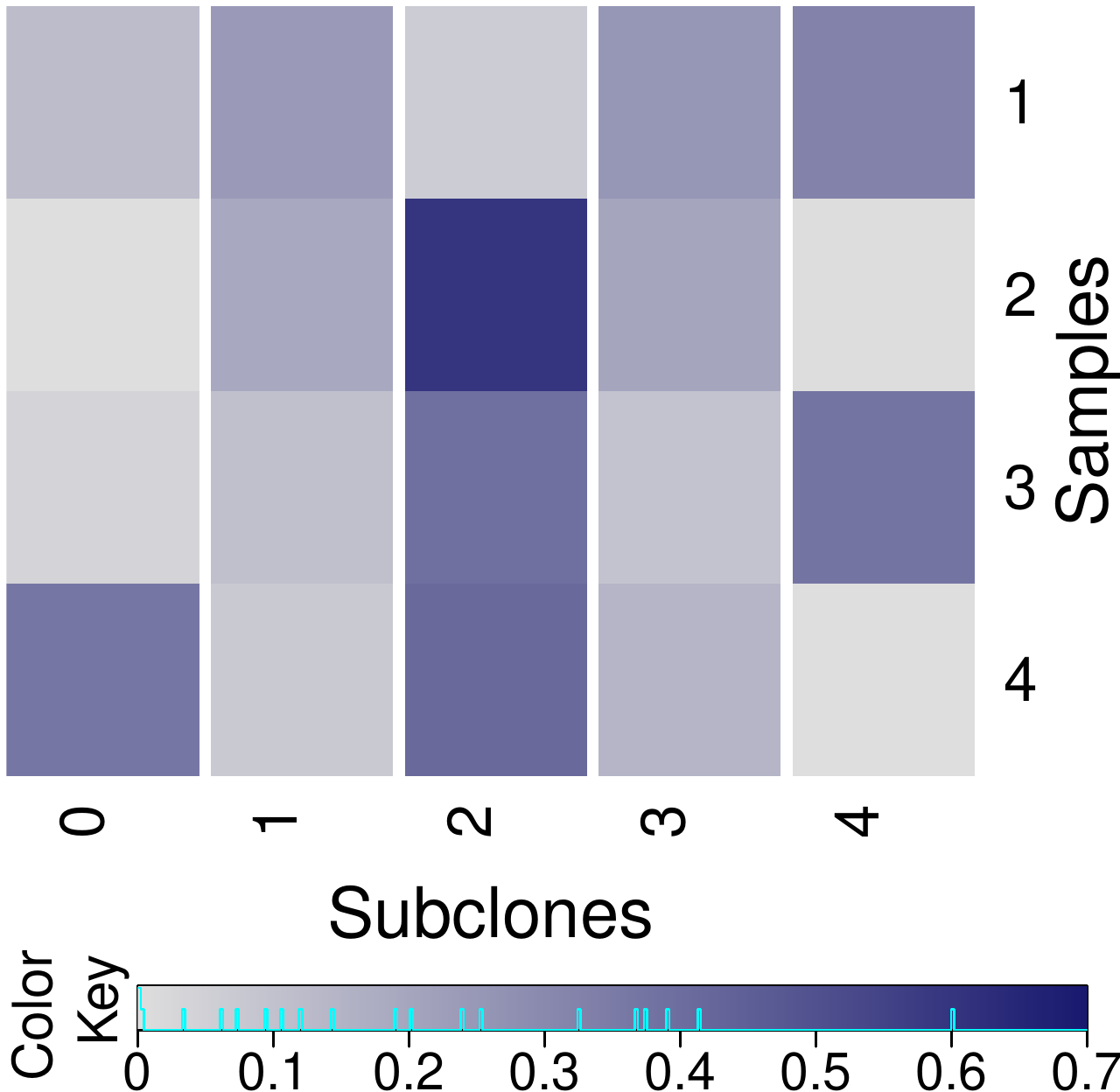}
\caption{$\what_{\BC}$}
\end{subfigure}
\begin{subfigure}[t]{.4\textwidth}
\centering
\includegraphics[width=\textwidth]{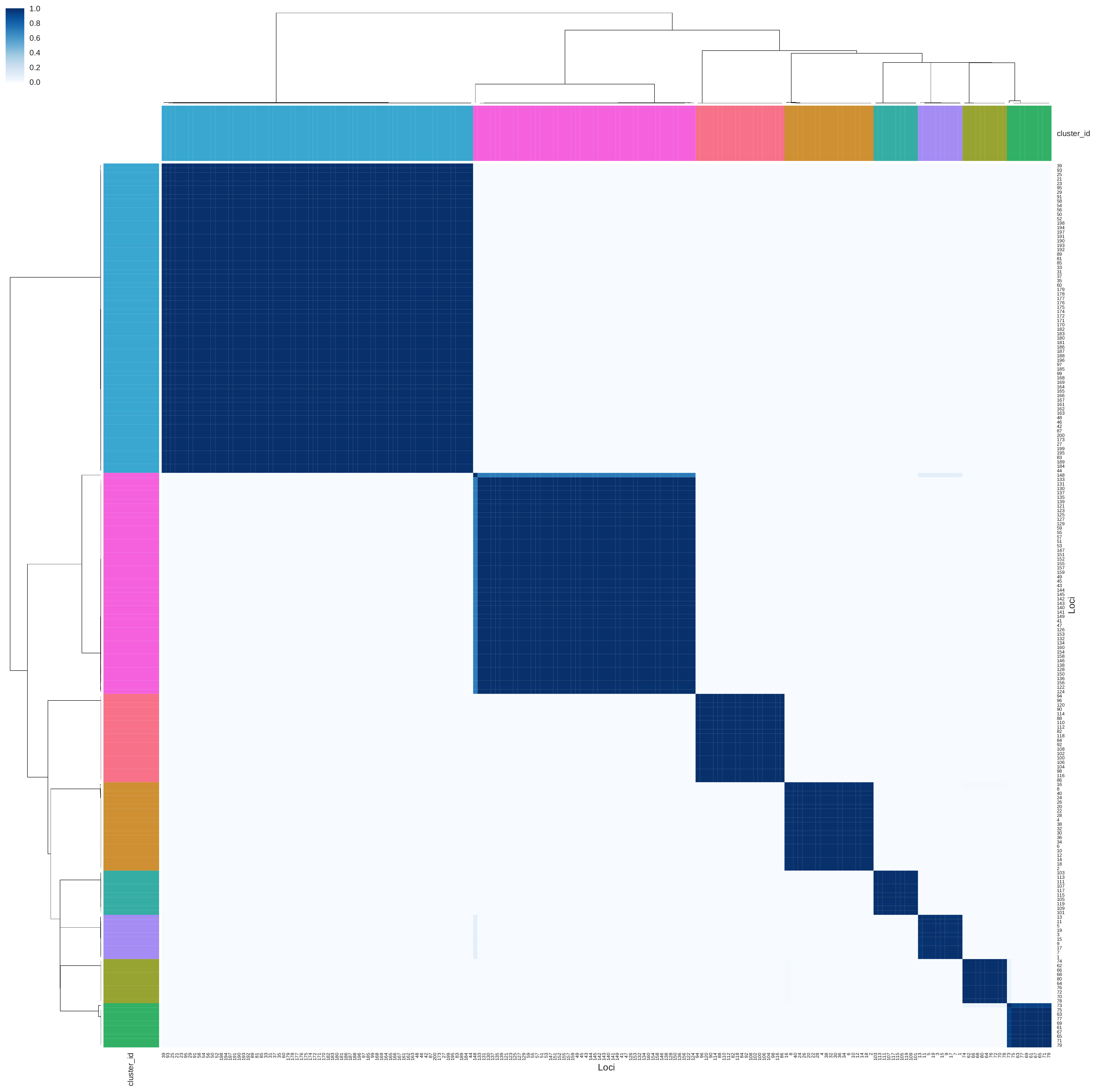}
\caption{Posterior similarity matrix}
\end{subfigure}
\begin{subfigure}[t]{.48\textwidth}
\centering
\includegraphics[width=\textwidth]{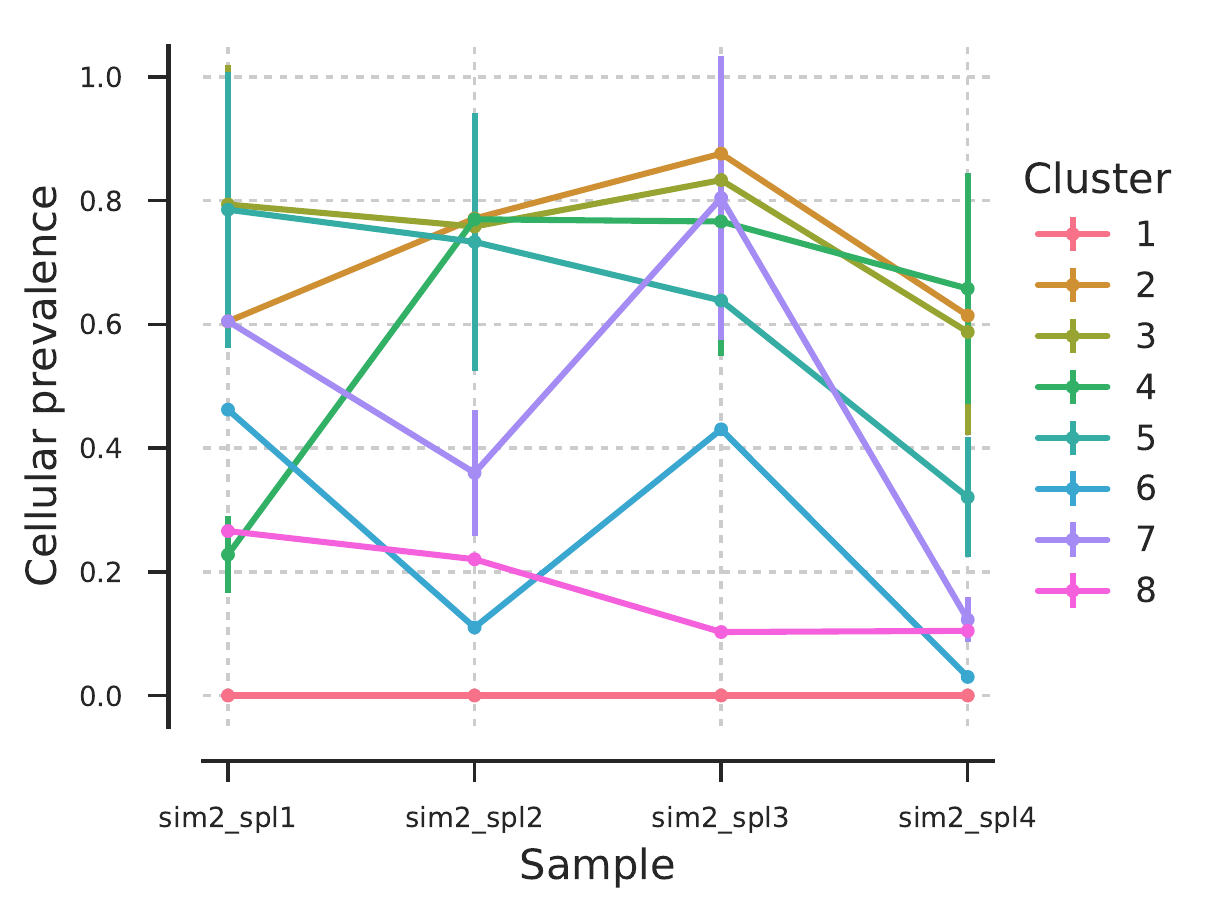}
\caption{Cellular prevalence}
\end{subfigure}
\end{center}
\caption{Simulation 2. Posterior inference under BayClone (a, b, c) and PyClone (d, e).}
\label{fig:sim2_BC}
\end{figure}

For comparison, we again fit the same simulated data with BayClone and
PyClone.  
\underline{BayClone} chooses the model with 4 subclones, which still
recovers the truth.  However, using only SNV data, BayClone can not
see the connection between adjacent SNVs, and inference fails to
recover $\bw_{\text{BC}}\true$ and therefore $\bZ_{\text{BC}}\true$,
even approximately.  
\underline{PyClone} infers 8 clusters for the 200 loci, which
reasonably recovers the truth. However, since the underlying subclone
structure is more complex, the PyClone cellular prevalence is not
directly comparable to PairClone outputs.

\subsection{Simulation 3}
\label{app:sim3}
\begin{figure}[h!]
\begin{center}
\begin{subfigure}[t]{.325\textwidth}
\centering
\includegraphics[width=\textwidth]{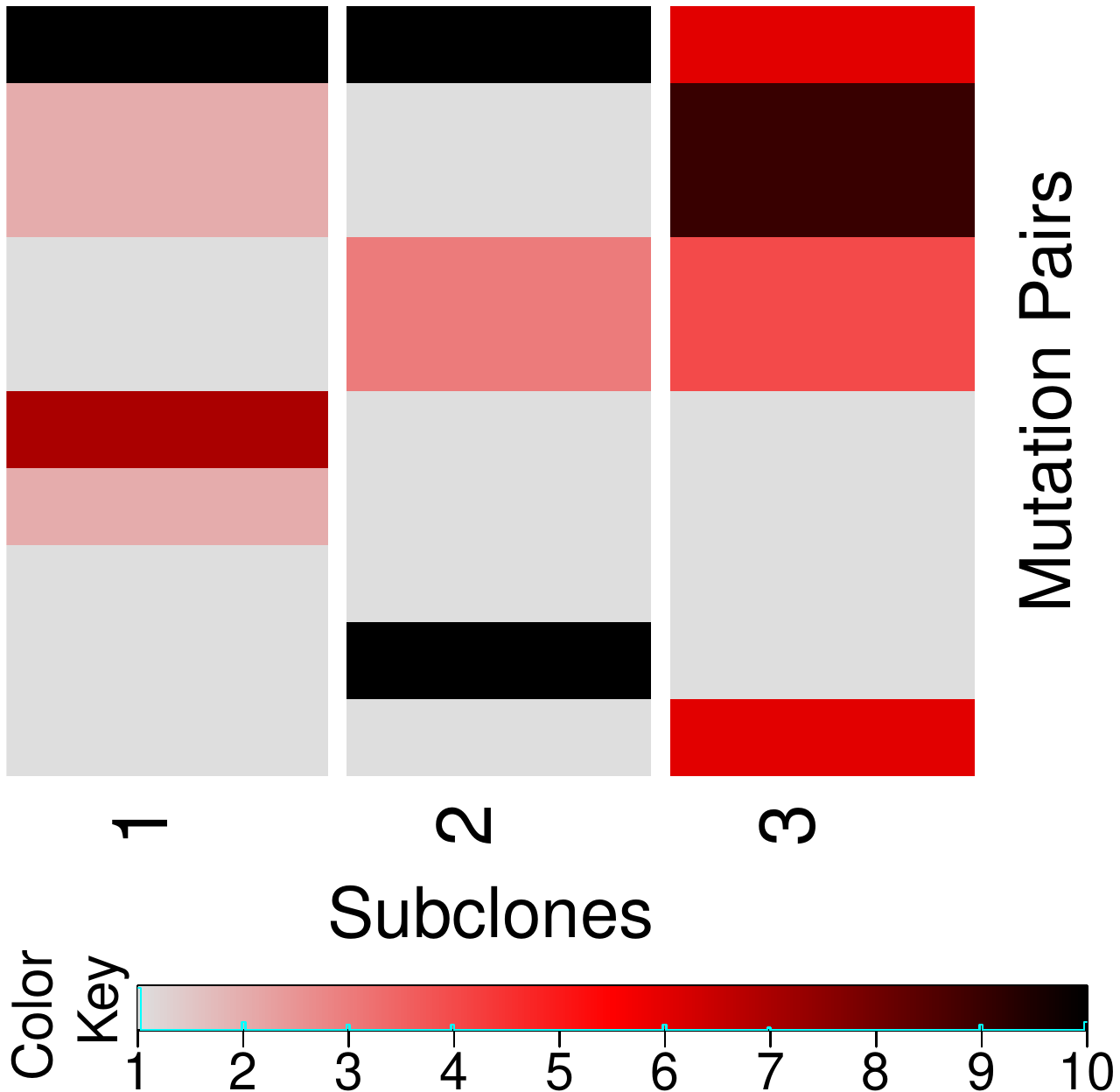}
\caption{$\bZ\true$}
\end{subfigure}
\begin{subfigure}[t]{.325\textwidth}
\centering
\includegraphics[width=\textwidth]{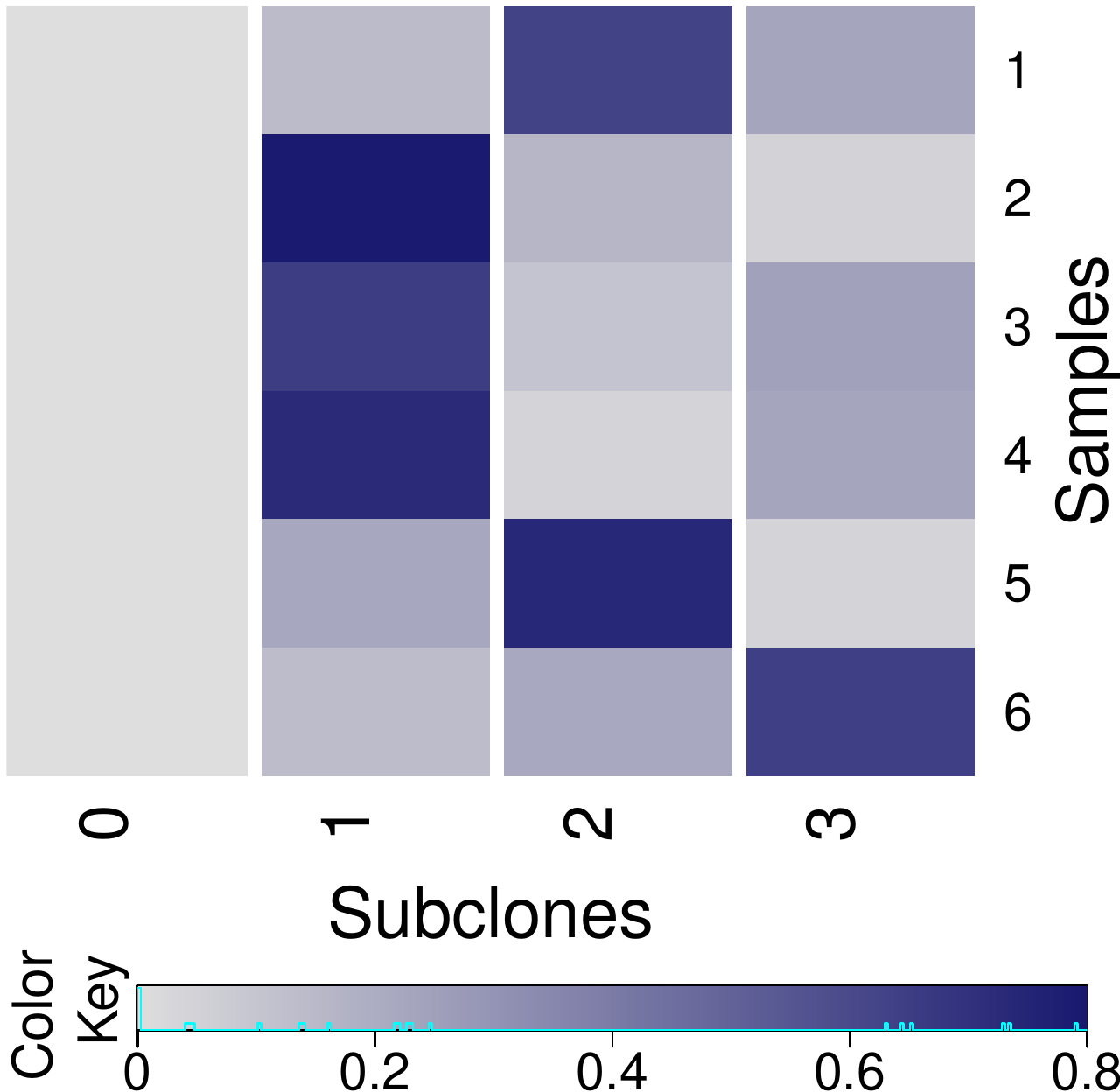}
\caption{$\bw\true$}
\end{subfigure}
\begin{subfigure}[t]{.325\textwidth}
\centering
\includegraphics[width=\textwidth]{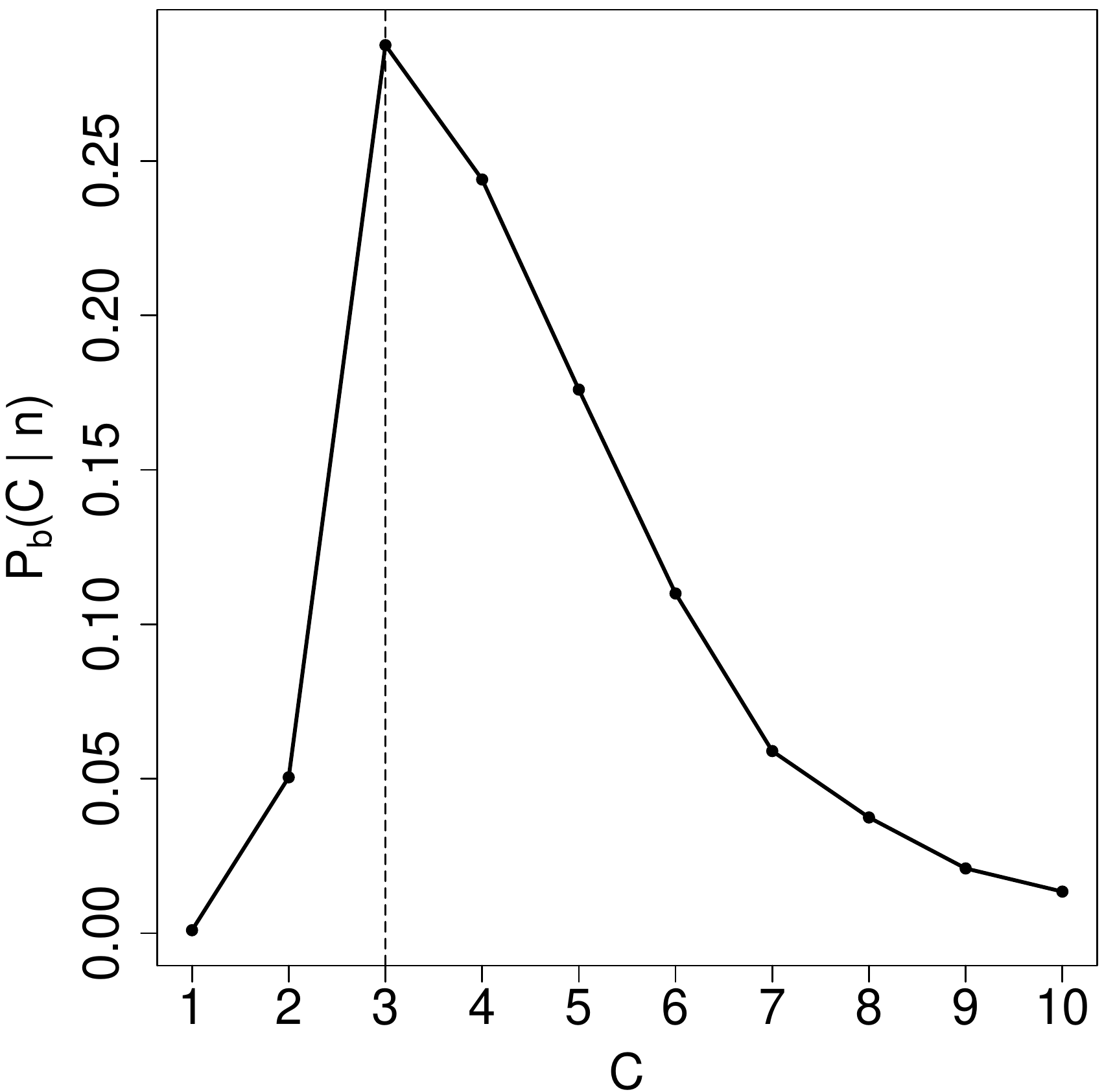}
\caption{$p_b(C \mid \bn'')$}
\end{subfigure}
\begin{subfigure}[t]{.325\textwidth}
\centering
\includegraphics[width=\textwidth]{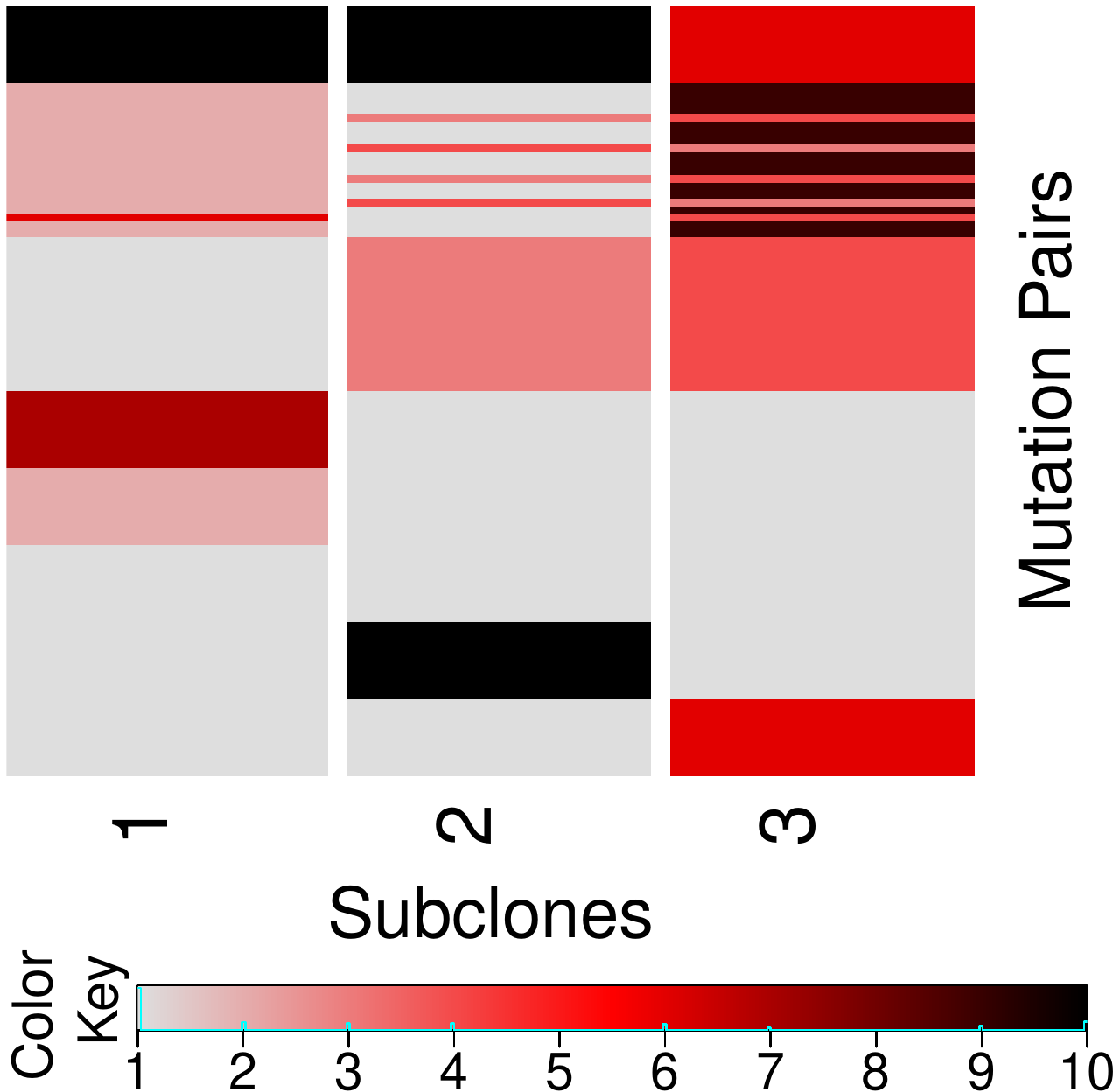}
\caption{$\Zhat$}
\end{subfigure}
\begin{subfigure}[t]{.325\textwidth}
\centering
\includegraphics[width=\textwidth]{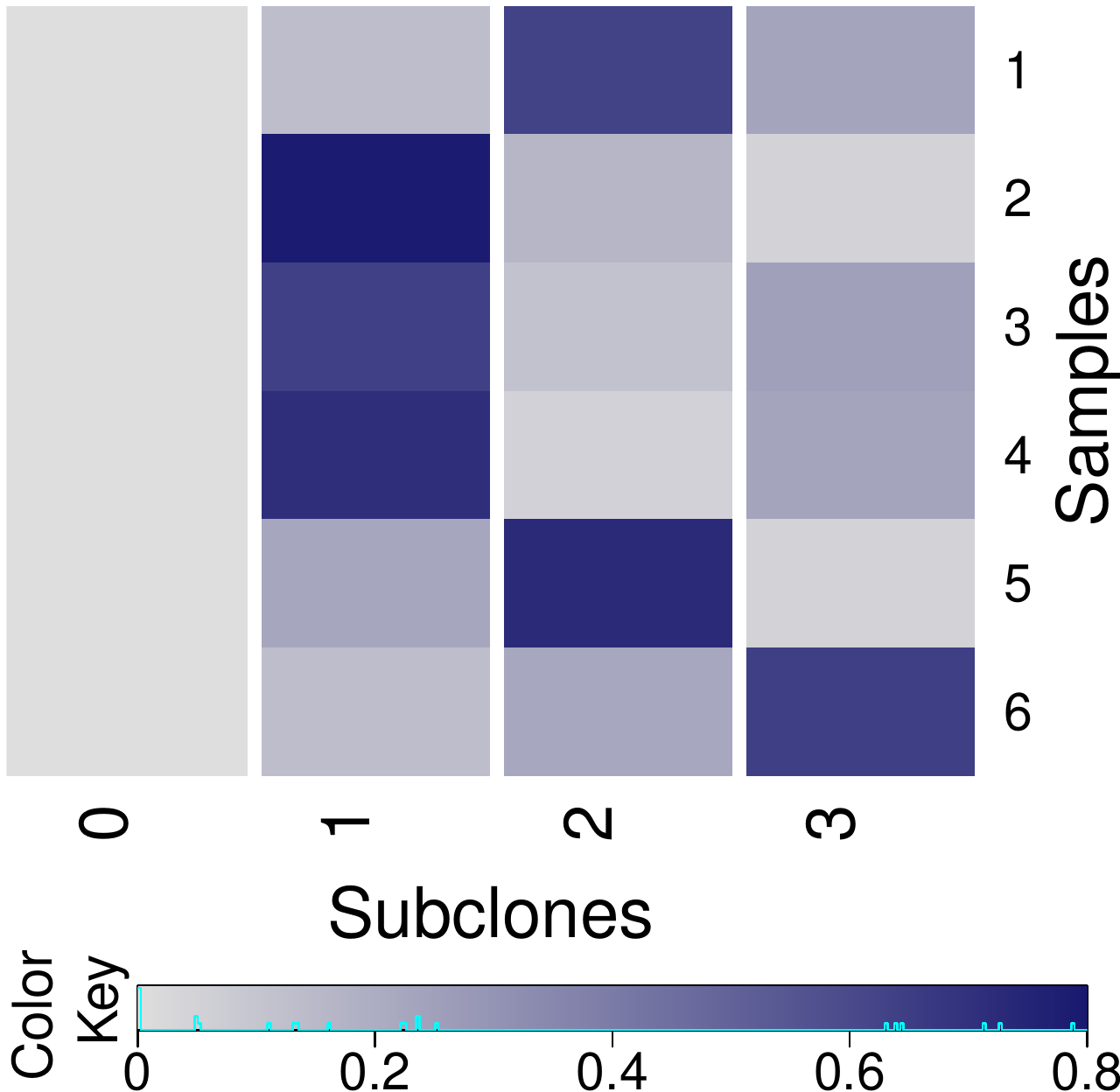}
\caption{$\what$}
\end{subfigure}
\begin{subfigure}[t]{.325\textwidth}
\centering
\includegraphics[width=\textwidth]{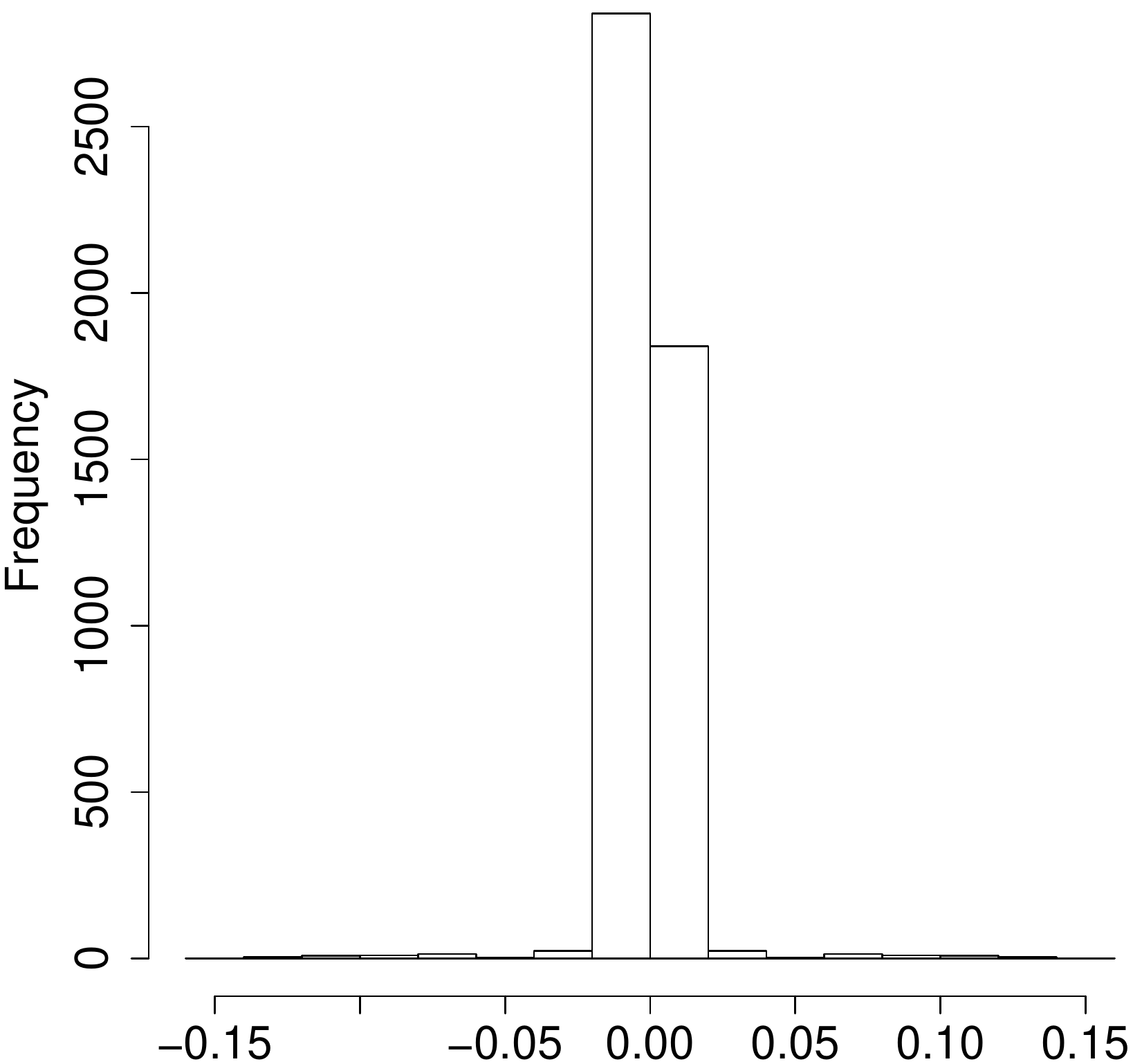}
\caption{Histogram of $(\ptkghat - p_{tkg}\true)$}
\end{subfigure}
\end{center}
\caption{Simulation 3. Simulation truth $\bZ\true$ and $\bw\true$ (a, b), and posterior inference under PairClone (c, d, e, f).}
\label{app:figsim3}
\end{figure}

In the last simulation we use  
$T = 6$ samples with $C^{\text{TRUE}} = 3$ and latent subclones.  
We still consider $K = 100$ mutation pairs.
The subclone matrix $\bZ\true$ is shown in Figure~\ref{app:figsim3}(a). For
each sample $t$, we generate the subclone proportions from
$\bw_t\true \sim \Dir(0.01, \sigma(14, 6, 3))$, where $ \sigma (14,
6, 3)$ is a random permutation of $(14, 6, 3)$. The proportions
$\bw\true$ are shown in
Figure~\ref{app:figsim3}(b).  The parameters $\brho\true$ and
$N_{tk}$ are generated using the same approach as before, and we use the same $v_{tk2}$ and $v_{tk3}$ as in Simulation 2.
Finally, we calculate $\{p_{tkg}\true\}$ and generate read counts
$n_{tkg}$ from equation \eqref{eq:multi} similar to simulation 1.

We fit the model with the same hyperparameter and the same MCMC
tuning parameters as in simulation 1.  We now use a smaller test
sample size, i.e., a smaller fraction 
$b$   in the transdimensional MCMC.  
See Section \ref{app:sec:calib} for a discussion.

Figure \ref{app:figsim3}(c) shows $p_b(C \mid \bn'')$, with the
posterior mode $\Chat = 3$ recovering the truth.
Figures~\ref{app:figsim3}(d, e) show $\Zhat$ and $\what$.
Comparing with panels (a) and (b) we can see an almost perfect
recovery of the truth. 
Figure~\ref{app:figsim3}(f) shows a histogram of
the residuals $(\ptkghat - p_{tkg}\true)$. The plot indicates a good model fit.

\begin{figure}[h!]
\begin{center}
\begin{subfigure}[t]{.325\textwidth}
\centering
\includegraphics[width=\textwidth]{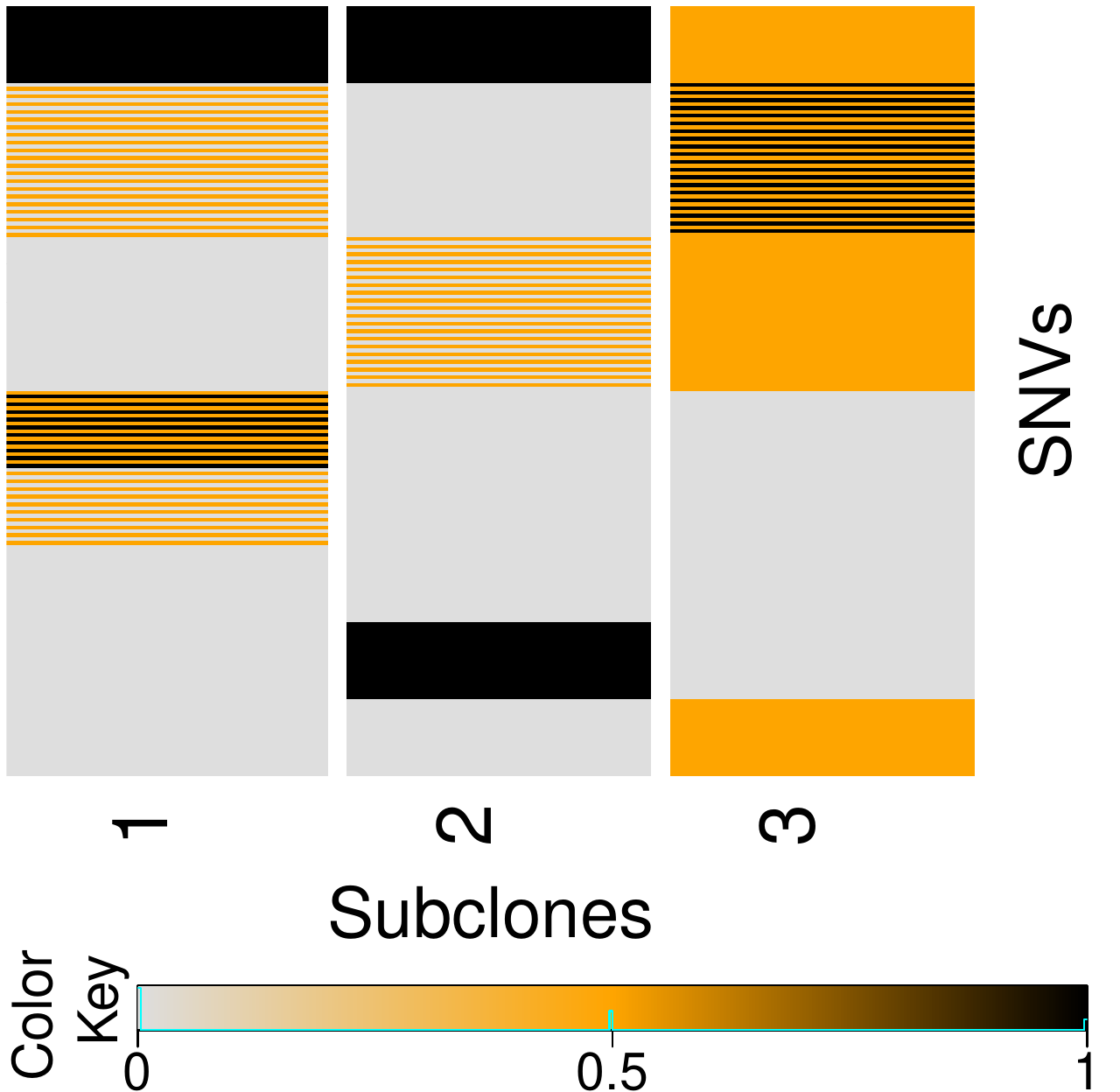}
\caption{$\bZ_{\BC}\true$}
\end{subfigure}
\begin{subfigure}[t]{.325\textwidth}
\centering
\includegraphics[width=\textwidth]{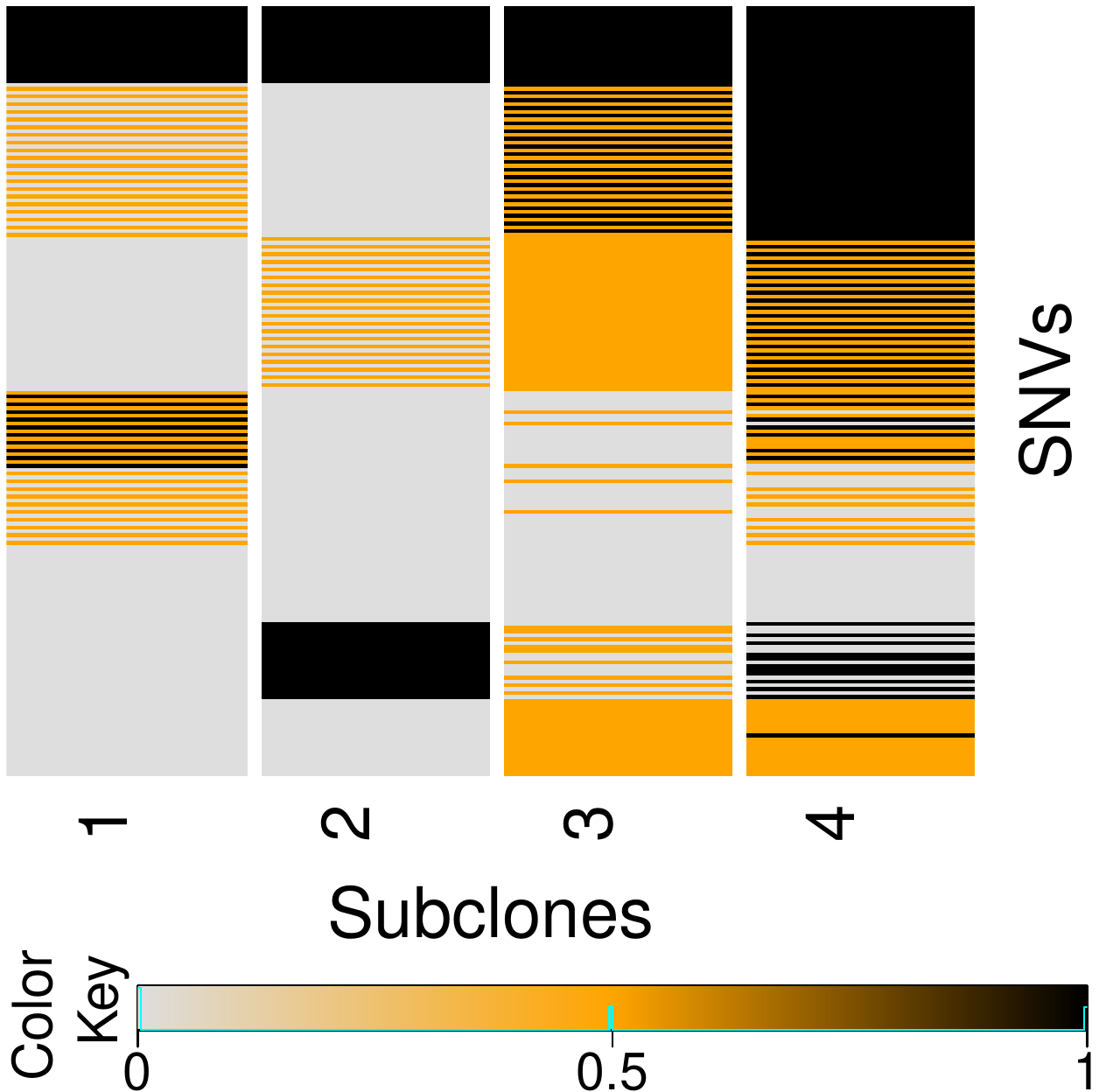}
\caption{$\Zhat_{\BC}$}
\end{subfigure}
\begin{subfigure}[t]{.325\textwidth}
\centering
\includegraphics[width=\textwidth]{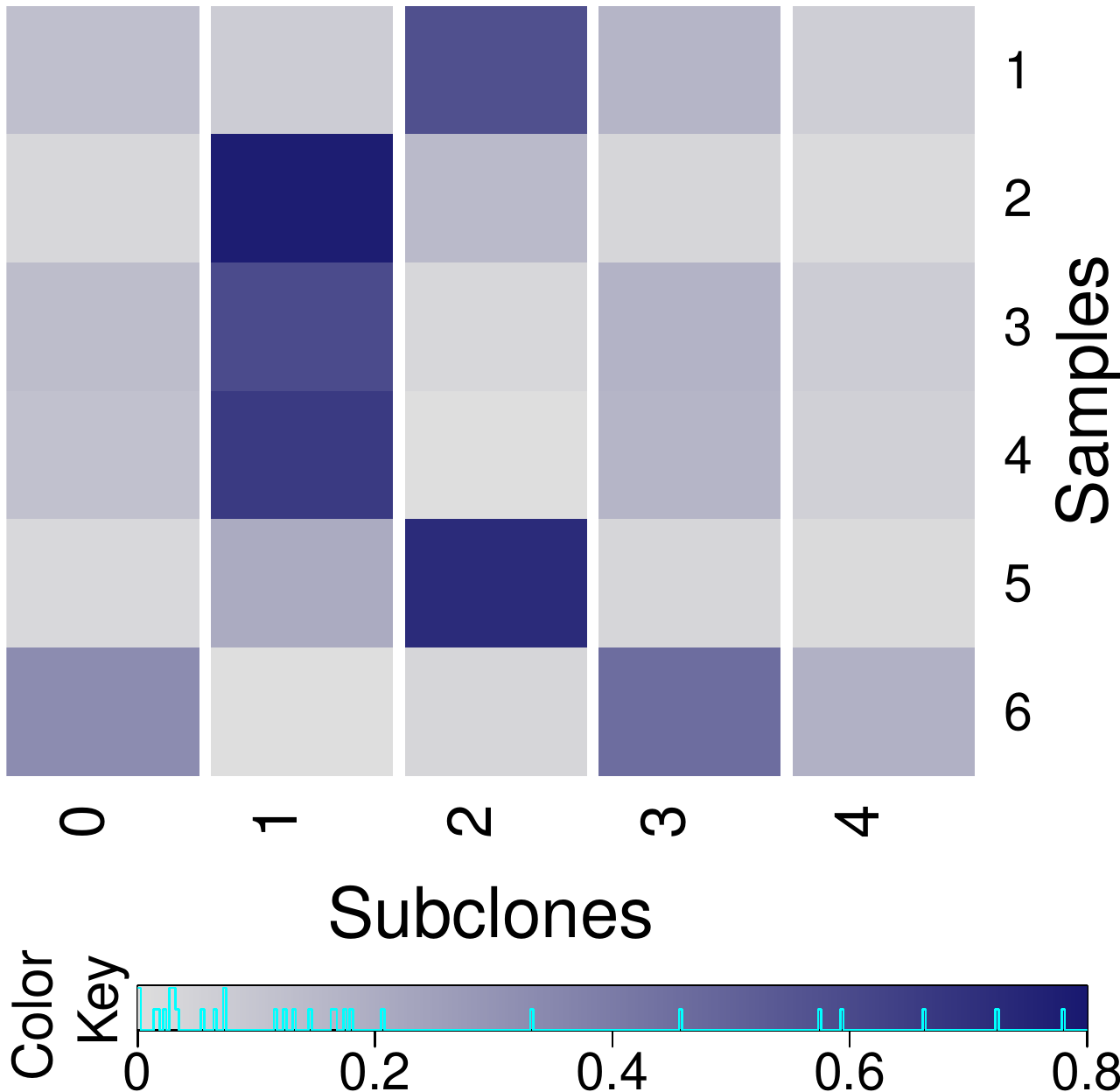}
\caption{$\what_{\BC}$}
\end{subfigure}
\begin{subfigure}[t]{.4\textwidth}
\centering
\includegraphics[width=\textwidth]{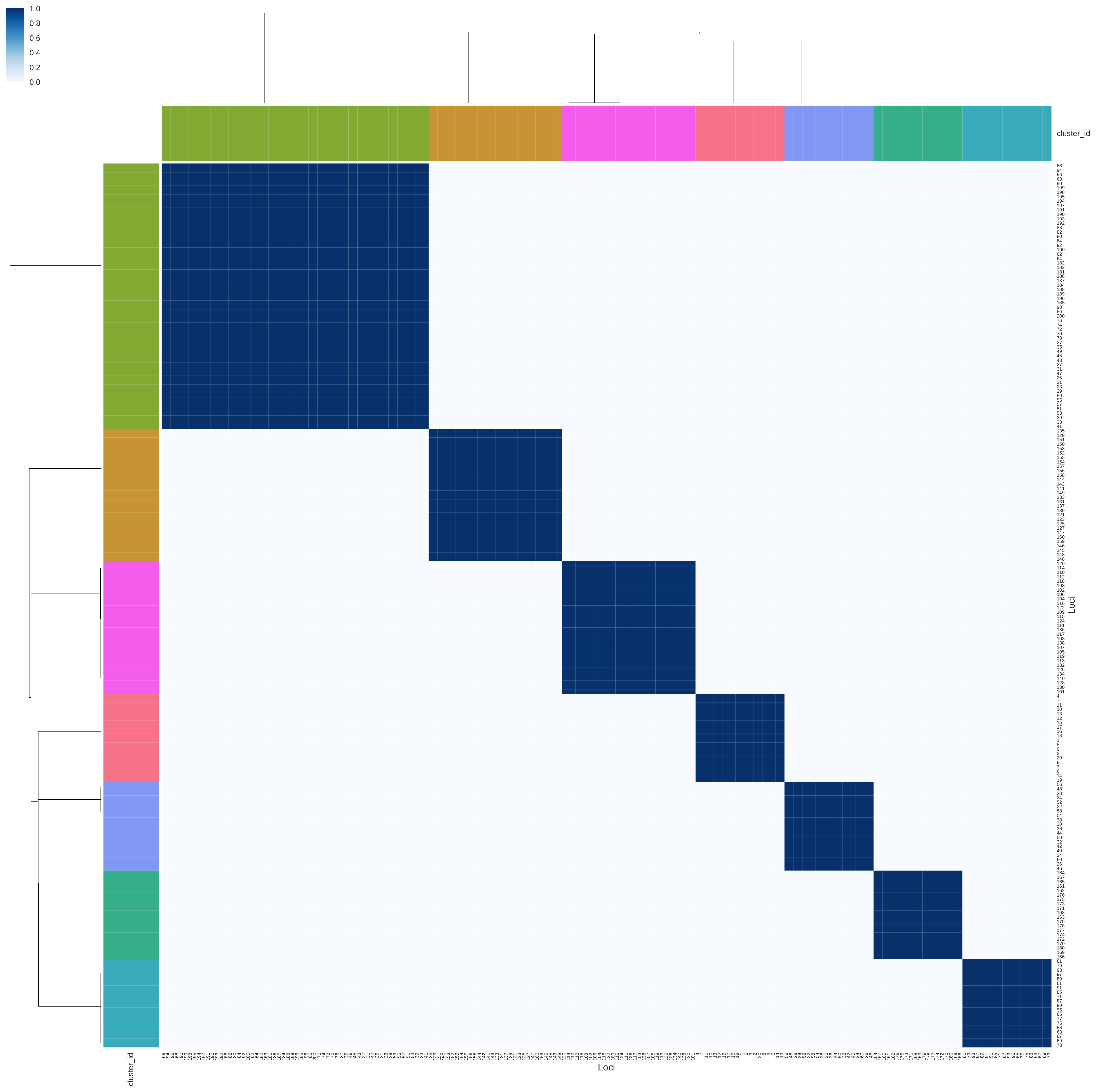}
\caption{Posterior similarity matrix}
\end{subfigure}
\begin{subfigure}[t]{.48\textwidth}
\centering
\includegraphics[width=\textwidth]{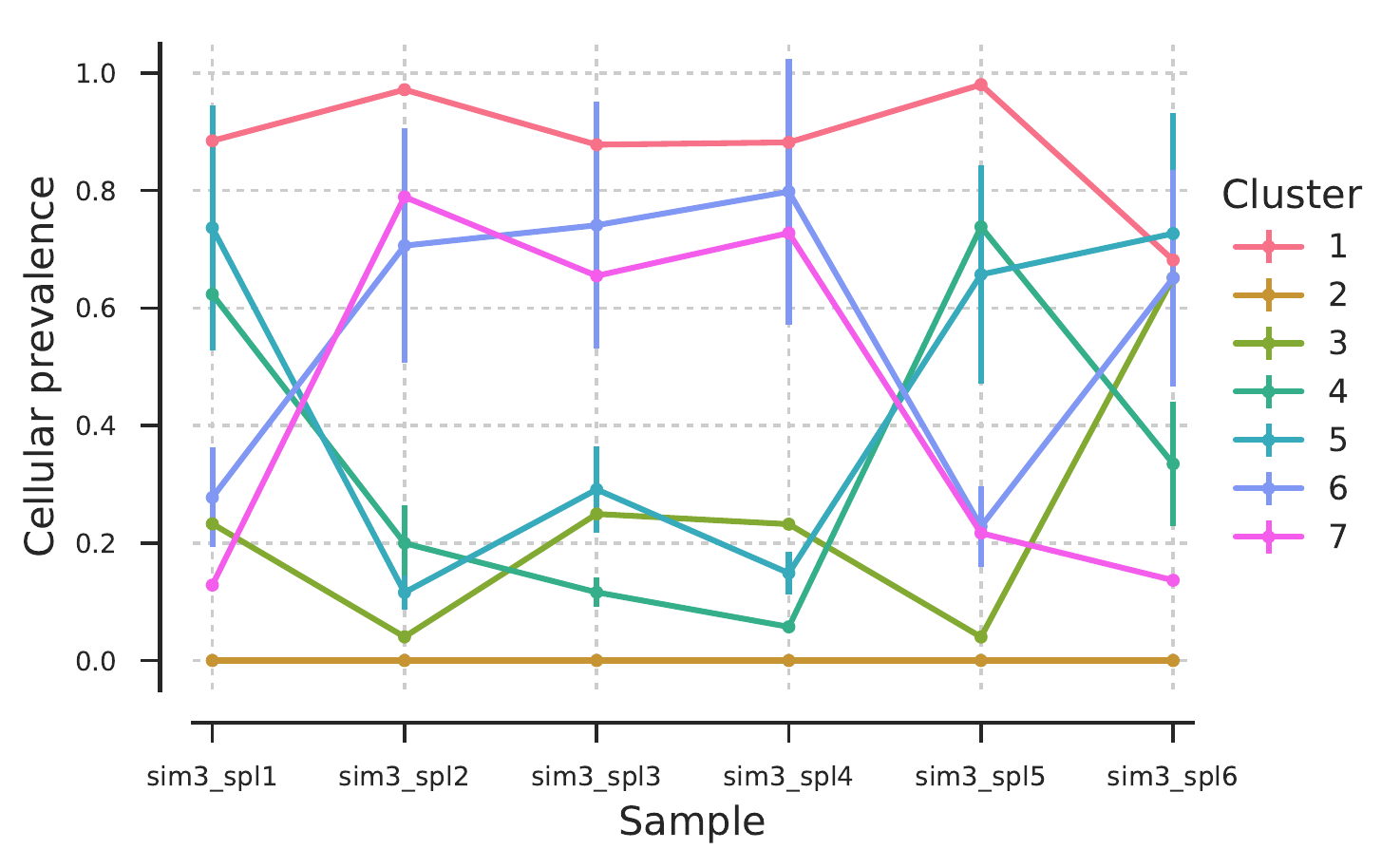}
\caption{Cellular prevalence}
\end{subfigure}
\end{center}
\caption{Simulation 3. Posterior inference under BayClone (a, b, c) and PyClone (d, e).}
\label{fig:sim3_BC}
\end{figure}

We again compare with inference under BayClone and PyClone.
In this case, \underline{BayClone} chooses the model with 4 subclones,
failing to recover the truth.
\underline{PyClone} infers 7 clusters for the 200 loci, which
reasonably recovers the truth, but the result is still not directly
comparable.

\section{PairClone Extensions}
\label{sec:pipeline}

\subsection{Incorporating Marginal Read Counts}
\label{sec:mrc}
Most somatic mutations are not part of the paired reads that we
use in PairClone. 
We refer to these single mutations as SNVs (single
nucleotide variants) and consider the following simple extension to
incorporate marginal counts for SNVs in PairClone.
We introduce a new $S \times C$ matrix $\bZ^{S}$ to represent the
genotype of the $C$ subclones for these additional SNVs.
To avoid confusion, we denote the earlier $K \times C$ subclone matrix
by $\bZ^{P}$ in this section.
The $(s,c)$ element of $\bZ^S$ reports the  genotype of SNV $s$ in
subclone $c$, with $z^{S}_{sc} \in \{ 0, 0.5, 1\}$ denoting
homozygous wild-type ($0$), heterozygous variant
($0.5$), and homozygous variant ($1$), respectively.
The $c$-th column of $\bZ^{P}$ and $\bZ^{S}$ together define 
subclone $c$. We continue to assume copy number 
neutrality in all SNVs and mutation pairs
(we discuss an extension to incorporating subclonal copy number
variations in the next subsection).
The marginal read counts are easiest
incorporated in the PairClone model by recording them as right (or left)
missing reads (as described in Section \ref{sec:splmodel})
for hypothetical pairs, $k=K+1,\ldots,k+S$. 
Let $\tN_{ts}$ and $\tn_{ts}$ denote
the total count and the number of reads bearing a variant
allele, respectively, for SNV $s$ in sample $t$.
Treating $s$ as a mutation pair $k=K+s$ with missing second read, we
record $n_{tk8}=\tn_{ts}$, $n_{tk1}=\ldots=n_{tk7}=0$ and
$N_{tk}=\tN_{ts}$. 
We then proceed as before, now with $K+S$ mutation pairs. 
Inference reports an augmented $(K+S) \times C$ subclone matrix
$\tilde{\bZ}^P$. We record the first $K$ rows of
$\tilde{\bZ}^P$ as $\bZ^P$, and transform the remaining $S$ rows
to $\bZ^S$ by only recording the genotypes of the observed loci.

\begin{figure}[h!]
\begin{center}
\begin{subfigure}[t]{.65\textwidth}
\centering
Simulation truth
\end{subfigure}
\begin{subfigure}[t]{.325\textwidth}
\centering
~~\vspace{3mm}
\end{subfigure} \\
\begin{subfigure}[t]{.325\textwidth}
\centering
\includegraphics[width=\textwidth]{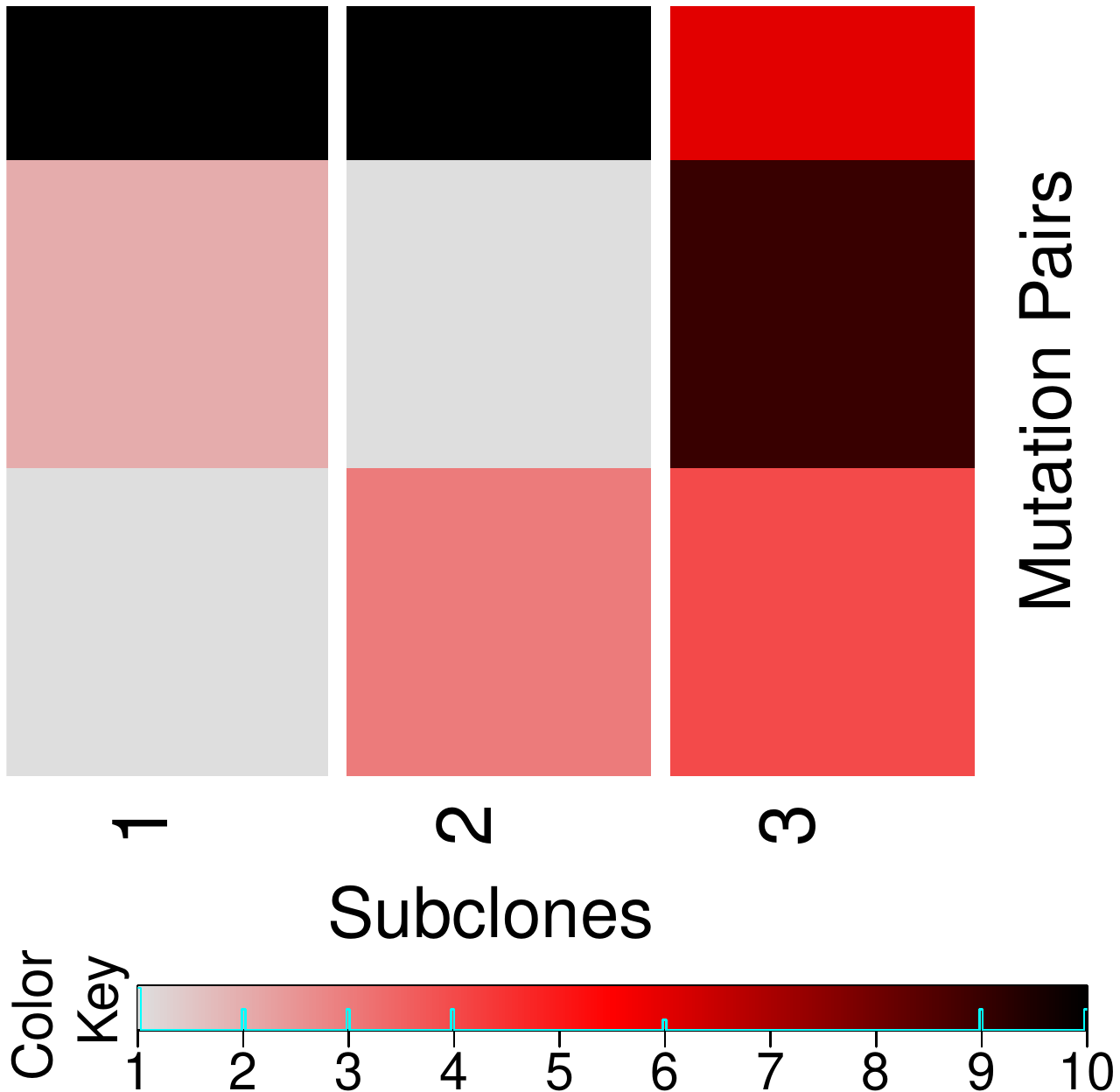}
\caption{$\bZ^{P, \text{TRUE}}$}		
\end{subfigure}
\begin{subfigure}[t]{.325\textwidth}
\centering
\includegraphics[width=\textwidth]{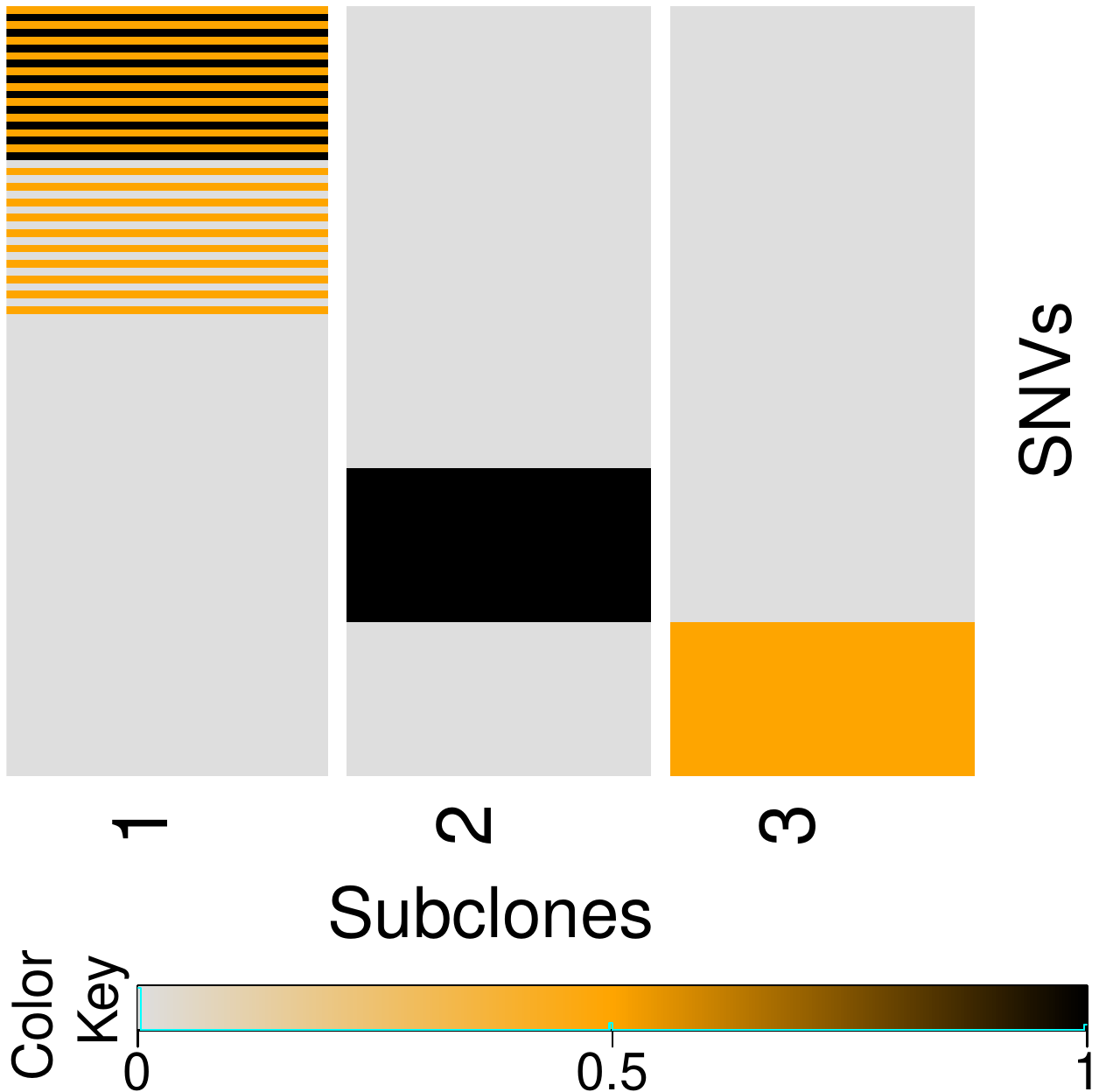}
\caption{$\bZ^{S, \text{TRUE}}$}		
\end{subfigure}
\begin{subfigure}[t]{.325\textwidth}
\centering
\includegraphics[width=\textwidth]{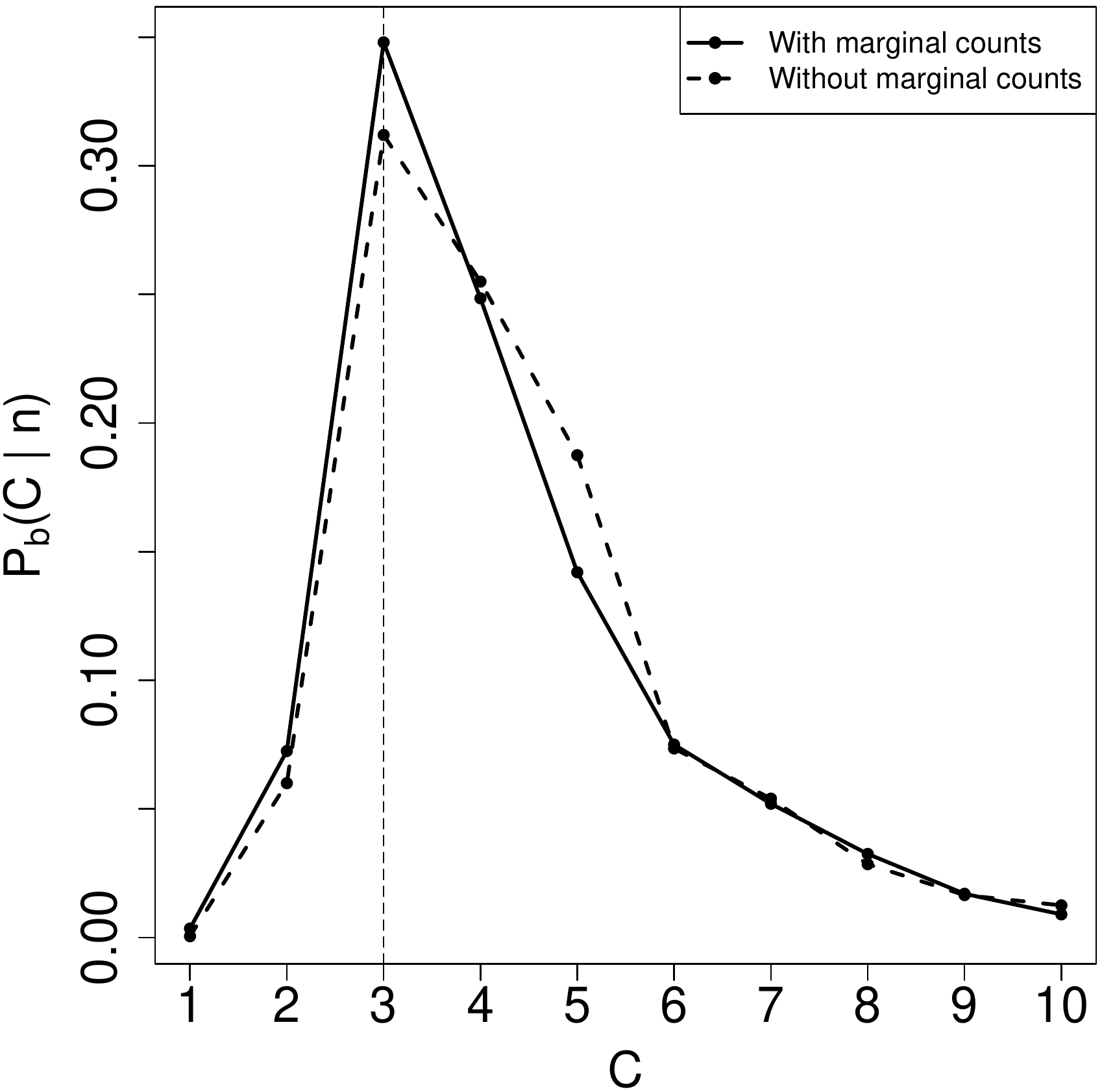}
\caption{$p_b(C \mid \bm n'')$}	
\vspace{8mm}
\end{subfigure}
\begin{subfigure}[t]{.65\textwidth}
\centering
Pairs and SNVs
\end{subfigure}
\begin{subfigure}[t]{.325\textwidth}
\centering
Pairs only \vspace{3mm}
\end{subfigure}  \\
\begin{subfigure}[t]{.325\textwidth}
\centering
\includegraphics[width=\textwidth]{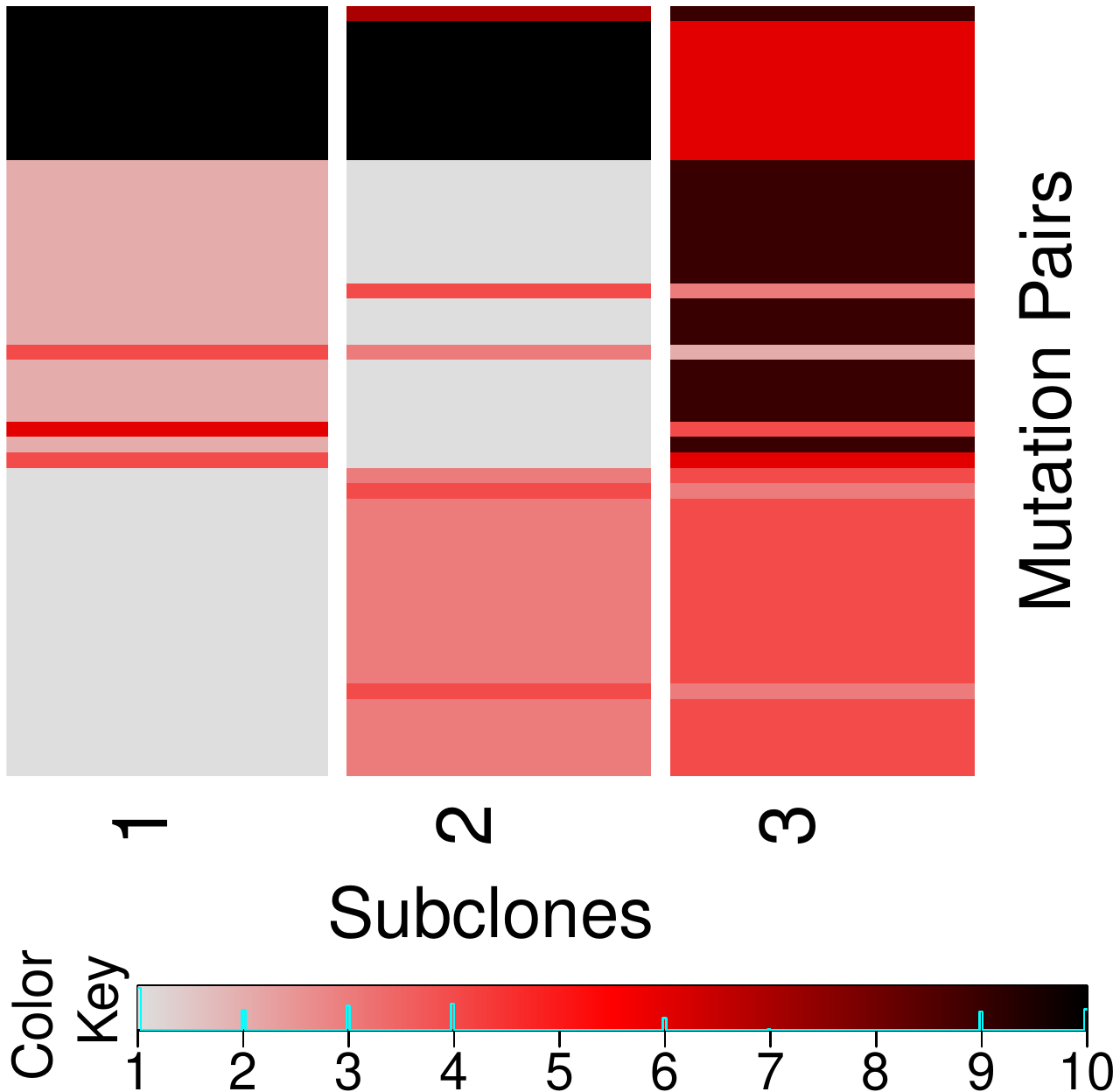}
\caption{$\Zhat^{P}$}		
\end{subfigure}
\begin{subfigure}[t]{.325\textwidth}
\centering
\includegraphics[width=\textwidth]{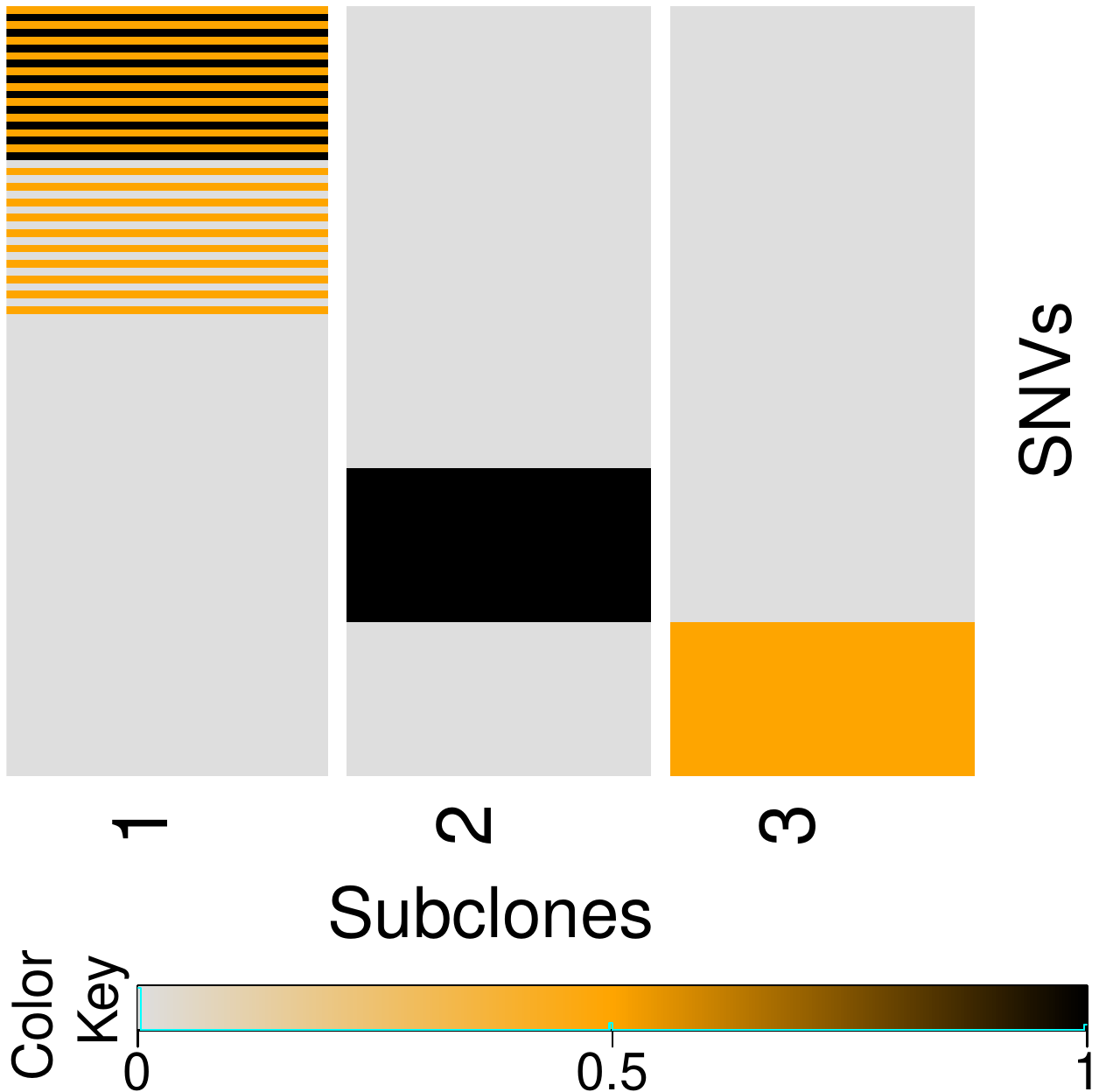}
\caption{$\Zhat^{S}$}		
\end{subfigure}
\begin{subfigure}[t]{.325\textwidth}
\centering
\includegraphics[width=\textwidth]{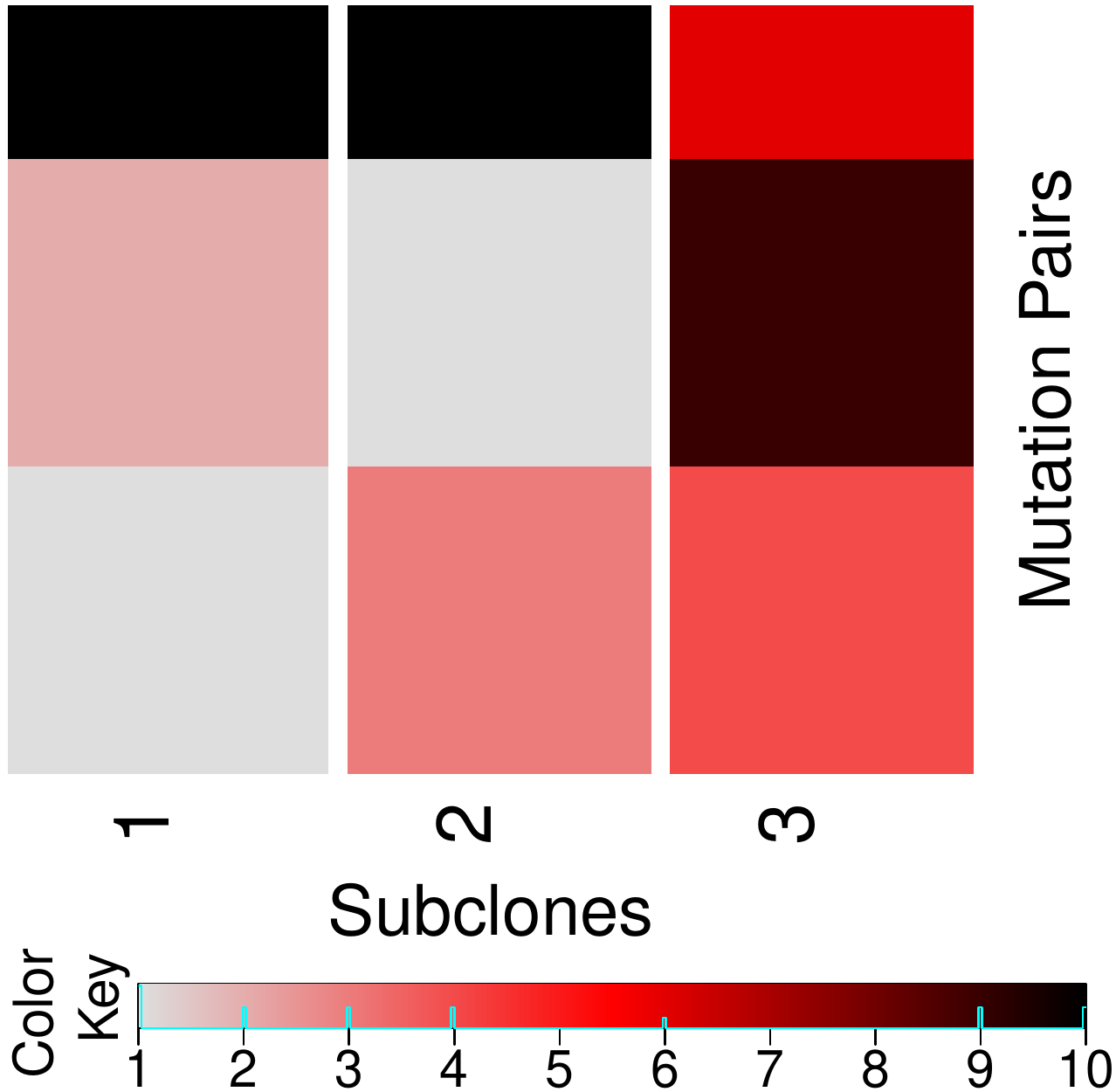}
\caption{$\Zhat_{\text{wom}}^{P}$}		
\end{subfigure}
\end{center}
\caption{Summary of simulation results using additional marginal read counts. Simulation truth $\bZ^{P, \text{TRUE}}$ and $\bZ^{S, \text{TRUE}}$ (a, b), posterior inference with marginal read counts incorporated (c, d, e), and posterior inference without marginal read counts (c, f).}
\label{fig:sim_extension}
\end{figure}

We evaluate the proposed modeling approach
with a simulation study. The simulation setting is the same as
simulation 3 in Section \ref{sec:simulation},   except that we discard
the phasing information of mutation pairs $51-100$ and only record
their marginal read counts. 
Figure \ref{fig:sim_extension}(a)--(f) summarizes the simulation results.
Panels (a, b) show the simulation truth for the mutation pairs and SNVs, respectively. 
Panel (c) shows the posterior $p(C \mid \bn'')$ and panels (d, e) show the estimated 
genotypes $\Zhat^P$ and $\Zhat^S$. Inference for 
the weights $w_{tc}$ recovers the
simulation truth (not shown). 
The result compares favorably to inference under BayClone (Figure
\ref{fig:sim3_BC}(b)), due to the additional phasing information for the first 50 mutation pairs.

For a direct evaluation of the information in the additional marginal
counts we also evaluate posterior inference with only the first 50
mutation pairs, shown in Figure \ref{fig:sim_extension} (c, f). 
Comparison with Figure \ref{fig:sim_extension} (c, d) shows that the additional
marginal counts do not noticeably improve inference on tumor
heterogeneity.

\subsection{Incorporating Tumor Purity}
\label{sec:purity}
Usually, tumor samples are not pure in the sense that they contain
certain proportions of normal cells. Tumor purity refers to the
fraction of tumor cells in a tumor sample. To explicitly model tumor
purity, we introduce a normal subclone, the proportion of which in
sample $t$ is denoted by $\wts$, $t = 1, \ldots, T$. The normal
subclone does not possess any mutation  (since we only
consider somatic mutations). 
The tumor purity for sample $t$ is thus $(1 - \wts)$. The normal
subclone is denoted by $\bz_*$, with $\bz_{k*} = \bz^{(1)}$ for all
$k$. 
The remaining subclones are still denoted by
$\bz_c$, $c = 1, \ldots, C$, with proportion $w_{tc}$ 
in sample $t$, and $\sum_{c = 0}^C w_{tc} + \wts = 1$.
 
The probability model needs to be slightly modified to accommodate the
normal subclone. The sampling model remains unchanged as
\eqref{eq:multi}. Same for the prior models for $\bZ$, $\brho$ and $C$.
We only change the construction of $\tp_{tkg}$ and 
$p(\bw)$ as follows. 
With a new normal subclone, the probability of observing a short read
$\bh_g$ becomes 
$ 
   \tp_{tkg} = \sum_{c = 1}^C w_{tc}\,A(\bh_g, \bz_{kc}) +  \wts \,
   A(\bh_g, \bz^{(1)}) +  w_{t0} \, \rho_g,
$
based on the same generative model described in Section \ref{sec:prior}.
Let $\tilde{w}_{tc} = w_{tc} / (1 - \wts)$.
We use a Beta-Dirichlet prior,
$\wts \stackrel{iid}{\sim} \Be(d_1^{*}, d_2^{*})$,
and $\tilde{\bw}_{t} \stackrel{iid}{\sim} \Dir(d_0, d,\ldots, d).$ 
An informative prior for $\wts$ could be based on
an estimate from a purity caller, for
example, \cite{van2010allele} or \cite{carter2012absolute}.

\begin{figure}[h!]
\begin{center}
\begin{subfigure}[t]{.325\textwidth}
\centering
\includegraphics[width=\textwidth]{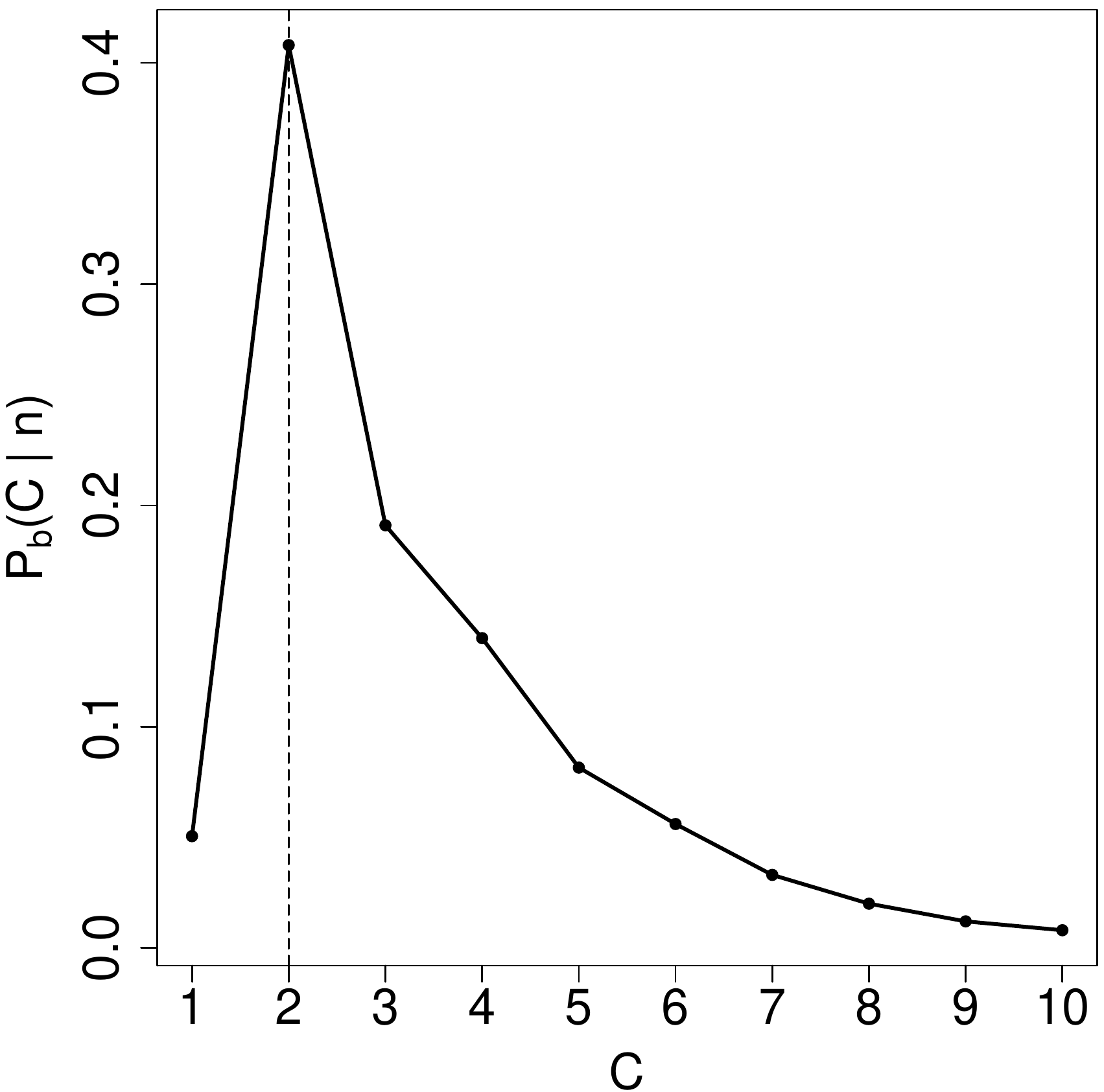}
\caption{$p_b(C \mid \bm n'')$}		
\end{subfigure}
\begin{subfigure}[t]{.325\textwidth}
\centering
\includegraphics[width=\textwidth]{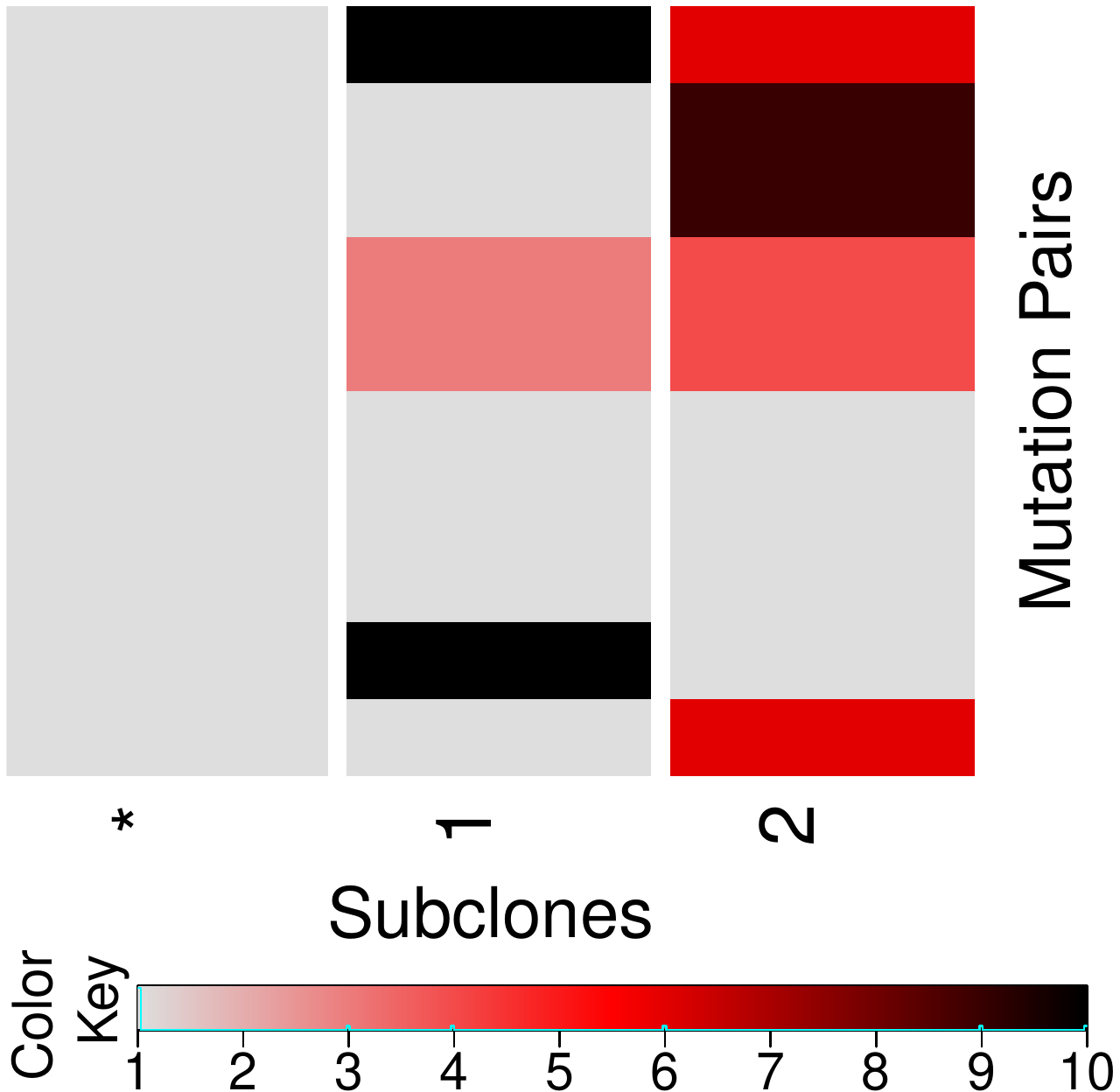}
\caption{$\Zhat$}		
\end{subfigure}
\begin{subfigure}[t]{.325\textwidth}
\centering
\includegraphics[width=\textwidth]{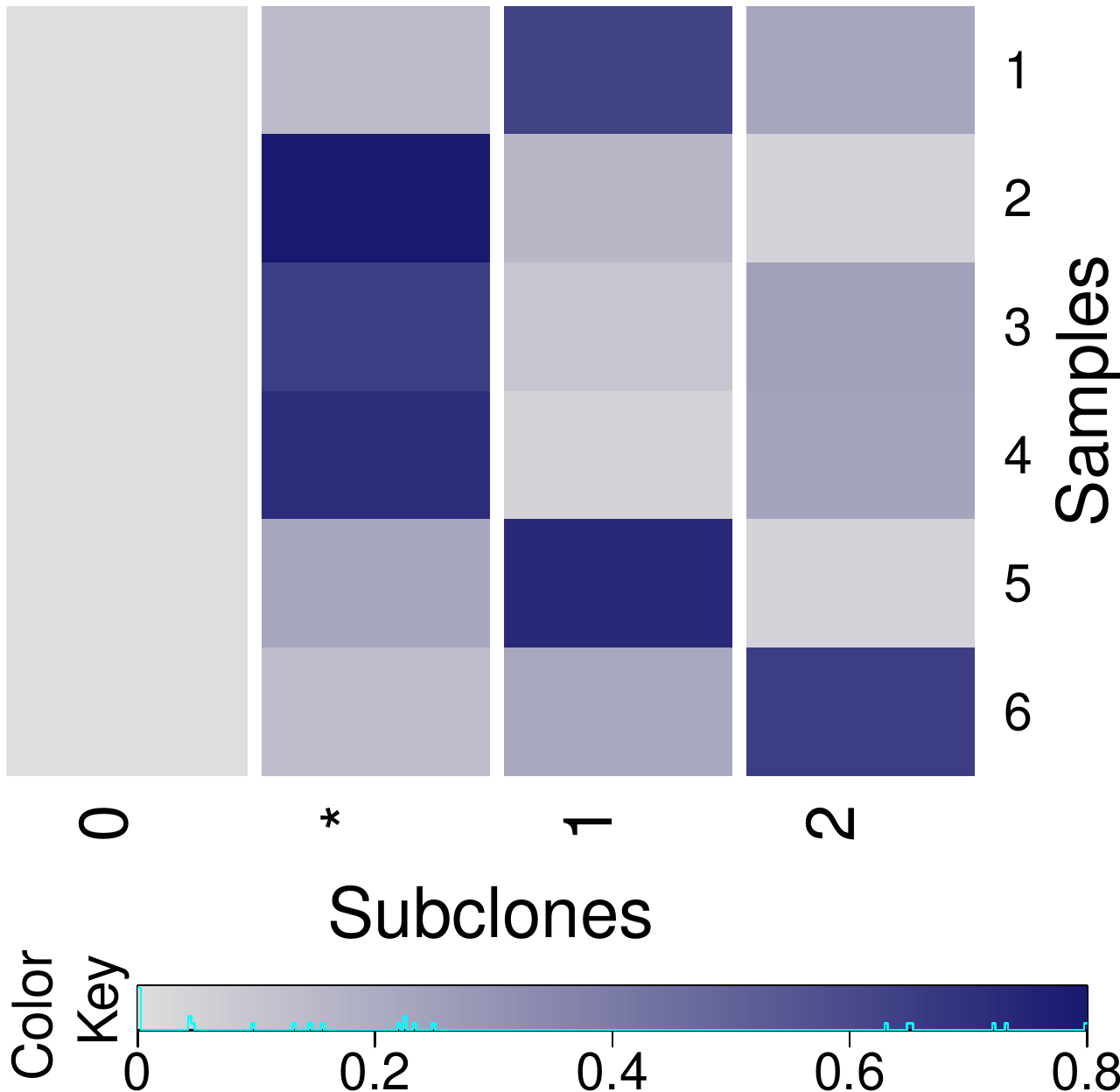}
\caption{$\what$}		
\end{subfigure}
\end{center}
\caption{Summary of simulation results with tumor purity incorporated.}
\label{app:figsimpurity}
\end{figure}

We evaluate the modified model with a simulation study. The simulation setting is the same as simulation 3 in Section \ref{sec:simulation},  except that we substitute the first subclone with a normal subclone. 
We use exactly the same hyperparameters  as those in simulations 2 and 3, and in addition we take $d_1^* = d_2^* = 1$.
Figure \ref{app:figsimpurity} summarizes inference results.
Columns in panels (b) and (c) marked with ``*'' correspond to the
normal subclone.  Panel (a) shows $p_b(C \mid \bm n'')$.
Posterior inference recovers the simulation truth, with 
posterior mode $\Chat = 2$.
Panel (b) shows $\Zhat$. Comparing with subclones 2 and 3 in
Figure \ref{app:figsim3}(a) we find a good recovery of the simulation
truth.
Panel (c) shows $\what$, which can be compared with Figure \ref{app:figsim3}(b). 

\subsection{Incorporating Copy Number Changes}
\label{sec:cnv}
Tumor cells not only harbor sequence mutations such as SNVs and
mutation pairs, they often undergo copy number changes and produce
copy number variants (CNVs). Genomic regions with CNVs 
have copy number $\ne 2$. 
We briefly outline an extension of
PairClone that includes CNVs in the inference, but do not implement inference under this
model, as it would require significantly more complex modeling.
In addition to $\bZ$ which describes sequence variation we 
introduce a $K \times C$ matrix $\bL$ to represent subclonal copy
number variation with $\ell_{kc}$ reporting the copy number for
mutation pair $k$ in subclone $c$. 
We use $\bL$ to augment the sampling model to include the total
read count $N_{tk}$.
Earlier in \eqref{eq:multi}, the multinomial sample size $N_{tk}$ was
considered fixed. We now add a sampling model. Following
\cite{lee2016bayesianjrssc} we assume
$$
  N_{tk} \mid \phi_t, M_{tk} \sim \text{Poisson}(\phi_t M_{tk} / 2)
$$
Here, $\phi_t$ is the expected number of reads in sample $t$ under
copy-neutral conditions, and
$M_{tk}$ is a weighted average copy number across subclones,
$$
   M_{tk} = \sum_{c=1}^C w_{tc} \ell_{kc} + w_{t0} \ell_{k0}.
$$
The last term $w_{t0} \ell_{k0}$ accounts for noise and
artifacts, where $w_{t0}$ and $\ell_{k0}$ are the population frequency
and copy number of the background subclone, respectively. We assume no
CNVs for the background subclone, that is, $\ell_{k0} = 2$ for all $k$. 
We complete the model with a prior $p(\bL)$. 
Assuming $\ell_{kc} \in \{0,\ldots,Q\}$, i.e., a maximum copy number $Q$, 
we use another instance of a finite cIBP. For each
column of $\bL$, we introduce $\bm \pi_c = (\pi_{c0}, \pi_{c1},
\ldots, \pi_{cQ})$ and assume $p(\ell_{kc} = q) = \pi_{cq}$, again
with a Beta-Dirichlet prior for $\bm \pi_c$. 

Recall the construction of $\tp_{tkg}$ in \eqref{pprior1},
including in particular the generative model.
This generative model is now updated to include the varying $\ell_{tc}$.
To generate a short read for mutation pair $k$, we
first select a subclone $c$ from which the read arises, using the
population frequencies $w_{tc} \ell_{kc} / \sum_{c = 0}^C w_{tc} \ell_{kc}$
for sample $t$. Next we select with probability $z_{kcj} / \ell_{kc}$ one
of the four possible alleles, $\bh_g$, $g=1,2,3$ or $4$, 
where we now use $\bz_{kc} = (z_{kcj}, j = 1, \ldots, 4)$ to denote
numbers of alleles having genotypes $00$, $01$, $10$ or $11$, and
$\sum_j z_{kcj} = \ell_{kc}$. 
In the case of left (or right) missing
locus we observe $\bh_g$, $g = 5$ or $6$ (or $g = 7$ or $8$),
corresponding to the observed locus of the chosen allele, similar to before.
In summary, the probability of observing a short read $\bh_g$ can be
written as 
\begin{align*}
\tp_{tkg} =  \sum_{c = 0}^C  \left[ \frac{w_{tc} \ell_{kc}}{\sum_{c =
  1}^C w_{tc} \ell_{kc} + w_{t0} \ell_{k0}} \cdot \frac{A(\bh_g,
  \bz_{kc})}{\ell_{kc}} \right] = \frac{\sum_{c=0}^C w_{tc} A(\bh_g,
  \bz_{kc})}{M_{tk}}, 
\end{align*}
where $A( \cdot )$ corresponds to the described generative model.

\section{Lung Cancer Data}
\label{sec:realdata}
\subsection{Using PairClone}
We apply PairClone to analyze whole-exome in-house data.
Whole-exome sequencing data is generated from four
($T = 4$) surgically dissected tumor samples taken from a single patient
diagnosed with lung adenocarcinoma. The resected tumor is divided into
two portions. One portion is flash frozen and another portion is
formalin fixed and paraffin embedded (FFPE). Four different samples
(two from each portion) are taken. DNA is extracted from all 
four samples. Agilent SureSelect v5+UTR probe kit (targeting coding
regions plus UTRs) is used for exome capture. The exome library is
sequenced in paired-end fashion on an Illumina HiSeq 2000
platform. About 60 million reads are obtained in FASTQ file format,
each of which is 100 bases long. We map paired-end reads to the human
genome (version HG19)~\citep{church2011modernizing} using
BWA~\citep{li2009fast} to generate BAM files for each individual
sample. After mapping the mean coverage of the samples is around $70$
fold. We call variants using UnifiedGenotyper from GATK
toolchain~\citep{mckenna2010genome} and generate a single VCF file for
all of them. A total of nearly $115,000$ SNVs and small indels are
called within the exome coordinates.

Next, using \texttt{LocHap}~\citep{sengupta2016ultra} we find mutation
pair positions, the number of alleles
and number of reads mapped to them.
\texttt{LocHap} searches for multiple SNVs that are scaffolded by
the same pair-end reads,  that is,  they  can be recorded on one
paired end read. We refer to such sets of multiple SNV's as
local haplotypes (LH). When more than two genotypes are exhibited by
an LH, it is called a LH variant (LHV). Using individual BAM files
and the combined VCF file, \texttt{LocHap} generates four individual output
file in HCF format~\citep{sengupta2016ultra}. An HCF file contains LHV
segments with two or three SNV positions. In this analysis, we are
only interested in mutation pair, and therefore filter out all the LHV
segments consisting of more than two SNV locations.
We restrict our analysis to copy number neutral regions. 
To further improve data quality, we drop all LHVs where two
SNVs are very close to each other (within, say, $50$ bps) or close to
any type of structural variants such as indels. We also remove
those LHVs where either of the SNVs is mapped with strand bias by most
reads, or either of the SNVs is mapped towards the
end of the most aligned reads. 
Finally, we only consider mutation pairs that have strong
evidence of heterogeneity.
Since LHVs exhibit $>2$ genotypes in the short reads, by
definition they are somatic mutations.

  At the end of this process, $69$ mutation
pairs are left and we record the read data from HCF files for the
analysis.  In addition, in the hope of utilizing more information
from the data, we randomly choose 69 un-paired SNVs and include them
in the analysis. 
Since in practice, tumor samples often include contamination with 
normal cells, we incorporate inference for tumor purity as described
in Section \ref{sec:purity}. 
 We run MCMC simulation for $30,000$ iterations, discarding the first 
$10,000$ iterations as initial burn-in and 
keeping
every 10th MCMC sample. 
We set the hyperparameter exactly  as in the simulation study 
of Section \ref{sec:purity}. 

\begin{figure}[h!]
\begin{center}
\begin{subfigure}[t]{.3\textwidth}
\centering
\includegraphics[width=\textwidth]{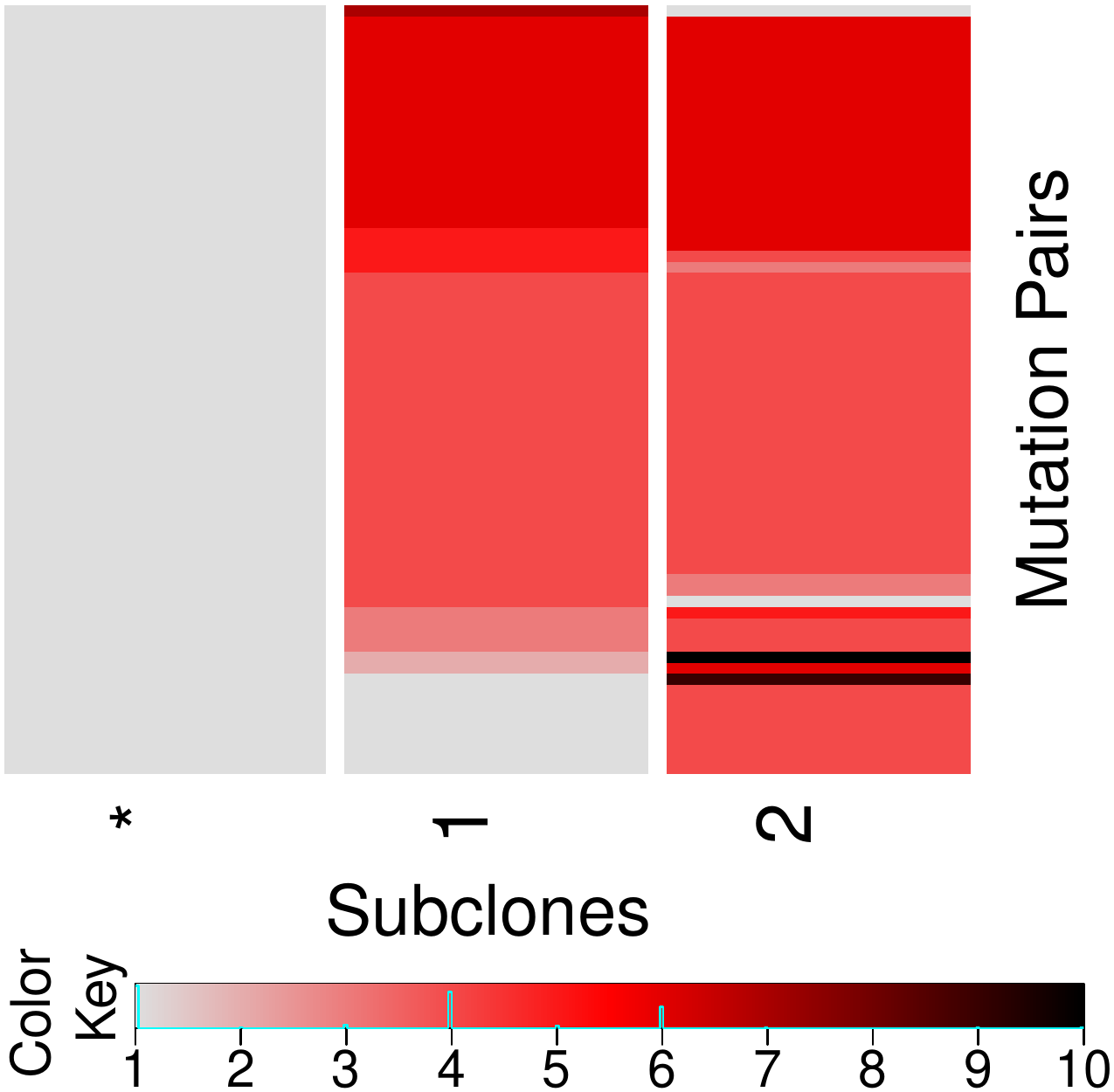}
\caption{$\Zhat^P$}		
\end{subfigure}
\begin{subfigure}[t]{.3\textwidth}
\centering
\includegraphics[width=\textwidth]{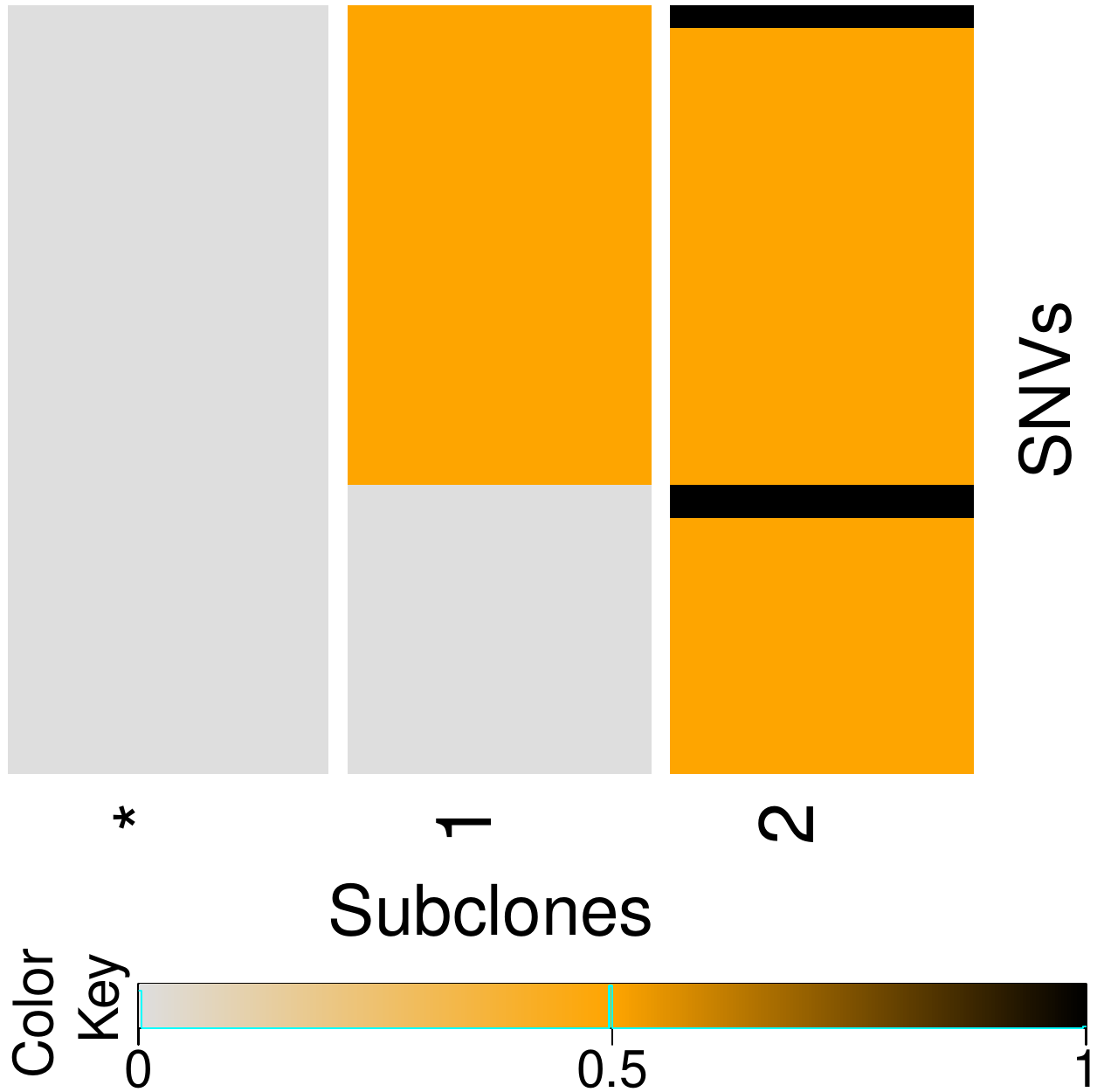}
\caption{$\Zhat^S$}		
\end{subfigure}
\begin{subfigure}[t]{.3\textwidth}
\centering
\includegraphics[width=\textwidth]{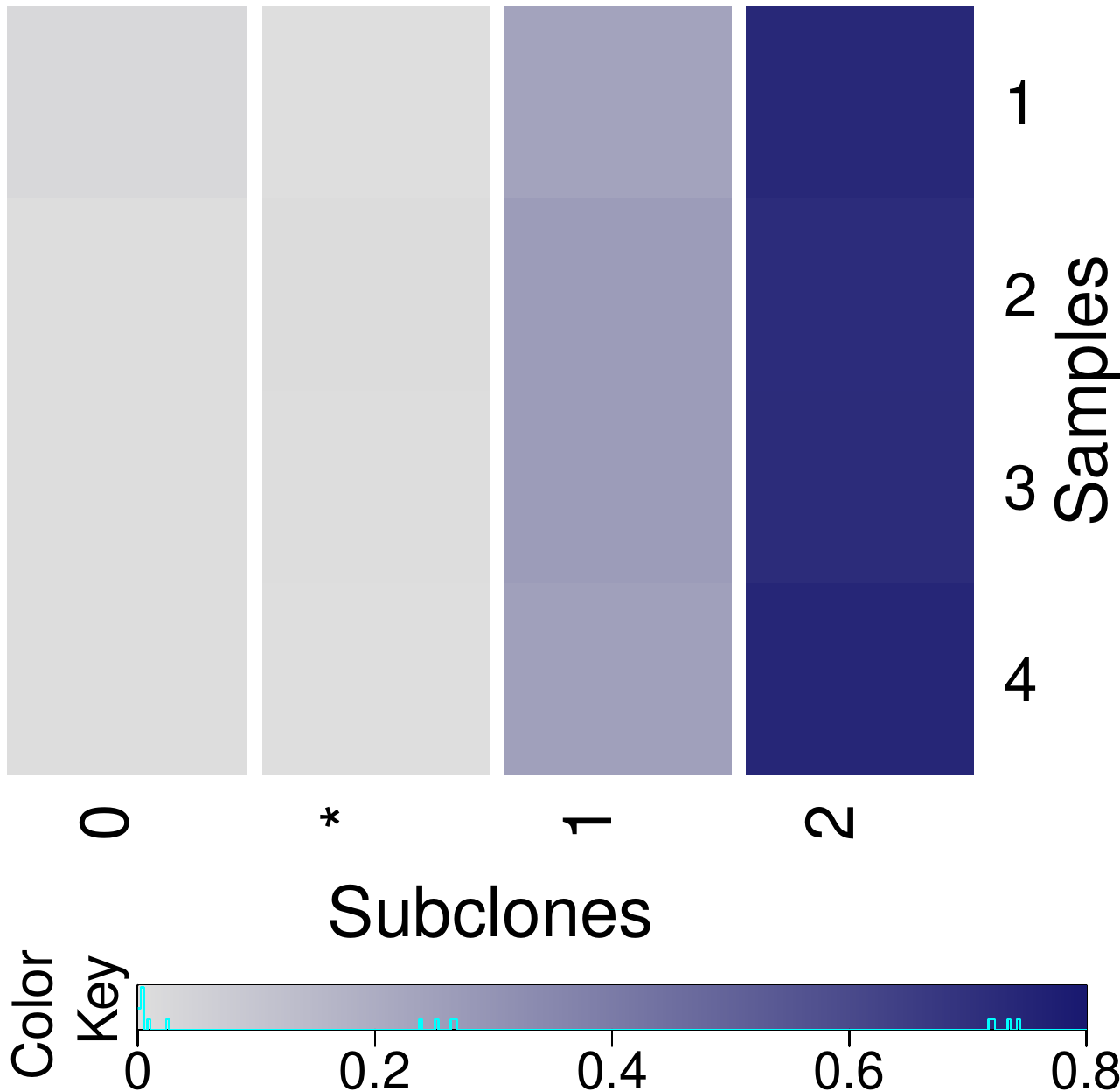}
\caption{$\what$}		
\end{subfigure}
\begin{subfigure}[t]{.3\textwidth}
\centering
\includegraphics[width=\textwidth]{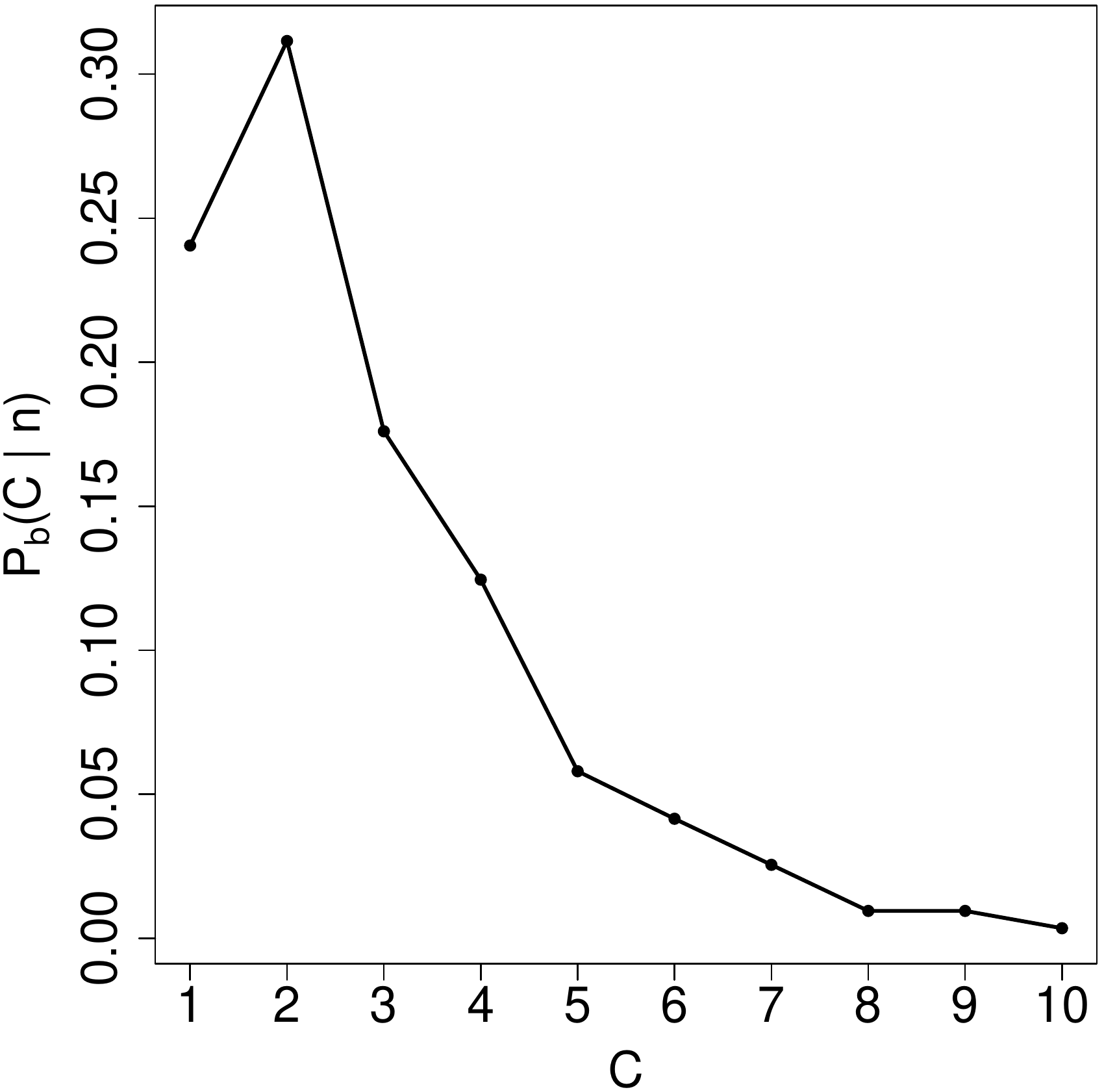}
\caption{$p_b(C \mid \bn'')$}	
\label{app:figlungC}	
\end{subfigure}
\hspace{6mm}\begin{subfigure}[t]{.3\textwidth}
\centering
\includegraphics[width=\textwidth]{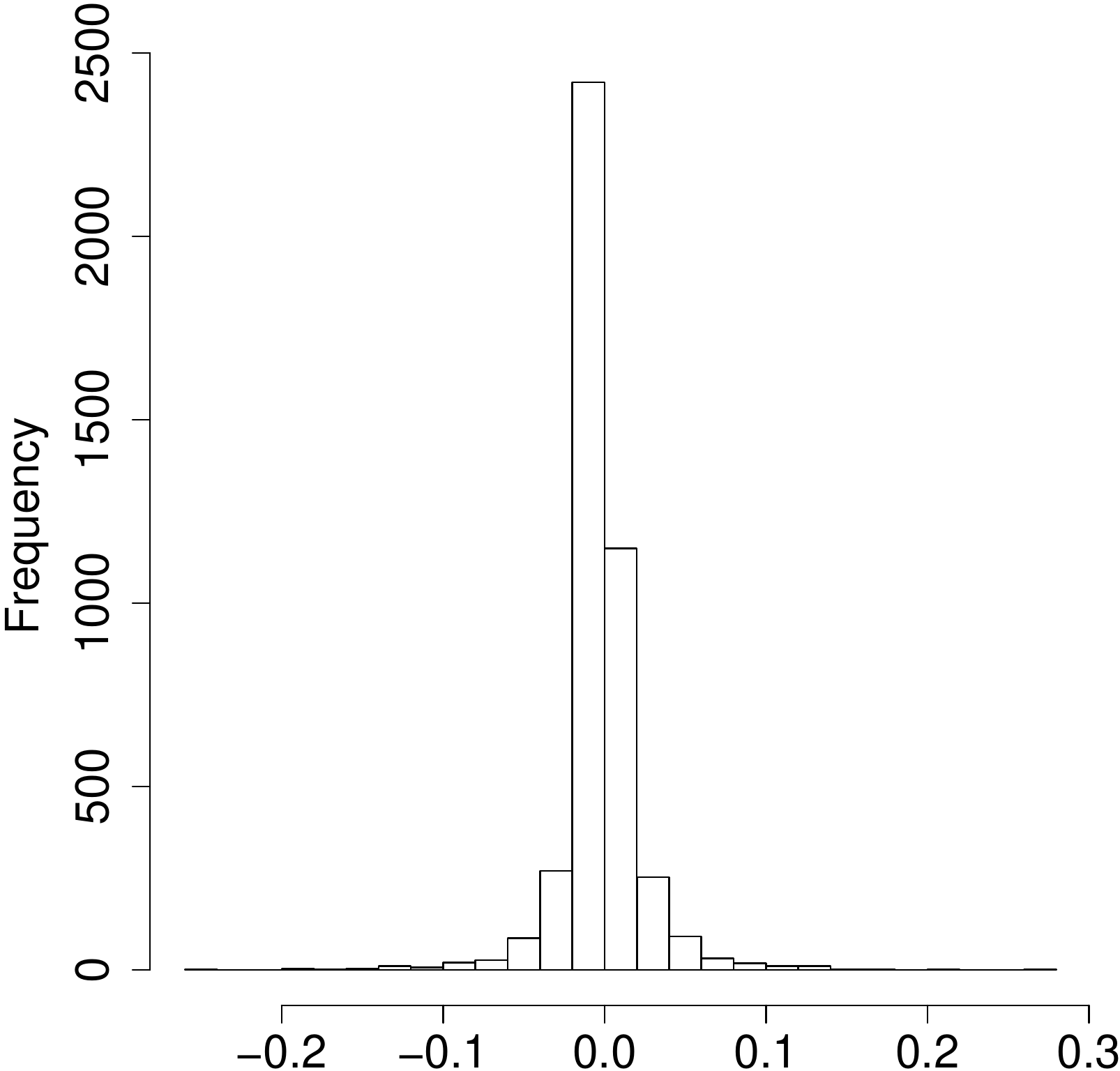}
\caption{Histogram of $(\ptkghat -\pbar_{tkg})$}		
\label{app:figlungresid}
\end{subfigure}
\end{center}
\caption{Lung cancer. Posterior inference under PairClone.}
\label{fig:lung}
\end{figure}

\paragraph{Results}
The posterior distribution $p_b(C \mid \bn'')$ (Figure \ref{app:figlungC}) reports
$p_b(C \mid \bn'') = 0.24$, $0.31$, $0.17$ and $0.12$ for $C = 1$, $2$, $3$ and $4$, respectively, and then quickly drops below $0.1$, with posterior mode $\Chat = 2$. 
This means, excluding the effect of normal cell contamination, the tumor samples have two subclones. Figure \ref{fig:lung}(a, b) show the estimated subclone matrix $\Zhat^P$ and $\Zhat^S$ corresponding to mutation pairs and SNVs, respectively.  
The first column of $\Zhat^P$ and $\Zhat^S$ represents the normal
subclone.  The rows for both matrices are reordered for a better
display.  Figure \ref{fig:lung}(c)
shows the estimated subclone proportions $\what$ for the four samples.
The second column of $\what$ represents the proportions of normal
subclones in the four samples. The small values indicate high purity
of the tumor samples.
The similar proportions across the four
samples reflect the spatial proximity of the samples. 
Furthermore, excluding a few exceptions that might be due to model
mis-fitting, the subclones form a simple phylogenetic tree: 
$*\rightarrow 1 \rightarrow 2$.  Subclones 1 and 2 share a large portion
of common mutations, while subclone 2 has some private
mutations that are missing in subclone 1.

For informal model checking we inspect a histogram of realized residuals (Figure \ref{app:figlungresid}). To define residuals, we calculate estimated multinomial probabilities $\{\ptkghat\}$ according to $\Zhat$, $\what$ and empirical values of $\{v_{tk1}, v_{tk2}, v_{tk3}\}$.
Let $\pbar_{tkg} = n_{tkg} / N_{tk}$. The figure plots the residuals $(\ptkghat- \pbar_{tkg})$.
The resulting histogram of residuals is centered around zero with little mass beyond $\pm 0.04$, indicating a good model fit.

\subsection{Using SNVs only}

\begin{figure}[h!]
\begin{center}
\begin{subfigure}[t]{.22\textwidth}
\centering
\includegraphics[width=\textwidth]{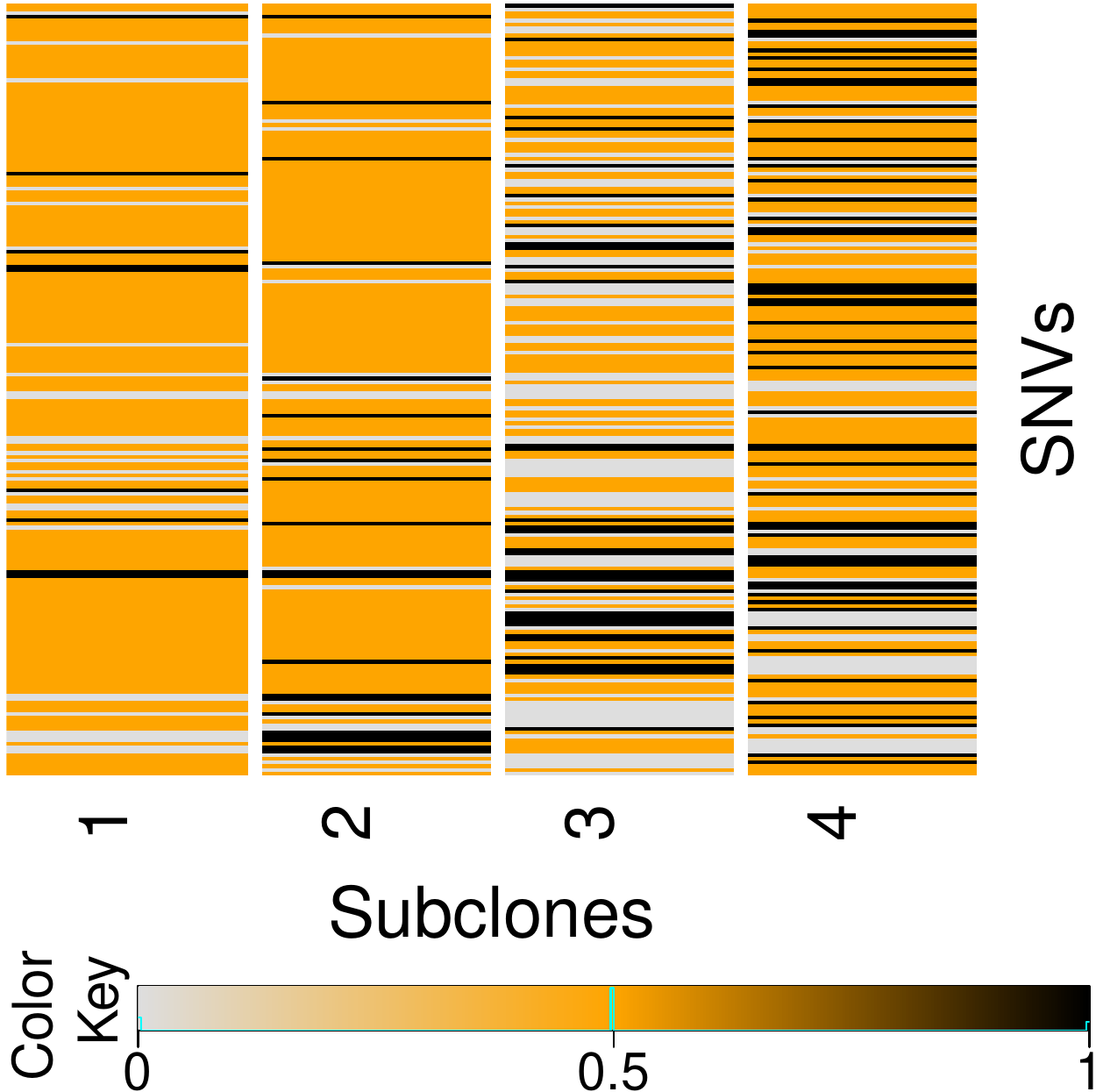}
\caption{$\Zhat_{\text{BC}}$}		
\end{subfigure}
\begin{subfigure}[t]{.22\textwidth}
\centering
\includegraphics[width=\textwidth]{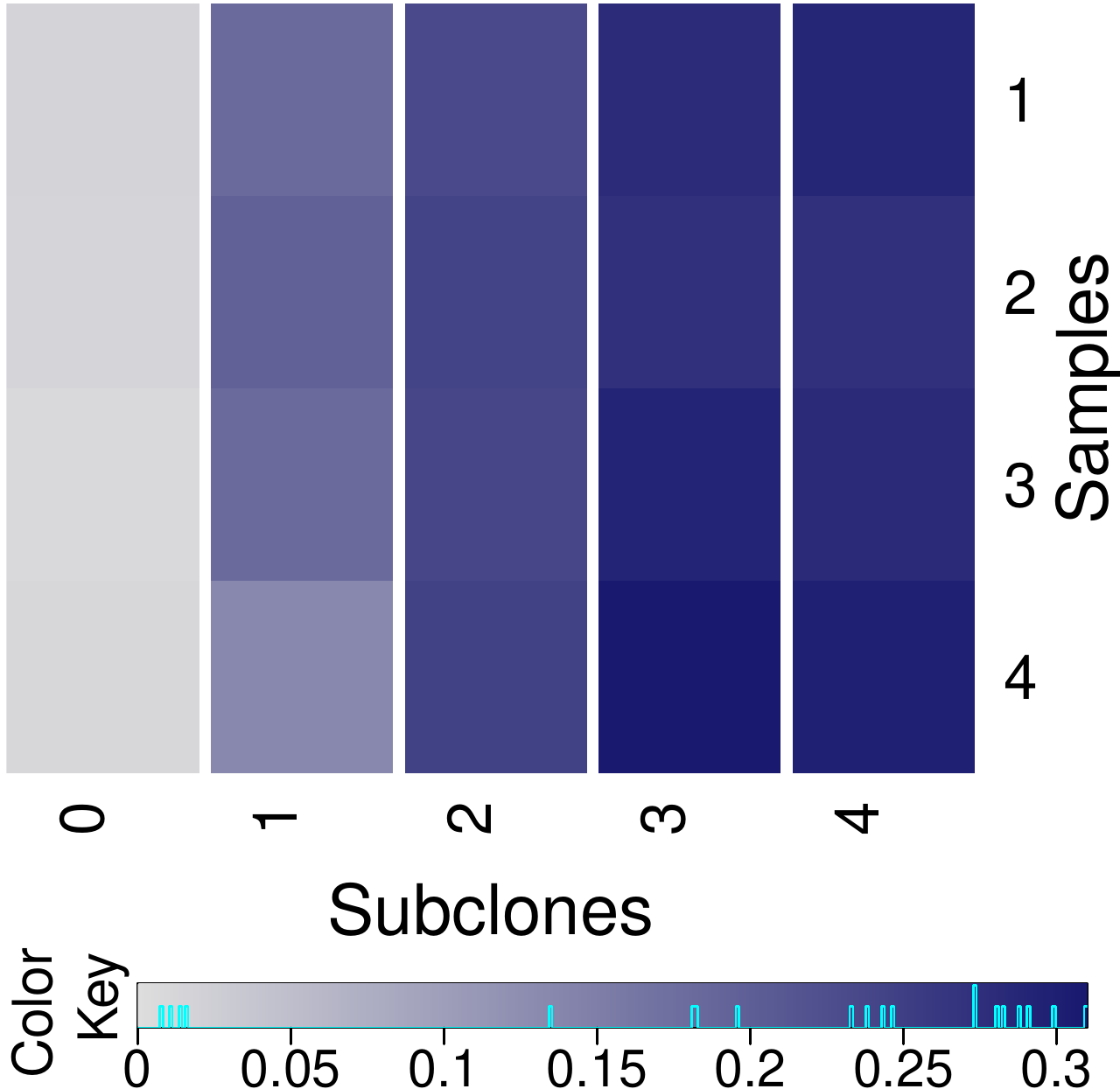}
\caption{$\what_{\text{BC}}$}		
\end{subfigure}
\begin{subfigure}[t]{.25\textwidth}
\centering
\includegraphics[width=\textwidth]{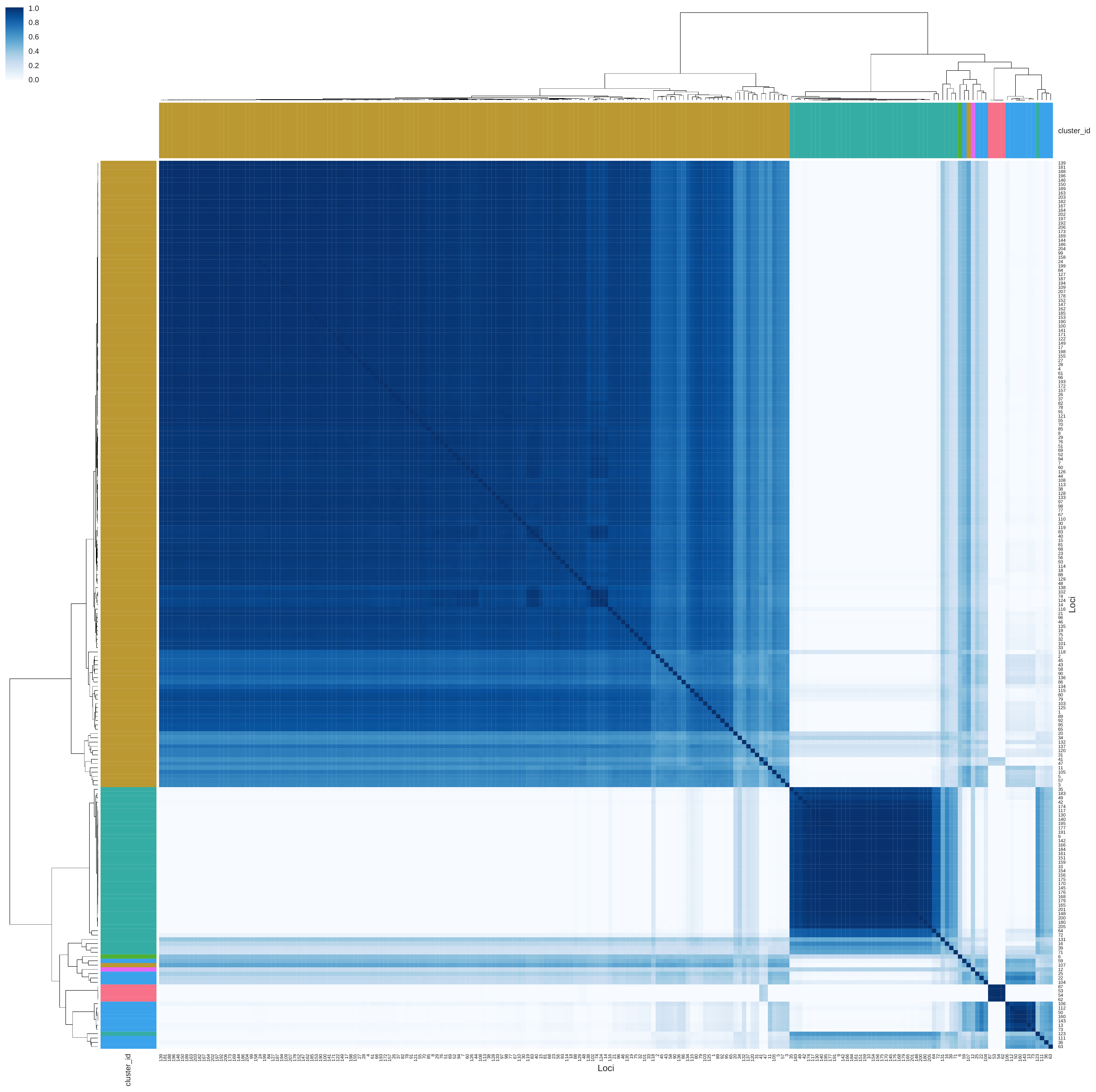}
\caption{Clustering matrix}		
\end{subfigure}
\begin{subfigure}[t]{.27\textwidth}
\centering
\includegraphics[width=\textwidth]{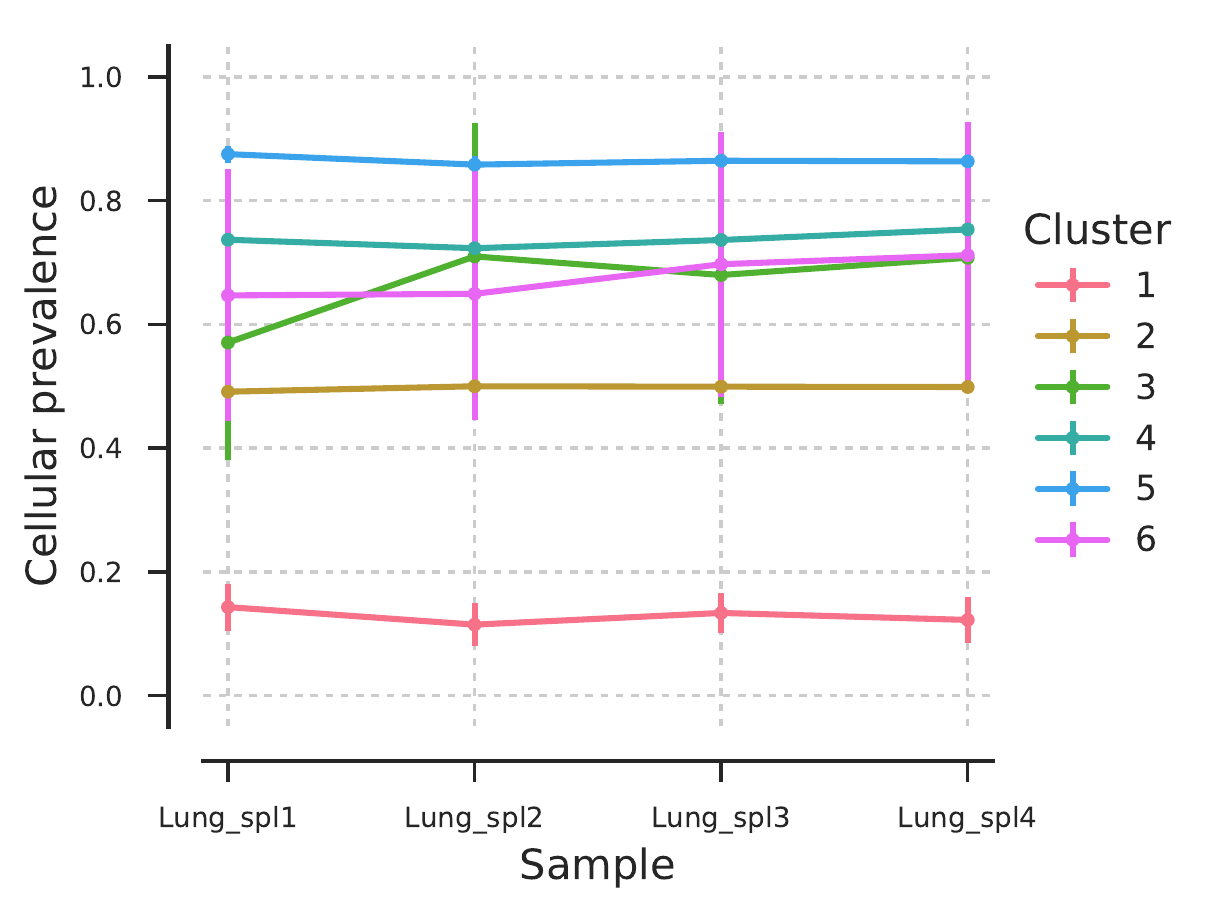}
\caption{Cellular prevalence}		
\end{subfigure}
\end{center}
\caption{Lung cancer. Posterior inference under BayClone (a, b) and PyClone (c, d).}
\label{fig:lungBC}
\end{figure}

For comparison, we also run BayClone and PyClone on the same dataset. Using the log pseudo marginal likelihood (LPML), BayClone reports $\Chat=4$ subclones.
The estimated subclone matrix in
BayClone's format is shown in Figure \ref{fig:lungBC}(a), with the
rows reordered in the same way as in Figure \ref{fig:lung}(a, b).
In light of the earlier simulation results we believe
that the inference under PairClone is more reliable.
Figure \ref{fig:lungBC}(b) shows the estimated subclone proportions
under BayClone. 
Figure \ref{fig:lungBC}(c) shows the estimated clustering of the SNV
loci under PyClone (the color coding along the axes).  
PyClone identifies 6 different clusters. 
The largest cluster (shown in brown) corresponds to loci that have heterozygous variants in
both subclones 1 and 2, the second-largest cluster (shown in blueish green) corresponds to loci
that have homozygous wild types in subclone 1 and homozygous
variants in subclone 2, and the other smaller clusters represent other
less common combinations. The clusters match with clustering of
rows of $\Zhat^P$ and $\Zhat^S$. 
PyClone does not immediately give inference on subclones, but combing clusters with similar cellular prevalence across samples one is able to conjecture subclones. In this sense, PyClone gives similar result compared with PairClone.
Finally, Figure \ref{fig:lungBC}(d) displays PyClone's estimated
cellular prevalences of clusters across different samples. 
The estimated subclone proportions and cellular prevalences across the
four samples remain very similar also under the BayClone and PyClone
output, which strengthens our inference that the four samples possess the same subclonal profile, each with two subclones. 


\begin{figure}[h!]
\begin{center}
\begin{subfigure}[t]{.4\textwidth}
\centering
\includegraphics[width=\textwidth]{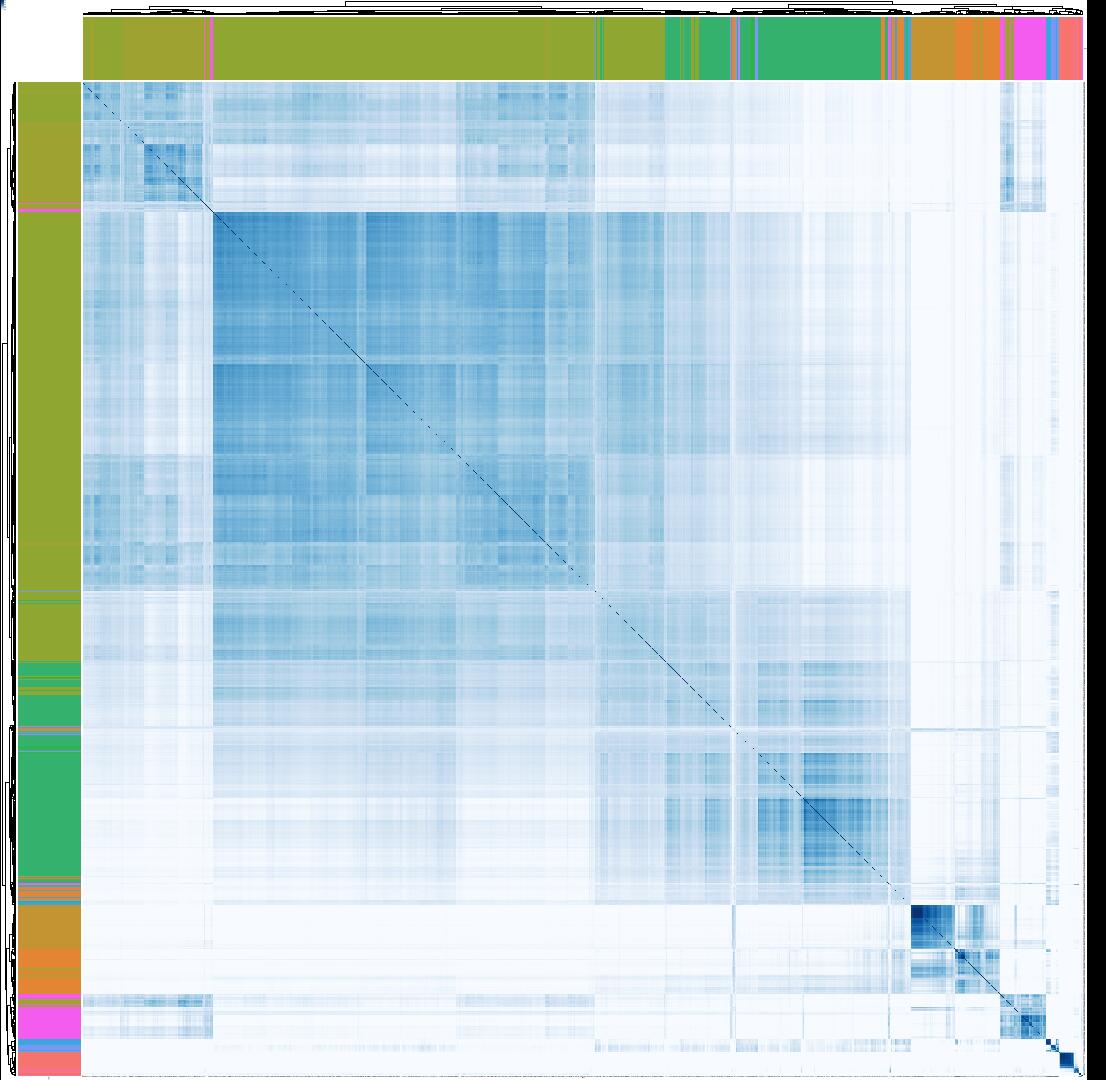}
\caption{Clustering matrix}		
\end{subfigure}
\hspace{7mm}\begin{subfigure}[t]{.45\textwidth}
\centering
\includegraphics[width=\textwidth]{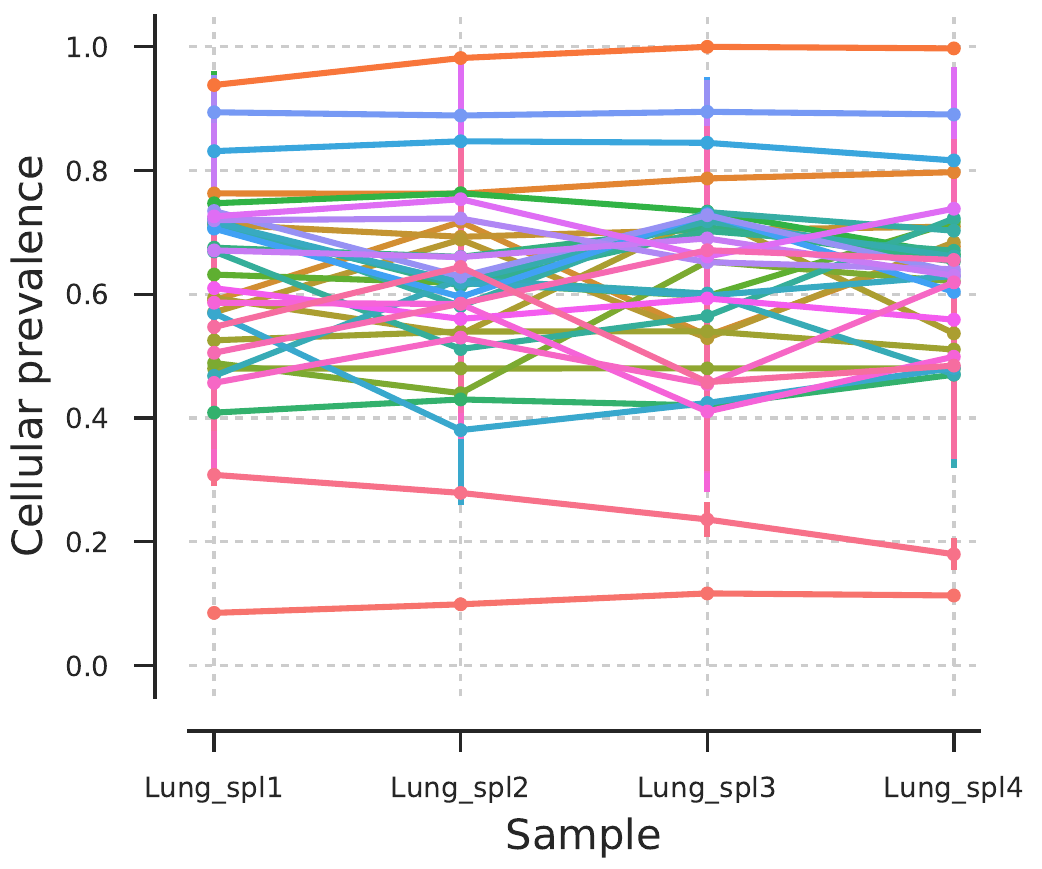}
\caption{Cellular prevalence}		
\end{subfigure}
\end{center}
\caption{Lung cancer. Posterior inference under PyClone using 1800 SNVs. PyClone inferred 34 clusters with two major clusters (olive and green) and many small noisy clusters (other colors).}
\label{fig:lung1800}
\end{figure}

For another comparison, we run PyClone with a much larger number of SNVs ($S = 1800$, which include the 69 pairs and 69 SNVs we ran analysis before) to evaluate the information gain by using additional marginal counts. The results are summarized in Figure \ref{fig:lung1800}, with panel (a) showing the estimated clustering of the 1800 SNVs. PyClone reports 34 clusters. The two largest clusters (olive and green clusters) in Panel (a) match with the two largest clusters (brown and bluish green clusters) in Figure \ref{fig:lungBC}(c) and also corroborate the two subclones inferred by PairClone.
In addition, PyClone infers lots of noisy tiny clusters using 1800 SNVs, which we argue model only noise. 
In summary, this comparison shows the additional marginal counts do not noticeably improve inference on tumor heterogeneity, and modeling mutation pairs is a reasonable way to extract useful information from the data.

\section{Discussion}
\label{sec:conclusion}

We can significantly enrich our understanding of cancer development by
using high throughput NGS data to infer co-existence of subpopulations
which are genetically different across tumors and within a single
tumor (inter and intra tumor heterogeneity, respectively).
In this paper,
we have presented a novel feature allocation model for reconstructing such subclonal structure 
using mutation pair data.
Proposed inference explicitly models overlapping mutation pairs.
We have shown that more accurate inference can be obtained
using mutation pairs data compared to using only marginal counts for
single SNVs.
Short reads mapped to mutation pairs can provide direct evidence for
heterogeneity in the tumor samples. In this way the proposed approach
is more reliable than methods for subclonal reconstruction that rely
on marginal variant allele fractions only. 

The proposed model is easily extended for data where an
LH segment consists of more than two SNVs. We can easily accommodate
$n$-tuples instead of pairs of SNVs by increasing the number
of categorical values ($Q$) that the entries in the $\bZ$
matrix can take. There are several more interesting
directions of extending the current model.
For example, one could account for the potential
phylogenetic relationship among subclones (i.e the columns in the $\bZ$
matrix). Such extensions would enable one to
infer mutational timing and allow the reconstruction of tumor evolutionary
histories.

The proposed model characterizes tumor heterogeneity with SNV data. Other genetic variants including CNV could also provide important information about tumor heterogeneity. However, incorporating such data increases the complexity in modeling. Like other existing methods we chose to focus on the SNV data, keeping the development of models for other genetic variants as a research topic of it own. A future direction is to develop computationally efficient and reliable approaches to incorporating other genetic variants, presumably by utilizing other available data in tumor heterogeneity.

Lastly, we focus on statistical inference using bulk sequencing
data on tumor samples. Alternatively, biologists can apply single-cell
sequencing on each tumor cell and study its genome one by one. This is
a gold standard that can examine tumor heterogeneity at the single-cell
level. However, single-cell sequencing is still expensive and cannot
scale up. Also, many bioinformatics and statistical challenges are
unmet in analyzing single-cell sequencing data.

\chapter{A Bayesian Treed Feature Allocation Model for Tumor Subclone Phylogeny Reconstruction Using Mutation Pairs}
\label{chap:PairCloneTree}

We present a latent feature allocation model to reconstruct a tumor
phylogenetic tree and corresponding tumor heterogeneity. Similar to
most current methods for inference on tumor heterogeneity, we consider
data from next-generation sequencing. Unlike most methods that use
information in short reads mapped to single nucleotide variants
(SNVs), we consider subclone reconstruction using pairs of two
proximal SNVs that can be mapped by the same short reads. A key part
of the inference model is a phylogenetic tree prior that is used to
construct a dependent prior on tumor cell subpopulations.  The use of
the tree structure in the prior greatly strengthens inference. Only
subclones that conform with an a priori plausible phylogenetic tree
are assigned non-negligible probability. The proposed Bayesian
framework implies posterior distributions on the number of subclones,
their genotypes, cellular proportions, and the phylogenetic tree
spanned by the inferred subclones. The proposed method is validated
against different sets of simulated and real-world data using single
and multiple tumor samples. An open source software package is
available at \url{http://www.compgenome.org/pairclonetree}.

\section{Introduction}
\label{sec:intro}
Tumor cells emerge from a Darwinian-like selection among
multiple competing subpopulations of cells
\citep{nowell1976clonal,bonavia2011heterogeneity,
marusyk2012intra}. 
During
tumorigenesis, through sequential clonal expansion and
selection cells acquire distinct mutations.
This process leads to genetically divergent
subpopulations of cells, also known as 
subclones
\citep{navin2010tumor,Gerlinger2012intratumor,nik2012life,
bignell2010signatures,bozic2010accumulation,raphael2014identifying}.
Reconstructing the subclones and their evolutionary relationship
could help investigators to identify driver mutations that emerge early in
the development or during the progression period.
Such results provide insight about targeted
therapies
\citep{aparicio2013implications,papaemmanuil2011somatic,varela2011exome,
stephens2012landscape}. 

A recent surge of genetic sequencing data makes it
possible to investigate tumor subclonal architecture in
detail \citep{oesper2013theta,
strino2013trap,fischer2014high,
miller2014sciclone,roth2014pyclone,
jiao2014inferring,deshwar2015phylowgs,zare2014inferring,
sengupta2015bayclone,marass2017phylogenetic}. 
We will discuss details of some \citep{marass2017phylogenetic, jiao2014inferring,deshwar2015phylowgs} later in Section \ref{sec:sim}, after we have introduced the required notation.
Latest developments of next generation sequencing (NGS) technology enabled 
researchers to develop a variety of techniques that are broadly
known as subclonal reconstruction. 
One of the aims is to deconvolute observed genomic data
from a tumor into constituent signals  corresponding to 
various subclones and to reconstruct their relationship in a
phylogeny. 
In most methods the reconstruction is based on
short reads that are mapped to single nucleotide variants
(SNVs) (few methods also consider somatic copy number aberrations,
SCNA).  
SNV-based subclone calling methods utilize variant allele fractions
(VAFs), that is, the fractions of alleles (or short reads) at
each locus that carry mutations. 
Since humans are diploid, 
the VAFs of short reads for a homogeneous cell population
should be 0, 0.5 or 1.0 for any locus in copy number neutral (copy
number = 2) regions and after adjusting for tumor purity.
VAFs different from 0, 0.5 or 1.0 
are therefore evidence for heterogeneity. 
Based on this idea, existing SNV-based
subclone calling methods either cluster
mutations \citep{miller2014sciclone,roth2014pyclone,
jiao2014inferring,deshwar2015phylowgs},
or use latent feature allocation methods to infer the subclone
genotypes and their proportions
\citep{zare2014inferring,sengupta2015bayclone,marass2017phylogenetic}.
All are based on observed VAFs. 

\subsection{Main Idea}
\begin{figure}[h!]
\centering
\includegraphics[width=\textwidth]{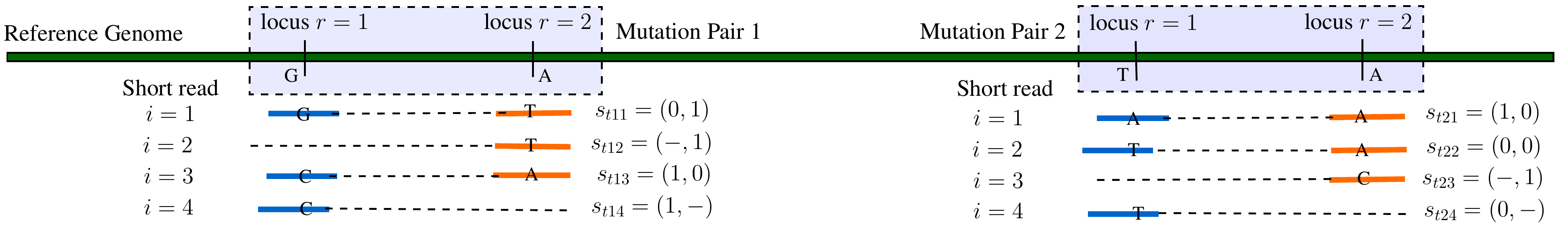}
\caption{Short reads data from mutation pairs
  using NGS. Here $s_{tki}$ denotes the $i$-th read
  for  the $k$-th mutation pair in sample $t$. 
  Each $s_{tki}$ is a 2-dimensional vector  which corresponds to the
  two proximal SNVs in the mutation pair, and each component of the
  vector takes values $0$, $1$ or -- representing wild type,
  variant or missing genotype, respectively.} 
\label{fig:lochap_data}
\end{figure}

We assume that the available data are from $T$ ($T\geq1$) samples 
from a single patient and the main inference goal is
intra-tumor heterogeneity.
We present a novel approach to reconstruct tumor
subclones and their corresponding phylogenetic tree based on
mutation pairs. Here a mutation pair refers to a pair of proximal
SNVs on the genomes that can be simultaneously mapped by the same
paired-end short reads, with one SNV on each end.
In other words, mutation pairs can be phased by short reads. See
Fig. \ref{fig:lochap_data} for an illustration. 
Short reads mapped to only one of the SNV loci are treated as partially missing paired-end reads and are not excluded from our approach. Specifically, marginal SNV reads can be included in our analysis. See Section \ref{sc:sampling_model} for more details.
The idea of working with phased mutation pairs was introduced in 
Chapter \ref{chap:PairClone}.
We build on it and develop a novel and entirely
different inference approach by explicitly modeling the
underlying phylogenetic relationship.
That is, we model tumor heterogeneity based on a representation of
a phylogenetic tree of tumor cell subpopulations. A prior probability
model on such phylogenetic trees induces a dependent prior on
the mutation profiles of latent tumor cell subpopulations.
In particular, the phylogenetic tree of tumor cell subpopulations is
included as a 
random quantity in the Bayesian model.  Currently, we only consider
mutation pairs in copy neutral region i.e. copy number two.
The proposed inference aims
to reconstruct (i) subclones defined by the haplotypes across all the
mutation pairs, (ii) cellular proportion of each subclone, and (iii) a
phylogenetic tree spanned by the subclones.

Consider an NGS data set with $K$ mutation pairs shared
across all $T$ ($T \ge 1$) samples.
We assume that the samples are composed of $C$
homogeneous subclones. The number of subclones $C$ is unknown and
becomes one of the model parameters.
We use a $K \times C$ matrix $\bZ$ to represent
the subclones, in which each column of $\bZ$ represents a subclone and
each row represents a mutation pair. 
That is, the $(kc)$ element  $\bz_{kc}$ of the matrix corresponds to
subclone $c$ and mutation pair $k$.
Each $\bz_{kc}$ is itself again a matrix. It is a $2 \times 
2$ matrix that represents the genotypes of the two alleles of the
mutation pair. See Fig. \ref{fig:sc_evo}(b).
An important step in the model construction is that 
the columns (subclones) of $\bZ$ form a phylogenetic tree $\Tau$.
The tree encodes the parent-child relationship across the subclones. 
A detailed construction of the tree and a prior probability model of
$\Tau$ and $\bZ$ are introduced later.
Lastly, we denote $\bw_t = (w_{t1}, \ldots, w_{tC})$ the cellular
proportions of the subclones in sample $t$ where $0 < w_{tc} < 1$ for
all $c$ and $\sum_{c=0}^C w_{tc} = 1$.  

Using NGS data we infer $\Tau$, $C$, $\bZ$ and $\bw$ based on a simple idea that variant
reads can only arise from subclones with variant  alleles
consistent with an underlying phylogeny. We develop a 
treed {\em latent feature allocation model} (LFAM) to
implement this reconstruction.
Mutation pairs are the objects of the LFAM, and
subclones are the latent features chosen by the mutation pairs
(in contrast to the phylogenetic Indian Buffet Process \citep{miller2012phylogenetic} which builds a tree structure
on objects, rather than features).
Note that subclone reconstruction based on LFAM allows 
overlapping mutations across subclones and therefore does not require 
the infinite sites assumption \citep{nik2012life}.
This is different from many existing cluster-based models in the
literature.  While LFAM attempts to directly infer genotypes of
all subclonal genomes, cluster-based models first infer SNV clusters
based on VAFs and then reconstruct subclonal genotypes based on the
clusters.

\paragraph{Advantage of using mutation pairs}
Mutation pairs contain
phasing information that improves the accuracy of subclone
reconstruction.  If two nucleotides reside on the same short
read, 
we know that they must appear in the same DNA strand in a subclone. For
example, consider a scenario with one mutation pair and two
subclones. Suppose the reference genome allele is (A, G) for that
mutation pair, with the notion that A and G are phased by the same
DNA strand.  Suppose the two subclones have diploid genomes at
the two loci and the genotypes for both DNA strands are  
((C, G), (A, T)) for subclone $c=1$, and ((C, T), (A, G)) for
$c=2$. 
Since in NGS short reads are
generated from a single DNA strand, 
short reads could be any of the four haplotypes
(C, G), (A, T), (C, T) or (A, G) for this mutation pair. 
If indeed relative large counts of short reads with each haplotype are observed, one can
reliably infer that there are heterogeneous cell subpopulations
in the tumor sample.
In contrast, if we ignore
the phasing information and only consider the (marginal) VAFs for
each SNV, then the observed VAFs for both SNVs are 0.5, which could be heterogeneous mutations from a single cell population.
%
In this paper, we leverage the power of using mutation pairs over single
SNVs to incorporate partial phasing information in our model. 
We assume that mutation pairs and their mapped short reads counts have
been obtained using a bioinformatics pipeline, such as 
\texttt{LocHap}
\citep{sengupta2016ultra}. Our aim is to use short reads mapping data
on mutation pairs to reconstruct tumor subclones and their
phylogeny.

\paragraph{Difference from traditional phylogenetic tree}
Phylogenetic trees are usually used to approximate perfect phylogeny
for a fixed number of
haplotypes \citep{gusfield1991efficient,bafna2003haplotyping,pe2004incomplete}.
Most methods lack assessment of tree uncertainties and report a
single tree estimate. Also,  methods based on SNVs 
put the observed mutation profile of SNV at the leaf nodes.
This is natural if the splits in the tree create subpopulations
that acquire or do not acquire a new mutation (or set of mutations).
In contrast, we define a tree with all descendant nodes differing from
the parent node by some mutations. 
That is, all node, including interior nodes, correspond to
tumor cell subpopulations.
See details below. 
For clarification we note that 
the prior structure in our model is
different from the phylogenetic Indian Buffet Process
(pIBP) \citep{miller2012phylogenetic}, which models phylogeny of the
objects rather than the features.\\

\subsection{Representation of Subclones}
\label{sec:rep_subclone}
\begin{figure}[h!]
\centering
\begin{subfigure}[t]{.53\textwidth}
\centering
\includegraphics[width=\textwidth]{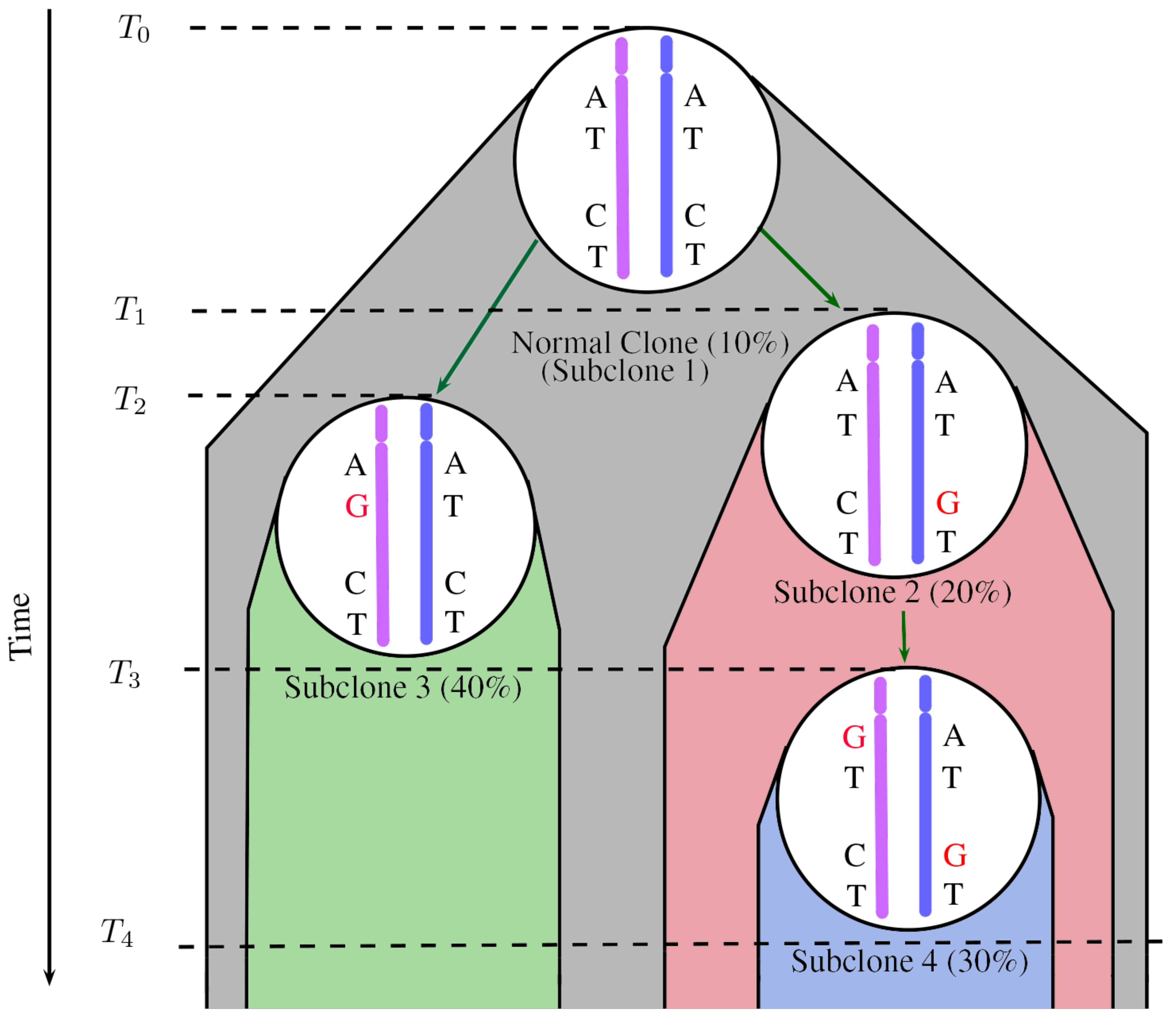}
\caption{}
\end{subfigure}
\begin{subfigure}[t]{.46\textwidth}
\centering
\includegraphics[width=\textwidth]{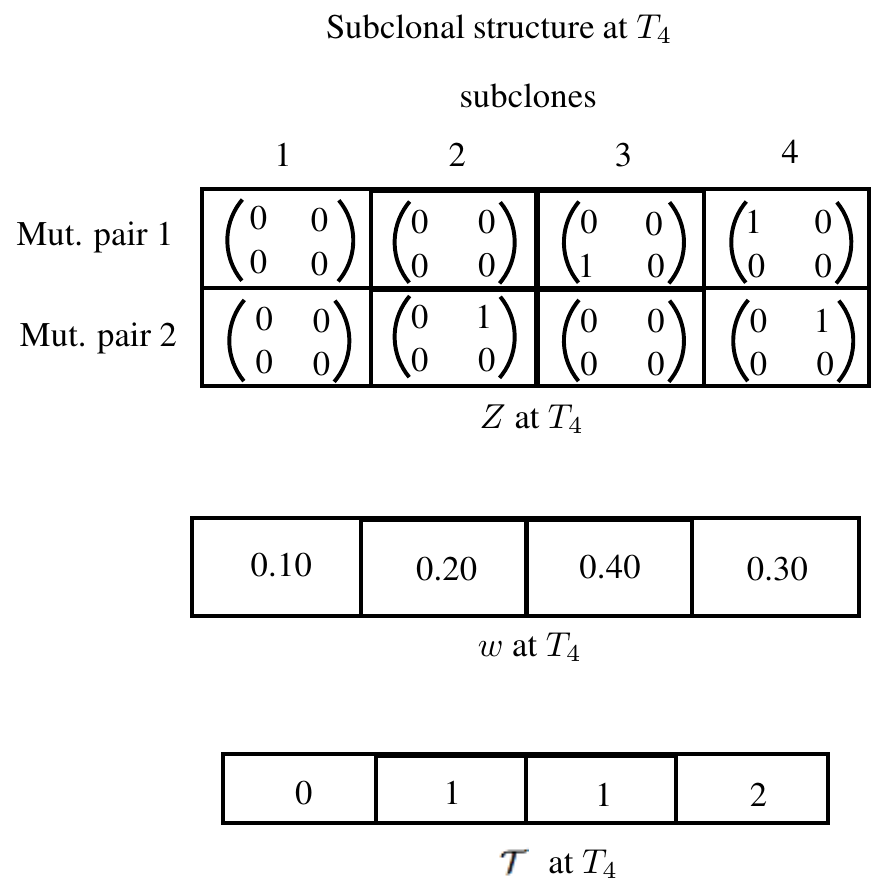}
\caption{}
\end{subfigure}
\caption{Schematic of subclonal evolution and subclone
  structure. Panel (a) shows the evolution of subclones over time.
  Panel (b) shows the subclonal structure at $T_4$ with 
  genotypes $\bZ$, cellular proportions $\bw$ and parent vector $\Tau$. 
  For each mutation pair $k$ and subclone $c$, 
  the entry $\bz_{kc}$ of $\bZ$ is a $2\times2$ matrix corresponding to
  the arrangement in the figure in panel (a), that is, 
  with alleles in the two columns, and SNVs in the rows.}
\label{fig:sc_evo}
\end{figure}

Fig.~\ref{fig:sc_evo} presents a stylized example of temporal
evolution of a tumor, starting from time $T_0$ and evolving until time
$T_4$ with the normal clone (subclone $c=1$) and three tumor subclones
($c=2,3,4$).
Each tumor subclone is marked by two mutation pairs with distinct
somatic mutation profiles. In Fig.~\ref{fig:sc_evo} the true
phylogenetic tree is plotted connecting the stylized subclones.  The
true population frequencies of the subclones are marked in
parentheses. In panel (b) subclone genomes, their
population frequencies and the phylogenetic relationship are
represented by $\bZ$, $\bw$, and $\Tau$. The entries of $\Tau$ 
report for each subclone the index of the parent subclone
(with $\Tau_1=0$ for the root clone $c=1$).

Suppose $K$ mutation pairs with $C$ subclones are
present. The subclone phylogeny can be visualized with a rooted tree
with $C$ nodes.  We use a $C$-dimensional {\it parent} vector $\Tau$ 
to encode the parent-child relationship of a
tree, where $\Tau_c = \Tau[c] = j$ means that subclone $j$ is
the parent of subclone $c$. The parent vector uniquely
defines the topology of a rooted tree.  We assume that the tumor
evolutionary process always starts from the normal clone, indexed
by $c = 1$. The normal clone does not have a parent, and we denote
it by $\Tau_1 = 0$. For example, the parent vector representation of
the subclone phylogeny in Fig. \ref{fig:sc_evo} is $\Tau = (0, 1, 1,
2)$.

We use the $K \times C$ matrix $\bZ$ to represent the subclone
genotypes. Each column of $\bZ$ defines a subclone, and each row of
$\bZ$ corresponds to a mutation pair.  The entry $\bz_{kc}$ records
the genotypes for mutation pair $k$ in subclone $c$.  Since each
subclone has two alleles $j = 1, 2$, and each mutation pair has two
loci $r = 1, 2$, the entry $\bz_{kc}$ is itself a $2 \times 2$ matrix,
i.e. $\bz_{kc} = (\bz_{kcj}, j = 1, 2)$ and $\bz_{kcj} = (z_{kcjr}, r
= 1, 2)$,
\begin{equation*}
\bz_{kc} = (\bz_{kc1}, \bz_{kc2}) =  
\begin{bmatrix}
   \left(\begin{array}{c}
        z_{kc11}  \\
        z_{kc12}  \\
  \end{array}\right) 
   \left(\begin{array}{c}
        z_{kc21}  \\
        z_{kc22}  \\
  \end{array}\right) 
\end{bmatrix}
\end{equation*}
where 
$\left(\begin{array}{c} 
         z_{kc11}  \\ z_{kc12}  \\ 
 \end{array}\right)$ 
and 
$\left(\begin{array}{c} 
         z_{kc21}  \\ z_{kc22}  \\ 
\end{array}\right)$ 
represent mutation pairs of allele $1$ and allele $2$, respectively.
Theoretically, each $z_{kcjr}$ can be any one of the four nucleotide
sequences, A, C, G, T. However, at a single locus, the probability of
having more than two sequences is negligible since it would require
the same locus to be mutated twice throughout the life span of the
tumor, which is extremely unlikely.
Therefore, we assume $z_{kcjr}$ can only take two possible values,
with $z_{kcjr} = 1$ (or $0$) indicating that the corresponding locus has a
mutation (or does not have a mutation) compared to the reference
genome, respectively.  
For example, in Fig. \ref{fig:sc_evo}, we have $K = 2$ mutation pairs
and $C = 4$ subclones.
For mutation pair $k = 2$ in subclone $c = 4$, the allele $j=1$
harbors no mutation, while the allele $j=2$ has a mutation at the
first locus $r=1$, which translates to $\bz_{24} = (00, 10)$ (writing
00 as a shorthand for $(0,0)^T$, etc.). 
Altogether, $\bz_{kc}$ can take $2^4 = 16$ possible
values $\bz_{kc} \in \{(00, 00), (00, 01), \ldots, (11,11) \}$. Since
we do not have phasing information across mutation pairs, the
$\bz_{kc}$ values having mirrored columns  lead to exactly the same data
likelihood and thus are indistinguishable. 
Therefore, we reduce the number of
possible values of $\bz_{kc}$ to $Q = 10$. We list them below for
further reference: \\
$\bz^{(1)} = (00, 00)$, $\bz^{(2)} = (00, 01)$,
$\bz^{(3)} = (00, 10)$, $\bz^{(4)} = (00, 11)$, $\bz^{(5)} = (01,
01)$, $\bz^{(6)} = (01, 10)$, $\bz^{(7)} = (10, 10)$, $\bz^{(8)} =
(01, 11)$, $\bz^{(9)} = (10, 11)$ and $\bz^{(10)} = (11, 11)$. \\
We assume that the normal subclone has no mutation, $\bz_{k1} = \bz^{(1)}$ for
all $k$, indicating all mutations are somatic. 
In addition to these $C$ true subclones, we introduce a background
subclone, indexed as $c=0$ and without biological meaning, to account
for experimental noise and tiny subclones that are not detectable
given the sequencing depth. 
We assume that the background subclone is a random mixture of all
possible genotypes. See more discussion in Section~\ref{sc:sampling_model}.

Finally, we introduce notation for mixing proportions.
Suppose $T$ tissue samples are dissected from the same patient. We
assume that the samples are admixtures of $C$ subclones, each sample
with a different set of mixing proportions (population
frequencies). We use a $T \times (C+1)$ matrix $\bw$ to record the
proportions, where $w_{tc}$ represents the population frequencies
of subclone $c$ in sample $t$, $0 < w_{tc} < 1$ and $\sum_{c=0}^C
w_{tc} = 1$. The proportions $w_{t1}$ denotes the proportion of
normal cells contamination in sample $t$. 

The rest of the paper is organized as follows: Section~\ref{sec:model}
and Section~\ref{sec:post} describe the latent feature allocation
model and posterior inference, respectively. Section~\ref{sec:sim}
presents two simulation studies. Section~\ref{sec:real} reports
analysis results for an actual experiment.  We conclude with a
discussion in Section~\ref{sec:disc}.

\section{The PairCloneTree Model}
\label{sec:model}

\subsection{Sampling Model}
\label{sc:sampling_model}

Let $\bN$ be a $T \times K$ matrix with $N_{tk}$ representing read
depth for mutation pair $k$ in sample $t$. It records the number of
times any locus of the mutation pair is covered by sequencing reads
(see Fig.~\ref{fig:lochap_data}). 
Let $\bs_{tki} = \left(s_{tkir}, r = 1, 2 \right)$ be a specific short
read where $r = 1, 2$ index the two loci in a mutation pair, $i = 1,
2, \ldots, N_{tk}$. We use $s_{tkir} = 1$ (or $0$) to denote a
variant (reference) sequence at the read, 
compared to the reference genome. An important feature of the data is
that read $i$ may not overlap with locus $r$.
We use $s_{tkir} = -$ to represent the missing sequence on the
read. Reads that do not overlap with either of the two loci are not
included in the model as they do not contribute any information about
the mutation pair. In summary, $\bs_{tki}$ can take $G = 8$
possible values, 
$$
   \bs_{tki} \in \{\bs^{(1)}, \ldots, \bs^{(8)} \} =
       \{00, 01, 10, 11, -0, -1, 0-, 1- \}.
$$
Among all $N_{tk}$ reads, let
$n_{tkg} = \sum_i I \left(\bs_{tki} = \bs^{(g)} \right)$ be the number
of short reads having genotype $\bs^{(g)}$. As illustrated in
Fig.~\ref{fig:lochap_data} out of total 4 reads ($N_{t1} = 4$), we
have $n_{t12} = 1, n_{t13} = 1,n_{t16} = 1$ and $n_{t18}=1$.

We assume a multinomial sampling model for the observed read counts
\begin{align*}
   (n_{tk1}, \ldots, n_{tk8}) \mid N_{tk} \sim \Mn(N_{tk}; p_{tk1}, \ldots, p_{tk8}),
\end{align*}
where $p_{tkg}$ is the probability of observing a short read $\bs_{tki}$
with genotype $\bs^{(g)}$. Later we link $p_{tkg}$ with the
underlying subclone structures.

If desired, it is straightforward to incorporate data for marginal SNV reads in the model. These reads can be treated as, without loss of generality, right missing reads, i.e. $s_{tki2} = -$. In this case, $n_{tk1} = \ldots = n_{tk6} = 0$, and the multinomial sampling model reduces to a binomial model. The addition of marginal SNV counts does not typically improve inference. See more details in Chapter \ref{chap:PairClone}.

\paragraph{Construction of $p_{tkg}$}
For a short read $\bs_{tki}$, depending on whether it covers both loci or only one locus, we consider three cases: (i) a read covers both loci, taking values $\bs_{tki} \in \{\bs^{(1)}, \ldots, \bs^{(4)} \}$ (complete read); (ii) a read covers the second locus, taking values $\bs_{tki} \in \{\bs^{(5)}, \bs^{(6)} \}$ (left missing read); and (iii) a read covers the first locus, taking values $\bs_{tki} \in \{\bs^{(7)}, \bs^{(8)} \}$ (right missing read). 
Let $v_{tk1}, v_{tk2}, v_{tk3}$ denote the probabilities of observing a short read satisfying cases (i), (ii) and (iii), respectively. Conditional on cases (i), (ii) or (iii), let $\tp_{tkg}$ be the conditional probability of observing $\bs_{tki} = \bs^{(g)}$. We have $p_{tkg} = v_{tk1} \, \tp_{tkg}, g = 1, \ldots,
4$, $p_{tkg} = v_{tk2} \, \tp_{tkg}, g = 5, 6$, and $p_{tkg} = v_{tk3}
\, \tp_{tkg}, g = 7, 8$. We assume non-informative missingness and do not make inference on $v$'s, so they remain constants in the likelihood. 

We express $\tp_{tkg}$ in terms of $\bZ$ and $\bw$ based on the following generative model.
Consider a sample $t$. To generate a short read, we first select a subclone $c$ with probability $w_{tc}$.  Next we select with probability $0.5$ one of the two 
alleles $j= 1 , 2$. Finally, we record the read $\bs^{(g)}$, $g=1,2,3$ or $4$, corresponding to the chosen allele $\bz_{kcj}=(z_{kcj1},z_{kcj2})$.
In the case of left (or right) missing locus we observe $\bs^{(g)}$, $g=5$
or $6$ (or $g=7$ or $8$), corresponding to the observed locus of the chosen allele.
Reflecting these three generative steps, we denote the probability of
observing a short read $\bs^{(g)}$ from subclone $c$ that bears sequence $\bz_{kcj}$ by 
\begin{equation}
 A(\bs^{(g)}, \bz_{kc}) =
   \sum_{j = 1}^2 0.5\,\times\,I(s^{(g)}_1 = z_{kcj1})\, I(s^{(g)}_2=z_{kcj2}),
 \label{eq:A-PCT}
\end{equation}
with the understanding that $I(- = z_{kcjr}) \equiv 1$ for missing reads.
Implicit in \eqref{eq:A-PCT} is the restriction
$A(\bs^{(g)}, \bz_{kc}) \in \{0, 0.5, 1\}$, depending on the arguments.

Finally, using the conditional probabilities $A(\cdot)$ we obtain the marginal probability
of observing a short read $\bs^{(g)}$ from the tumor sample $t$ with $C$ subclones with cellular proportions $\{ w_{tc}\}$ as
\begin{equation}
   \tp_{tkg} = \sum_{c = 1}^C w_{tc}\,A(\bs^{(g)}, \bz_{kc}) + w_{t0} \, \rho_g.
   \label{eq:ptkg}
\end{equation}
The first term in Eq.~\ref{eq:ptkg} states that the probability of observing a short read with genotype $\bs^{(g)}$ is a weighted sum of the $A$'s across all the subclones.  Here $w_{t0} \rho_g$ stands for the probability of observing
$\bs^{(g)}$ due to random noise. It can be thought of as a
background subclone with weight  $w_{t0}$, which is a random mixture
of four genotypes 00, 01, 10 and 11 with proportions $\rho_g$. We
assume the random noise does not differ across different mutation
pairs, thus $\rho_g$ does not have an index $k$. Note that
$\rho_1+\ldots + \rho_4= \rho_5+\rho_6 = \rho_7+\rho_8 = 1$. Again,
the background subclone  (denoted by $c=0$) has no biological meaning
and is only used to account for noise and artifacts in the NGS data
(sequencing errors, mapping errors, etc.).

\subsection{Prior Model}
We construct a hierarchical prior model, starting
with $p(C)$, then a prior on the tree for a given number of nodes,
$p(\Tau \mid C)$, and  finally a prior on the subclonal genotypes
given the phylogenetic tree $\Tau$.

\paragraph{Prior for $C$ and $\Tau$}  
We assume a geometric prior for the number of subclones,
$p(C) = (1 - \alpha)^{C - 1} \alpha$, $C \in \{1, 2, 3,
\ldots\}$. Conditional on $C$, the prior on the tree, 
$p(\Tau \mid C)$ is as in~\cite{chipman1998bayesian}. For a tree with $C$ nodes, we let 
\begin{align*}
p(\Tau \mid C) \propto \prod_{c = 1}^C (1 + \eta_c)^{-\beta}, 
\end{align*}
where $\eta_c$
is the depth of node $c$, or the number of generations between node
$c$ and the normal subclone $1$. The prior penalizes deeper trees and
thus favors parsimonious representation of subclonal structure. 


\paragraph{Prior for $\bZ$} 
The subclone genotype matrix $\bZ$ can be thought of as a feature
allocation for categorical matrices. The mutation pairs are the
objects, and the subclones are the latent features chosen by the
objects. Each feature has 10 categories corresponding to the $Q=10$ different
genotypes. Conditioning on $\Tau$ the 
subclone genotype matrix needs to introduce dependence across features
to reflect the assumed phylogeny.
We construct a prior for $\bZ$ based on the following
generative model.  
We start from a normal subclone denoted by $\bz_{\cdot 1} = \bm 0$. 
Now consider a subclone $c>1$ and defined by $\bz_{\cdot c}$. 
The subclone preserves all mutations from its parent $\bz_{\cdot \Tau_c}$,
but also gains a Poisson number of new mutations. We assume the new
mutations randomly happen at the unmutated loci of the parent
subclone. A formal description of prior of $\bZ$ follows.

For a subclone $c$, let $\ell_{kc} = \sum_{j, r} z_{kcjr}$ denote the
number of mutations in mutation pair $k$, and let
$\mathcal{L}_c = \{k: \ell_{kc} < 4 \}$ denote the mutation pairs
in subclone $c$ that have less than four mutations. 
Let $m_{kc} = \ell_{kc} - \ell_{k \Tau_c}$ denote the number of new mutations that
mutation pair $k$ gains compared to its parent, and let $m_{\cdot c} =
\sum_k m_{kc}$. 
We assume
(i) The child subclone should acquire at least one additional mutation
compared with its parent (otherwise subclone $c$ would be identical
to its parent $\Tau_c$).
(ii) If the parent has already acquired all four mutations for a
given $k$, then the child can not gain any more new mutation.   
That is, if $\ell_{k \Tau_c} = 4$, then $m_{kc} = 0$.  
(iii) Each mutation pair can gain at most one additional mutation in
each generation, $m_{kc} \in \{0, 1\}$.  
Based on these assumptions,  given a parent subclone
$\bz_{\cdot \Tau_c}$, we construct a child subclone $\bz_{\cdot c}$ as follows. 
Let $\MM_c = \{k: m_{kc} = 1 \}$ be the set of   mutation   pairs in
subclone $c$  where new mutations are gained.
Let $\Choose(\LL, m)$ denote
a uniformly chosen subset of $\LL$ of size $m$, and let
$X \sim \tpois(\lambda; [a, b])$ represent a Poisson distribution with mean
$\lambda$, truncated to $a \leq X \leq b$. We assume
\begin{align}
 m_{\cdot c} \mid \bz_{.\Tau_c},\Tau,C &\sim \tpois(\lambda; [1, |\mathcal{L}_{\Tau_c}| ]), \nonumber\\
  \MM_c \mid m_{.c},\bz_{.\Tau_c},\Tau,C &\sim \Choose(\LL_{\Tau_c}, m_{\cdot c}).
\label{eq:prior_z}
\end{align}
The lower bound and upper bound of the truncated Poisson reflect
assumptions (i) and (ii) respectively. 
Also, Eq.~\ref{eq:prior_z} implicitly captures assumption (iii).    

Next, for a mutation pair that gains one new mutation, we assume the
new mutation randomly arises in any of the unmutated loci in the
parent subclone.
Let $\ZZ_{kc} = \{(j, r): z_{kcjr} = 0 \}$,  and let 
$\unif(A)$ denote a uniform distribution over the set $A$.
We first choose
\begin{align*}
(j^*, r^*) \mid \bz_{.\Tau_c},\Tau,C \sim \unif(\ZZ_{k \Tau_c}),
\end{align*}
and then set $z_{kcj^* r^*} = 1$. 
So we have 
\begin{align*}
p(\bZ \mid \Tau, C) \propto \prod_{c=2}^C \tpois(m_{\cdot c}; [1, |\mathcal{L}_{\Tau_c}| ]) . \frac{1}{\left(\begin{array}{c} |\mathcal{L}_{\Tau_c}| \\ m_{\cdot c} \\ 
 \end{array}\right)} . \prod_{k\in \mathcal{M}_c} \frac{1}{|\ZZ_{k \Tau_c}|}.
\end{align*}

\paragraph{Prior for $\bw$ and $\brho$}
We design $p(\bw)$ in such a manner that we could put an informative prior for
$w_{t1}$ if a reliable estimate for tumor purity is available based
on some prior bioinformatics pipeline 
(e.g.~\cite{van2010allele,carter2012absolute}). 
Recall that $c=1$ is the normal subclone, i.e.,
$w_{t1}$ is the normal subclone proportion, and
that $\sum_{{c=0},{c\neq 1}}^C w_{tc} + w_{t1} = 1$. We assume a
Beta-Dirichlet prior on $\bw$ such that,
\begin{equation*}
  w_{t1} \sim \Be(a_p,b_p);\quad \mbox{and} \;\;
  \frac{w_{tc}}{(1-w_{t1})} \sim \Dir (d_0,d,\cdots,d),
\end{equation*}
where $c = 0, 2, 3,\cdots,C$. 
We set $d_0 << d$ as $w_{t0}$ is only a correction term to account for background noise and model mis-specification term.

The model is completed with a prior for $\brho = \{\rho_g\}$. We
consider complete read, left missing read and right missing read
separately, and assume
\begin{align*}
  \rho_{g_1} \sim \Dir(d_1, \ldots, d_1); \quad
  \rho_{g_2} \sim  \Dir(2d_1, 2d_1); \quad
  \rho_{g_3} \sim  \Dir(2d_1, 2d_1),
\end{align*}
where $g_1=\{1,2,3,4\}$, $g_2=\{5,6\}$ and $g_3 = \{7,8\}$.

\section{Posterior Inference}
\label{sec:post}
Let $\bx = (\bZ, \bw, \brho)$ denote the unknown parameters except for
the number of subclones $C$ and the tree $\Tau$. Markov chain Monte
Carlo (MCMC) simulation from the posterior $p(\bx
\mid \bn, \Tau, C)$ is used to implement posterior inference.
Gibbs sampling transition probabilities are used
to update $\bZ$, and Metropolis-Hastings transition probabilities are
used to update $\bw$ and $\brho$. For example, we update $\bZ$ by row
with 
\begin{multline*}
p(\bz_{k \cdot} \mid \bz_{-k \cdot}, \ldots) \propto \prod_{t = 1}^T \prod_{g = 1}^G \left[ \sum_{c = 1}^C w_{tc} \, A(\bh_g, \bz_{kc}) + w_{t0} \, \rho_g \right]^{n_{tkg}} \cdot \\
p(\bz_{k \cdot} \mid  \bz_{-k \cdot}, \Tau, C),
\end{multline*}
where $\bz_{k \cdot}$ is a row of $\bZ$ satisfying the phylogeny
$\Tau$.  

Since the posterior distribution $p(\bx \mid \bn, \Tau, C)$
is expected to be highly multi-modal, we utilize parallel
tempering \citep{geyer1991markov} to improve the mixing of the
chain. Specifically, we use OpenMP parallel computing
API \citep{dagum1998openmp} in C++, to implement a scalable parallel
tempering algorithm.

\paragraph{Updating $C$ and $\Tau$} 
In general, posterior MCMC on tree structures can be very
challenging to implement \citep{chipman1998bayesian,Deni:Mall:Smit:baye:1998}.
However, the problem here is manageable since plausible numbers for
$C$ constrain $\Tau$ to moderately small trees.
We assume that the number of nodes is {\it a priori} restricted to
$C_{\min} \leq C \leq C_{\max}$. Conditional on the number of
subclones $C$, the number of possible tree topologies is
finite.
Let $\mathscr{T}$ denote the (discrete) sample space of $(\Tau, C)$. 
Updating the values of $(\Tau, C)$ involves trans-dimensional MCMC.
At each iteration, we propose new values for $(\Tau, C)$ from a
uniform proposal, $q(\tTau, \tC \mid \Tau, C) \sim
\unif(\mathscr{T})$.

In order to search the space $\mathscr{T}$ for the number of subclones
and trees that best explain the observed data, we follow a
similar approach as in \cite{lee2015bayesian} and Chapter \ref{chap:PairClone} (motivated by
fractional Bayes' factor in \cite{ohagan1995}) that splits the data
into a training set and a test set. Recall that $\bn$ represents the
read counts data. We split $\bn$ into a training set
$\bn'$ with $n_{tkg}' = b n_{tkg}$, and a test set $\bn''$ with
$n_{tkg}'' = (1-b) n_{tkg}$.  
Let $p_b(\bx \mid \Tau, C) = p(\bx \mid
\bn', \Tau, C)$ be the posterior evaluated on the
training set only. We use $p_b$ in two instances. First, $p_b$ is used
as an informative prior instead of the original prior $p((\bx \mid
\Tau, C)$, and second, $p_b$ is used as a proposal distribution for
$\tbx$, $q(\tbx \mid \tilde{\Tau}, \tC) = p_b(\tbx \mid \tilde{\Tau},
\tC)$. Finally, the acceptance probability of proposal $(\tilde{\Tau},
\tC, \tbx)$ is evaluated on the test set.
Importantly, in the acceptance probability the (intractable) normalization
constant of $p_b$ cancels out, making this approach computationally
feasible.
\begin{multline*}
  p_{\text{acc}}(\Tau, C, \bx, \tTau, \tC,\tbx) = 1 \wedge
  \frac{p(\bn'' \mid \tbx, \tTau, \tC)}
       {p(\bn'' \mid \bx, \Tau, C)} \cdot 
  \frac{p(\tTau, \tC) \cancel{p_b(\tbx \mid \tTau, \tC)}}
       {p(\Tau, C)   \cancel{p_b(\bx  \mid \Tau, C  )}} \cdot \\
  \frac{q(\Tau, C \mid \tTau, \tC) \cancel{q(\bx \mid \Tau, C)}}
  {q(\tTau, \tC \mid \Tau, C) \cancel{q(\tbx \mid \tTau, \tC)}}.
\end{multline*}
Here we use $p_b$ as an informative proposal distribution for $\tbx$ to achieve a better mixing Markov chain Monte Carlo simulation with reasonable acceptance probabilities. Without the use of an informative proposal, the proposed new tree is almost always rejected because the multinomial likelihood with the large sample size is very peaked.
Under the modified prior $p_b(\cdot)$, the resulting conditional posterior on $\bx$ remains entirely unchanged, $p_b(\bx \mid \Tau, C, \bn) = p(\bx \mid \Tau, C, \bn)$ (Appendix \ref{app:sec:updatec}).

The described uniform tree proposal is in contrast to usual search algorithms for trees that generate proposals from neighboring trees.
The advantage of this kind of proposal is to ensure a reasonable acceptance probability. But such algorithms have an important drawback that they quickly gravitate towards a local mode and then get stuck. A possible approach to addressing this problem is to repeatedly restart the algorithm from different starting trees. See  \cite{chipman1998bayesian} for more details. Our uniform tree proposal combined with the data splitting scheme is another way to mitigate this challenge, efficiently searching the tree space while keeping a reasonable acceptance probability.

\paragraph{Point estimates for parameters} 
All posterior inference is contained in the posterior
distribution for $\bx, C$ and $\Tau$.
For example, the marginal posterior distribution of $C$ and $\Tau$
gives updates posterior probabilities for all possible values of $C$
and $\Tau$. It is still useful to report point estimates. We use the
posterior modes $(\Chat, \Tauhat)$ as point estimates for $(C, \Tau)$,
and conditional on $\Chat$ and $\Tauhat$, 
we use the maximum a posteriori (MAP) estimator as an estimation for
the other parameters. 
The MAP is approximated as the MCMC sample with highest posterior
probability.
Let $\{ \bx^{(l)}, l = 1, \ldots, L \}$ be a set
of MCMC samples of $\bx$, and 
\begin{align*}
\hat{l} = \argmax_{l \in \{1, \ldots, L\}} \; p(\bn \mid \bx^{(l)}, \Tauhat, \Chat) \, p(\bx^{(l)} \mid \Tauhat, \Chat).
\end{align*}
We report point estimates as $\Zhat = \bZ^{(\hat{l})}$, $\what =
\bw^{(\hat{l})}$ and $\hat{\brho} = \brho^{(\hat{l})}$.

\section{Simulation Studies}
\label{sec:sim}
We present two simulation studies to assess the proposed approach. We simulate single sample and multi-sample data with different read depths to test the performance of our model in different scenarios. In both simulation studies, we generate hypothetical read count data for $K = 100$ mutation pairs, which is a typical number of mutation pairs in a tumor sample. 
However, if needed, a much larger number of SNVs could be included in the model, with the only limiting concern being computational efficiency, which remains a challenge for all current methods.

\subsection{Simulation 1}
In the first simulation study, we consider $T = 1$ sample, which is
the case for most real-world tumor cases due to the challenge in
obtaining multiple samples from a patient.  However, this does not
rule out meaningful inference. As we will show, with good read
depth, the simulation truth can still be recovered. 
Note that the relevant sample size is not the number of tissue
samples, but closer to the number of reads, which is large even for
$T=1$. 
  
\begin{figure}[h!]
\begin{center}
\begin{subfigure}[t]{.325\textwidth}
\centering
\includegraphics[width=\textwidth]{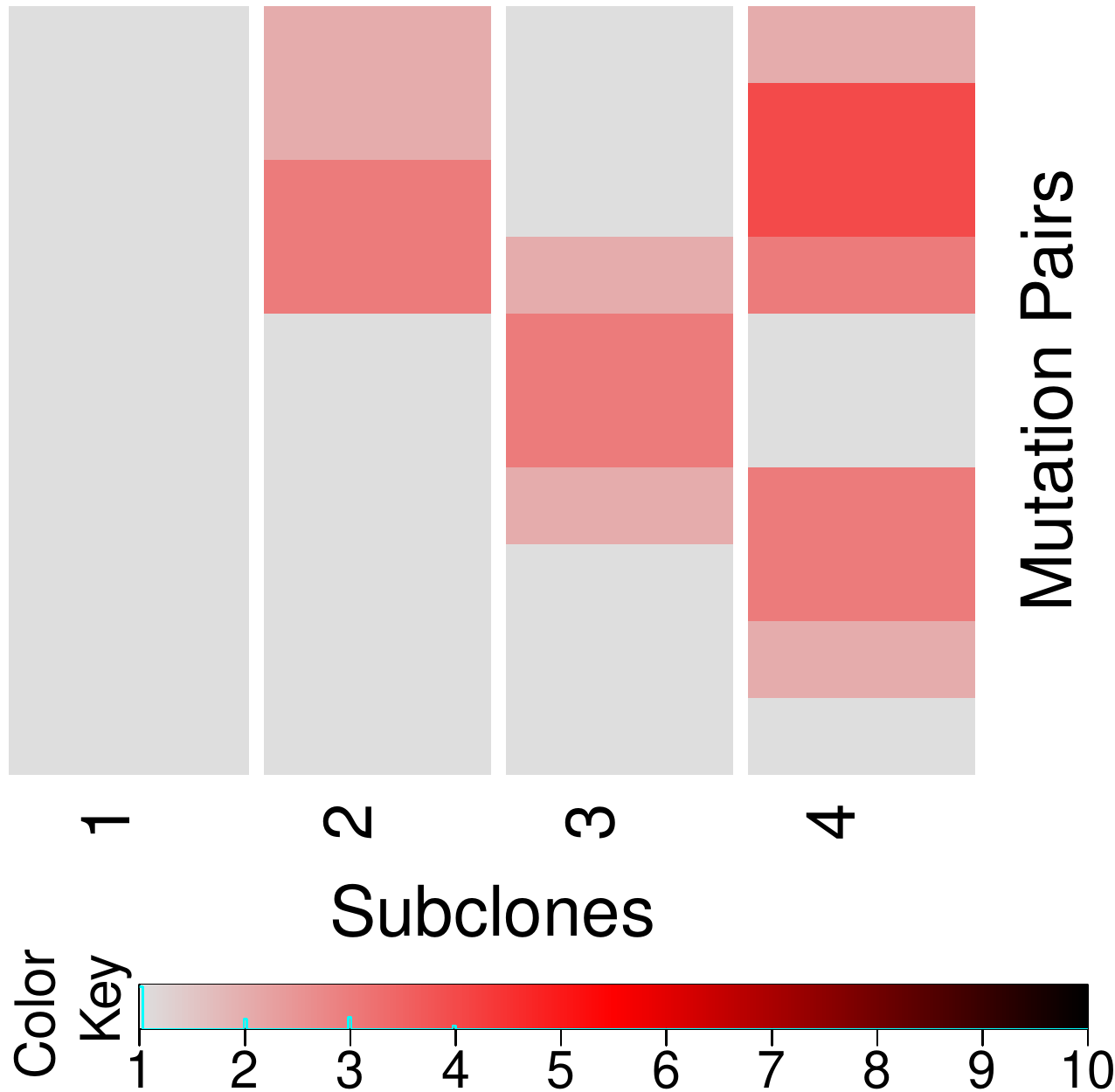}
\caption{$\bZ$}
\end{subfigure}
\begin{subfigure}[t]{.325\textwidth}
\centering
\includegraphics[width=\textwidth]{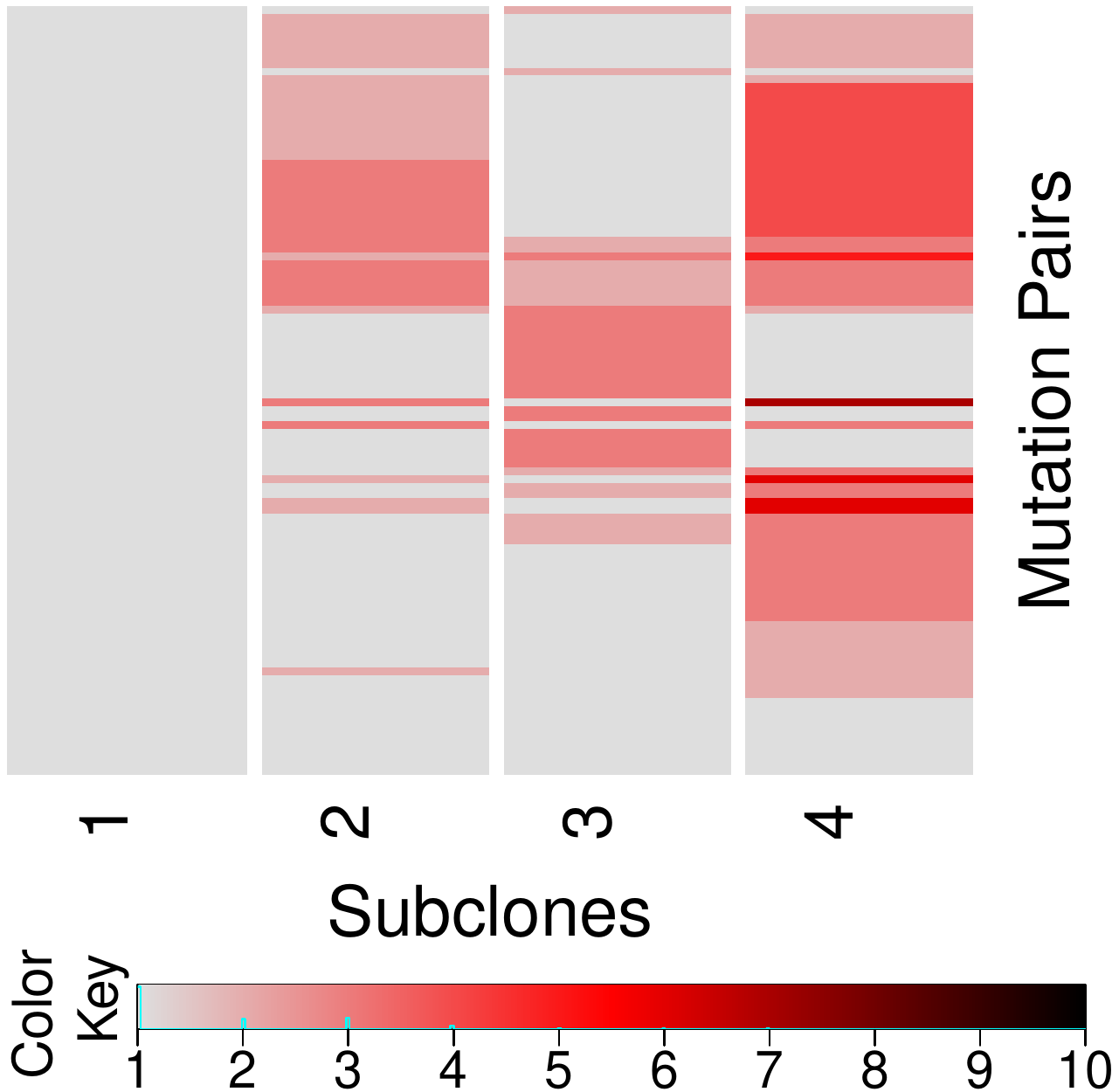}
\caption{$\Zhat_{\text{500x}}$}
\end{subfigure}
\begin{subfigure}[t]{.325\textwidth}
\centering
\includegraphics[width=\textwidth]{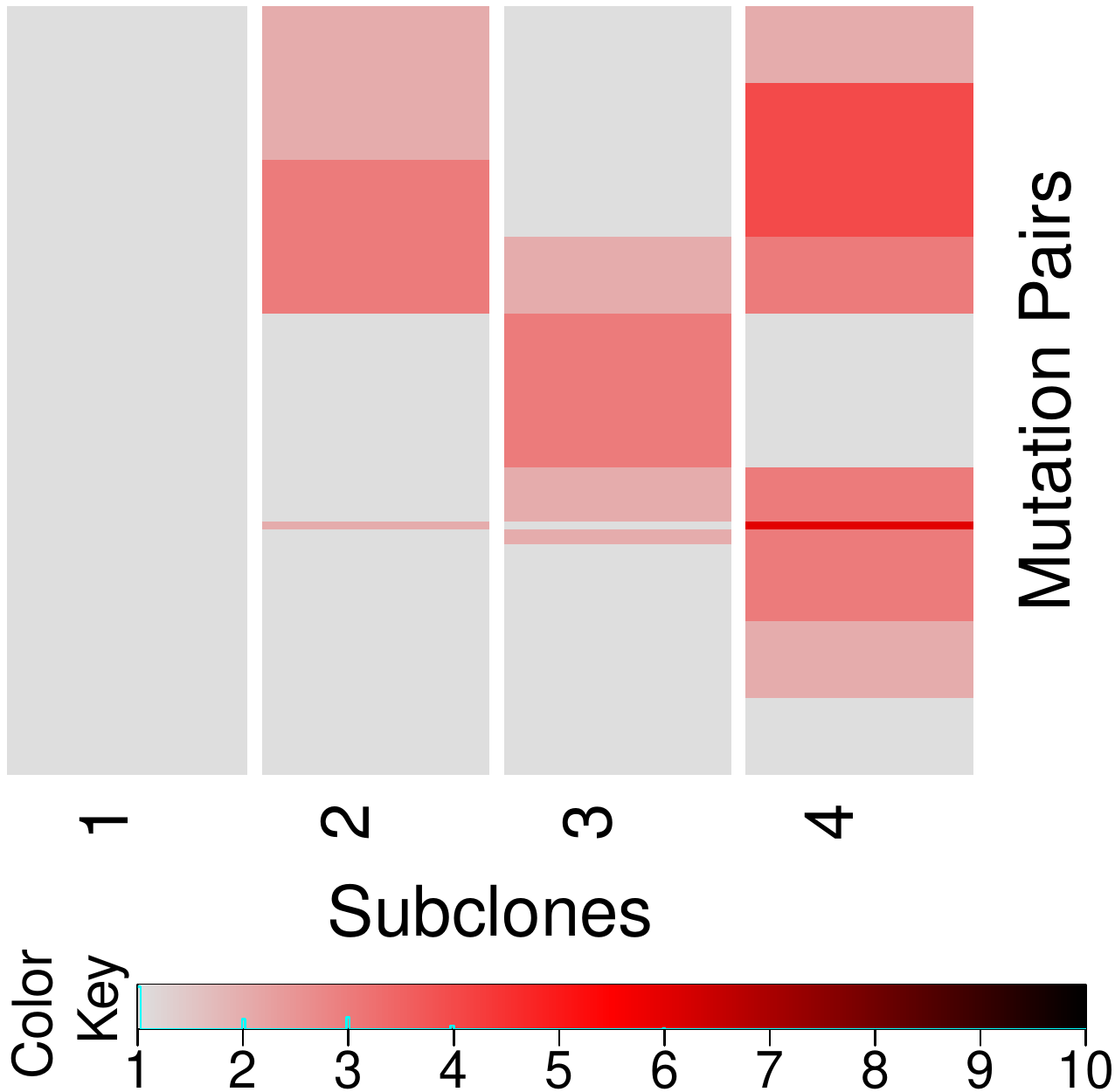}
\caption{$\Zhat_{\text{2000x}}$}
\end{subfigure}
\begin{subfigure}[t]{.29\textwidth}
\centering
\includegraphics[width=.9\textwidth]{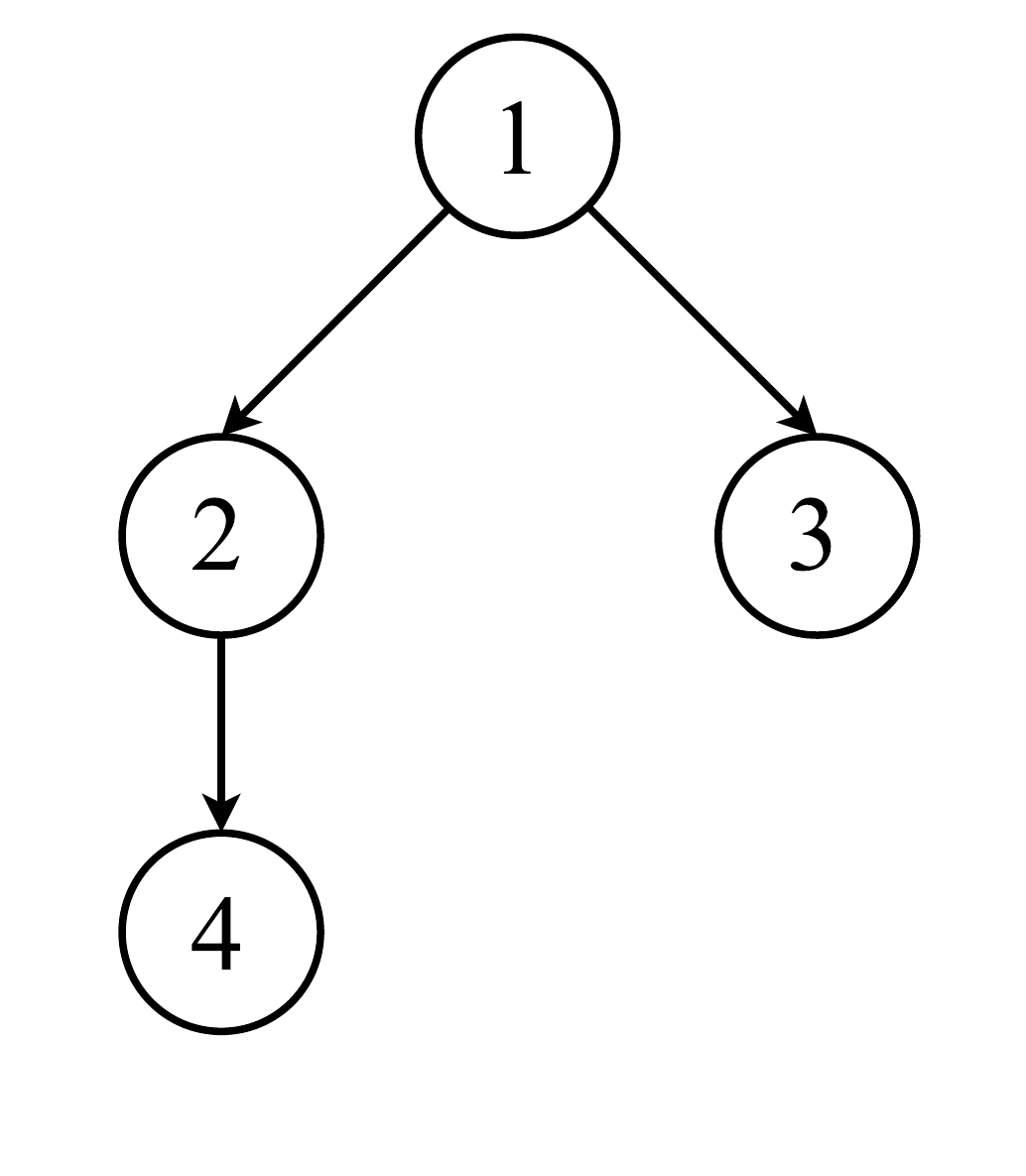}
\caption{True phylogeny}
\end{subfigure}
\hspace{8mm}
\begin{subfigure}[t]{.36\textwidth}
\centering
\includegraphics[width=\textwidth]{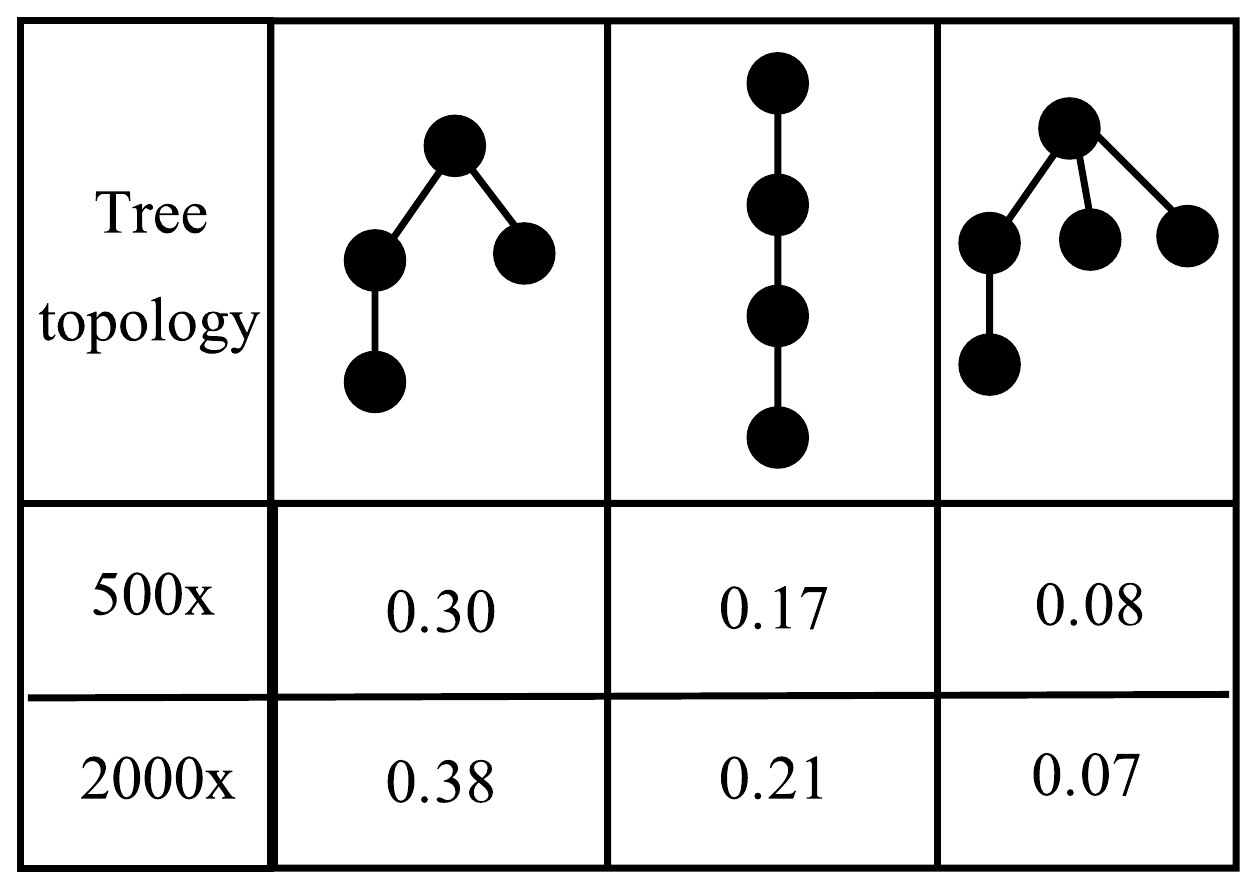}
\caption{Posterior prob. of tree topology}
\end{subfigure}
\end{center}
  \caption{Simulation 1. Simulation truth $\bZ$ (a) and phylogeny (d), and posterior
    inference under PairCloneTree (b, c, e). }
  \label{fig:sim1}
\end{figure}

We consider $K = 100$ mutation pairs and assume
a simulation truth with $C = 4$ latent
subclones. Fig.~\ref{fig:sim1}(a) and (d) show the true
underlying subclonal genotypes and phylogeny, respectively. 
We use a heatmap to show the subclone matrix $\bZ$, where colors from
light gray to red to black are used to represent genotypes $\bz^{(1)}$
to $\bz^{(10)}$. The subclone weights are simulated from $\Dir(0.01,
\sigma(15, 10, 8, 5))$, where $\sigma(15, 10, 8, 5)$ stands for a
random permutation of the four numbers.
For the single sample in this simulation we get $\bw = (0.000$, $0.135$, $0.169$, $0.470$, $0.226)$.
The noise factor $\brho$ is
generated from its prior with $d_1 = 1$. In order to mimic a typical
rate of observing left (or right) missing reads, we set $v_{tk2} =
v_{tk3} = 0.25$, for $k = 1, \ldots, 50$, and $v_{tk2} = v_{tk3} =
0.3$, for $k = 51, \ldots, 100$. For the read depth $N_{tk}$, we consider
two scenarios. 
In the first scenario, we consider 500x depth and
generate $N_{tk} \sim \du([400, 600])$; in the second scenario, we
consider 2000x depth and generate $N_{tk} \sim \du([1900, 2100])$.
While these read depth values are impossible from existing
whole-genome sequencing technology, they are available from
whole-exome or targeted sequencing experiments.   

We fit the model with the following hyperparameters: $\alpha = 0.5$, $\beta = 0.5$, $d = 0.5$, $d_0 = 0.03$, $d_1 = 1$, where the values of $\alpha$ and $\beta$ imply mild penalty for deep and bushy trees \citep{chipman1998bayesian}, and other hyperparameters are generic non-informative choices.  We set $a_p = d, b_p = d_0 + (C-1)d$ for given $C$ as a non-informative prior choice and set $\lambda = 2K/C$ to express our prior belief that about half of the mutations occur uniformly at each generation. We use $C_{\min} = 2$ and $C_{\max} = 5$ as the range of $C$, since the vast majority of the methods in the literature show
that even though a tumor sample could possess thousands to millions of SNVs, the number
of inferred subclones usually is in the low single digit.
Empirically, we choose the training set fraction as $b = 0.95$, as it performs well in all simulation studies.
We run a total of 8000 MCMC simulations. Discarding the first 3000 draws as initial burn-in, we have a Monte Carlo sample with 5000 posterior draws. 


Posterior inference with 500x read depth is summarized in
Fig.~\ref{fig:sim1}(b, e). Fig.~\ref{fig:sim1}(e) shows 
the top three tree topologies and corresponding posterior
probabilities.
The posterior mode
recovers the true phylogeny. Fig.~\ref{fig:sim1}(b) shows the
estimated genotypes with 500x read depth, conditional on the posterior modes $(\Chat, \Tauhat)$. 
Some mismatches are due to
the single sample and limited read depth. The estimated subclone
proportions are 
$\what = (0.000$, $0.073$, $0.171$, $0.517$, $0.239)$,
which agrees with the truth.

Posterior inference with 2000x read depth is summarized in
Fig.~\ref{fig:sim1}(c, e). 
The posterior mode recovers the true
phylogeny. Fig.~\ref{fig:sim1}(c) shows the estimated genotypes. 
The simulation  shows how larger read depths improve posterior
accuracy and improve the power of recovering the latent structure. 
In particular, this shows that even with a single sample, with
reasonable read depth, the truth can still be recovered. The estimated
subclone proportions are 
$\what = (0.000$, $0.127$, $0.168$, $0.477$, $0.228)$.

\subsection{Comparison with Cloe and PhyloWGS}
\begin{figure}[h!]
\begin{center}
\begin{subfigure}[t]{.325\textwidth}
\centering
\includegraphics[width=\textwidth]{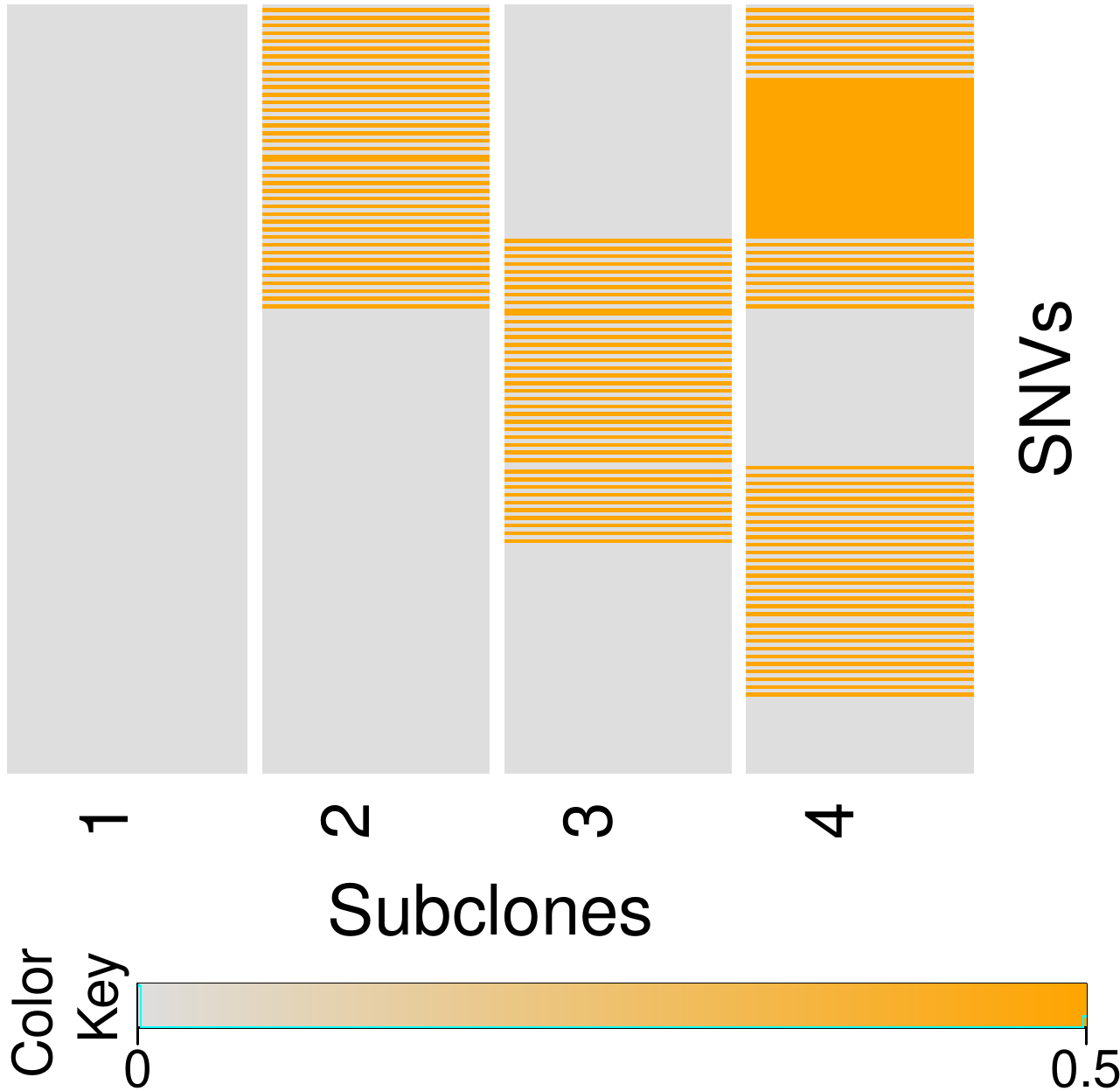}
\caption{$\bZ^{\text{Cloe}}$}
\label{fig:sim1_Z_Cloe}
\end{subfigure}
\begin{subfigure}[t]{.325\textwidth}
\centering
\includegraphics[width=\textwidth]{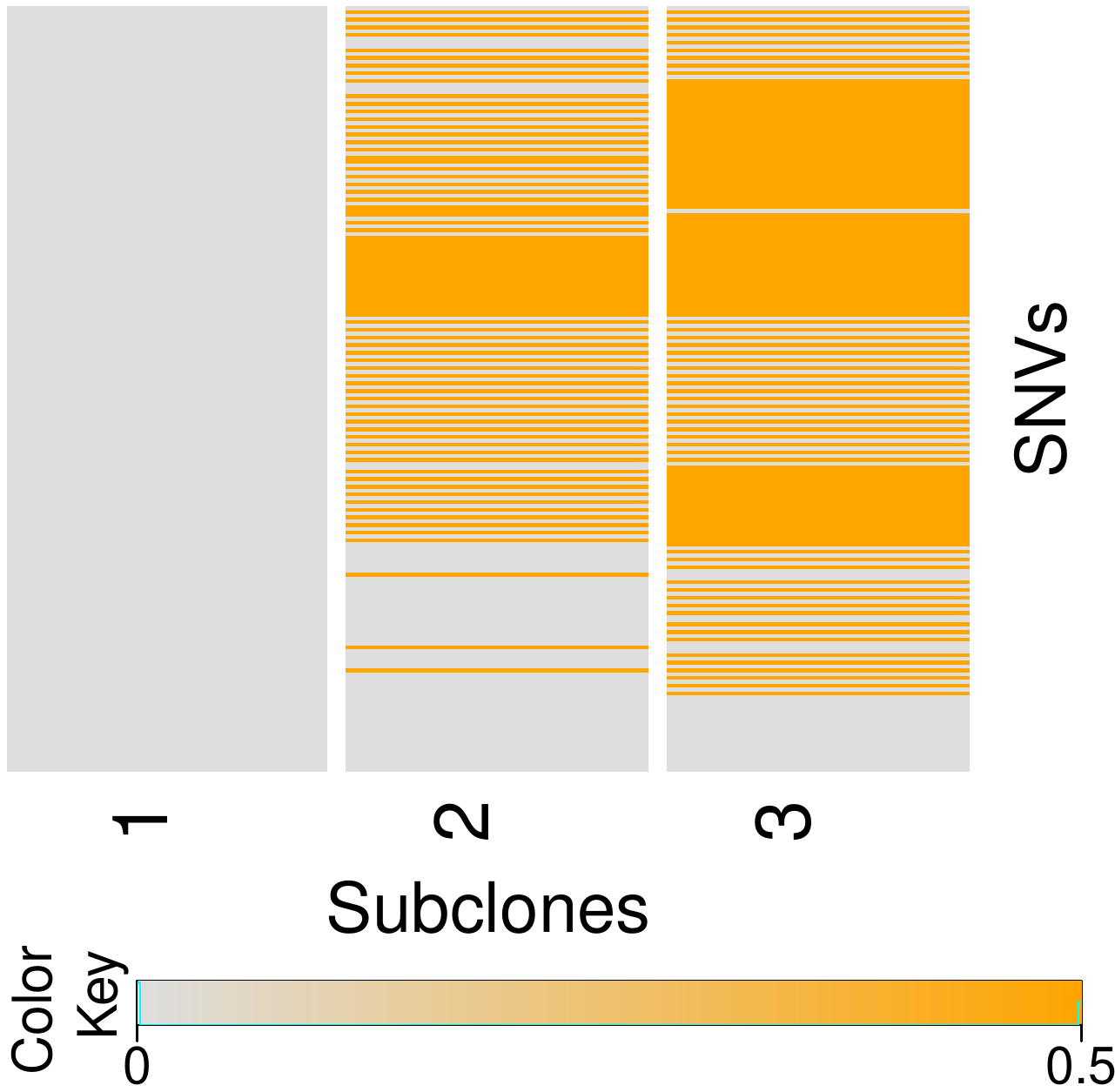}
\caption{$\Zhat_{\text{500x}}^{\text{Cloe}}$}
\label{fig:sim1_Zhat_Cloe}
\end{subfigure}
\begin{subfigure}[t]{.325\textwidth}
\centering
\includegraphics[width=\textwidth]{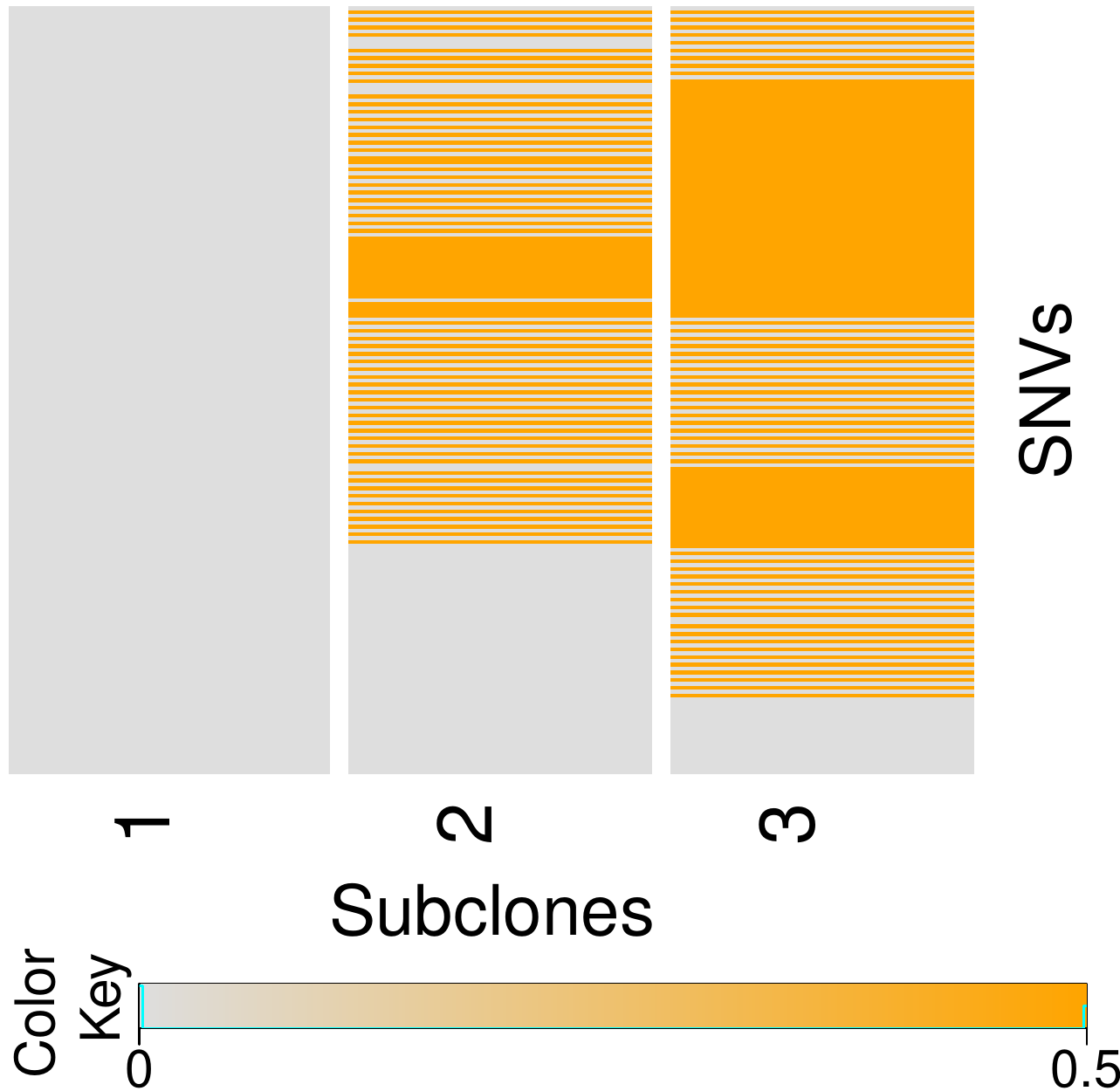}
\caption{$\Zhat_{\text{500x}}^{\text{PWGS}}$}
\label{fig:sim1_Zhat_PWGS}
\end{subfigure}
\end{center}
  \caption{Simulation 1. Simulation truth $\bZ^{\text{Cloe}}$ (a), and posterior
    inference under Cloe (b) and PhyloWGS (c). }
  \label{fig:sim1_compare}
\end{figure}

There is no other subclone calling method based on
paired-end read data that also infers phylogeny. 
We therefore compare with other similar model-based
approaches. In particular, we use Cloe \citep{marass2017phylogenetic} and 
PhyloWGS \citep{jiao2014inferring, deshwar2015phylowgs}
for inference with the same simulated data. 
These two methods also use highly structured Bayesian nonparametric priors and MCMC simulations for posterior inference.
Both methods take mutant read counts and read depths for SNVs as input.
Therefore, we discard the phasing information in
mutation pairs and only record marginal counts for SNVs as the input.
The simulation truth in Cloe and PhyloWGS's format is shown in Fig. \ref{fig:sim1_Z_Cloe}.
The orange color means a heterozygous mutation at the corresponding SNV locus.

\underline{Cloe} infers clonal genotypes and phylogeny based on a similar feature allocation model.
We run Cloe with the default hyperparameters for the same number of
8000 iterations with the first 3000 draws as initial burn-in. After that
we carry out model selection for $C$ with $2 \leq C \leq 5$.  For the 500x read
depth data, based on MAP estimate, Cloe reports 3 subclones with
phylogeny $1 \rightarrow 2 \rightarrow 3$, and the subclone genotypes
are shown in Fig. \ref{fig:sim1_Zhat_Cloe} with subclone proportions $\what^{\text{Cloe}} = (0.569$, $0.218$, $0.213)$. 
For the 2000x read depth data, Cloe infers 2 subclones (genotypes not
shown).

\underline{PhyloWGS}, on the other hand, infers clusters of mutations and phylogeny. One can then make phylogenetic analysis to conjecture subclones and genotypes.
Let $\tilde{\phi}_i$ denote the fraction of cells with a variant allele at
locus $i$. The $\tilde{\phi}_i$'s are latent quantities related to the observed VAF for each SNV. PhyloWGS infers the phylogeny by clustering SNVs with
matching $\tilde{\phi}_i$'s under a tree-structured prior for the unique values $\phi_j$. In particular, they use the tree-structured stick breaking process (TSSB)
\citep{adams2010tree}. 
The TSSB implicitly defines a prior
on the formation of subclones, including the prior on $C$ and the number of
novel loci that arise in each subclone.
In contrast, PairCloneTree explicitly defines these
model features, allowing easier prior control on $C$ and
$\mathcal{M}_c$.
We run PhyloWGS with the default hyperparameters and 2500 iterations with a burn-in of 1000 samples. We only consider loci with VAF $> 0$ as the other loci do not provide information for PhyloWGS clustering.
We then report cluster sizes and phylogeny based on MAP estimate. For the 500x read
depth data, PhyloWGS reports 3 subclones with
phylogeny $0 \rightarrow 1 (77, 0.429) \rightarrow 2 (53, 0.218)$, where 0 refers to the normal subclone, and the numbers in the brackets refer to the cluster sizes and cellular prevalences. The conjectured subclone genotypes are shown in Fig. \ref{fig:sim1_Zhat_PWGS}, with subclone proportions $\what^{\text{PWGS}} = (0.571$, $0.211$, $0.218)$.
For the 2000x read
depth data, PhyloWGS reports 3 subclones with
phylogeny $0 \rightarrow 1 (80, 0.431) \rightarrow 2 (50, 0.227)$ (genotypes not
shown).

Inferences under Cloe and PhyloWGS do not entirely recover the truth. The reason is probably that the common mutations of subclones 2 and 4 ($\mathcal{M}_2$ with a cellular prevalence of $0.169+0.226$) have a similar cellular prevalence with the mutations of subclone 3 ($\mathcal{M}_3$ with a cellular prevalence of $0.470$). Here we abuse the notation slightly and let $\mathcal{M}_c$ denote the new mutations that subclone $c$ gained. Therefore, Cloe infers that $\mathcal{M}_2$ and $\mathcal{M}_3$ belong to the same subclone ($\mathcal{M}_2^{\text{Cloe}} \approx \mathcal{M}_2 \cup \mathcal{M}_3$ and $\mathcal{M}_3^{\text{Cloe}} \approx \mathcal{M}_4$). Similarly, PhyloWGS clusters $\mathcal{M}_2$ and $\mathcal{M}_3$ together.
Using more informative mutation pairs data, PairCloneTree is able to identify that $\mathcal{M}_2$ and $\mathcal{M}_3$ belong to different subclones.
The comparisons support the argument in Section \ref{sec:intro} that the
inclusion of phasing information from the paired-end read data
increases statistical power in recovering the underlying structure.
Note that PairCloneTree is based on a different sampling model and has a very different representation of $\bZ$. Therefore, there is no obvious way to quantify the three model's performance under the same scale.


\subsection{Simulation 2}
\begin{figure}[h!]
\begin{center}
\begin{subfigure}[t]{.325\textwidth}
\centering
\includegraphics[width=\textwidth]{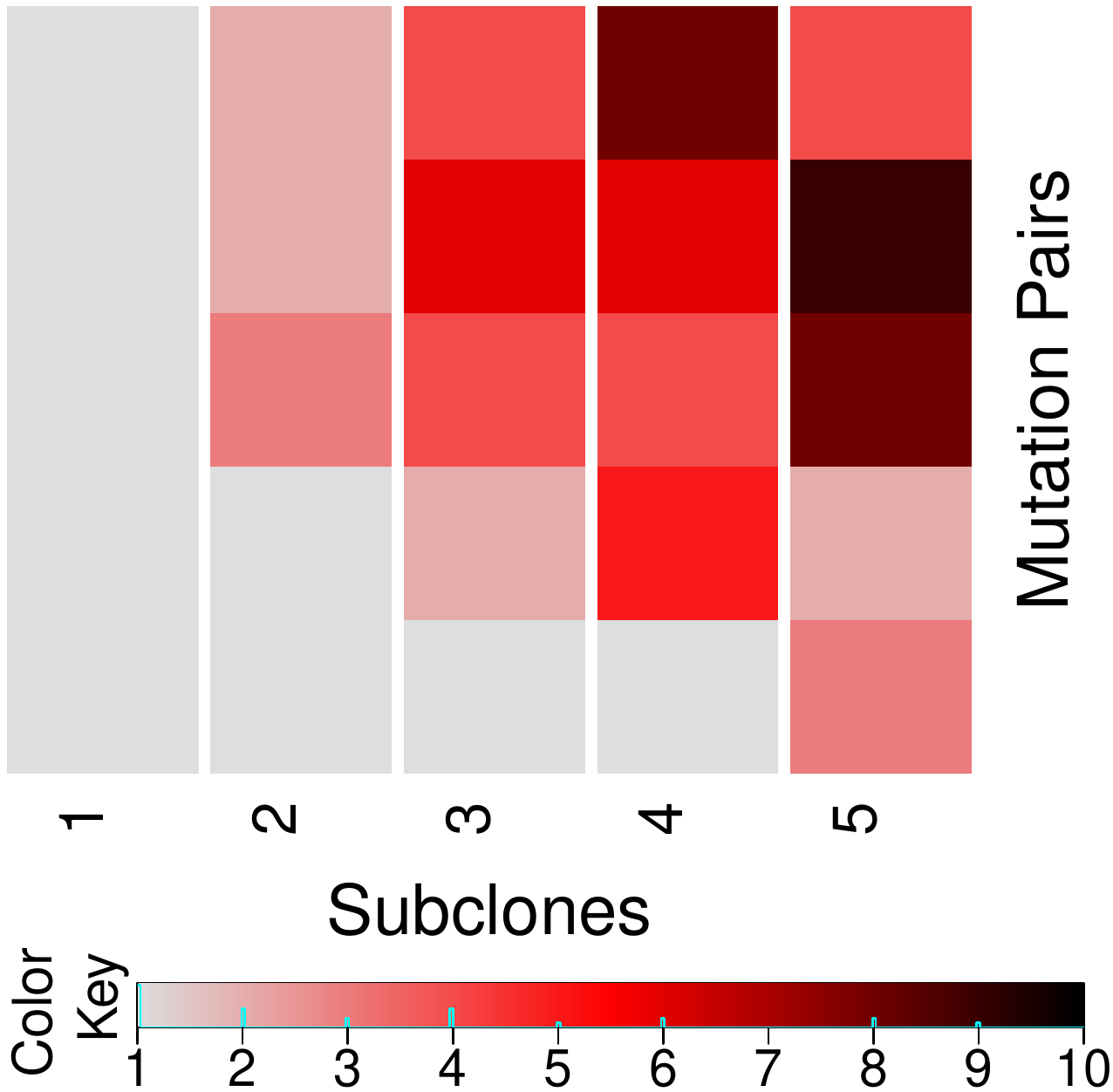}
\caption{$\bZ$}
\end{subfigure}
\begin{subfigure}[t]{.325\textwidth}
\centering
\includegraphics[width=\textwidth]{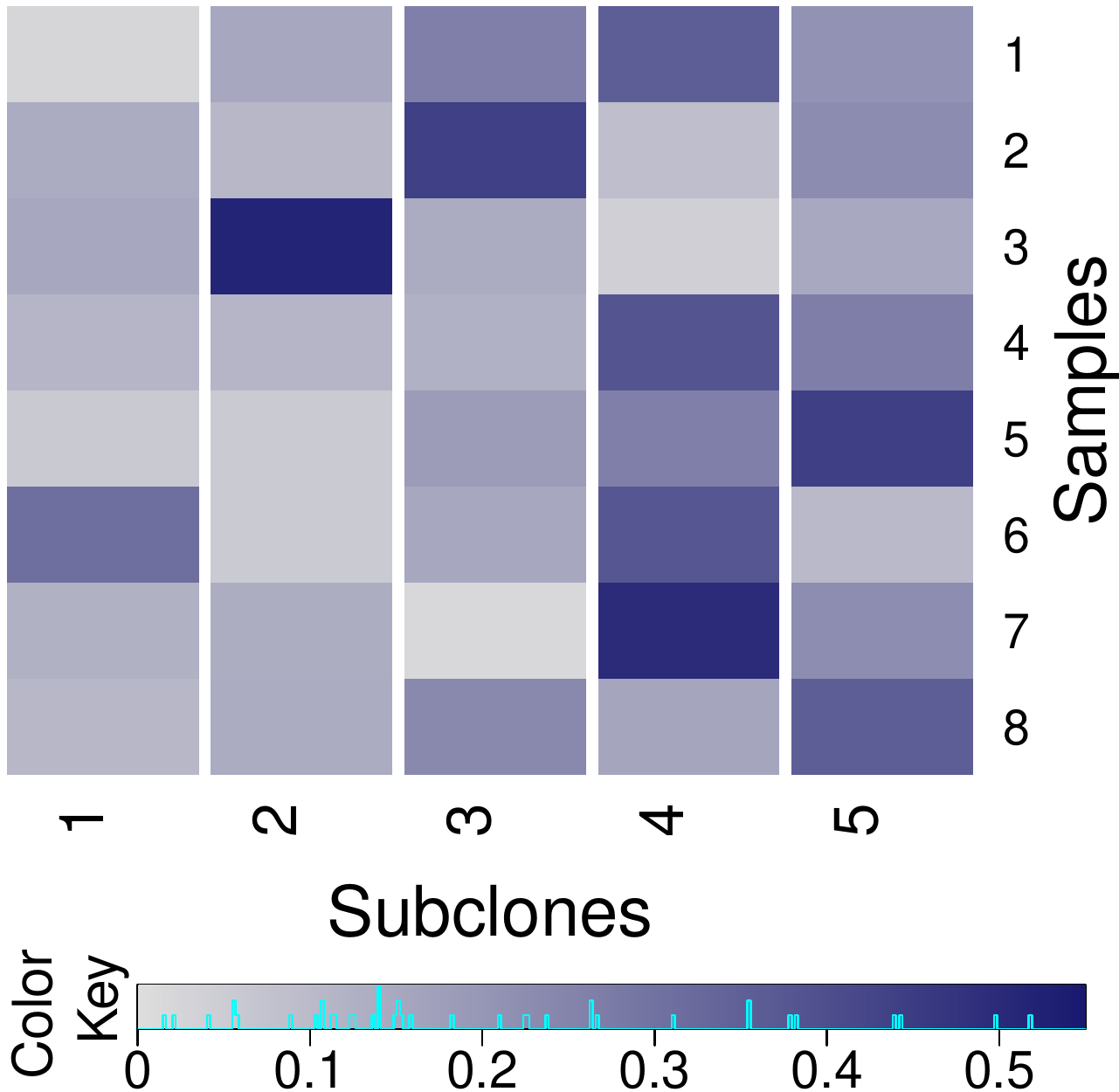}
\caption{$\bw$}
\end{subfigure}
\begin{subfigure}[t]{.325\textwidth}
\centering
\includegraphics[width=.75\textwidth]{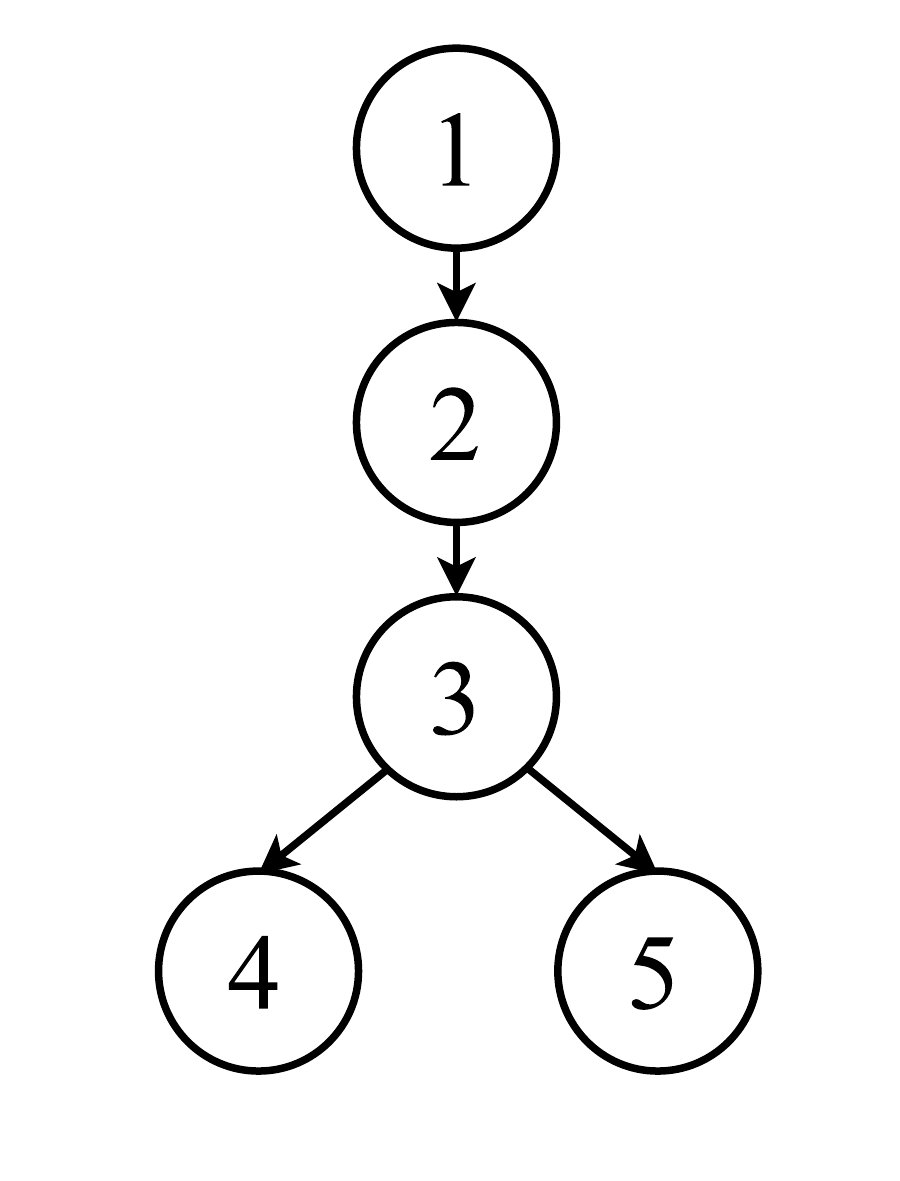}
\caption{True and est. phylogeny}
\end{subfigure}
\begin{subfigure}[t]{.325\textwidth}
\centering
\includegraphics[width=\textwidth]{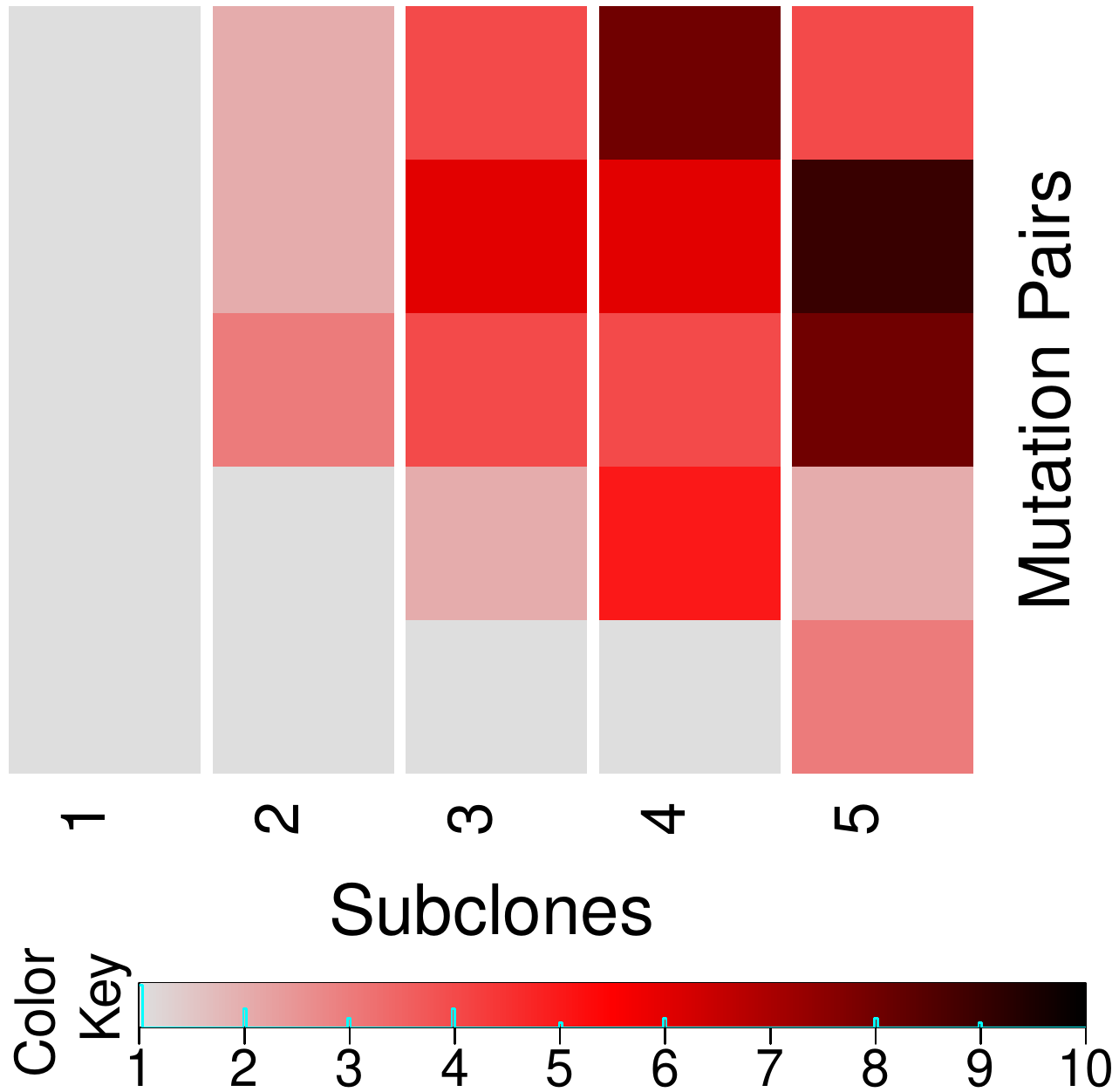}
\caption{$\Zhat$}
\end{subfigure}
\begin{subfigure}[t]{.325\textwidth}
\centering
\includegraphics[width=\textwidth]{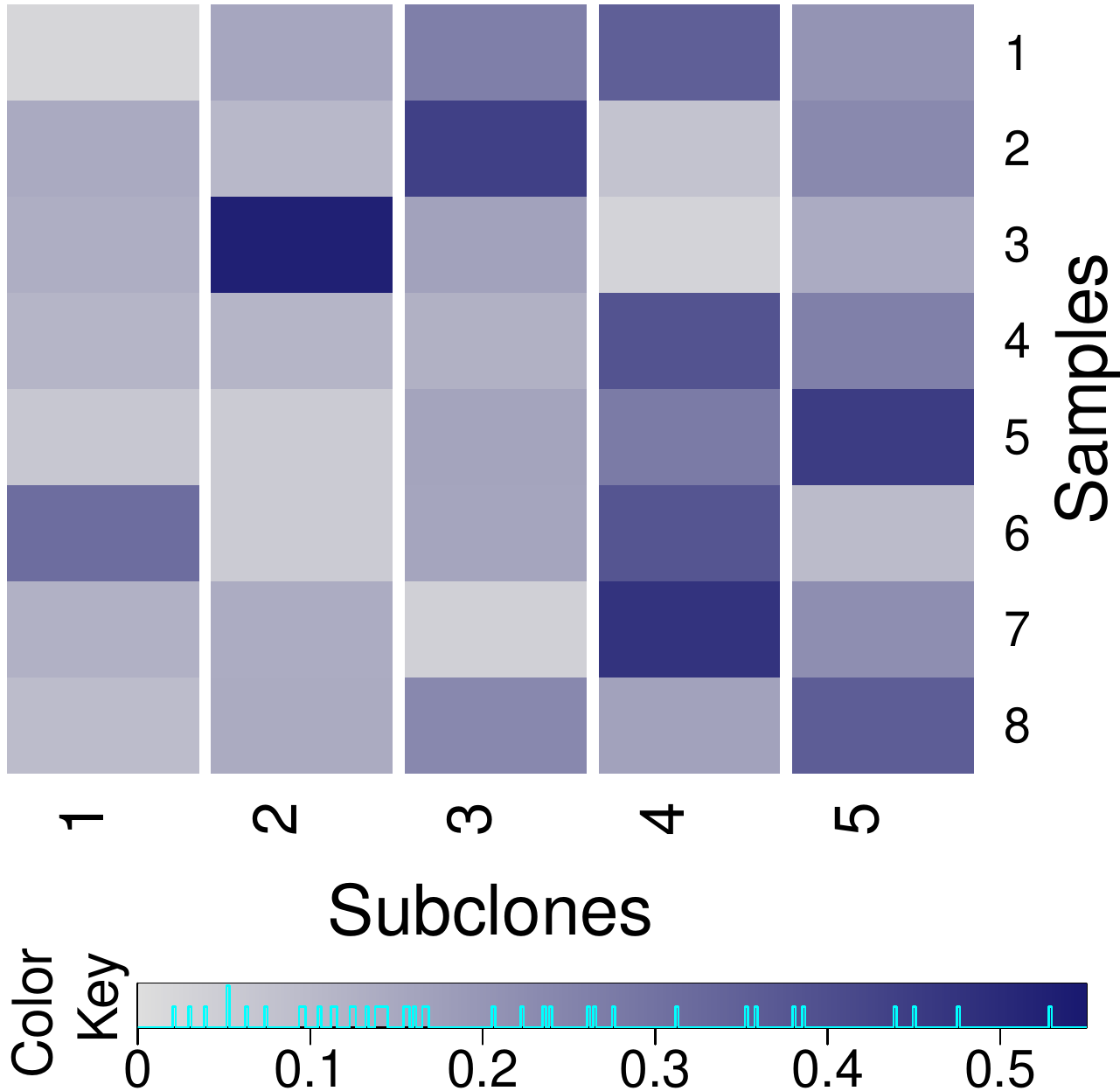}
\caption{$\what$}
\end{subfigure}
\begin{subfigure}[t]{.325\textwidth}
\centering
\includegraphics[width=\textwidth]{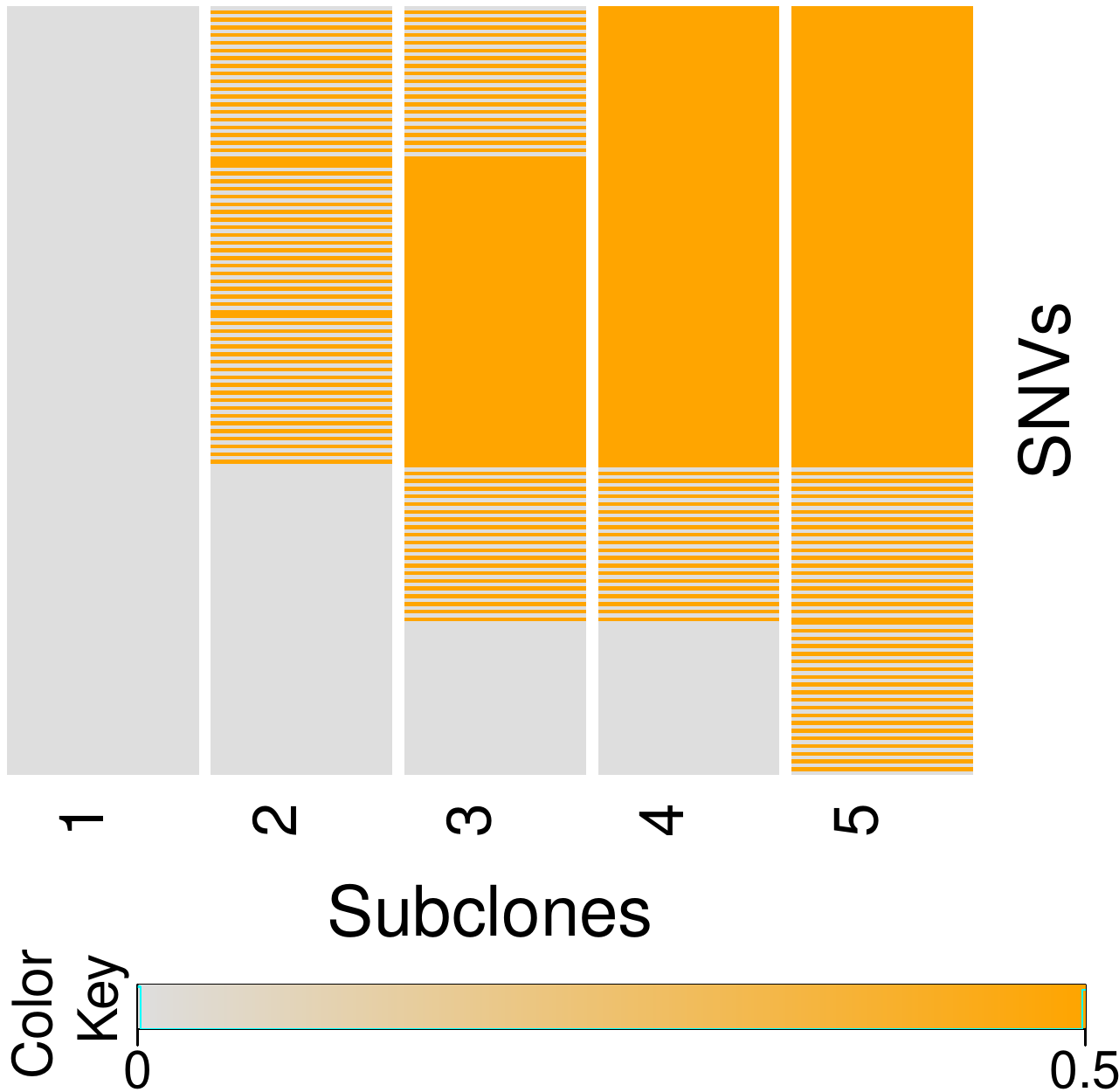}
\caption{$\Zhat^{\text{Cloe}}$}
\end{subfigure}
\end{center}
\caption{Simulation 2. Simulation truth $\bZ$ (a), $\bw$ (b) and phylogeny (c), and posterior inference under PairCloneTree (c, d, e) and  Cloe (f). }
\label{fig:sim2}
\end{figure}

In the second simulation, we evaluate the performance of the proposed approach
on multiple  samples. We still consider $K = 100$ mutation pairs, but
with a more complicated subclone structure, $C = 5$. We generate
hypothetical data for $T = 8$ samples. The subclone proportions in
each sample $t$ are generated from $\bw_t \sim \Dir(0.01, \sigma(25,
15, 10, 8, 5))$. Fig. \ref{fig:sim2}(a, b, c) show simulation truth
$\bZ$, $\bw$ and the phylogeny, respectively. We show $\bw$ in a
heatmap with light gray to deep blue scale. A darker blue color
indicates higher abundance of a subclone in a sample, while a lighter
gray color indicates lower abundance. 
The proportions of the background subclone $w_{t0}$'s are not shown as they only take tiny values, $w_{t0} < 10^{-3}$.
The average sequence depth for
the eight samples was about 500x.   

The hyperparameters are set to be the same as in simulation 1. We run
the same number of MCMC iterations. 

The true phylogeny is recovered with 100\% posterior
probability (Fig.~\ref{fig:sim2}(c)). Fig.~\ref{fig:sim2} (d, e) show the estimated genotypes
$\Zhat$ and subclone proportions $\what$. The truth is  exactly
recovered.  The simulation shows that with more
information from eight samples inference becomes quite reliable.

For comparison we again run Cloe and PhyloWGS on this data.
Cloe correctly infers the number of subclones, and the estimated
subclone genotypes match the truth, shown in Fig. \ref{fig:sim2}
(f). However, Cloe infers the phylogeny as $ 1 \rightarrow 2
\rightarrow 3 \rightarrow 4 \rightarrow 5$. 
On the other hand, PhyloWGS infers the phylogeny as 
$
0 \rightarrow 1 
\begin{array}{c}
\rightarrow 2  \\
\rightarrow 3 
\end{array}
$ (details not shown).
Both methods approximate but still miss some detail in the simulation truth.

\section{Lung Cancer Data}
\label{sec:real}
We use whole-exome sequencing (WES) data generated from four ($T$ = 4)
surgically dissected tumor samples taken from a single patient
diagnosed with lung adenocarcinoma. DNA is extracted from all four
samples and exome library is sequenced on an Illumina HiSeq 2000
platform in paired-end fashion. Each of the read is 100 base-pair long
and coverage is 200x-400x. 
We use BWA \citep{li2009fast} and GATK's UniformGenotyper \citep{mckenna2010genome} for
mapping and variant calling, respectively. In order to find mutation
pair location along with their genotypes with number of reads
supporting them, we use a bioinformatics tool called
\texttt{LocHap} \citep{sengupta2016ultra}. This tool searches for two or
three SNVs that are scaffolded by the same reads. When the scaffolded
SNVs, known as local haplotypes, exhibits more than two haplotypes, it
is known as local haplotype variant (LHV). Using the individual BAM
and VCF files \texttt{LocHap} finds a few hundreds LHVs on  
average in a WES sample. We select LHVs with two SNVs as we are interested in mutation pairs only. On those LHVs, we run 
the bioinformatics filters suggested by \texttt{LocHap} to keep the mutation
pairs with high calling quality. 
We focus our analysis in copy number neutral regions. In the end, we get 
69 mutation pairs for the sample and record the read count data
from \texttt{LocHap}'s output.

\begin{figure}[h!]
\begin{center}
\begin{subfigure}[t]{.345\textwidth}
\centering
\includegraphics[width=\textwidth]{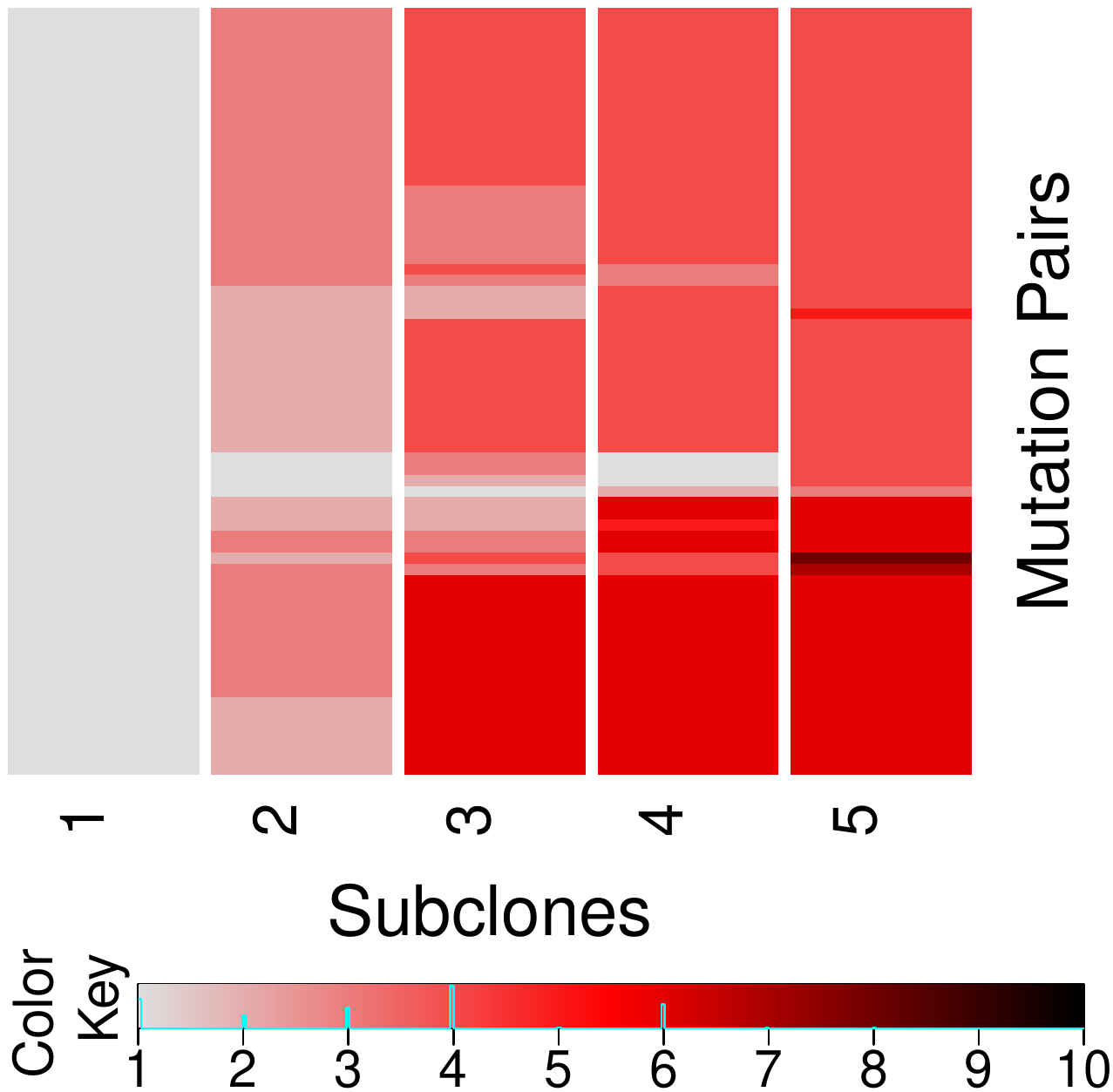}
\caption{$\Zhat$}
\end{subfigure}
\begin{subfigure}[t]{.345\textwidth}
\centering
\includegraphics[width=\textwidth]{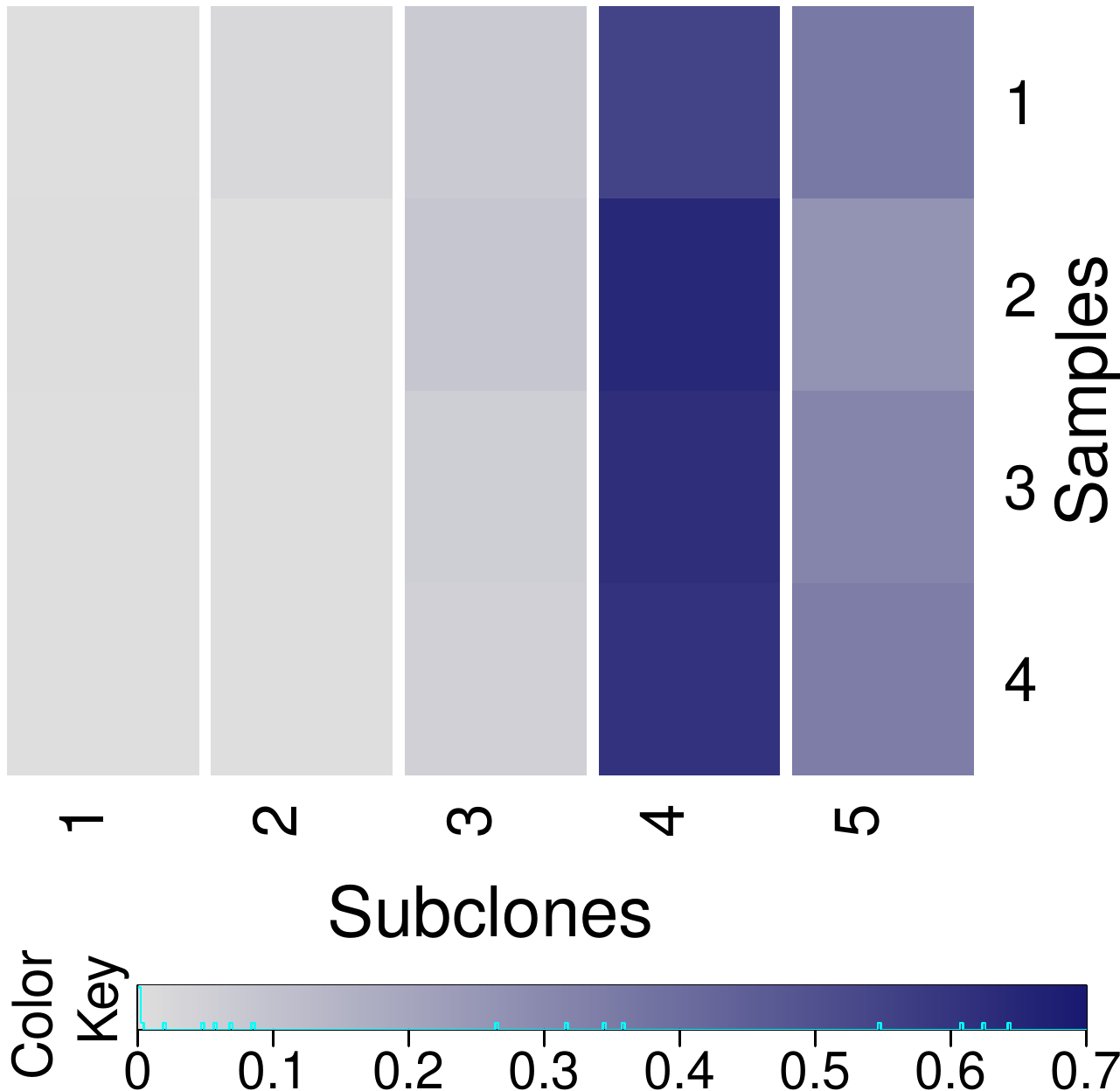}
\caption{$\what$}
\end{subfigure}
\begin{subfigure}[t]{.29\textwidth}
\centering
\includegraphics[width=.65\textwidth]{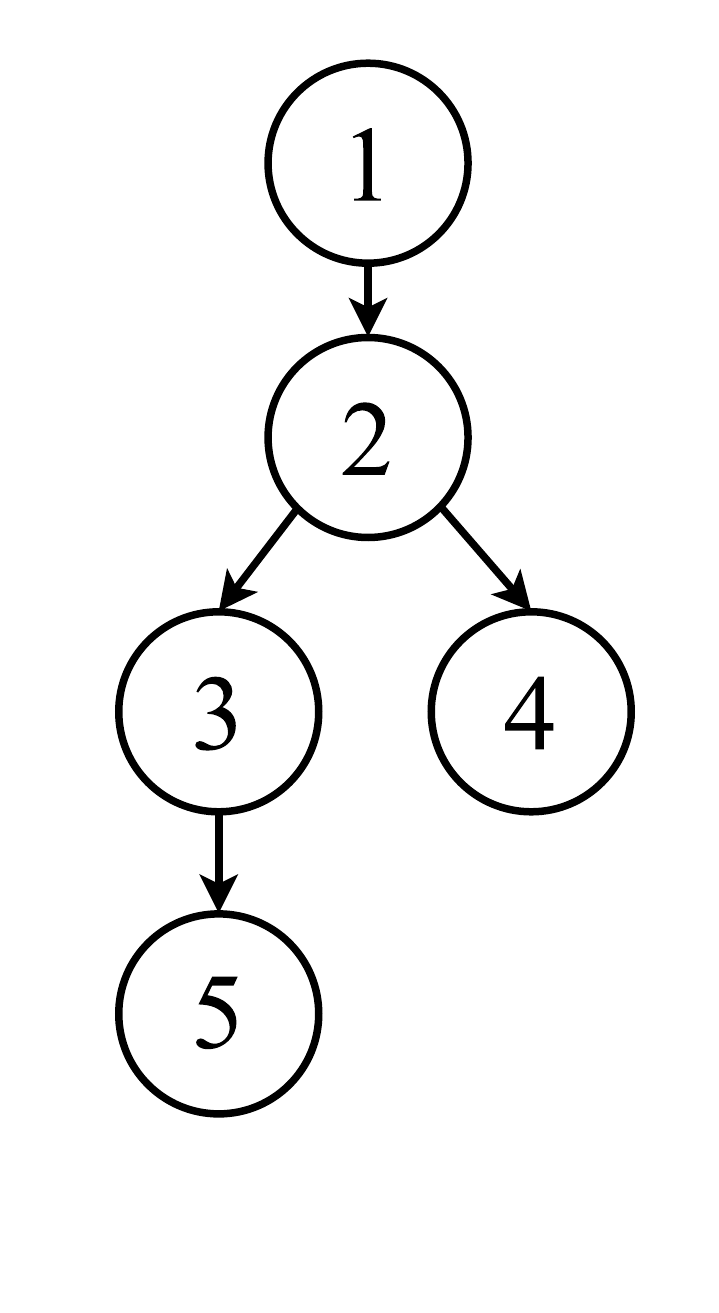}
\caption{Estimated phylogeny}
\end{subfigure}
\begin{subfigure}[t]{.45\textwidth}
\centering
\includegraphics[width=\textwidth]{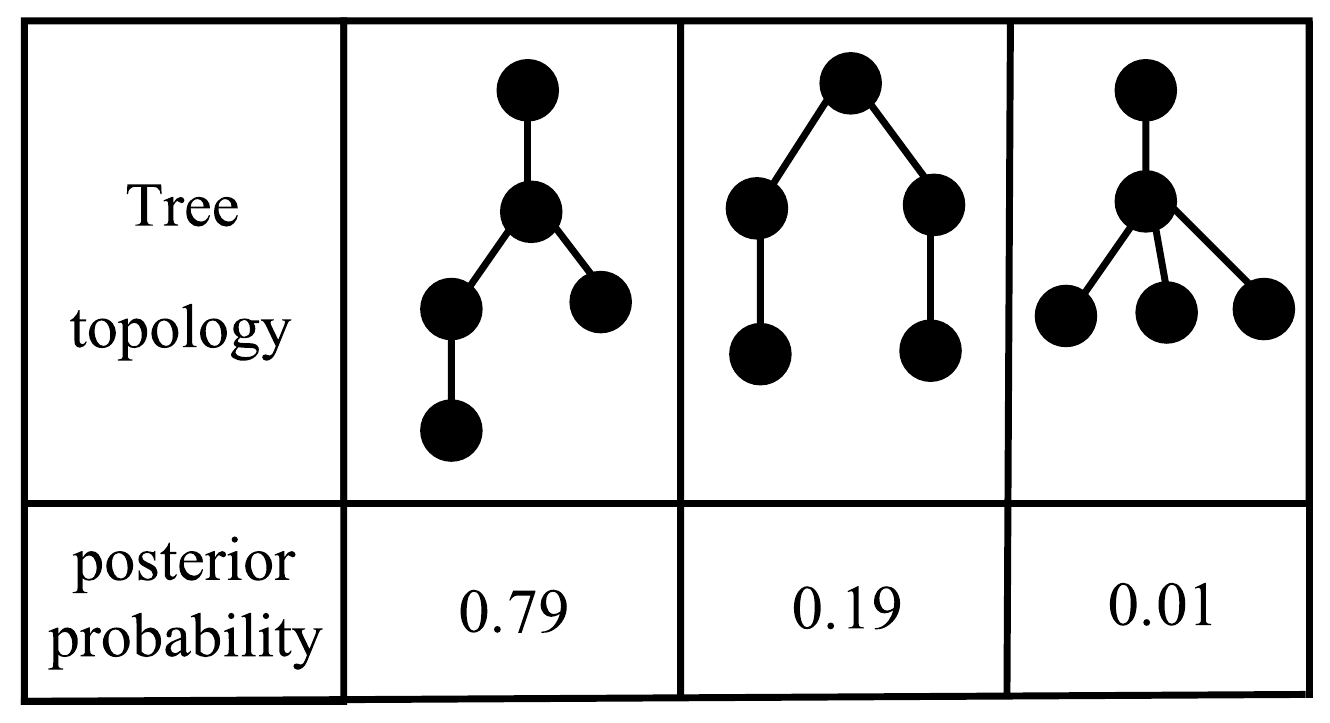}
\caption{Posterior prob. of tree topology}
\end{subfigure}
\hspace{7mm}
\begin{subfigure}[t]{.325\textwidth}
\centering
\includegraphics[width=\textwidth]{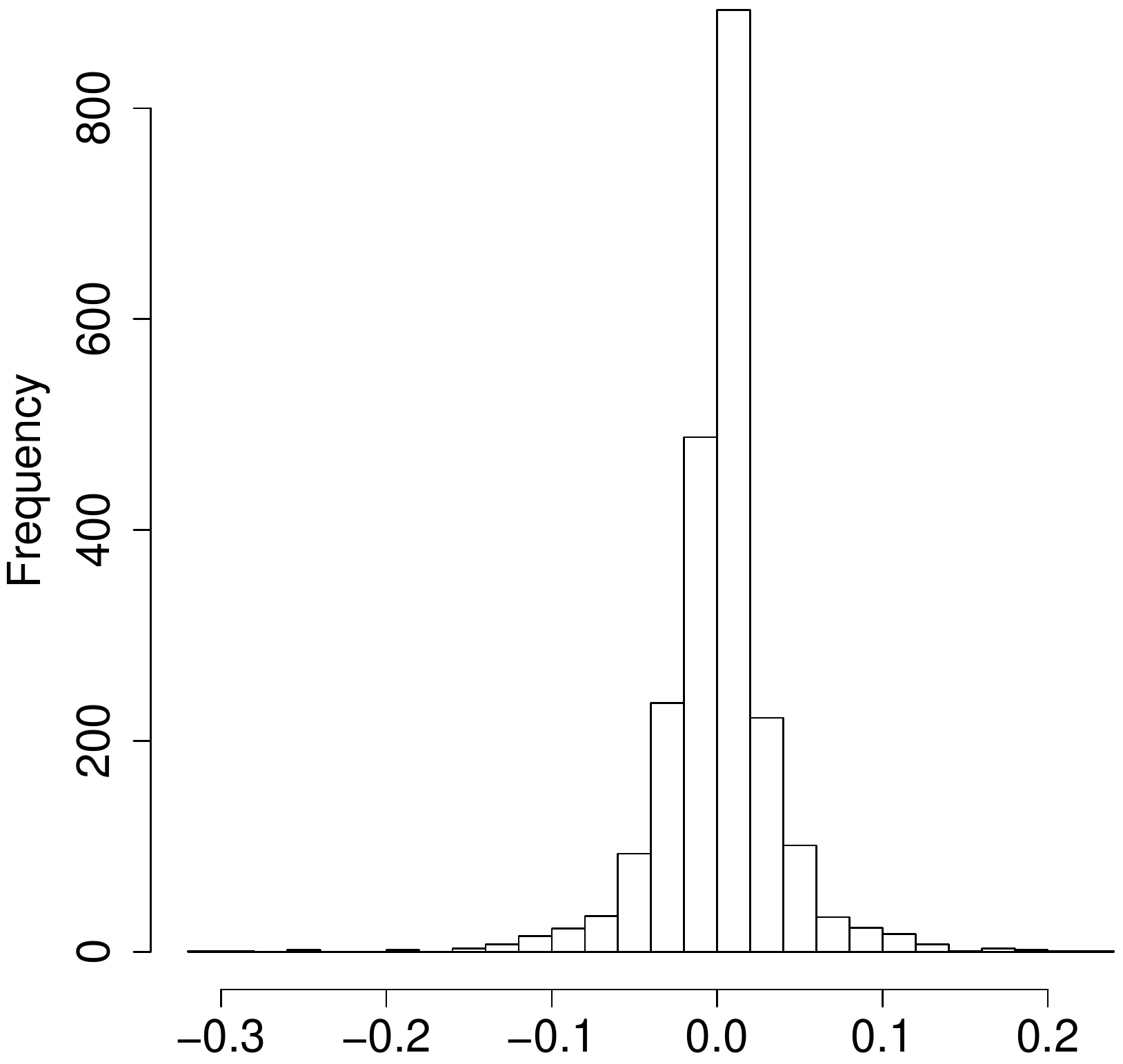}
\caption{Histogram of $(\phat_{tkg} - \bar{p}_{tkg})$}
\end{subfigure}
\end{center}
\caption{Posterior inference with PairCloneTree for lung cancer data set.}
\label{fig:lungPCT}
\end{figure}

We use the same hyperparameters and MCMC 
setting as in the simulations. Fig. \ref{fig:lungPCT} (d) shows some of the the
posterior probabilities of the subclone phylogeny. The posterior mode is
shown in Fig. \ref{fig:lungPCT} (c) with $C=5$
subclones. Fig. \ref{fig:lungPCT} (a, b) show the estimated subclone
genotypes $\Zhat$ and cellular proportions $\what$, respectively ($\hat{w}_{t0} < 4 \times 10^{-3}$ and are not shown). The
rows for $\Zhat$ are reordered for better display. The cellular
proportions of the subclones show strong similarity across the 4
samples, indicating homogeneity of the samples. 
This is expected as the samples are dissected from proximal sites. Subclone 1,
which is the normal subclone, takes a small proportion in all 4
samples, indicating high purity of the tumor samples. Subclones 2
and 3 are also included in only small proportions.
They have almost vanished in the samples. 
However, as parents of subclones 4 and 5, respectively, they are
important for the reconstruction of the subclone phylogeny.
Subclones 4 and 5 are the two main subclones. They
share a large proportion of common mutations, but each one has some
private mutations, consistent with the estimated tree.
Finally,  Fig. \ref{fig:lungPCT} (e) shows a histogram
of residuals, where we calculate empirical values $\bar{p}_{tkg} =
n_{tkg} / N_{tk}$ and plot the difference $(\phat_{tkg} -
\bar{p}_{tkg})$. The residuals are centered at zero with little
variation, indicating a good model fit.

\begin{figure}[h!]
\begin{center}
\begin{subfigure}[t]{.345\textwidth}
\centering
\includegraphics[width=\textwidth]{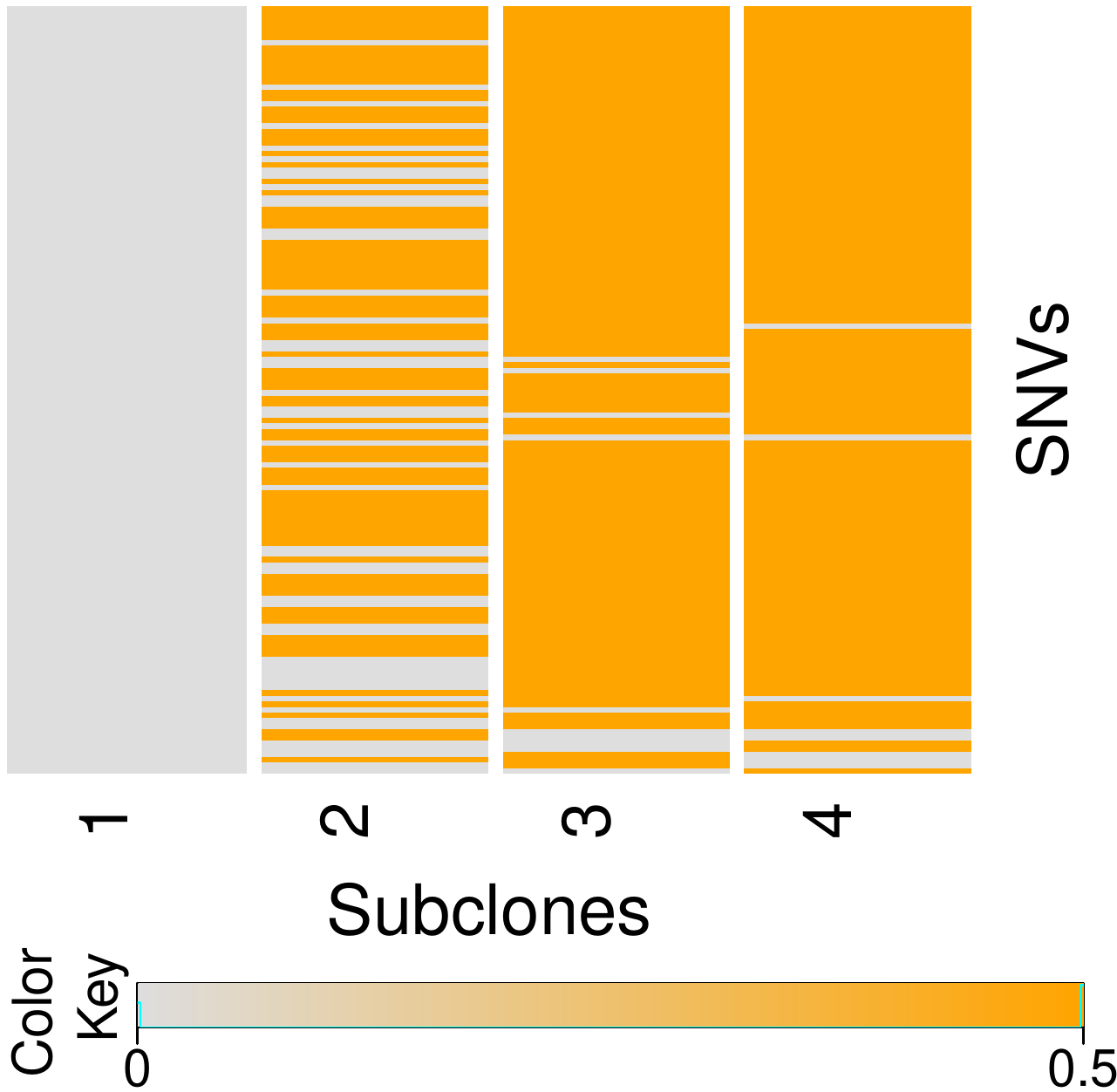}
\caption{$\Zhat^{\text{Cloe}}$}
\end{subfigure}
\begin{subfigure}[t]{.345\textwidth}
\centering
\includegraphics[width=\textwidth]{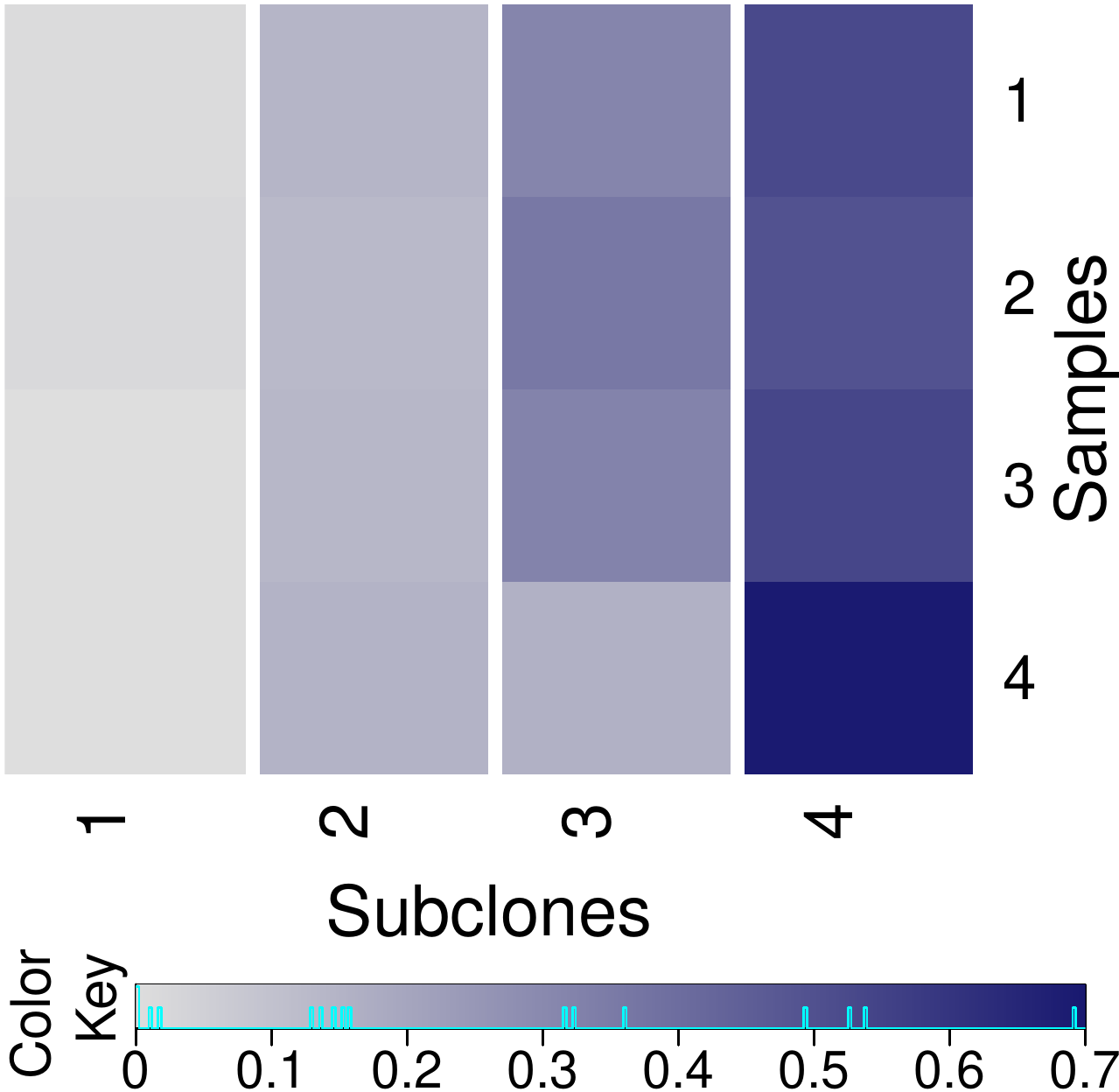}
\caption{$\what^{\text{Cloe}}$}
\end{subfigure}
\hspace{1mm}
\begin{subfigure}[t]{.6\textwidth}
\centering
\vspace{5mm}
\scalebox{0.75}{
\begin{tabular}{|l|c|c|c|c|}
\hline
\backslashbox{Clusters \\ (sizes)}{Samples} & 1 & 2 & 3 & 4 \\\hline
\qquad \quad 0 (0) & 1.0 & 1.0 & 1.0 & 1.0  \\\hline
\qquad \quad 1 (93) & 0.977 & 0.990 &  0.989 & 0.980 \\\hline
\qquad \quad 2 (32) & 0.854 & 0.816 & 0.832 & 0.852 \\\hline
\qquad \quad 3 (6) & 0.532 & 0.431 & 0.473 & 0.596 \\\hline
\qquad \quad 4 (3) & 0.433 & 0.408 & 0.445 & 0.386 \\\hline
\qquad \quad 5 (3) & 0.252 & 0.180 &  0.229 & 0.152 \\\hline
\qquad \quad 6 (1) & 0.112 & 0.108 & 0.056 & 0.127 \\\hline
\end{tabular}
}
\caption{Cellular prevalence (PhyloWGS)}
\end{subfigure}
\end{center}
\caption{Posterior inference with Cloe (a, b) and PhyloWGS (c) for lung cancer data set.}
\label{fig:lungPCT_compare}
\end{figure}

For comparison, we run Cloe 
and PhyloWGS on the same data set with default hyperparameters.
Cloe infers four subclones with phylogeny $ 1 \rightarrow 2 \rightarrow 3 \rightarrow 4$. 
Fig. \ref{fig:lungPCT_compare} (a, b) show the estimated genotypes $\Zhat^{\text{Cloe}}$ and cellular proportions $\what^{\text{Cloe}}$, respectively. 
PhyloWGS estimates 6 clusters (and a cluster 0 for normal subclone) of the SNVs with phylogeny
$$
0  \rightarrow 1 \rightarrow 2 
\begin{array}{cccc}
\rightarrow  &  3 &  \rightarrow & 4  \\
\rightarrow & 5 &  \rightarrow & 6 
\end{array} .
$$
Fig. \ref{fig:lungPCT_compare} (c) summarizes the cluster sizes and cellular prevalences. In light of the earlier simulation studies we believe that the inference under PairCloneTree is more reliable.
Cloe and PhyloWGS outputs confirm that the four samples have similar proportions 
of all the subclones, indicating little inter-sample heterogeneity.  
Also, Cloe and PhyloWGS infer very small normal cell proportion, 
which corroborates PairCloneTree's finding that the tumor samples have high purity.

\section{Discussion}
\label{sec:disc}

In this work, using a treed LFAM we infer subclonal
genotypes structure for 
mutation pairs, their cellular proportions and the phylogenetic
relationship among subclones. This is the first attempt to generate a
subclonal phylogenetic structure using mutation pair data. We show 
that more accurate inference can be obtained using mutation
pairs data compared to using only marginal counts for single SNVs. 
The model can be easily extended to incorporate more than two
SNVs. Another way of extending the model is to encode mutation times
inside the length of the edges of phylogenetic tree.
 
The major motivation for accurate estimation of heterogeneity in tumor
is personalized medicine. The next step towards this goal is to
use varying estimates of subclonal proportions across patients to
drive adaptive treatment allocation.

Currently the heterogeneity is measured
mostly with SNV and CNA data. However, structural variants (SVs) like
deletion, duplication, inversion, translocation and other large genome
rearrangement arguably provide more 
accurate \citep{fan2014towards} VAF estimation, which is the key input
for characterizing the heterogeneity. Therefore incorporation of SVs
into the current model could significantly improve the outcome of
tumor heterogeneity analysis.
Recently, in~\cite{brocks2014intratumor} the authors 
attempted to explain the intratumor
heterogeneity in DNA methylation and copy-number pattern by a unified
evolutionary process. So the current genome centric definition of
tumor heterogeneity could be extended by incorporation of methylation,
DNA mutation, and RNA expression data in an integromics model.

Finally in the era of big data it is important to factor computation
into the research effort, and build 
efficient computational models that could handle millions of
SNVs. Linear response variational Bayes \citep{giordano2015linear} or
MAD-Bayes \citep{broderick2013mad,xu2015mad} methods could be
considered as alternative computational strategies to tackle the
problem.

\chapter{A Nonparametric Bayesian Approach to Dropout in Longitudinal Studies with Auxiliary Covariates}
\label{chap:bnpmis}
We develop a nonparametric Bayesian approach to missing outcome data in longitudinal studies in the presence of auxiliary covariates. In the presence of auxiliary covariates, we consider a joint model for the full data response, missingness and auxiliary covariates. 
We include auxiliary covariates to ``move'' the missingness ``closer'' to missing at random (MAR).
In particular, we specify a nonparametric Bayesian model for the observed data via Gaussian process priors and Bayesian additive regression trees. These model specifications allow us to capture non-linear and non-additive effects, in contrast to existing parametric methods. We then separately specify the conditional distribution of the missing data response given the observed data response, missingness and auxiliary covariates (i.e. the extrapolation distribution) using identifying restrictions. We introduce meaningful sensitivity parameters that allow for a simple sensitivity analysis. Informative priors on those sensitivity parameters can be elicited from subject-matter experts. We use Monte Carlo integration to compute the full data estimands. Performance of our approach is assessed using simulated datasets. Our methodology is motivated by, and applied to, data from a clinical trial on treatments for schizophrenia.

\section{Introduction}

In longitudinal clinical studies, the research objective is often to make inference on a subject's full data response conditional on covariates that are of primary interest; for example, to calculate the treatment effect of a test drug at the end of a study.
The vector of responses for a research subject is often incomplete due to dropout.
Dropout is typically non-ignorable \citep{rubin1976inference, daniels2008missing} and in such cases the joint distribution of the full data response and missingness needs to be modeled. In addition to the covariates that are of primary interest, we would often have access to some \emph{auxiliary covariates} (often collected at baseline) that are not desired in the model
for the primary research question. Such variables can often provide information about the missing responses and missing data mechanism. For example, missing at random (MAR) \citep{rubin1976inference} might only hold conditionally on auxiliary covariates \citep{daniels2008missing}. In this setting, auxiliary covariates should be incorporated in the joint model as well, but we should proceed with inference unconditional on these auxiliary covariates.

The full data distribution can be factored into the observed data distribution and the extrapolation distribution \citep{daniels2008missing}. The observed data distribution can be identified by the observed data, while the extrapolation distribution cannot. Identifying the extrapolation distribution relies on untestable assumptions such as parametric models for the full data distribution or identifying restrictions \citep{linero2017general}. Such assumptions can be indexed by unidentified parameters called \emph{sensitivity parameters} \citep{daniels2008missing}.
The observed data do not provide any information to estimate the sensitivity parameters.
Under the Bayesian paradigm, informative priors can be elicited from subject-matter experts and be placed on those sensitivity parameters. 
Finally, it is desirable to conduct a \emph{sensitivity analysis} \citep{daniels2008missing, national2011prevention} to assess the sensitivity of inferences to such assumptions.
The inclusion of auxiliary covariates can ideally reduce the extent of sensitivity analysis that is needed for drawing accurate inferences.

In this paper, we propose a Bayesian nonparametric model for the joint distribution of the full data response, missingness and auxiliary covariates. We use identifying restrictions to identify the extrapolation distribution and introduce sensitivity parameters that are meaningful to subject-matter experts and allow for a simple sensitivity analysis.

\subsection{Missing Data in Longitudinal Studies}

Literature about longitudinal missing data 
with non-ignorable dropout 
can be mainly divided into two categories: likelihood-based and moment-based (semiparametric). Likelihood-based approaches include selection models (e.g. \citealp{heckman1979sample, diggle1994informative, molenberghs1997analysis}), pattern mixture models (e.g. \citealp{little1993pattern, little1994class, hogan1997mixture}) and shared-parameter models (e.g. \citealp{wu1988estimation, follmann1995approximate, pulkstenis1998model, henderson2000joint}). These three types of models differ from how the joint distribution of the response and missingness is factorized. Likelihood-based approaches often make strong parametric model assumptions to identify the full data distribution.
For a comprehensive review see, for example, \cite{daniels2008missing} or \cite{little2014statistical}.
Moment-based approaches, on the other hand, typically specify a semiparametric model for the marginal distribution of the response, and a semiparametric or parametric model for the missingness conditional on the response.
Moment-based approaches are in general more robust to model misspecification since they make minimal distributional assumptions.
See, for example, \cite{robins1995analysis, rotnitzky1998semiparametric, scharfstein1999adjusting, tsiatis2007semiparametric, tsiatis2011improved}.

There are several recent papers under the likelihood-based paradigm that are relevant to our approach, such as \cite{wang2010bayesian, linero2015flexible, linero2017nonparametric, linero2017general}. 
These papers specify Bayesian nonparametric models for the observed data distribution, and thus have similar robustness to semiparametric approaches. 
However, existing approaches do not utilize information from auxiliary covariates.
In the presence of auxiliary covariates, \cite{daniels2014fully} model longitudinal binary responses using a parametric model under ignorable missingness.
Our goal is to develop a nonparametric Bayesian approach to longitudinal missing data 
with non-ignorable dropout that also allows for incorporating auxiliary covariates.
As mentioned earlier, the reason to include auxiliary covariates is that we anticipate it will make the missingness ``closer'' to MAR.

\subsection{Notation and Terminology}
We introduce some notation and terminology as follows. 
Consider the responses for a subject $i$ at $J$ time points.  
Let $\bm Y_i = (Y_{i1}, \ldots, Y_{iJ})$ be the vector of longitudinal outcomes that was planned to be collected, 
$\bm \bY_{ij} = (Y_{i1}, \ldots, Y_{ij})$ be the history of outcomes through the first $j$ times, 
and $\bm \tY_{ij} = (Y_{i,j+1}, \ldots, Y_{iJ})$ be the future outcomes after time $j$. 
Let $S_i$ denote the dropout time or dropout pattern, which is defined as the last time a subject's response is recorded, i.e. $S_i = \max \{j: Y_{ij} \text{ is observed} \}$.
Missingness is called \emph{monotone} if $Y_{ij}$ is observed for all $j \leq S_i$, and missingness is called \emph{intermittent} if $Y_{ij}$ is missing for some $j < S_i$.
For monotone missingness, $S_i$ captures all the information about missingness. In the following discussion, we will concern ourselves with monotone missingness. Dropout is called \emph{random} \citep{diggle1994informative} if the dropout process only depends on the observed responses, i.e. the missing data are MAR; dropout is called \emph{informative} if the dropout process also depends on the unobserved responses, i.e. the missing data are missing not at random (MNAR).
We denote by $\bX_i$ the covariates that are of primary interest, and
$\bV_i = (V_{i1}, \ldots, V_{iQ})$ the $Q$ auxiliary covariates that are not of primary interest. Those auxiliary covariates should be related to the outcome and missingness. The observed data for subject $i$ is  $(\bm \bY_{iS_i}, S_i, \bV_i, \bX_i)$, and the full data is $(\bm Y_{i}, S_i, \bV_i, \bX_i)$. In general, we are interested in expectation of the form $E[t(\bm Y_i) \mid \bX_i]$, where $t$ denotes some functional of $\bm Y_i$. 
Finally, denote by $p(\by, s, \bv \mid \bx, \bomega)$ the joint model for the full data response, missingness and auxiliary covariates conditional on the covariates that are of primary interest, where $\bomega$ represents the parameter vector.

\subsection{The Schizophrenia Clinical Trial}
\label{sec:sctdata}
Our work is motivated by a multicenter, randomized, double-blind clinical trial on treatments for schizophrenia. For this clinical trial, the longitudinal outcomes are the positive and negative syndrome scale (PANSS) scores \citep{kay1987positive}. The outcomes are collected at $J = 6$ time points corresponding to baseline, day 4 after baseline, and weeks 1, 2, 3 and 4 after baseline. The possible dropout patterns are $S_i = 2, 3, 4, 5, 6$. 
The covariate of primary interest is treatment, with $X_i = 1, 2$ or $3$ corresponding to test drug, active control or placebo, respectively. In addition, we have access to $Q = 7$ auxiliary covariates including age, onset (of schizophrenia) age, height, weight, country, sex and education level. 

\begin{figure}[h!]
\begin{center}
\includegraphics[width=\textwidth]{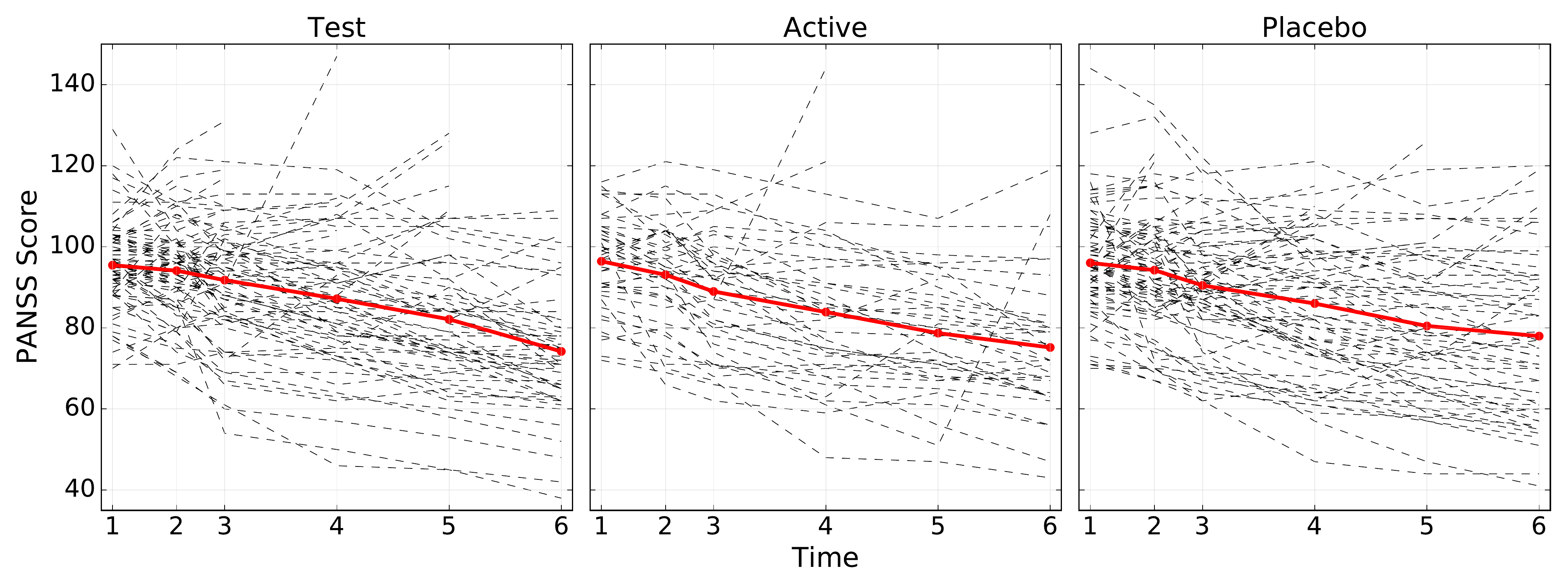}
\end{center}
\caption{Trajectories of individual responses (dashed black lines) and mean responses (thick red lines) over time for the active control, placebo and test drug arms.}
\label{fig:bnpmis_traj}
\end{figure}

The dataset consists of $N = 204$ subjects, with 45 subjects for the active control arm, 78 subjects for the placebo arm, and 81 subjects for the test drug arm. 
Figure \ref{fig:bnpmis_traj} shows the individual trajectories and mean responses over time for the three treatment arms.
The dropout rates are $33.3\%$, $20.0\%$ and $25.6\%$ for the test drug, active control and placebo arms, respectively.   
Table \ref{tbl:pattern_summary} shows the detailed dropout rates for each dropout pattern.
Subjects drop out for a variety of reasons. Some reasons including adverse events (e.g. occurrence of side effects), pregnancy and protocol violation are thought to be random dropouts, while the other reasons such as disease progression, lack of efficacy, physician decision and withdraw by patient are thought to be informative dropouts. It is ideal to treat those reasons differently while making inference.
Table \ref{tbl:pattern_summary} shows the informative dropout rates for each dropout pattern. For the test, active and placebo arms,
the percentages of informative dropouts among all dropouts are $88.9\%$, $77.8\%$ and $100.0 \%$, respectively. The dataset has a few intermittent missing outcomes (1 for the test drug arm, 1 for the active control arm, and 2 for the placebo arm). We focus our study on monotone missingness and assume partial ignorability \citep{harel2009partial} for the few intermittent missing outcomes.

\begin{table}[h!]
\centering
\begin{tabular}{|c|c|c|c|c|c|}
\hline
               & $S_i = 2$ & $S_i = 3$ & $S_i = 4$ & $S_i = 5$ & Overall\\ \hline
Test      & 4.9 (3.7) & 12.3 (9.9) & 8.6 (8.6) & 7.4 (7.4) & 33.3 (29.6)  \\ \hline
Active & 2.2 (2.2) & 4.4 (2.2) & 8.9 (6.7) & 4.4 (4.4) & 20.0 (15.6) \\ \hline
Placebo        & 3.8 (3.8) & 5.1 (5.1)  & 11.5 (11.5) & 5.1 (5.1) & 25.6 (25.6) \\ \hline
\end{tabular}
\caption{Dropout rates (\%) for different dropout patterns in the three treatment arms, with informative dropout rates in parentheses.}
\label{tbl:pattern_summary}
\end{table}


The goal of this study is to estimate the change from baseline treatment effect,
\begin{align*}
r_x = \E[Y_{i6} - Y_{i1} \mid X_i = x].
\end{align*}
In particular, the treatment effect improvements over placebo, i.e. $r_1 - r_2$ and $r_3 - r_2$, are of interest.
This dataset was previously analyzed in \cite{linero2015flexible}, which took a Bayesian nonparametric approach as well, but did not utilize information from the auxiliary covariates.

\subsection{Overview}
We stratify the model by treatment, and suppress the treatment variable $x$ to simplify notation hereafter. 
The extrapolation factorization \citep{daniels2008missing} is
\begin{align*}
p(\by, s, \bv \mid \bomega) = p(\tby_{s} \mid \bby_s, s, \bv, \bomegae) p(\bby_s, s, \bv \mid \bomegao),
\end{align*}
where the extrapolation distribution, $p(\tby_{s} \mid \bby_s, s, \bv, \bomegae)$, is not identified by the data in the absence of uncheckable assumptions or constraints on the parameter space. The observed data distribution $p(\bby_s, s, \bv \mid \bomegao)$ is identified and can be estimated nonparametrically.
We factorize the observed data distribution based on pattern-mixture modeling \citep{little1993pattern},
\begin{align}
p(\bby_s, s, \bv \mid \bomegao) = p(\bby_s \mid s, \bv, \bpi) p(s \mid \bv, \bm f) p(\bv \mid \bm \eta),
\label{eq:joint_model}
\end{align}
where we assume distinct parameters $\bomegao = (\bpi, \bm f, \bm \eta)$ parametrizing the response model, the missingness and the distribution of the auxiliary covariates, respectively. 

The remainder of this article is structured as follows. In Section \ref{sec:prob_model} we specify Bayesian (nonparametric) models for \eqref{eq:joint_model}.
In Section \ref{sec:extrapolation}, we use identifying restrictions to identify the extrapolation distribution. 
In Section \ref{sec:comp}, we describe our posterior inference and computation approaches.
In Section \ref{sec:simu}, we present simulation studies to validate our model and compare with results using other methods.
In Section \ref{sec:real_data}, we apply our method to a clinical trial on treatments for schizophrenia.
We conclude with a discussion in Section \ref{sec:discuss}.

\section{Probability Model for the Observed Data}
\label{sec:prob_model}
The model specification \eqref{eq:joint_model} brings two challenges:

1. For the models $p(\bby_s \mid s, \bv, \bpi)$ and $p(s \mid \bv, \bm f)$, it is unclear how the auxiliary covariates are related to the responses and dropout patterns. For example, the auxiliary covariates contain height and weight, which might not have a linear and additive effect on the responses. For example, the responses might have a linear relationship with the body mass index, which is calculated by $\text{weight} / \text{height}^2$.

2. For the model $p(\bby_s \mid s, \bv, \bpi)$, the observed patterns are sparse. For example, the dropout pattern $S_i = 2$ for the active control arm has only 1 observation.

To mitigate challenge 1, 
we specify nonparametric models for $p(\bby_s \mid s, \bv, \bpi)$ and $p(s \mid \bv, \bm f)$ via Gaussian process priors and Bayesian additive regression trees. Such models are highly flexible and robust to model misspecification.
To address challenge 2, we utilize informative priors such as autoregressive (AR) and conditional autoregressive (CAR) priors to share information across neighboring patterns. Detailed model specifications are as follows.

\subsection{Model for the Observed Data Responses Conditional on Pattern and Auxiliary Covariates}
We define the model for observed data responses conditional on drop out time and auxiliary covariates, i.e. $p(\bby_{s} \mid s, \bv, \bpi)$, as follows. The distribution $p(\bby_{s} \mid s, \bv, \bpi)$ can be factorized as
\begin{align}
p_s\left(\bby_{s} \mid  \bv , \bpi\right) = p_s (y_{s} \mid \bby_{s-1}, \bv, \bpi) \cdots p_s(y_{2} \mid \bby_{1}, \bv, \bpi) p_s(y_{1} \mid \bv, \bpi),
\label{eq:factorize_obs}
\end{align}
where the subscript $s$ corresponds to conditioning on dropping out pattern $S = s$.

We assume
\begin{align}
\left(Y_{j} \mid \bm \bY_{j-1} = \bby_{j-1}, S = s, \bV = \bv, \bpi \right) = 
\begin{cases}
a_0(\bv, j, s) + \varepsilon_{js}, \quad &j = 1; \\
a(\bv, j, s) + \bby_{j-1}^T \bPhi_{js} + \varepsilon_{js}, \quad &j \geq 2;
\end{cases}
\label{model_for_response}
\end{align}
where $j = 1, \ldots, s$; $s = 2, \ldots, J$. Here $a_0$ and $a$ are stochastic processes indexed by $\mathcal{U} = \mathcal{V} \times \mathcal{J}$, where $\mathcal{V}$ is the state space of $\bv$ and $\mathcal{J} \subset \{ 1, \ldots, J\}^2$ is the state space of $(j, s)$. Furthermore,   $\bPhi_{js}$ is the vector of lag coefficients for each time/pattern, and $\varepsilon_{js}$'s are Gaussian errors,
\begin{align*}
\varepsilon_{js} \sim N(0, \sigma_{js}^2).
\end{align*}
We place Gaussian process priors \citep{rasmussen2006gaussian} on $a$ and $a_0$,
\begin{align*}
a_0(\bv, j, s) &\sim \GP
 \left[ \mu_0(\bv, j, s), \, C_0(\bv, j, s ;  \bv', j', s') \right]; \\
a(\bv, j, s) &\sim \GP
 \left[ \mu(\bv, j, s), \, C(\bv, j, s ;  \bv', j', s') \right],
\end{align*}
with mean functions $\mu_0$, $\mu$: $\mathcal{U} \rightarrow \mathbb{R}$ and covariance functions $C_0$, $C$: 
$\mathcal{U} \times \mathcal{U} \rightarrow \mathbb{R}^+$. Specifically,
\begin{align}
\begin{split}
\mu_0(\bv, j, s) &= \bv^T \bbeta_{0s} + b_{js}; \\
\mu(\bv, j, s) &=  \bv^T \bbeta_{s} + b_{js} ,
\end{split}
\label{eq:GP_mean}
\end{align}
and

\begin{align}
\begin{split}
C_0(\bv, j, s ;  \bv', j', s') &= 
\kappa_0^2 \,  D_0(\bv, j, s ;  \bv', j', s') + 
 \tkappa_{0}^2 \, I(\bv, j, s ;  \bv', j', s');
 \\
C(\bv, j, s ;  \bv', j', s') &=
\kappa^2 \, D(\bv, j, s ;  \bv', j', s') + 
 \tkappa^2 \, I(\bv, j, s ;  \bv', j', s').
\end{split}
\label{eq:GP_cov}
\end{align}

We use two different stochastic processes $a_0$ and $a$ for $j = 1$ and $j \geq 2$. The reason is that for $j = 1$, $a_0$ represent the mean initial response with no past; for $j \geq 2$, $a$ represents the mean at subsequent thus with a measured past. In the mean functions \eqref{eq:GP_mean}, $\bbeta_{0s}$ and  $\bbeta_{s}$ are the vectors of regression coefficients of the auxiliary covariates, and $b_{js}$ is the time/pattern specific intercepts. 
In the covariance functions \eqref{eq:GP_cov}, $D_0(a; b)$ and $D(a; b)$ are the exponential distances between $a$ and $b$, defined by
\begin{align*}
D_0(\bv, j, s ;  \bv', j', s')  &= \exp \bigg[- \frac{\| \tilde{\bv}- \tilde{\bv}' \|_2^2}{2 h_{v0}^2} - \frac{|\tj - \tj'|}{h_{j0}} - \frac{|\ts - \ts'|}{h_{s0}} \bigg], \\
D(\bv, j, s ;  \bv', j', s')  &= \exp \bigg[- \frac{\| \tilde{\bv}- \tilde{\bv}' \|_2^2}{2 h_{v}^2} - \frac{|\tj - \tj'|}{h_{j}} - \frac{|\ts - \ts'|}{h_{s}} \bigg].
\end{align*}
Here $\kappa_0^2$, $h_{v0}$, $h_{j0}$, $h_{s0}$, $\tkappa_{0}^2$, 
$\kappa^2$, $h_{v}$, $h_{j}$, $h_{s}$, $\tkappa^2$ are the hyperparameters. The values $\tilde{\bv}$, $\tj$ and $\ts$ are standardized values for $\bv$, $j$ and $s$,
\begin{multline*}
\tilde{v}_{iq} = \frac{v_{iq} - \text{mean}(v_{\cdot q})}{\text{sd}(v_{\cdot q})},
\\
\tj_i = \frac{j_i - \text{min}(j_{\cdot})}{\text{max}(j_{\cdot}) - \text{min}(j_{\cdot})}, \quad \text{and} \; \;
\ts_i = \frac{s_i - \text{min}(s_{\cdot})}{\text{max}(s_{\cdot}) - \text{min}(s_{\cdot})}.
\end{multline*}
For categorical covariates, the distance between $\bv$ and $\bv'$ is calculated by counting the number of different values.
In addition, in \eqref{eq:GP_cov}, $I(a; b)$ is the Kronecker delta function that takes the value 1 if $a = b$ and 0 otherwise. 
The function $I(a; b)$ is used to introduce a small nugget for the diagonal covariances, which overcomes near-singularity of the covariance matrices and improves numerical stability.
The Gaussian processes flexibly model the relationship between auxiliary covariates and response and accounts for possibly non-linear and non-additive effects.

For the noise variance $\sigma_{js}^2$, we assume an inverse Gamma shrinkage prior,
\begin{align*}
\sigma_{js}^2 \mid g_{\sigma} \iidsim \IG(\lambda_{\sigma}, \lambda_{\sigma} g_{\sigma}), \quad j = 1, \ldots, s, \; s = 2, \ldots, J,
\end{align*}
with $\E (1/\sigma_{js}^2) = 1/g_{\sigma}$ and $\Var (1/\sigma_{js}^2) = 1 / \lambda_{\sigma} g_{\sigma}^2$. We put hyper-priors on $\lambda_{\sigma}$ and $g_{\sigma}$, 
\begin{align*}
\lambda_{\sigma} - 2 &\sim \IG(1, 1),\\
g_{\sigma} &\sim \text{Gamma}(1, 1).
\end{align*}

Next, we consider the parameters in the mean functions \eqref{eq:GP_mean}. We allow the regression coefficients of the auxiliary covariates to vary by pattern. 
However, it is typical to have sparse patterns. 
As a result, we consider an informative prior that assumes regression coefficients for neighboring patterns to be similar. In particular, 
we specify AR(1) type priors on $\bbeta_{0s}$ and $\bbeta_s$. For $\bbeta_s$, we assume
\begin{align*}
\bbeta \sim N \left[ X_{\beta} \tbbeta, \sigma_{\beta}^2 \Sigma_{\beta}(\rho) \right],
\end{align*}
where
\begin{align*}
\bbeta = \left( \begin{array}{c}
\bbeta_{2} \\
\bbeta_{3}\\
\vdots \\
\bbeta_{J}  \end{array} \right), \qquad
X_{\beta} = \left(\begin{array}{c}
I \\
I \\
\vdots \\
I  \end{array} \right), 
\end{align*}
and
\begin{align*}
\Sigma_{\beta}(\rho) = \frac{1}{1 - \rho^2} \left( \begin{array}{cccc}
I &  \rho I & \cdots & \rho^{J - 2} I \\
\rho I & I & \cdots & \rho^{J - 3} I \\
\vdots & \vdots &  & \vdots \\
\rho^{J - 2} I & \rho^{J - 3} I & \cdots & I  \end{array} \right).
\end{align*}
The prior on $\bbeta$ introduces three unknown hyperparameters $\tbbeta$, $\sigma_{\beta}^2$ and $\rho$. We specify diffuse normal, inverse Gamma and uniform priors, respectively,
\begin{align*}
\tbbeta \sim N(\bm 0, \delta_{\beta}^2 I), \quad
\sigma_{\beta}^2 \sim \IG (\lambda_1^{\beta}, \lambda_2^{\beta}), \quad
\rho \sim \Unif(0, 1).
\end{align*}
Similarly, for $\bbeta_{0s}$,
\begin{align*}
&\bbeta_0 \sim N \left[ X_{\beta} \tbbeta_0, \sigma_{\beta_0}^2 \Sigma_{\beta}(\rho_0) \right], \quad \text{with hyper-priors} \\
&\tbbeta_0 \sim N(\bm 0, \delta_{\beta_0}^2 I), \quad
\sigma_{\beta_0}^2 \sim \IG (\lambda_1^{\beta_0}, \lambda_2^{\beta_0}), \quad
\rho_0 \sim \Unif(0, 1).
\end{align*}

The time/pattern specific intercepts are given conditional autoregressive (CAR) type priors \citep{banerjee2014hierarchical, de2012bayesian} as we expect them to be similar for neighboring patterns/times.
Let $\bb_0 = (b_{12}$, $b_{13}$, $\ldots$,  $b_{1J})$ and $\bb = (b_{22}$; $b_{23}$, $b_{33}$; $\ldots $ ; $b_{2J}$, $\ldots$, $b_{JJ})$.
The potential neighbors of $b_{js}$ are $\{b_{j-1, s}, b_{j+1, s}, b_{j, s-1}, b_{j, s+1} \}$. Denote by $\mathcal{N}_{js}^b = \{(j', s'): b_{j's'} \text{~is neighbor of~} b_{js} \}$ and $N_{js}^b = |\mathcal{N}_{js}^b|$ which is the number of neighbors of $b_{js}$. The CAR type prior assigns conditional priors on $b_{js}$ given its neighbors, and under several regularity conditions the conditionals indicate a joint distribution.
In particular, we assume 
\begin{align*}
b_{js} \mid b_{-js} \sim N\left( \tilde{b} + \sum_{j's' \in \mathcal{N}_{js}^b} \frac{\gamma_b}{N_{js}^b} \left( b_{j's'} - \tilde{b} \right)  , \frac{\sigma_{b}^2}{N_{js}^b} \right),
\end{align*}
which induces a joint prior on $\bb$ of the form
\begin{align*}
\bb \sim N\left( \bone \tilde{b}, \sigma_{b}^2 (I - \gamma_b W_b)^{-1} \mathcal{N}_b \right),
\end{align*}
where 
\begin{align*}
(W_b)_{jsj's'} = 
\begin{cases}
1/N_{js}^b, \quad &\text{if $(j,s)$ and $(j', s')$ are neighbors}; \\
0, \quad &\text{otherwise},
\end{cases}
\end{align*}
$ \mathcal{N}_b = \text{diag}(1/N_{js}^b)$, $\tb$ is a mean parameter for $\bb$, $\sigma_{b}^2$ is a variance parameter and $\gamma_b$ is a spatial dependence parameter. Let $\left( e_1^{b} \right)^{-1}$ and $\left( e_2^{b} \right)^{-1}$ denote the max and min eigenvalues of $W_b$. 
To guarantee that $I - \gamma_b W_b$ is positive definite, $\gamma_b$ is required to belong to $(e_2^{b}, e_1^{b})$. Furthermore, it is not unreasonable to assume the spatial correlation is positive, i.e. $0 < \gamma_b <  e_1^{b}$. 
We put hyper-priors on $\tb$, $\sigma_{b}^2$ and $\gamma_b$,
\begin{align*}
\tb \sim N(0, \delta_{b}^2),   \quad
\sigma_b^2 \sim \IG(\lambda_1^{b}, \lambda_2^{b}), \quad
\gamma_b \sim \Unif(0, e_1^{b}).
\end{align*}
Similarly, for $\bb_0$, we assume
\begin{align*}
&\bb_0 \sim N\left( \bone \tilde{b}_0, \sigma_{b_0}^2 (I - \gamma_{b_0} W_{b_0})^{-1} \mathcal{N}_{b_0} \right);  \quad \text{with hyper-priors} \\
&\tb_0 \sim N(0, \delta_{b_0}^2), \quad
\sigma_{b_0}^2 \sim \IG(\lambda_1^{b}, \lambda_2^{b}), \quad
\gamma_{b_0} \sim \Unif(0, e_1^{b_0}).
\end{align*}

We then consider the parameters in the covariance functions \eqref{eq:GP_cov}. We put inverse Gamma priors on $\kappa_0^2$ and $\kappa^2$,
\begin{align*}
\kappa_0^2   \sim \IG(\lambda_1^{\kappa_0}, \lambda_2^{\kappa_0}), \quad \kappa^2 \sim \IG(\lambda_1^{\kappa}, \lambda_2^{\kappa}).
\end{align*}
For simplicity, we fix the length scales $h_{v0}$, $h_{j0}$, $h_{s0}$, $h_{v}$, $h_{j}$ and $h_{s}$. For example, in practice, we set 
$h_{v0}^2 = h_{v}^2 = Q$ to introduce moderate correlation between the responses of two subjects with similar $\bV$'s; we set $h_{j0} = h_j = 5$, $h_{s0} = h_s = 5$ to introduce strong correlation between the responses of one subject measured at two different time points. We also fix $\tkappa_{0}^2$ and $\tkappa^2$ at small values, e.g. $\tkappa_{0}^2 = \tkappa^2 = 0.01$.

We complete the model with a prior for the lag coefficients. For each time/pattern, we break $\bPhi_{js}$ into three parts: $\bPhi_{js} = (\bphi_{3js}, \phi_{2js}, \phi_{1js})$, $\phi_{1js}, \phi_{2js}$ and $\bphi_{3js}$ correspond to lag-1 response, lag-2 response and higher-order lag responses, respectively. 
We put a CAR type prior on $\bphi_1 = \{ \phi_{1js} \}$, similar to the priors on $b_{js}$, 
\begin{align*}
&\bphi_1 \sim N \left(  \bone \tphi_1, \sigma_{\phi_1}^2 (I - \gamma_{\phi_1} W_{\phi_1})^{-1} \mathcal{N}_{\phi_1}  \right);  \quad \text{with hyper-priors}\\
&\tphi_1 \sim N(1, \delta_{\phi_1}^2), \quad 
\sigma_{\phi_1}^2 \sim \IG(\lambda_1^{\phi_1}, \lambda_2^{\phi_1}), \quad
\text{and} \;\; \gamma_{\phi_1} \sim \Unif(0, e_1^{\phi_1}).
\end{align*}

For $\phi_{2js}$ and $\bphi_{3js}$ we simply put normal priors with more prior mass around 0 to indicate the prior belief that higher-order lags have less impact on current response. Specifically,
\begin{alignat*}{2}
&\phi_{2js} \sim N(0, \sigma_{\phi_2}^2 ), \quad
&&\sigma_{\phi_2}^2 \sim \IG(\lambda_1^{\phi_2}, \lambda_2^{\phi_2}); \\
&\bphi_{3js} \sim N(0, \sigma_{\phi_3}^2 I), \quad
&&\sigma_{\phi_3}^2 \sim \IG(\lambda_1^{\phi_3}, \lambda_2^{\phi_3}),
\end{alignat*}
with $\lambda_1^{\phi_2} > \lambda_2^{\phi_2}$ and $\lambda_1^{\phi_3} > \lambda_2^{\phi_3}$.

\subsection{Model for the Pattern Conditional on Auxiliary Covariates}
We model the hazard of dropout at time $j$ with Bayesian additive regression trees (BART) \citep{chipman2010bart},
\begin{align*}
p(S = j \mid S \geq j, \bv, \bm f) = F_N(f_j(\bv)),
\end{align*}
where $F_N$ denotes the standard normal cdf (probit link), and $f_j(\bv)$ is the sum of tree models from BART.  The BART model captures complex relationships between auxiliary covariates and dropout including interactions and nonlinearities. We use the default priors for $f_j(\cdot)$ given in \cite{chipman2010bart}.

\subsection{Model for the Auxiliary Covariates}
We use a Bayesian bootstrap \citep{rubin1981bayesian} prior for the distribution for $\bv$. Suppose $\bv$ can only take the $N$ discrete values that we observed, $\bV \in \{ \bv_1, \ldots,  \bv_N\}$. The probability for each is 
\begin{align}
p(\bV  = \bv_i \mid \bm \eta ) = \eta_i,
\label{eq:bb}
\end{align}
where $\sum_{i=1}^N \eta_i = 1$. We place a Dirichlet distribution prior on $\bm \eta$,
\begin{align*}
(\eta_1, \ldots, \eta_N) \sim \text{Dir}(d_{\eta}, \ldots, d_{\eta}).
\end{align*}

\section{The Extrapolation distribution}
\label{sec:extrapolation}
The extrapolation distribution for our setting can be sequentially factorized as
\begin{multline}
p_s(\tby_{s}  \mid \bby_{s}, \bv, \bomegae) = p_s(y_{s+1} \mid \bby_{s}, \bv, \bomegae) \cdot \\
p_s(y_{s+2} \mid \bby_{s+1}, \bv, \bomegae) \cdots
p_s(y_{J} \mid \bar{\bm y}_{J-1}, \bv, \bomegae).
\label{extrapolation_distribution}
\end{multline}

The extrapolation distribution is not identified by the observed data. To identify the extrapolation distribution, we use identifying restrictions that express the extrapolation distribution as a function of the observed data distribution; 
see \cite{linero2017general} for a comprehensive discussion. For example, missing at random (MAR) \citep{rubin1976inference} is a joint identifying restriction that completely identifies the extrapolation distribution.
It is shown in \cite{molenberghs1998monotone} that MAR is equivalent to the available case missing value (ACMV)  restriction in the pattern mixture model framework. The same statement is true when conditional on $\bV$, in which case MAR is referred to as auxiliary variable MAR (A-MAR) \citep{daniels2008missing}. ACMV sets 
\begin{align*}
p_k(y_j \mid \bby_{j - 1}, \bv, \bomegae) = p_{\geq j}(y_j \mid \bby_{j - 1}, \bv, \bpi),
\end{align*}
for $k < j$ and $ 2 \leq j < J$, where the subscript $\geq j$ corresponds to conditioning on $S \geq j$. 

When the missingness is not at random, a partial identifying restriction \citep{linero2017general} is the missing non-future dependence (NFD) assumption  \citep{kenward2003pattern}. NFD states that the probability of dropout at time $j$ depends only on $\bby_{j + 1}$. Similarly, when conditional on $\bV$, auxiliary variable NFD (A-NFD) sets
\begin{align*}
p(S = j \mid \bby_J, \bv , \bomega) = p(S = j \mid \bby_{j+1}, \bv , \bomega).
\end{align*}
Within the pattern-mixture framework, NFD is equivalent to the non-future missing value (NFMV) restriction \citep{kenward2003pattern}. Under A-NFD, we have
\begin{align}
p_k(y_j \mid \bar{\bm y}_{j - 1}, \bv, \bomegae) = p_{\geq j - 1}(y_j \mid \bar{\bm y}_{j - 1}, \bv, \bpi),
\label{bnpmis:NFMV}
\end{align}
for $k < j - 1$ and $ 2 < j \leq J$. NFMV leaves one conditional distribution per incomplete pattern unidentified: $p_s(y_{s+1} \mid \bby_{s}, \bv)$. To identify $p_s(y_{s+1} \mid \bby_{s}, \bv)$, we assume a location shift $\tau_{s+1}$ \citep{daniels2000reparameterizing},
\begin{align}
\left[Y_{s+1} \mid \bm \bY_{s}, S = s, \bV, \bomega \right] \eqind
\left[ Y_{s+1} + \tau_{s+1} \mid \bm \bY_{s}, S \geq s+1, \bV, \bomega \right],
\label{location_shift}
\end{align}
where $\eqind$ denotes equality in distribution, and $\tau_{s+1}$ measures the deviation of the unidentified distribution $p_s(y_{s+1} \mid \bby_{s}, \bv)$ from ACMV. In particular, ACMV holds when $\tau_{s+1} = 0$; $\tau_{s+1}$ is  a \emph{sensitivity parameter} \citep{daniels2008missing}. 
To help calibrate the magnitude of $\tau_{s+1}$, we set
\begin{align}
\left[ \tau_{s+1} \mid \bY_{s} = \bby_s, \bV = \bv \right] = \tilde{\tau} \cdot \Delta_{s+1}(\bby_{s}, \bv),
\label{sensparam}
\end{align}
where $\Delta_{s+1}(\bby_{s}, \bv)$ is the standard deviation of $p_s(y_{s+1} \mid \bby_{s}, \bv)$ under ACMV, and $\tilde{\tau}$ represents the number of standard deviations that $p_s(y_{s+1} \mid \bby_{s}, \bv)$ is deviated from ACMV. Importantly, note that, based on the calibration, for a fixed $\tilde{\tau}$ we would have a smaller $\Delta$ using auxiliary covariates and thus a smaller deviation from ACMV, in comparison to unconditional on $\bV$. 

\section{Posterior Inference and Computation}
\label{sec:comp}

\subsection{Posterior Sampling for Observed Data Model Parameters}
We use a Markov chain Monte Carlo (MCMC) algorithm to draw samples from the posterior $\bw_O^{(l)} \iidsim p(\bw_O \mid \{ \bby_{is_i}, s_i, \bv_i \}_{i=1}^N )$, $l = 1, \ldots, L$. 
Note that we use distinct parameters $\bpi, \bm f, \bm \eta$ for $p(\bby_s \mid s, \bv, \bpi)$, $p(s \mid \bv, \bm f)$ and $p(\bv \mid \bm \eta)$, and the parameters are also a priori independent, $p(\bpi, \bm f, \bm \eta) = p(\bpi) p(\bm f) p(\bm \eta)$. Therefore, the posterior distribution of $\bw_O$ can be factored as
\begin{multline*}
p \left ( \bw_O \mid \{ \bby_{is_i}, s_i, \bv_i \}_{i=1}^N \right) = p \left (\bpi \mid \{ \bby_{is_i}, s_i, \bv_i  \}_{i=1}^N \right)  \\
p \left (\bm f \mid \{ s_i, \bv_i \}_{i=1}^N \right ) p \left (\bm \eta  \mid \{ \bv_i \}_{i=1}^N \right ),
\end{multline*}
and posterior simulation can be conducted independently for $\bpi$, $\bm f$ and $\bm \eta$.
Gibbs transition probabilities are used to update $\bpi$ (details in Appendix \ref{app-bnpmis-mcmc}), the R package \texttt{BayesTree} \citep{package:bayestree} is used to update $\bm f$, and $\bm \eta$ is updated by directly sampling from its posterior $\bm \eta \mid \{ \bv_i \}_{i=1}^N \sim \text{Dir}(1 + d_{\eta}, \ldots, 1 + d_{\eta})$.

\subsection{Computation of Expectation of Functionals of Full-data Responses}
Our interest lies in the expectation of functionals of $\by$, given by
\begin{align}
E[t(\by)] &= \int_{\by} t(\by) p(\by) d\by \nonumber\\
&=  \int_{\by} t(\by) \left[ \sum_s \int_v p_s(\tby_s \mid \bby_s, \bv) p_s(\bby_s \mid \bv)  p(s \mid \bv) p(\bv) d\bv \right] d\by.
\label{computation}
\end{align}

Once we have obtained posterior samples $\{ \bw_O^{(l)}, l = 1, \ldots, L \}$, the expression \eqref{computation} can be computed by Monte Carlo integration.  Since the desired functionals are functionals of $\by$, computing \eqref{computation} involves sampling pseudo-data based on the posterior samples. We note that this is an application of G-computation \citep{robins1986new, scharfstein2014global, linero2015flexible} within the Bayesian paradigm (see Algorithm \ref{G-computation}).

\begin{center}
\begin{minipage}{\textwidth}
\begin{algorithm}[H]
\caption{G-computation}
\label{G-computation}
\begin{algorithmic}[1]
\For{$l$  in $1, \ldots, L$ }
\For{$m$  in $1, \ldots, M$ }
\State 1. Draw $\bV^* = \bv^* \sim p(\bv^* \mid \bm \eta^{(l)})$
\State 2. Draw $S^* = s^* \sim p(s^* \mid \bv^*, \bm f^{(l)})$
\State 3. Draw $\bm \bY_s^* = \bby_s^* \sim p(\bby_s^* \mid s^*, \bv^*, \bpi^{(l)})$
\State 4. Draw $\bm{\tilde{Y}}_s^*  = \tby_s^* \sim p(\tby_s^* \mid \bby_s^*, s^*, \bv^*, \bomegae^{(l)})$
\State 5. Set $\bm Y^{*(m, l)} = (\bm \bY_s^*, \bm{\tilde{Y}}_s^*)$
\EndFor
\EndFor\\
\Return{$(1 / ML) \cdot \sum_{m,l} t \left[ \bm Y^{*(m, l)} \right]$}
\end{algorithmic}
\end{algorithm}
\end{minipage}
\end{center}

In detail, for step 1, we draw $\bV^* = \bv_i$ with probability 
$p(\bV = \bv_i \mid \bm \eta^{(l)} ) = \eta_i^{(l)}$. For step 2, we draw $S^*$ 
by sequentially sampling from $R \sim \bernoulli [p(S^* = j \mid S^* \geq j, \bv)]$. If $R = 1$, take $S^* = j$; otherwise proceed with $p(S^* = j+1 \mid S^* \geq j+1, \bv)$, $j = 2, \ldots, J$. 
For step 3, we first draw $y_1^* \sim N \left(a_{0}(\bv^*, 1, s^*), \sigma_{1s^*}^2 \right)$ 
and then sequentially draw $y_j^* \sim N \left(a ( \bv^*, j, s^*) + \bby_{j-1}^T \bPhi_{js^*}, \sigma_{js^*}^2  \right)$, $j = 2, \ldots, s^*$ as in \eqref{eq:factorize_obs}, 
where $a_{0} (\bv^*, 1, s^*)$ and  
$a ( \bv^*, j, s^* )$ are generated by GP prediction rule \citep{rasmussen2006gaussian}. For step 4, we sequentially draw $y_j^*$ for $j = s^*+1, \ldots, J$ as in \eqref{extrapolation_distribution} from the unidentified distributions, now identified using identifying restrictions. When the ACMV restriction is specified, step 4 involves sampling from a distribution of $p_{\geq j }(y_j \mid \bar{\bm y}_{j - 1}, \bv)$, where
\begin{align}
p_{\geq j }(y_j \mid \bar{\bm y}_{j - 1}, \bv) = \sum_{k = j}^J 
\alpha_{kj}(\bar{\bm y}_{j - 1},  \bv) \; 
p_k(y_j \mid  \bar{\bm y}_{j - 1}, \bv),
\label{sample_NFD}
\end{align}
and
\begin{align*}
\alpha_{kj}(\bar{\bm y}_{j - 1},  \bv) &= p(S = k \mid \bar{\bm y}_{j - 1},  S \geq j, \bv) \\
&= \frac{p(\bar{\bm y}_{j - 1} \mid S = k, \bv)p(S = k \mid S \geq j, \bv)}{\sum_{k=j}^J p(\bar{\bm y}_{j - 1} \mid S = k, \bv)p(S = k \mid S \geq j, \bv)}.
\end{align*}
The distribution in \eqref{sample_NFD} is a mixture distribution over patterns. 
We sample from \eqref{sample_NFD} by first 
drawing $K = k$ with probability $\alpha_{kj}$, $k = j, \ldots, J$, 
then drawing a sample from $p_k(y_j \mid  \bar{\bm y}_{j - 1}, \bv)$. When the NFMV restriction is specified, step 4 also involves sampling from a distribution $p_{\geq j-1 }(y_j \mid \bar{\bm y}_{j - 1}, \bv)$, where
\begin{multline*}
p_{\geq j -1}(y_j \mid \bar{\bm y}_{j - 1}, \bv) 
= \alpha_{j-1, j-1}(\bar{\bm y}_{j - 1},  \bv) \; 
p_{j-1}(y_j \mid  \bar{\bm y}_{j - 1}, \bv) +  \\
[1-\alpha_{j-1, j-1}(\bar{\bm y}_{j - 1},  \bv)]
p_{\geq j }(y_j \mid \bar{\bm y}_{j - 1}, \bv).
\end{multline*}
Sampling from $p_{\geq j-1 }(y_j \mid \bar{\bm y}_{j - 1}, \bv)$ is done by first sampling $Y_j^* \sim p_{\geq j} (y_j \mid  \bar{\bm y}_{j - 1}, \bv)$ as in \eqref{sample_NFD}. Then draw $R \sim \bernoulli [\alpha_{j-1, j-1}]$. If $R = 1$, apply the location shift \eqref{location_shift}, otherwise, retain $Y_j^*$. See Appendix \ref{app-bnpmis-gcomp} for more details of steps 3 and 4.


\section{Simulation Studies}
\label{sec:simu}
We conduct several simulation studies similar to the data example to assess the operating characteristic of our proposed model. We simulate responses for $J = 6$ time points and fit our model to estimate the 
change from baseline treatment effect, i.e. $\E[Y_J - Y_1]$. We take $\kappa_0^2   \sim \IG(10, 1)$ and $\kappa^2 \sim \IG(10, 1)$ to shrink the nonparametric model towards a simple linear regression model, and set the other prior and hyperprior parameters at standard noninformative choices. See Appendix \ref{app-bnpmis-sim} for exact values.
For comparison, we consider two alternatives: (1) a parametric model that consists of a linear regression model for $p_s(y_j \mid \bby_{j-1}, \bv)$, a sequential logit model for $p(s \mid \bv)$, and a Bayesian bootstrap model for $p(\bv)$, as in Equations \eqref{sim1:mu}, \eqref{sim1:s} and \eqref{eq:bb}, respectively; (2) a parametric model without $\bV$ that consists of a linear regression model for $p_s(y_j \mid \bby_{j-1})$ and a Bayesian bootstrap model for $p(s)$.
We use noninformative priors for the two parametric models. 
For each simulation scenario below, we generate 500 datasets with $N = 200$ subjects per dataset.

\subsection{Performance Under MAR}
\label{sec:simuMAR}
We first evaluate the performance of our model under the ACMV restriction (MAR). 
Since this restriction completely identifies the extrapolation distribution, this simulation study validates the appropriateness of our observed data model specification.
We consider the following three simulation scenarios.
\paragraph{Scenario 1.} 
We test the performance of our approach when the data are generated from a simple linear pattern-mixture model to assess loss of efficiency from using an unnecessary complex modeling approach.
For each subject, we first simulate $Q = 4$ auxiliary covariates from a  multivariate normal distribution 
\begin{align}
\bV \iidsim N(\bm 0, \Sigma_{vv}).\label{sim1:v}
\end{align}
We then generate dropout time using a sequential logit model
\begin{align}
\logit \, P(S = s \mid S \geq s, \bV) &= \zeta_s + \bV^T \bxi_s. \label{sim1:s}
\end{align}
Next, we generate $\bm \bY_{s}$ from
\begin{align}
\left(Y_{j} \mid \bm \bY_{j-1}, S = s, \bV \right) &\sim N \left(\mu_{js}(\bm \bY_{j-1}, \bV), \sigma_{js}^2 \right), \quad \text{for $j = 1, \ldots, s$} \nonumber\\
\text{where} \quad \mu_{js}(\bm \bY_{j-1}, \bV) &= 
\begin{cases}
\bV^T \bbeta_{0s} + b_{js}  \quad &\text{if $j = 1$} \\
\bV^T \bbeta_s + b_{js} + \bm \bY_{j-1}^T \bPhi_{js} \quad &\text{if $j \geq 2$}
\end{cases} \label{sim1:mu}
\end{align}
Finally, the distribution of $\bm \tY_{s}$ is specified under the ACMV restriction (for calculating the simulation truth of the mean estimate).

The parameters in \eqref{sim1:v}, \eqref{sim1:s} and \eqref{sim1:mu} are chosen by fitting the model to the test drug arm of the schizophrenia clinical trial (after standardizing the responses and the auxiliary covariates with mean 0 and standard deviation 1). See Appendix \ref{app-bnpmis-sim} for details.

\paragraph{Scenario 2.} 
We consider a scenario where the covariates and the responses have more complicated structures in order to test the performance of our model when linearity does not hold. 
For simplicity, for each subject, we simulate $Q = 3$ auxiliary covariates from
$\bV \iidsim N(\bm 0, \Sigma_{vv})$.
The responses and drop out times are generated in the same way as in scenario 1, but we include interactions and nonlinearities by replacing $\bV$ in Equations \eqref{sim1:s} and \eqref{sim1:mu} with $\tilde{\bV} = (V_1, V_2, V_3, V_1 \times V_2, V_1 \times V_3, V_2 \times V_3, V_1^2, V_2^2, V_3^3)$. The regression coefficients $\bxi_s$, $\bbeta_{0s}$ and $\bbeta_{s}$ change accordingly. See Appendix \ref{app-bnpmis-sim} for further details.

\paragraph{Scenario 3.} 
We consider a scenario with a very different structure from our model formulation.  In particular, we consider a 
lag-1 selection model with a mixture model for the joint distribution of $\bm Y$ and $\bV$. 
We generate
\begin{align}
&K \sim \mbox{Categorical}(\bm \pi), \nonumber\\
&\Omega^{(K)} \sim \mathcal{W}^{-1}\left( (\nu - J - Q - 1) \Omega_0^{(K)}, \nu \right), \nonumber\\
&\left( \begin{array}{c}
\bm{Y} \\
\bV  \end{array} \right) \mid K \sim  N\left[ \bm{\mu}^{(K)}, \Omega^{(K)}\right], \label{eq:normal_mixture}\\
&\logit \, P(S = s \mid S \geq s, \bm{Y}, \bV) = \zeta_s + \ell_s Y_{s} + \bV^T \bxi_s, \nonumber
\end{align}
where $\mathcal{W}^{-1}\left( (\nu - \dim(\Omega_0) - 1) \Omega_0, \nu \right)$ is an inverse-Wishart distribution with precision parameter $\nu$ and mean $\Omega_0$.
See \cite{linero2015flexible} for further details on this type of model. Formulating a joint distribution as in \eqref{eq:normal_mixture} allows us to impose complicated relationships between $\bm{Y}$ and $\bV$ \citep{muller1996bayesian}. We consider $Q = 3$ auxiliary covariates and $5$ mixture components.  We assume $\bm{\mu}^{(K)}$ and $\Omega_0^{(K)}$ correspond to a linear model of $(\bm Y \mid \bV)$ and have the form
\begin{align*}
\bm{\mu}^{(K)} = \left( \begin{array}{c}
\bm{\mu}_y^{(K)} \\
\bm 0  \end{array} \right), \quad
\Omega_0^{(K)} = \left( \begin{array}{cc}
\Sigma_{yy}^{(K)} & \Sigma_{yv}^{(K)} \\
\Sigma_{vy}^{(K)} & \Sigma_{vv} \end{array} \right).
\end{align*}
In particular, we generate $\bm{\mu}^{(K)}$ and $\Omega_0^{(K)}$ according to \cite{linero2015flexible}
by fitting the mixture model to the active control arm of the schizophrenia clinical trial. See Appendix \ref{app-bnpmis-sim} for further details.

\begin{table}[h!]
\centering
\begin{tabular}{crrrr}
\toprule
Model        & \multicolumn{1}{c}{Bias}    & \multicolumn{1}{c}{CI width}  & \multicolumn{1}{c}{CI coverage} & \multicolumn{1}{c}{MSE} \\ \hline
          & \multicolumn{4}{c}{Scenario 1} \\
NP & -0.010(0.004)   &  0.294(0.002)    & 0.910(0.012)        & 0.008(0.000)  \\ 
LM & -0.005(0.004)   &  0.379(0.001)    & 0.969(0.007)        & 0.008(0.000)  \\ 
No $\bV$ & 0.004(0.004)   &  0.385(0.002)    & 0.969(0.007)        & 0.008(0.000)  \\ 
          & \multicolumn{4}{c}{Scenario 2} \\
NP &  0.038(0.010)   &  0.915(0.004)    &     0.933(0.011)    &     0.058(0.004) \\ 
LM &  0.197(0.010)   &  0.954(0.004)    &     0.855(0.015)    &     0.097(0.005) \\  
No $\bV$ & 0.289(0.010)   &  1.009(0.004)    & 0.794(0.017)        & 0.138(0.007)  \\ 
          & \multicolumn{4}{c}{Scenario 3} \\
NP &   0.001(0.007)  & 0.668(0.002) &   0.953(0.009)  & 0.028(0.002)  \\ 
LM &   0.008(0.007)  &  0.705(0.002) &   0.968(0.008)  & 0.028(0.002) \\
No $\bV$ &   0.026(0.007)  &  0.707(0.002) &   0.964(0.008)  & 0.028(0.002) \\ \bottomrule
\end{tabular}
\caption{Summary of simulation results under MAR. Values shown are posterior means, with Monte Carlo standard errors in parentheses. NP,  LM and No $\bV$ represent the proposed nonparametric model, the linear regression model with auxiliary covariates and the linear regression model without auxiliary covariates, respectively. Coverage of $95\%$ credible intervals. }
\label{tbl:simu_MAR}
\end{table}

The simulation results are summarized in Table \ref{tbl:simu_MAR}. 
For scenario 1, the true data generating model is the linear regression model with $\bV$.
The three models have similar performance in terms of MSE. The $95\%$ credible interval of the nonparametric model has a frequentist coverage rate less than $95\%$ due to the 
prior information, i.e., the Gaussian process priors and the AR/CAR priors, being quite strong and the sample size ($N = 200$) being relatively small. Therefore, the Bayesian credible interval is unlikely to have the expected frequentist coverage. The linear regression model ignoring $\bV$ does not perform worse than the one including $\bV$. The reason is probably that the (linear) effects of different $\bV$'s on $t(\bm Y)$ cancel out in the integration \eqref{computation}.
For scenario 2, the true data generating model does not match any of the three models used for inference.
The nonparametric model significantly outperforms the parametric linear regression models in all aspects. The result suggests that when the model is misspecified, the nonparametric model has much more robust performance. We also note that when $\bm Y$ and $S$ do not have a linear relationship with $\bV$, ignoring $\bV$ results in more significant bias than including $\bV$ (even mistakenly).
For scenario 3, the true data generating model is a mixture of linear regression models but has a different parameterization with the three models used for inference.
The three models again have similar performance. For a pattern-mixture model, the marginal distribution of the responses  $\bm Y$ is a mixture distribution over patterns, which explains the good performance of the three models. For all the three scenarios, the nonparametric model always gives narrower credible intervals and has lower bias, in particular versus the model without auxiliary covariates.  

In summary, the nonparametric approach loses little when the corresponding parametric model holds, 
and it significantly outperforms the other approaches when the model used for inference is misspecified.
The simulation results suggest that the nonparametric approach is more favorable compared with the parametric approaches and accommodates complex mean models.

\subsection{Performance Under MNAR}
To assess the sensitivity of our model to untestable assumptions for the extrapolation distribution, we fit our model to simulated data under an NFD restriction \eqref{bnpmis:NFMV}. We consider simulation scenarios 2 and 3 as in Section \ref{sec:simuMAR}, where the simulation truth is still generated under MAR. 
We complete our model with a location shift (Equations \eqref{location_shift} and \eqref{sensparam}). Recall that the sensitivity parameter $\tilde{\tau}$ measures the deviation of our model from MAR, and the simulation truth corresponds to $\tilde{\tau} = 0$.
The sensitivity parameter $\tilde{\tau}$ is given four different priors: $\Unif(-0.75, 0.25)$, $\Unif(-0.5, 0.5)$, $\Unif(-0.25, 0.75)$, $\Unif(0, 1)$.
All the four priors contain the simulation truth.
Compared to fixing the value of $\tilde{\tau}$, using a uniform prior conveys uncertainty about the identifying restriction. For example, using a point mass prior $\tilde{\tau} = 0$ implies MAR with no uncertainty, while using a prior such that $\E[\tilde{\tau}] = 0$ and $\Var[\tilde{\tau}] > 0$ implies MAR with uncertainty.

\begin{table}[h!]
\centering
\begin{tabular}{crrrrr}
\toprule
Model &  \multicolumn{1}{c}{$\E(\tilde{\tau})$}  & \multicolumn{1}{c}{Bias}    & \multicolumn{1}{c}{CI width}  & \multicolumn{1}{c}{CI coverage} & \multicolumn{1}{c}{MSE} \\ \hline
   &       &    \multicolumn{4}{c}{Scenario 2} \\
NP & -0.25 & -0.046(0.010)  &  0.979(0.004)  &    0.968(0.008)  &   0.051(0.003)\\
      &  0      &  0.039(0.010)   &  0.994(0.004)  &    0.966(0.008)  &   0.052(0.003) \\
      &  0.25 &  0.129(0.010)   &  1.016(0.004)  &    0.942(0.010)  &   0.068(0.004) \\
      &  0.5   &  0.224(0.010)   &  1.036(0.004)  &    0.874(0.014)  &   0.103(0.006) \\ 
LM & -0.25 & 0.081(0.010)    &  1.056(0.004)  &    0.966(0.008)  &   0.056(0.004) \\
      &  0      & 0.189(0.010)    &  1.075(0.004)  &    0.923(0.012)  &   0.086(0.005) \\
      &  0.25 & 0.301(0.010)    &  1.101(0.005)  &    0.841(0.016)  &   0.142(0.007) \\
      &  0.5   & 0.418(0.010)    &  1.132(0.005)  &    0.700(0.020)  &   0.228(0.009) \\
No $\bV$  & -0.25 & 0.155(0.009)   &  1.135(0.005)  &    0.961(0.008)  &   0.069(0.004) \\
                 &  0      & 0.281(0.009)   &  1.161(0.005)  &    0.889(0.014)  &   0.125(0.006) \\
                 &  0.25 & 0.411(0.009)   &  1.191(0.005)  &    0.771(0.018)  &   0.216(0.008) \\
                &  0.5    & 0.546(0.010)   &  1.223(0.005)  &    0.583(0.021)  &   0.348(0.011) \\
       &          &    \multicolumn{4}{c}{Scenario 3} \\
NP  & -0.25 &  -0.049(0.007)   &  0.691(0.002)  &    0.944(0.010)  &   0.032(0.002)\\
       &  0      &  -0.008(0.007)   &  0.695(0.002)  &    0.972(0.007)  &   0.030(0.002) \\
       &  0.25 &   0.033(0.008)   &  0.696(0.002)  &    0.963(0.008)  &   0.032(0.002) \\
       &  0.5   &   0.076(0.008)   &  0.703(0.002)  &    0.929(0.011)  &   0.037(0.002) \\ 
LM  & -0.25 &  -0.042(0.007)   &  0.725(0.002)  &    0.961(0.008)  &   0.031(0.002) \\
       &  0      &   -0.001(0.007)  &  0.728(0.002)  &    0.980(0.006)  &   0.030(0.002) \\
       &  0.25 &    0.042(0.008)  &  0.734(0.002)  &    0.972(0.007)  &   0.032(0.002) \\
       &  0.5   &    0.085(0.008)  &  0.741(0.002)  &    0.948(0.010)  &   0.038(0.002) \\
No $\bV$ & -0.25 & -0.047(0.007)   &  0.751(0.002)  &    0.972(0.007)  &   0.031(0.002) \\
                &  0      &  0.015(0.007)   &  0.761(0.002)  &    0.987(0.005)  &   0.029(0.002) \\
                &  0.25 &  0.079(0.007)   &  0.768(0.002)  &    0.966(0.008)  &   0.036(0.002) \\
                &  0.5   &  0.144(0.008)   &  0.783(0.003)  &    0.909(0.012)  &   0.052(0.003) \\ \bottomrule
\end{tabular}
\caption{Summary of simulation results under MNAR. Values shown are posterior means, with Monte Carlo standard errors in parentheses. NP, LM and No $\bV$ represent the proposed nonparametric model, the linear regression model with auxiliary covariates and the linear regression model without auxiliary covariates, respectively. Coverage of $95\%$ credible intervals. The values of $\E(\tilde{\tau})$, $-0.25$, $0$, $0.25$ and $0.5$, correspond to prior specifications $\Unif(-0.75, 0.25)$, $\Unif(-0.5, 0.5)$, $\Unif(-0.25, 0.75)$ and $\Unif(0, 1)$, respectively.}
\label{tbl:simu_MNAR}
\end{table}

The simulation results are summarized in Table \ref{tbl:simu_MNAR}. 
When the sensitivity parameter $\tilde{\tau}$ is centered at the correct value 0, the nonparametric model significantly outperforms the parametric linear regression models under scenario 2 and performs as well as the parametric linear regression models under scenario 3. 
Comparing with the simulation results under MAR (Table \ref{tbl:simu_MAR}), the use of a uniform prior for $\tilde{\tau}$ induces more uncertainty on inference according to the wider credible intervals.
We also note that, when $\tilde{\tau}$ is not centered at 0, 
the models using $\bV$ still perform better than the model not using $\bV$. This is due to the calibration of the location shift (Equations \eqref{location_shift} and \eqref{sensparam}). For the same $\tilde{\tau}$ we would have a smaller deviation from ACMV using $\bV$ compared to not using $\bV$.
This property makes the missingness ``closer'' to MAR and reduces the extent of sensitivity analysis with the inclusion of $\bV$.

\section{Application to the Schizophrenia Clinical Trial}
\label{sec:real_data}

We implement inference under the proposed model for data from the  schizophrenia clinical trial described in Section \ref{sec:sctdata}. Recall the quantity of interest is the change from baseline
treatment effect, $r_x = E[Y_{i6} - Y_{i1} \mid X_i = x]$, where $x = 1, 2$ or $3$ correspond
to treatments under active control, placebo or test drug, respectively. We are particularly interested in the treatment effect improvements over placebo, i.e. $r_1 - r_2$ and $r_3 - r_2$.

\subsection{Comparison to Alternatives and Assessment of Model Fit}
We first compare the fit  among the proposed model and alternatives. 
We consider a linear regression model with auxiliary covariates and a linear regression model without auxiliary covariates, as we have used in the simulation studies. We use the log-pseudo marginal likelihood (LPML) as the model selection criteria. LPML is defined by
\begin{align*}
\text{LPML} = \frac{1}{N} \sum_{i=1}^N \log (\text{CPO}_i).
\end{align*}
Here $\text{CPO}_i$ is the conditional predictive ordinate \citep{geisser1979predictive} for observation $i$, 
\begin{align*}
\text{CPO}_i = p \left( \bar{\bm Y}_{iS_i}, S_i, V_i \mid  \{ \bar{\bm Y}_{i'S_{i'}}, S_{i'}, V_{i'} \}_{i'=1, i' \neq i}^{N} \right).
\end{align*}
LPML can be straightforwardly estimated using posterior samples $\{ \bomegao^{(l)}, l = 1, \ldots, L \}$ \citep{gelfand1994bayesian}. A model with higher LPML is more favorable compared to models with lower LPMLs.
We fit the three models to the data and calculate the LPML by taking the summation of the LPML under each treatment arm. The results are summarized in Table \ref{tbl:real_LPML}. The proposed nonparametric model had the largest LPML over the linear regression models with and without auxiliary covariates. This is not surprising in light of the earlier simulation results.
We also compare inferences on treatment effect improvements over placebo under the MAR assumption using the three models. The results are summarized in Table \ref{tbl:real_LPML}.


\begin{table}[h!]
\centering
\begin{tabular}{cc@{\hskip 0.3in}c@{\hskip 0.3in}c}
\toprule
Model        & LPML    & $r_1 - r_3$  & $r_2 - r_3$  \\ \hline
NP      & -31.97   &  0.90(-5.56, 7.63)      & -7.09(-15.75, 0.98)         \\ 
LM      & -32.61   &  -1.11(-7.96, 5.73)     & -6.99(-14.61, 0.54)         \\ 
No V  &  -32.71  &  -1.86(-8.53, 4.86)    &  -8.10(-15.44, -0.98)         \\ 
\bottomrule
\end{tabular}
\caption{Comparison of LPML (the second column) and inference results under MAR (the third and fourth columns). NP, LM and No V represent the proposed model, linear regression model with auxiliary covariates and  linear regression model without auxiliary covariates, respectively. For the inference results under MAR, values shown are posterior means, with $95\%$ credible intervals in parentheses.}
\label{tbl:real_LPML}
\end{table}

\begin{figure}[h!]
\begin{center}
\includegraphics[width=0.95\textwidth]{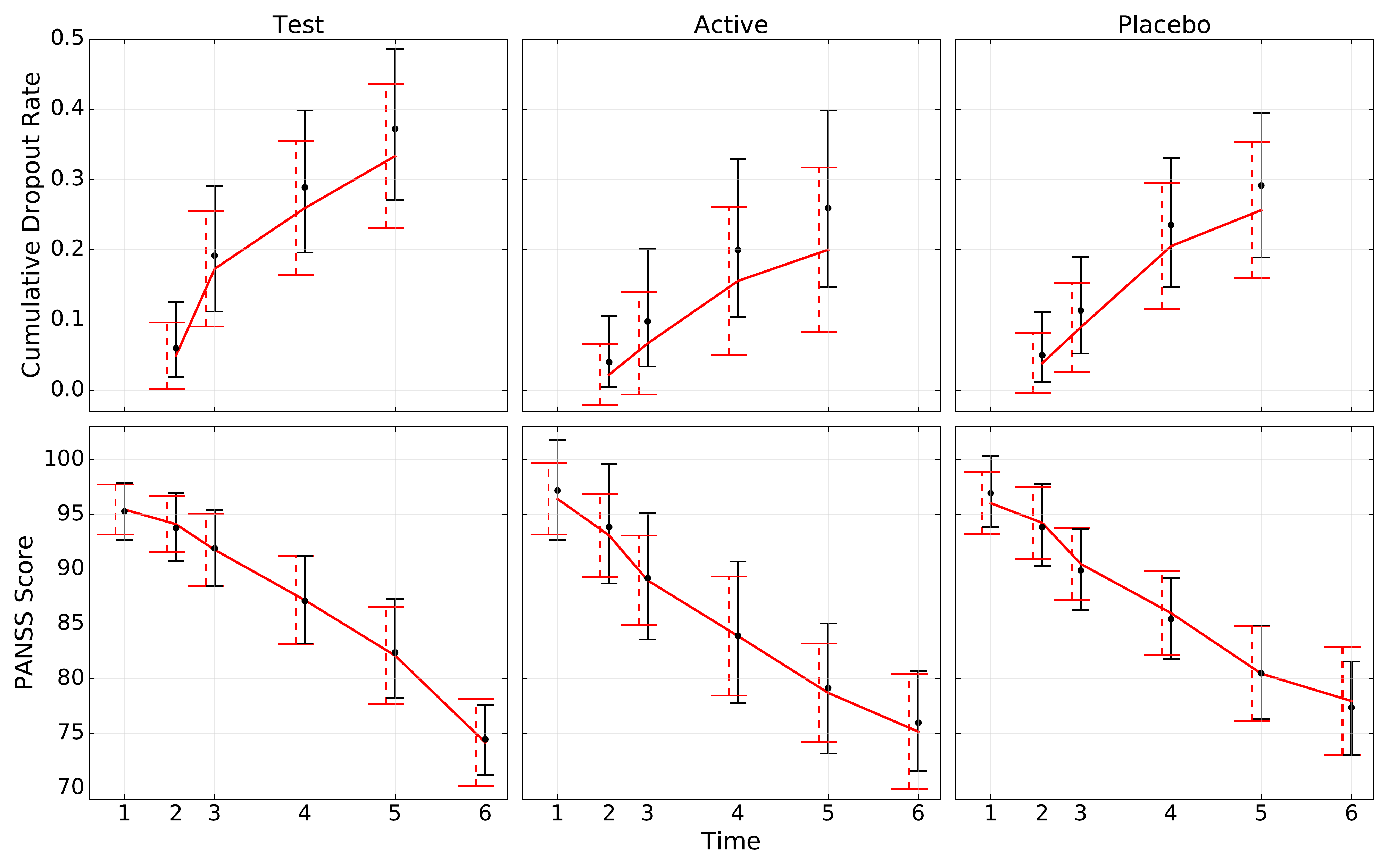}
\end{center}
\caption{Cumulative dropout rates (top) and observed-data means (bottom) over time obtained from the model versus the ones obtained from the empirical distribution.  The solid red line represents the empirical values, black dots represent the posterior means, red dashed error bars represent frequentist $95 \%$ confidence intervals, and black solid error bars represent the model's $95 \%$ credible intervals. }
\label{fig:bnpmis_real_emp}
\end{figure}

Next, we assess the ``absolute'' goodness of fit of the proposed model.
We estimate the cumulative dropout rates and observed-data means at each time point and under each treatment using the proposed model.
That is,
\begin{align*}
&p(S \leq j \mid x) = \int p(S \leq j \mid \bv, x) p(\bv \mid x) d\bv, \quad \text{and} \\
&\E(Y_j \mid S \geq j, x) = \int \E(Y_j \mid S \geq j, \bv, x) p(\bv \mid S \geq j, x) d\bv.
\end{align*}
We then compare those estimates with results obtained from the empirical distribution (not using information from auxiliary covariates). The comparison is shown in Figure \ref{fig:bnpmis_real_emp}. Despite some small differences that are due to the use of information from auxiliary covariates and prior information, there is no evidence for lack of fit. 

\subsection{Inference}
A large portion of subjects dropout for reasons that suggest the missing data are MNAR (see Section \ref{sec:sctdata}). To identify the extrapolation distribution, we make the NFD assumption \eqref{bnpmis:NFMV}. Recall that the NFD assumption leaves one conditional distribution per incomplete pattern unidentified: $p_s(y_{s+1} \mid \bby_{s}, \bv, x)$. To better identify $p_s(y_{s+1} \mid \bby_{s}, \bv, x )$, rather than simply assuming a location shift \eqref{location_shift}, we make use of information regarding the type of dropout.
Let $Z_i = 1$ or $0$ denote subject $i$ drops out for informative or noninformative reasons, respectively. 
We model $Z$ conditional on observed data responses, pattern, auxiliary covariates and treatment with a logistic regression, 
\begin{align*}
\logit P(Z = 1 \mid \bar{\bm Y}_s, S = s, \bV, X = x) = \zeta_{sx} + \bar{\bm Y}_s^T \bm \ell_{sx} + \bV^T \bm \xi_{sx}.
\end{align*}
An alternative nonparametric choice for modeling $P(Z \mid \bar{\bm Y}_S, S, \bV, X)$ is BART.  Since the sample size of $Z$ (i.e. number of dropout subjects for each pattern and each treatment) is small, we find that the simpler logistic regression model suffices here.

The indicator $Z$ is used to help identify $p_s(y_{s+1} \mid \bby_{s}, \bv, x )$. We assume
\begin{multline}
\left[Y_{s+1} \mid \bm \bY_{s}, S = s, \bV, X, \bomega \right] \eqind \\
P(Z=1 \mid \bm \bY_{s}, S = s, \bV, X) \cdot 
\left[ Y_{s+1} + \tau_{s+1} \mid \bm \bY_{s}, S \geq s+1, \bV, X, \bomega \right] + \\
P(Z=0 \mid \bm \bY_{s}, S = s, \bV, X)  \cdot \left[ Y_{s+1}  \mid \bm \bY_{s}, S \geq s+1, \bV, X, \bomega \right],
\label{NFD_real}
\end{multline}
which is a mixture of an ACMV assumption and a location shift. The idea is that, if a subject drops out for a reason associated with MAR, we impute the next missing value under ACMV; otherwise, we impute the next missing value by applying a location shift.
The sensitivity parameter $\tau_{s+1}$ is interpretable to subject-matter experts. 
Suppose two hypothetical subjects A and B have the same auxiliary covariates and histories up to time $s$, and suppose subject B drops out for an informative reason at time $s$ while subject A remains on study. Then, the response of subject B at time $(s + 1)$ is stochastically identical to the response of subject A at time $(s + 1)$ after applying the location shift $\tau_{s+1}$. 
As the prior for $\tau_{s+1}$, we assume $\tau_{s+1} \geq 0$ as we expect subject B would have a higher PANSS score at time $(s + 1)$ than subject A. 
The magnitude of $\tau_{s+1}$ is calibrated as in Equation  \eqref{sensparam},
\begin{align}
\left[ \tau_{s+1} \mid \bY_{s} = \bby_s, \bV = \bv, X = x \right] = \tilde{\tau}_x \cdot \Delta_{s+1, x}(\bby_{s}, \bv).
\label{sensparam_real}
\end{align}
We assume a uniform prior on  $\tilde{\tau}_x$, $\tilde{\tau}_x \sim \Unif(0, 1)$, as it is thought unlikely that the deviation from ACMV would exceed a standard deviation \citep{linero2015flexible}.

\begin{figure}[h!]
\begin{center}
\includegraphics[width=0.95\textwidth]{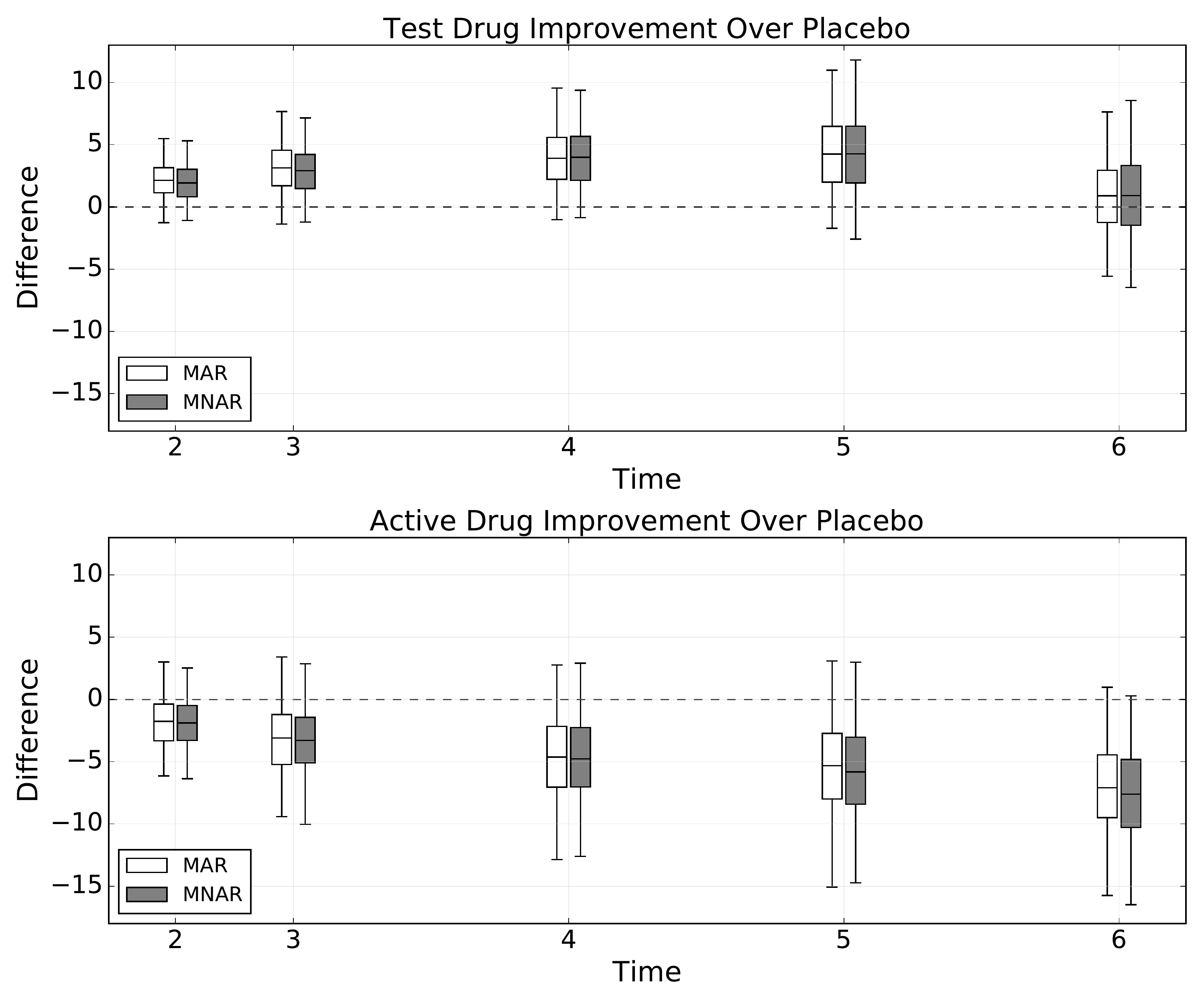}
\end{center}
\caption{Change from baseline treatment effect improvements of the test drug (top) and active drug (bottom) over placebo over time. Smaller values indicate more improvement compared to placebo. The dividing line within the boxes represents the posterior mean, the bottom and top of
the boxes are the first and third quartiles, and the ends of the whiskers show the 0.025 and 0.975 quantiles.}
\label{fig:bnpmis_real_improvement}
\end{figure}

Figure \ref{fig:bnpmis_real_improvement} summarizes change from baseline treatment effect improvements of the test drug and active drug over placebo. We implement inference under both the MAR and the mixture of MAR/MNAR (Equations \eqref{NFD_real} and \eqref{sensparam_real}) assumptions.
For the test drug arm, the treatment effect improvement $r_1 - r_3$ has posterior mean $0.90$ and $95\%$ credible interval $(-5.56, 7.63)$ under MAR, and posterior mean $0.91$ and $95\%$ credible interval $(-6.47, 8.54)$ under MNAR. There is no evidence that the test drug has better performance than placebo.
The MNAR assumption has little effect on inference on $r_1 - r_3$, since the test and placebo arms have similar informative dropout rates.
For the active drug arm, the treatment effect improvement $r_2 - r_3$ has posterior mean $-7.09$ and $95\%$ credible interval $(-15.75, 0.98)$ under MAR, and posterior mean $-7.61$ and $95\%$ credible interval $(-16.49, 0.28)$ under MNAR. There appears to be some evidence that the active drug has better treatment effect than placebo, especially under MNAR.
The MNAR assumption makes the difference between the active and placebo arms more significant, since the active arm has a smaller informative dropout rate.

\subsection{Sensitivity Analysis}
\begin{figure}[h!]
\begin{center}
\includegraphics[width=0.95\textwidth]{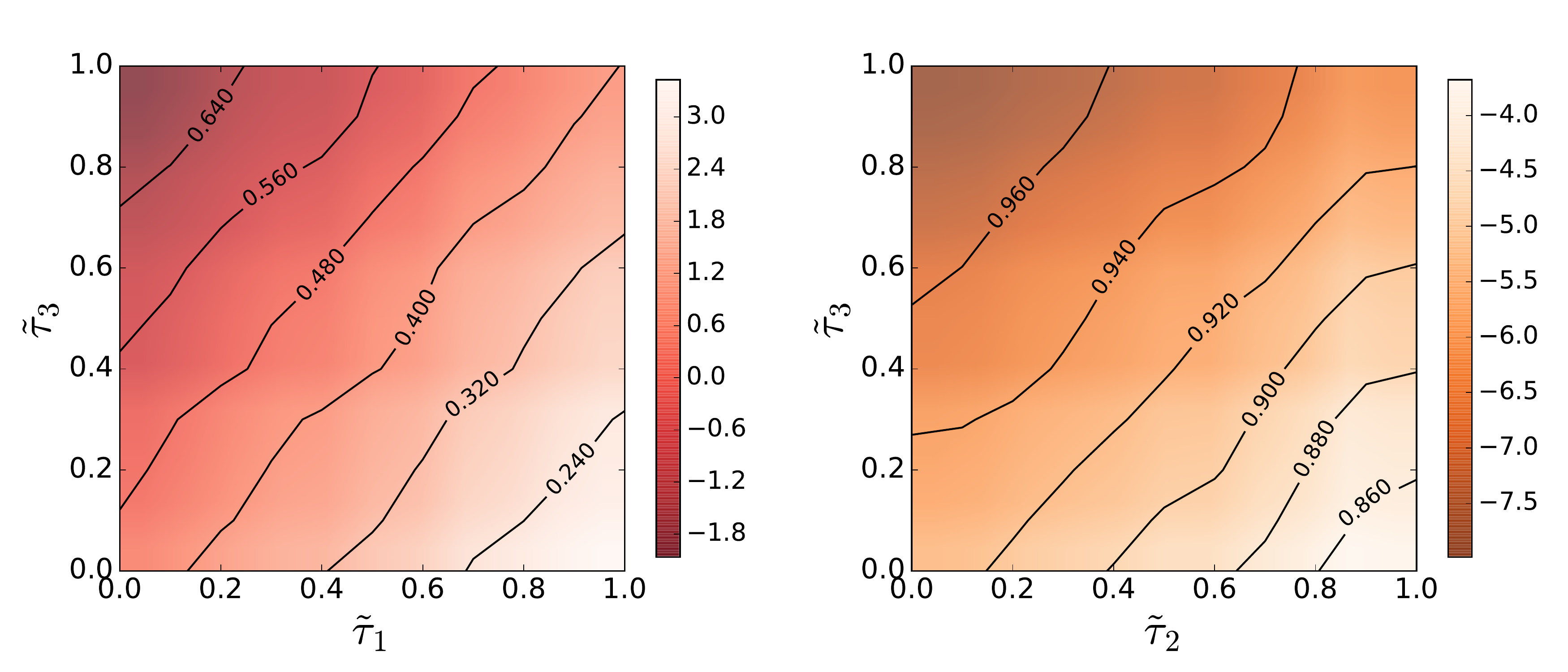}
\end{center}
\caption{Contour plots showing inferences on treatment effect improvements $r_1 - r_3$ (left) and $r_2 - r_3$ (right) for different choices of the sensitivity parameters along the $[0, 1]$ grid. The colors represent posterior means of $r_x - r_3$, where a deeper color indicates more improvement compared to placebo. The black lines show posterior probabilities of $r_x - r_3 < 0$. }
\label{fig:bnpmis_real_SA}
\end{figure}

To assess the sensitivity of inferences on treatment effect improvements ($r_1 - r_3$ and $r_2 - r_3$) to the informative priors on the sensitivity parameters ($\tilde{\tau}_1$, $\tilde{\tau}_2$ and $\tilde{\tau}_3$), we consider a set of point-mass priors for each $\tilde{\tau}_x$ along the $[0, 1]$ grid. Figure \ref{fig:bnpmis_real_SA} summarizes how inferences on $r_1 - r_3$ and $r_2 - r_3$ change for different choices of $\tilde{\tau}_1$, $\tilde{\tau}_2$ and $\tilde{\tau}_3$. Considering a significance level of 0.05. The sensitivity analysis corroborates our conclusion that there is no evidence that the test drug has better performance than placebo. For all the choices of $\tilde{\tau}_1$ and $\tilde{\tau}_3$, the posterior probability of $r_1 - r_3 < 0$ does not reach the 0.95 cutoff. On the other hand, the sensitivity analysis shows that there is some evidence that the active drug is superior than placebo. For all the combinations of $\tilde{\tau}_2$ and $\tilde{\tau}_3$, the posterior probability of $r_2 - r_3 < 0$ is greater than 0.84. For most favorable values of $\tilde{\tau}_2$ and $\tilde{\tau}_3$, the posterior probability of $r_2 - r_3 < 0$ is greater than 0.95, although it only occurs when $\tilde{\tau}_2$ is substantially smaller than $\tilde{\tau}_3$.
In summary, for all the choices of $\tilde{\tau}_x$, we do not reach substantially different results, which improves our confidence on the previous conclusions.
Finally, inferences on $r_1 - r_3$ and $r_2 - r_3$ under the uniform prior $\tilde{\tau}_x \sim \Unif(0, 1)$ are roughly the same as inferences under the point-mass prior $\tilde{\tau}_x = 0.5$, which means using the uniform prior does not induce much more uncertainty in this scenario.

\section{Discussion}
\label{sec:discuss}
In this work, we have developed a nonparametric Bayesian approach to monotone missing data with non-ignorable missingness in the presence of auxiliary covariates.
Under the extrapolation factorization, we flexibly model the observed data distribution and specify the extrapolation distribution using identifying restrictions.
We have shown the inclusion of auxiliary covariates in the model could in general improve the accuracy of inferences and reduce the extent of sensitivity analysis. 
We have also shown more accurate inferences can be obtained by using the proposed nonparametric Bayesian approach compared to using more restrictive parametric approaches.

In the model for the observed responses conditional on past responses, pattern and auxiliary covariates, we have assumed the effect of past responses on current response is linear (Equation \eqref{model_for_response} case $j \geq 2$). To make the model more flexible, we could include past responses in the index set of the stochastic process $a( \cdot )$, i.e. to model 
$\left(Y_{j} \mid \bm \bY_{j-1}, S, \bV, \bpi \right) = 
a(\bar{\bm Y}_{j-1}, \bV, j, S) + \varepsilon_{js}$. 
This way the model could account for possible nonlinearity and interactions in the past responses. However, this modeling approach is complicated by the fact that for different time $j$, the dimension of past responses $\bar{\bm Y}_{j-1}$ is different, so we leave it as an extension of this work. A possible compromise could be including only  lag-1 response in $a( \cdot )$.

The computation complexity of the Gaussian process is cubic in the number of data points. The problem is manageable in our application since the schizophrenia clinical trial dataset only contains $204$ subjects. When a much larger number of subjects is considered, several methods have been proposed to tackle the computational bottleneck of the GP (see \citealp{banerjee2008gaussian, banerjee2013efficient, hensman2013gaussian, datta2016hierarchical}).
To identify the extrapolation distribution under NFD, we assume a location shift. Alternatively, we can consider exponential tilting \citep{rotnitzky1998semiparametric, birmingham2003pattern}.

A possible extension of our work is to consider continuous time dropout. The Gaussian process is naturally suitable for the continuous case.  Another extension would be more flexible incorporation of auxiliary covariates beyond the mean. 
Another possible future direction is to extend our method to non-monotone missing data without imposing the partial ignorability assumption. In the setting of binary outcomes, our method can be extended by using a probit link. 

\chapter{Future Work}
\label{chap:future}

In the preceding chapters, we have developed nonparametric Bayesian models for biomedical data analysis. In particular, we have presented a novel feature allocation model for tumor subclone reconstruction using mutation pair data, a treed feature allocation model for tumor subclone phylogeny reconstruction using mutation pair data, and a nonparametric Bayesian approach to monotone missing data with auxiliary covariates in longitudinal studies. We have shown how inferences under our proposed models compare favorably
with inferences under existing methods, and significantly outperform inferences under existing methods in certain scenarios.

As we have mentioned in the discussion section in each chapter, there are several directions for future works. For the tumor heterogeneity problem, the proposed models can be extended for data where a local haplotype segment consists of more than two SNVs. We can accommodate $n$-tuples instead of pairs of SNVs by increasing the number of categorical values ($Q$) that the entries of $\bZ$ can take. 
The current model measures tumor heterogeneity with single nucleotide variants (SNVs) data in copy number neutral regions. We have discussed possibilities to incorporate copy number variants (CNVs) and 
have specific research plans to formally incorporate CNVs into the model and software. 
Also, structural variants (SVs) such as deletion, duplication, inversion, translocation and other large genome rearrangement provide more information for characterizing tumor heterogeneity. Utilizing information from SVs is another direction of characterizing tumor heterogeneity. The major motivation of tumor subclone reconstruction is application to precision medicine. The reconstructed tumor subclones can be used as basis for adaptive Bayesian clinical trial design. Finally, we plan to develop computation efficient algorithms to handle large numbers (e.g. millions) of SNVs. 

For the missing data problem, 
our model focuses on monotone missing data with discrete time dropout.
A possible extension of this work is to consider continuous time dropout. The Gaussian process is suitable for this case. Another possible future direction is to extend our method to non-monotone missing data without imposing the partial ignorability assumption. In the setting of binary outcomes, our method can be extended by using a probit link.
For large numbers of subjects, existing methods \citep{banerjee2008gaussian, banerjee2013efficient, hensman2013gaussian, datta2016hierarchical} that tackle the computational bottleneck of the Gaussian process can be employed in our application.






%
%
\appendices
\index{Appendices@\emph{Appendices}}%

\chapter{Appendix for Chapter \ref{chap:PairClone}}
\section{MCMC Implementation Details}
\label{app:sec:mcmc}
We first introduce $\theta_{tc}$ as an unscaled abundance level of
subclone $c$ in sample $t$. Assume $\theta_{t0} \sim \text{Gamma}(d_0,
1)$ and $\theta_{tc} \mid C \sim \text{Gamma}(d, 1)$. Let $w_{tc} =
\theta_{tc} / \sum_{c' = 0}^C \theta_{tc'}$, then $\bw_{t} \sim
\text{Dirichlet}(d_0, d, \ldots, d)$. We make inference on $\bm
\theta$ instead of $\bw$ as the value of $\bm \theta$ is not
restricted in a $C$-simplex. Similarly, we introduce $\rho_g^*$ as an
unscaled version of $\rho_g$. We let $\rho_g^* \sim \text{Gamma}(d_1,
1)$ and $\rho_g = \rho_g^* / \sum_{g' = 1}^4  \rho_{g'}^* $ for $g =
1, \ldots, 4$,  $\rho_g^* \sim \text{Gamma}(2d_1, 1)$ and $\rho_g =
\rho_g^* / \sum_{g' = 5}^6  \rho_{g'}^* $ for $g = 5, 6$, and
$\rho_g^* \sim \text{Gamma}(2d_1, 1)$ and $\rho_g = \rho_g^* /
\sum_{g' = 7}^8  \rho_{g'}^* $ for $g = 7, 8$. 

Conditional on $C$, the posterior distribution for the other parameters is given by
\begin{align*}
p(\bZ, \bm \pi, \bm \theta, \bm \rho^* \mid \bm n, C) \propto
&\prod_{t = 1}^T \prod_{k = 1}^K \prod_{g = 1}^G \tp_{tkg}^{n_{tkg}}
\times \prod_{c = 1}^C \prod_{q = 1}^Q \pi_{cq}^{m_{cq}} \times \\ 
&\prod_{c = 1}^C \left[ \pi_{c1}^{1 - 1} (1 - \pi_{c1})^{\alpha/C - 1}
  \cdot \prod_{q = 2}^Q \tilde{\pi}_{cq}^{\beta - 1} \right] \times 
\nonumber \\ 
&\prod_{t = 1}^T \left[ \theta_{t0}^{d_0 - 1} e^{- \theta_{t0}}
  \prod_{c = 1}^C \left( \theta_{tc}^{d - 1} e^{- \theta_{tc}}
  \right)\right] \times \\
&\prod_{g = 1}^4 \left( \rho_g^{*d_1 - 1} e^{-\rho_g^{*}} \right)
\cdot \prod_{g = 5}^8 \left( \rho_g^{*2d_1 - 1} e^{-\rho_g^{*}}
\right) .
\end{align*}
where $m_{cq} = \sum_{k = 1}^{K} I(\bz_{kc} = \bz^{(q)})$ counts the number of mutation pairs in subclone $c$ having genotype $\bz^{(q)}$.
\paragraph{Updating $\bZ$.} We update $\bZ$ by
sampling each $\bz_{kc}$ from: 
\begin{multline*}
p(\bz_{kc} = \bz^{(q)} \mid \ldots) \propto \\
\prod_{t = 1}^T \prod_{g = 1}^G \left[ \sum_{c' = 1, c' \neq c}^C w_{tc'} \, A(\bh_g, \bz_{kc'}) + w_{tc} \, A(\bh_g, \bz^{(q)}) + w_{t0} \, \rho_g \right]^{n_{tkg}} \cdot \pi_{cq}  
\end{multline*}

\paragraph{Updating $\bm \pi$.} The posterior distribution for $\bm \pi$ is
\begin{align*}
p(\bm \pi \mid \ldots) &\propto \prod_{c = 1}^C \left[ \left( \prod_{q = 1}^Q \pi_{cq}^{m_{cq}} \right) \cdot \pi_{c1}^{1 - 1} (1 - \pi_{c1})^{\alpha/C - 1} \cdot \prod_{q = 2}^Q \tilde{\pi}_{cq}^{\beta - 1} \right] \\
&= \prod_{c = 1}^C \left[\pi_{c1}^{m_{c1} + 1 - 1} (1 - \pi_{c1})^{K - m_{c1} + \alpha/C - 1} \cdot \prod_{q = 2}^Q \tilde{\pi}_{cq}^{m_{cq} + \beta - 1} \right].
\end{align*}
For each $c = 1, \ldots, C$, we update $\bm \pi_c$ by sampling from
\begin{align*}
\pi_{c1} \mid \ldots &\sim \text{Beta}(m_{c1} + 1, K - m_{c1} + \alpha / C), \\
(\tilde{\pi}_{c2}, \ldots, \tilde{\pi}_{cQ}) \mid \ldots &\sim \text{Dirichlet}(m_{c2} + \beta, \ldots,  m_{cQ} + \beta),
\end{align*}
and transforming by $(\pi_{c2}, \ldots, \pi_{cQ}) = (1 - \pi_{c1}) \cdot (\tilde{\pi}_{c2}, \ldots, \tilde{\pi}_{cQ})$.

\paragraph{Updating $\bm \theta$.} We update each $\theta_{tc}$ sequentially. For $c = 1, \ldots, C$, 
\begin{align*}
p(\theta_{tc} \mid \ldots) \propto \prod_{k = 1}^K \prod_{g = 1}^G \left[ \sum_{c = 1}^C w_{tc} \, A(\bh_g, \bz_{kc}) + w_{t0} \, \rho_g \right]^{n_{tkg}} \cdot \theta_{tc}^{d - 1} e^{-\theta_{tc}}.
\end{align*}
A Metropolis-Hastings transition probability
is used to update $\theta_{tc}$. At each
iteration, we propose a new $\tilde{\theta}_{tc}$ (on the log scale) by
$\log(\tilde{\theta}_{tc}) \sim N( \log\theta_{tc}, 0.2)$, and evaluate
the acceptance probability by 
$$
p_{\text{acc}}(\theta_{tc},
\tilde{\theta}_{tc}) = 1 \wedge \left[ \left( p(\tilde{\theta}_{tc}  \mid
  \ldots) \, p(\theta_{tc} \mid \tilde{\theta}_{tc}) \right) \middle/  \left( p(\theta_{tc} \mid \ldots) \, p(\tilde{\theta}_{tc} \mid \theta_{tc}) \right) \right].
$$
The term  $p(\theta_{tc} \mid \tilde{\theta}_{tc}) / p(\tilde{\theta}_{tc} \mid \theta_{tc}) = \tilde{\theta}_{tc} / \theta_{tc}$ takes into account the Jacobian of the log transformation.
For $c = 0$, the
only difference is to substitute $d$ with $d_0$. 

\paragraph{Updating $\bm \rho^*$.} We update each $\rho_{g}^*$ sequentially. For $g = 1, \ldots, 4$,
\begin{align*}
p(\rho_g^* \mid \ldots) \propto  \prod_{t = 1}^T \prod_{k = 1}^K \prod_{g = 1}^G \left[ \sum_{c = 1}^C w_{tc} \, A(\bh_g, \bz_{kc}) + w_{t0} \, \rho_g \right]^{n_{tkg}} \cdot \rho_g^{*d_1 - 1} e^{-\rho_g^{*}}.
\end{align*}
A Metropolis-Hastings transition probability
is used to update $\rho_g^*$. At each iteration, we propose a new
$\tilde{\rho}_{g}^*$ (on the log scale) by 
$\log(\tilde{\rho}_{g}^*) \sim N(\log{\rho_g^*}, 0.1)$,
and evaluate the acceptance probability by
$$
p_{\text{acc}}(\rho_g^*, \tilde{\rho}_{g}^*) = 1 \wedge \left[
 \left( p(\tilde{\rho}_{g}^* \mid \ldots) \, p(\rho_g^* \mid \tilde{\rho}_{g}^*) \right) /  \left( p(\rho_g^* \mid \ldots) \, p(\tilde{\rho}_{g}^* \mid \rho_g^* ) \right) \right]. 
$$
The term  $p(\rho_g^* \mid \tilde{\rho}_{g}^*) / p(\tilde{\rho}_{g}^* \mid \rho_g^* ) = \tilde{\rho}_{g}^* / \rho_g^*$ takes into account the Jacobian of the log transformation.
For $g = 4, \ldots, 8$, the only difference is to substitute
$d_1$ with $2 d_1$. 

\paragraph{Parallel tempering.} Parallel tempering (PT) is a MCMC technique first proposed by \cite{geyer1991markov}. A good review can be found in \cite{liu2008monte}.
PT is suitable for sampling from a multi-modal state space. It helps
the MCMC chain to move freely among local modes which is desired in
our application, and to create a better mixing Markov chain.

\begin{algorithm}
\caption{Parallel Tempering}
\label{PTalgorithm}
\begin{algorithmic}[1]
\State Draw initial state $(\bm x_1^{(0)}, \ldots, \bm x_I^{(0)})$ from appropriate distributions
\For{$l$  in $1, \ldots, L$ }
\State Draw $u \sim \text{Uniform}(0, 1)$
\If{$u \leq u_0$}
\State Conduct the parallel step: update every $\bm x_i^{(l)}$ to $\bm x_i^{(l+1)}$ via respective MCMC scheme
\Else
\State Conduct the swapping step: draw $i \sim \text{Discrete-Uniform}(1, \ldots, I - 1)$, propose a swap between $\bm x_i^{(l)}$ and $\bm x_{i+1}^{(l)}$, accept the swap with probability
\begin{align*}
\min\left\{  1, \frac{\pi_i(\bm x_{i+1}^{(l)}) \pi_{i+1}(\bm x_{i}^{(l)})}{\pi_i(\bm x_{i}^{(l)}) \pi_{i+1}(\bm x_{i+1}^{(l)})} \right\}
\end{align*}
\EndIf
\EndFor
\end{algorithmic}
\end{algorithm}

To sample from the target distribution $\pi(\bm x)$, we consider a
family of distributions $\Pi = \{ \pi_i , i = 1, \ldots, I \}$,  where
$\pi_i(\bm x) \propto \pi(\bm x)^{1 / \Delta_i}$. Without loss of
generality, let $\Delta_I = 1$ and $\pi_I(\bm x) = \pi(\bm x)$.
Denote by $\mathcal{X}_i$ the state space of $\pi_i(\bm x)$. The PT
scheme is illustrated in Algorithm \ref{PTalgorithm}. 

In our application, we find by simulation that PT works well with $I = 10$ temperatures and  $\{ \Delta_1, \ldots, \Delta_{10} \} = \{  4.5, 3.2, 2.5, 2, 1.7, 1.5, 1.35, 1.2, 1.1, 1\}$. We therefore use this parameter setting for all the simulation studies as well as the lung cancer dataset.

\section{Updating $C$}
\label{app:sec:updatec}
For updating $C$, we split the data into a training set $\bn'$, and a
test set $\bn''$ with $n_{tkg}' = b n_{tkg}$ and $n_{tkg}'' = (1 -
b) n_{tkg}$. Let $p_b(\bx \mid C) = p(\bx \mid \bn', C)$ denote the
posterior of $\bx$ conditional on $C$ evaluated on the training set
only. We use $p_b$ in two occasions. First, we replace the original
prior $p(\bx \mid C)$ by $p_b(\bx \mid C)$, and second, we use $p_b$
as a proposal distribution of $\tbx$ as $q( \tbx \mid \tilde{C} ) =
p_b(\tbx \mid \tilde{C})$. 
We show that the use of the training sample posterior as proposal
and modified prior in equation (4) (original manuscript) implies an
approximation in the reported marginal posterior for $C$, but leaves
the conditional posterior for all other parameters (given $C$)
unchanged. 

We evaluate the acceptance probability of $\tilde{C}$ on the test data
by
\begin{align*}
p_{\text{acc}} (C, \bx, \tC, \tbx)
   &= 1 \wedge \frac{p(\bn'' \mid \tbx, \tC)}{p(\bn'' \mid \bx, C)} \cdot
                         \frac{p(\tC) p(\tbx \mid \bn', \tC) }{p(C)
                         p(\bx \mid \bn', C)} \cdot \frac{q(C \mid
                         \tC) q(\bx \mid C)}{q(\tC \mid C) q(\tbx \mid
                         \tC)} \\ 
 &= 1 \wedge \frac{p(\bn'' \mid \tbx, \tC)}{p(\bn'' \mid \bx, C)} \cdot
  \frac{p(\tC) }{p(C)} . 
\end{align*}
Under the model $p_b(\cdot)$   with the modified prior,  the implied conditional posterior on $\bx$ satisfies 
\begin{align*}
p_b(\bx \mid C, \bn) 
&= \frac{p_b(\bx \mid C) p(\bn'' \mid \bx, C)}{\int p_b(\bx \mid C) p(\bn'' \mid \bx, C) d \bx} \\
&= \frac{p(\bx \mid C) p(\bn' \mid \bx, C) p(\bn'' \mid \bx, C)}{\int p(\bx \mid C) p(\bn' \mid \bx, C) p(\bn'' \mid \bx, C) d \bx} \\
&= p(\bx \mid C, \bn), 
\end{align*}
which indicates the conditional posterior of $\bx$ remains entirely unchanged.
The implied marginal posterior on $C$ is
$p_b(C \mid \bn'') \propto p(C) \, p_b(\bn'' \mid C)$,
with the likelihood on the test data evaluated as $p_b(\bn'' \mid C) =
\int p(\bn'' \mid \bx, C) \, p_b(\bx \mid C) d\bx$.
The use of the prior $p_b( \tbx \mid \tC)$ is similar to the
construction of the fractional Bayes factor (FBF)
\citep{ohagan1995}. Let $\bu = \{ \bm \pi, \bw, \bm \rho\}$ denote the
parameters other than $\bZ$ 
and let $\umle$ denote the maximum likelihood estimate for
$\bu$.
We follow \cite{ohagan1995} to show that inference on $C$ is as if we
were making use of only a fraction $(1-b)$ of the data, with a
dimension penalty.
In short,
$$
  p_b(C \mid \bn'') \propto p(C) p(\bn \mid
    \umle, C)^{1 - b}\, b^{\; p_{C} / 2},
$$
approximately, where $\umle$ is the maximum likelihood estimate of $\bu$, and $p_C$ is the number of unconstrained parameters in $\bu$. To obtain this approximation, consider the marginal
sampling model under $p_b(\cdot)$, after marginalizing with respect to $\bx$:
\begin{align*}
p_b(\bn'' \mid C) &= \int p(\bn'' \mid \bx, C) p_b(\bx \mid C)  d \bx \\ 
&= \int p(\bn'' \mid \bx, C)  \,
  \frac{p(\bn' \mid \bx, C) p(\bx \mid C)}
       {\int p(\bn' \mid \bx, C) p(\bx \mid C) d\bx}\,
  d \bx   \\
&= \frac{\int p(\bn \mid \bx, C) p(\bx \mid C) d \bx}{\int p(\bn' \mid
  \bx, C) p(\bx \mid C) d \bx}.  
\end{align*}
Here we substituted 
the training sample posterior as (new) prior $p_b(\bx \mid C)$. 
The integration
includes a  marginalization with respect to the discrete $\bZ$,
\begin{align*}
  \int p(\bn \mid \bx, C) p(\bx \mid C) d \bx &= \int \sum_{\bZ} p(\bn \mid \bZ, \bu, C) p(\bZ \mid \bu, C) p(\bu \mid C) d\bu \\
  &= \int  p(\bn \mid \bu, C) p(\bu \mid C) d\bu,
\end{align*}
For the remaining real valued parameters $\bu$ we use an appropriate
one-to-one transformation (e.g. logit transformation) $\bu \mapsto
\tbu$, such that $\tbu$ is unconstrained.  To simplify notation we
continue to refer to the transformed parameter as $\bu$ only.
Next, under the binomial sampling model
$p(\bn' \mid \bx,C) \propto p(\bn \mid \bx,C)^b$, leading
to

\begin{align*}
p_b(\bn'' \mid C) &=
\frac{\int p(\bn \mid \bu, C) p(\bu \mid C) d \bu}
     {\int p(\bn' \mid \bu, C) p(\bu \mid C) d \bu} \\
     &=
 \underbrace{\frac{\left[\prod_{t, k} N_{tk}! / (n_{tk1}! \cdots
      n_{tkG}!)\right]^b}{\prod_{t, k} (bN_{tk})! / \left[ (bn_{tk1})!
      \cdots (bn_{tkG})!\right]}}_{m(\bn)} \cdot
\underbrace{\frac{\int p(\bn \mid \bu, C) p(\bu \mid C) d \bu}{\int
    p(\bn \mid \bu, C)^{b} p(\bu \mid C) d \bu}}_{h_b(\bn \mid C)},
\nonumber
\end{align*}

Let $m(\bn)$ and $h_b(\bn \mid C)$ denote the two factors.
The first, $m(\bn)$, is a constant term. And the second factor,
$h_b(\bn \mid C)$, has exactly the same form as equation (12) in
\cite{ohagan1995}, who shows
\begin{align*}
   h_b(\bn \mid C) \approx 
   p(\bn \mid \umle, C)^{1-b} b^{\; p_C / 2}
\end{align*}
Let $N = \sum_{t,k} N_{tk}$. 
The argument of
\cite{gelfand1994bayesian} (case (e)) suggests that the
error in this approximation is of order $O(1/N^2)$
(note that Gelfand and Dey use expansion around the M.A.P. while
O'Hagan uses expansions around the M.L.E.).
This establishes the stated 
approximation of the posterior
$
p_b(C \mid \bn'') \approx k\cdot p(C) p(\bn \mid \umle, C)^{1 - b}\,
b^{\; p_{C} / 2}$, approximately.

\section{Calibration of $b$}
\label{app:sec:calib}

The construction of an informative prior
$p_b(\bx \mid C) \equiv p(\bx \mid \bn', C)$
based on a training sample $\bn'$ is similar to the use of a training
sample in the construction of the fractional Bayes factor (FBF) of
\cite{ohagan1995}.
However, there is an important difference. In the FBF construction the
aim is to replace a noninformative prior in the evaluation of a Bayes
factor. A minimally informative prior $p_b$ with small $b$ suffices.
In contrast, here $p_b(\bx \mid C)$ is (also) used as proposal
distribution in the trans-dimensional MCMC. The aim is to construct a
good proposal that fits the data well and thus leads to good
acceptance probabilities and a well mixing Markov chain.
With the highly informative multinomial likelihood we find that we
need a large training sample, that is, large $b$.
In Appendix \ref{app:sec:updatec} we show that the effect of using
$p_b$ is that $p(C \mid \bn)$ is approximated by
$$
   p_b(C \mid \bn'') \propto
   p(C) p(\bn \mid \umle, C)^{1 - b} b^{\;   p_{C} / 2},
$$
where $\bu = \{\bm \pi, \bw, \bm \rho\}$ are the parameters other than
$\bZ$, $\umle$ is the maximum likelihood estimate of $\bu$, and $p_C$
is the number of unconstrained parameters in $\bu$.
Importantly, however, inference on other parameters, $p(\bx \mid
C,\bn)$, remains entirely unchanged.

\begin{figure}[h!]
\begin{center}
\includegraphics[scale = 0.43]{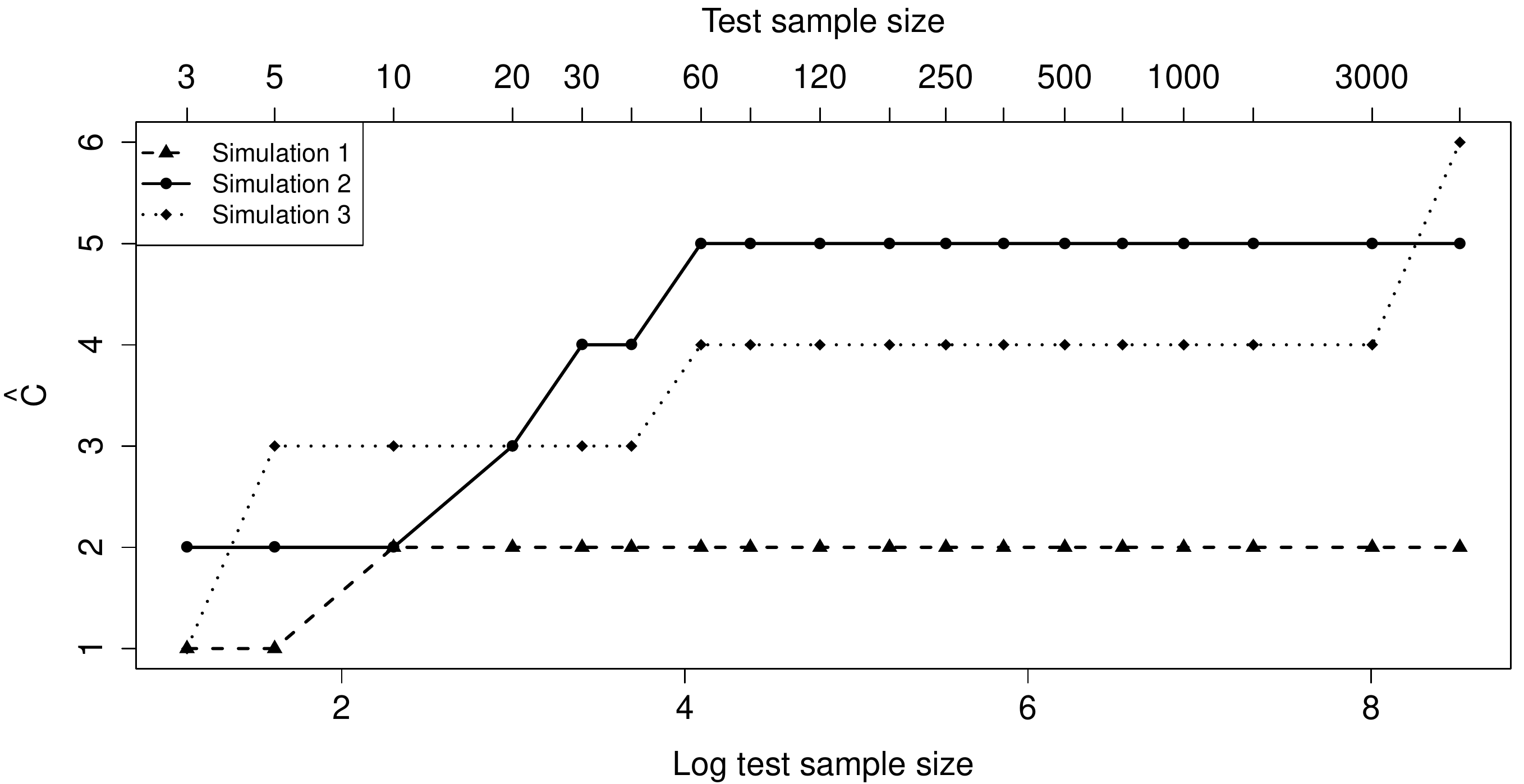}
\end{center}
\caption{Path plot of $\Chat$ with different test sample sizes for
   three simulations. The true number of subclones are 2, 4, and 3
  for simulations 1, 2, and 3, respectively. 
}
\label{fig:cpath}
\end{figure}

We therefore recommend to focus on inference for $C$ when calibrating
$b$. 
Carrying out simulation studies with single and multi-sample data, 
we find that the simulation truth for $C$ is best recovered with
a test sample size $(1-b) \sum_{t=1}^T \sum_{k=1}^K N_{tk} \approx 160 / T$, where
$N_{tk}$ is the total number of short reads mapped to mutation pair
$k$ in sample $t$.
For example, Figure \ref{fig:cpath} plots the posterior mode of $C$
against test sample sizes for simulated data in three simulations.
For multi-sample data we find (empirically, by simulation) that
the test sample size can be reduced, at a rate linear in
$T$. In summary we recommend to set $b$ to achieve a test sample
size around $160/T$. 
Following these guidelines, in our implementation in the previous
section, we used values $b = 0.992$ for simulation 1,  $b = 0.9998$
for simulation 2, and $b = 0.999911$ for simulation 3.

\section{Validation of the MCMC scheme}
 \paragraph{Validation of the correctness of the sampler.}  
We first use a scheme to validate the correctness of our MCMC sampler
in the style of \cite{geweke2004getting}. The joint density of the
parameters and observed data can be written as $p(\bx, \bn) = p( \bx)
p(\bn \mid \bx)$. Let $g$ be any function $g: \mathcal{X} \times
\mathcal{N} \rightarrow \mathbb{R}$ satisfying $\text{Var}[g(\bx,
\bn)] < \infty$, where $\mathcal{X}$ and $\mathcal{N}$ represent
sample spaces of $\bx$ and $\bn$, respectively. 
Denote by $\bar{g} = E[g(\bx, \bn) ]$, which can be evaluated
by independent Monte Carlo simulation from the joint distribution, or
in some cases might be known exactly as prior mean of functions of
parameters only. 
Alternatively, the same mean can be estimated by a different Markov
chain Monte Carlo scheme for the joint distribution, 
constructed by an initial draw $\bx^{(0)} \sim p(\bx)$, followed by
$\bn^{(l)} \sim p(\bn \mid \bx^{(l-1)})$, $\bx^{(l)} \sim q(\bx \mid
\bx^{(l-1)}, \bn^{(l)})$, and $g^{(l)} = g(\bn^{(l)}, \bx^{(l)})$ ,
for $l = 1, \ldots, L$. Under certain conditions, $\{\bx^{(l)},
\bn^{(l)}\}$ is ergodic with unique invariant kernel $p(\bx, \bn)$.
If the simulator is error-free, one should have
\begin{align}
 (\bar{g}^{(L)} - \bar{g}) / \left[ L^{-1} \hat{S}_g(0) \right]^{1/2}
  \xrightarrow{d} N(0, 1),
\label{eq:geweke}
\end{align}
where $\hat{S}_g(0)$ is consistent spectral density estimate for
$\{g^{(l)}, l = 1, \ldots, L \}$.  In our application, we take $g(\bx,
\bn) = w_{tc}$ and $p_{tkg}$. We set the number of samples $T = 4$,
and the number of mutation pairs $K = 80$. Since our inference on $C$
is not a standard MCMC, we fix $C = 3$ here and only consider $\bx =
\{ \bZ, \bm \pi, \bw, \brho\}$. Table \ref{tbl:convd1} shows the
statistic \eqref{eq:geweke} for five randomly selected $w_{tc}$ and
$p_{tkg}$. 
The recorded $z$-scores show no evidence for errors in the
simulator. 
\begin{table}[h!]
\begin{center}
\begin{tabular}{ | c | c | c |}
    \hline
    Test statistic & $z$-score & $p$-value \\ \hline
    $w_{12}$ & -0.4736149 & 0.6357745 \\ \hline
    $w_{43}$ & -1.441169 & 0.149537 \\ \hline
    $p_{1,23,3}$ & 0.9413715  & 0.3465145 \\ \hline
    $p_{3,60,7}$ & 1.388424  & 0.1650079 \\ \hline
    $p_{2,13,2}$ & -0.6051894  & 0.5450532 \\ \hline
\end{tabular}
\end{center}
\caption{Geweke's statistics and the corresponding $z$-scores and $p$-values.}
\label{tbl:convd1}
\end{table}
 
\paragraph{Convergence diagnostic.}
Next, we present some convergence diagnostics of our MCMC chain, including trace plots, autocorrelation plots, and test statistics described in \cite{geweke1991evaluating}. Those convergence diagnostics are based on the posterior distribution of parameters $p(\bx \mid \bn) \propto p(\bx) p(\bn \mid \bx)$. Let $g$ be any function $g: \mathcal{X} \rightarrow \mathbb{R}$, and $g^{(l)} = g(\bx^{(l)})$ where $\{\bx^{(l)}, l = 1, \ldots, L \}$ are samples from the posterior. Let
\begin{align*}
\bar{g}^{A}_L = L_{A}^{-1} \sum_{l = 1}^{L_A} g^{(l)}, \quad \quad \bar{g}^{B}_L = L_{B}^{-1} \sum_{l = l^*}^{L} g^{(l)} \; \; \; (l^* = L - L_B + 1),
\end{align*}
and let $\hat{S}_g^A(0)$ and $\hat{S}_g^B(0)$ denote consistent spectral density estimates for $\{g^{(l)}, l = 1, \ldots, L_A \}$ and $\{g^{(l)}, l = l^*, \ldots, L \}$, respectively. If the ratios $L_A / L$ and $L_B / L$ are fixed, with $(L_A + L_B) / L < 1$, then as $L \rightarrow \infty$, 
\begin{align*}
(\bar{g}^{A}_L - \bar{g}^{B}_L) / \left[ L_A^{-1} \hat{S}_g^A(0) + L_B^{-1} \hat{S}_g^B(0) \right]^{1/2} \xrightarrow{d} N(0, 1).
\end{align*}

\begin{figure}[h!]
\begin{center}
\begin{subfigure}[t]{.49\textwidth}
\centering
\includegraphics[width=\textwidth]{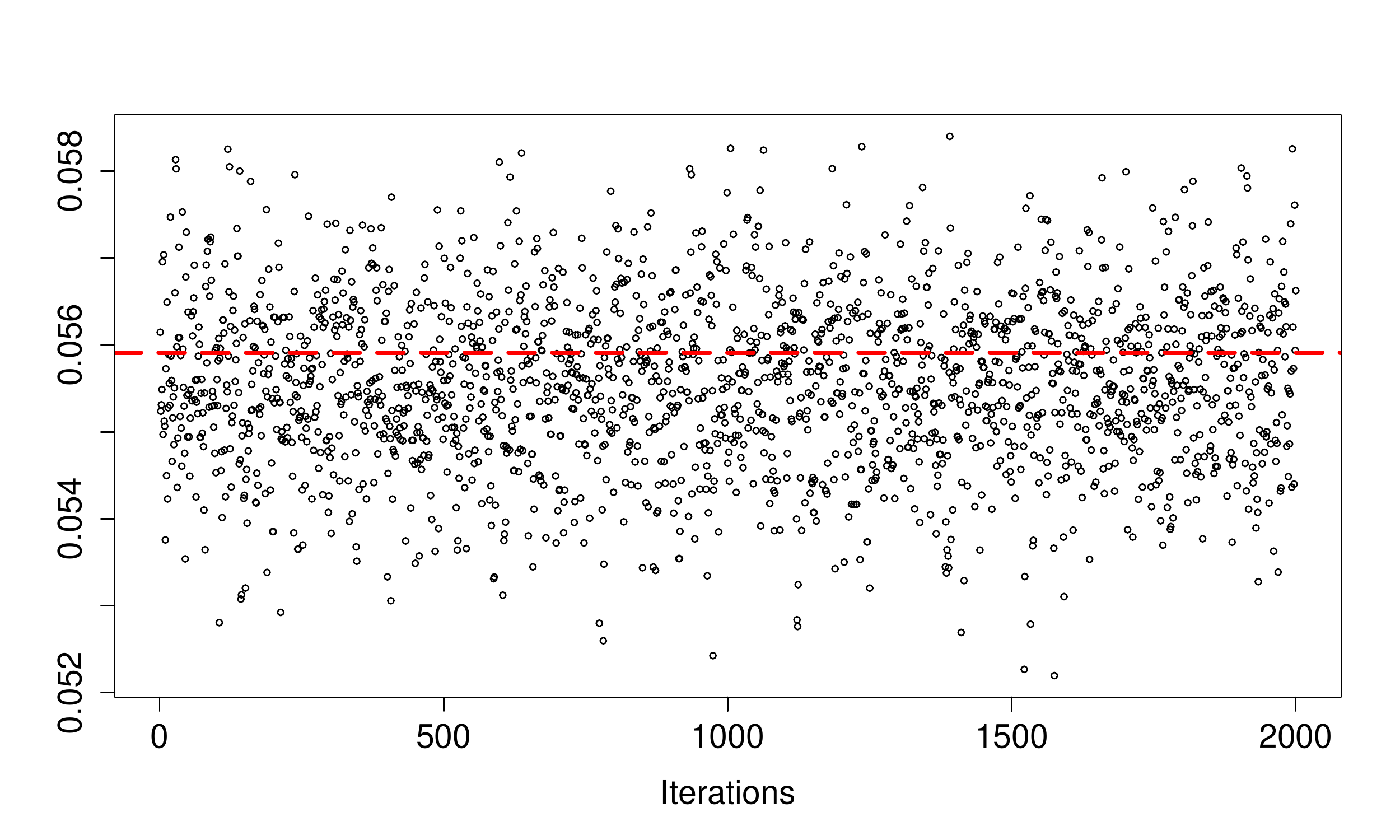}
\caption{Trace plot of $p_{1,5,2}$}
\end{subfigure}
\begin{subfigure}[t]{.49\textwidth}
\centering
\includegraphics[width=\textwidth]{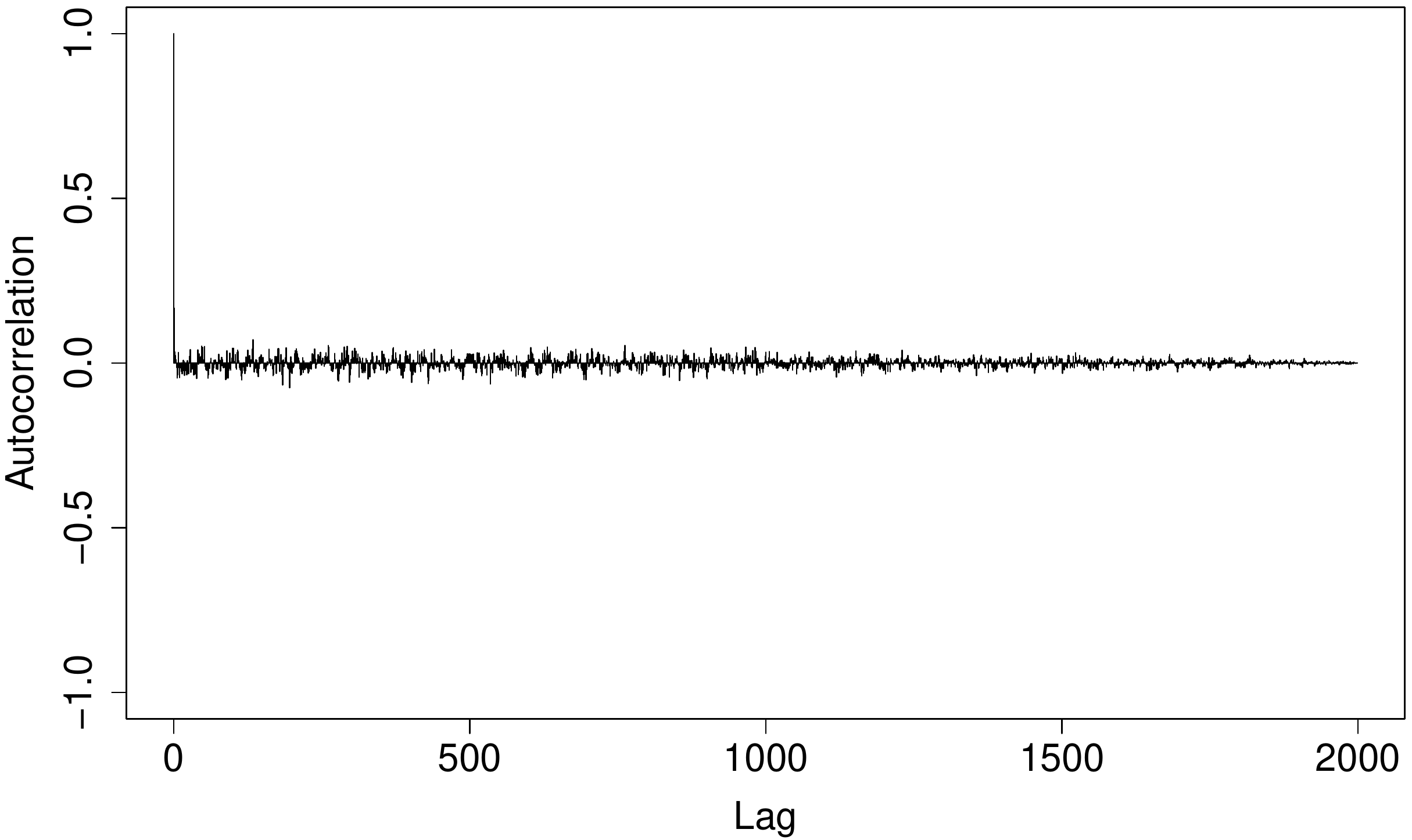}
\caption{Autocorrelation plot of $p_{1,5,2}$}
\end{subfigure}
\begin{subfigure}[t]{.49\textwidth}
\centering
\includegraphics[width=\textwidth]{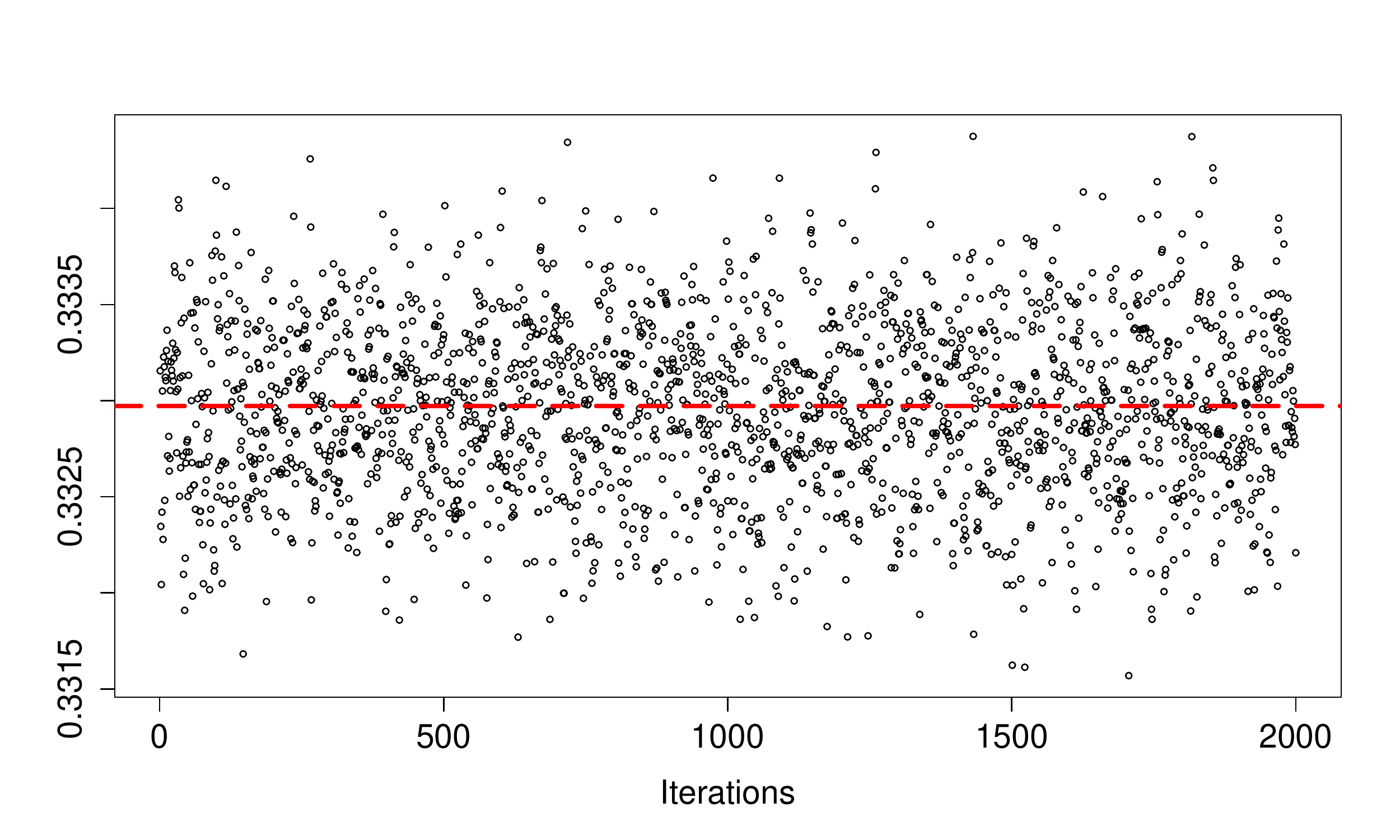}
\caption{Trace plot of $p_{3,68,7}$}
\end{subfigure}
\begin{subfigure}[t]{.49\textwidth}
\centering
\includegraphics[width=\textwidth]{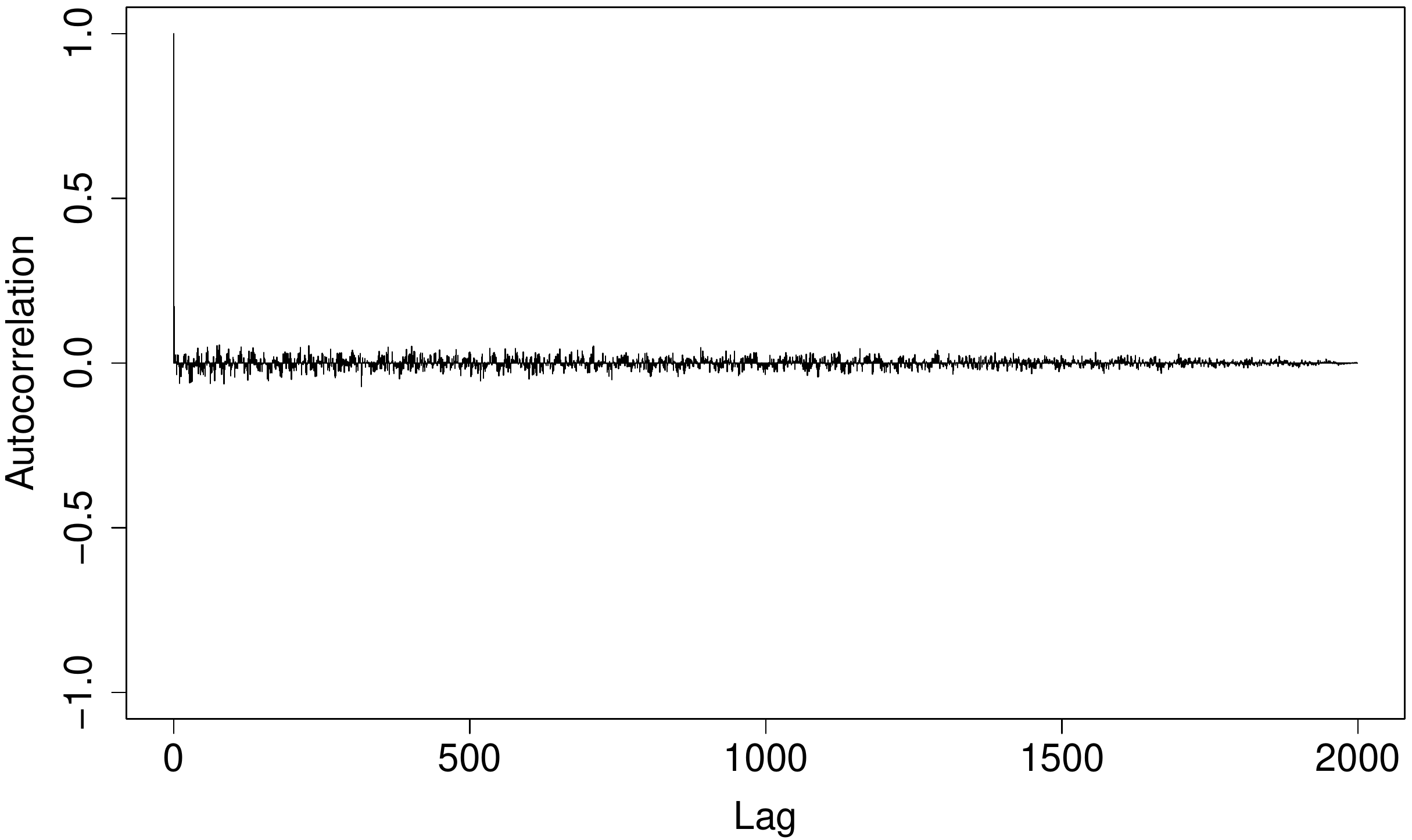}
\caption{Autocorrelation plot of $p_{3,68,7}$}
\end{subfigure}
\end{center}
\caption{Convergence check for Simulation 2.}
\label{fig:convd}
\end{figure}

In our application, a reasonable choice of $g$ is $g(\bx) = p_{tkg}(\bZ, \bw, \brho)$.
We use simulation 2 as an example, and show some plots and Geweke's statistics for some randomly chosen $p_{tkg}$.
Figure \ref{fig:convd}(a, c) shows the trace plot for $p_{tkg}$, with the red dashed line denoting the true value. The posterior samples are centered around the true value and symmetrically distributed. Figure \ref{fig:convd}(b, d) shows the autocorrelation plot for $p_{tkg}$. The autocorrelations between MCMC draws are small, indicating good mixing of the chain. Table \ref{tbl:convd2} shows the Geweke's statistics for five randomly selected $p_{tkg}$. The $p$-values for them are all greater than 0.05, representing those statistics pass the Geweke's diagnostic, and there is no strong evidence that the chain does not converge.

\begin{table}[h!]
\begin{center}
\begin{tabular}{ | c | c | c |}
    \hline
    Test statistic & $z$-score & $p$-value \\ \hline
    $p_{1,5,2}$ & 0.1748906 & 0.8611656 \\ \hline
    $p_{3,68,7}$ & -0.02609703 & 0.9791799 \\ \hline
    $p_{4,25,5}$ & 0.4454738  & 0.6559774 \\ \hline
    $p_{2,96,4}$ & -1.341994  & 0.179598 \\ \hline
    $p_{1,66,1}$ & -0.2727737  & 0.7850272 \\ \hline
\end{tabular}
\end{center}
\caption{Convergence check for Simulation 2.}
\label{tbl:convd2}
\end{table}

\chapter{Appendix for Chapter \ref{chap:bnpmis}}
\section{MCMC Implementation Details}
\label{app-bnpmis-mcmc}
We introduce some notation as follows. 
First considering the responses.
Denote by $N_s$ the number of subjects having dropout pattern $s$, $s = 2, \ldots, J$.
Let $\by_{js}$ denote the subjects' responses at time $j$ in pattern $s$, and $\bar{Y}_{js}$ denote the subjects' histories through the first $j$ times in pattern $s$, i.e.
\begin{align*}
\by_{js} &= (y_{1js}, y_{2js}, \ldots, y_{N_s, j, s} )^T, \\
\bY_{js} &= \left( \by_{1s}, \by_{2s}, \ldots, \by_{js} \right).
\end{align*}
Let $\by_{\vec 0}$ denote the initial responses (with no past) for all subjects, and $\by_{\vec}$ denote the subsequent responses (with measured pasts) for all subjects,
\begin{align*}
\by_{\vec 0} &= \left( \by_{12}^T, \by_{13}^T, \ldots, \by_{1J}^T \right)^T \\
\by_{\vec} &= \left( \by_{22}^T, \by_{23}^T, \by_{33}^T, \ldots, \by_{2J}^T, \ldots, \by_{JJ}^T \right)^T.
\end{align*}

We then consider the means and covariate matrices for the responses.
Let $\ba_{js}$ denote the vector of random variables (we abuse notation slightly, let $\ba_{js}$ include $\bY_{j-1, s} \bPhi_{js}$ when $j \geq 2$, to simplify notation),
\begin{align*}
\ba_{js} = 
\begin{cases}
\left( a_0(\bv_{1s}, j, s), \ldots,  a_0(\bv_{N_s, s}, j, s)  \right)^T, \quad &\text{if $j = 1$};\\
\left( a(\bv_{1s}, j, s), \ldots,  a(\bv_{N_s, s}, j, s)  \right)^T + \bY_{j-1, s} \bPhi_{js}, \quad &\text{if $j \geq 2$}.
\end{cases}
\end{align*}
The vector $\ba_{js}$ is the mean of $\by_{js}$.
Let $\ba_0$ and $\ba$ denote the vector of random variables, 
\begin{align*}
\ba_{0} &= \left( \ba_{12}^T, \ba_{13}^T, \ldots, \ba_{1J}^T \right)^T \\
\ba &= \left( \ba_{22}^T, \ba_{23}^T, \ba_{33}^T, \ldots, \ba_{2J}^T, \ldots, \ba_{JJ}^T \right)^T.
\end{align*}
Denote by
\begin{align*}
\Sigma_{y_0} &= \diag \left( \sigma_{12}^2 I_{N_2},  \ldots,  \sigma_{1J}^2 I_{N_J}  \right), \\
\Sigma_{y} &= \diag \left(
\sigma_{22}^2 I_{N_2} , \sigma_{23}^2 I_{N_3}, \sigma_{33}^2 I_{N_3}, \ldots, \sigma_{2J}^2 I_{N_J}, \ldots,  \sigma_{JJ}^2 I_{N_J}   \right).
\end{align*}
Thus, the likelihoods for the initial responses and subsequent responses are
\begin{align*}
\by_{\vec 0} \mid \ba_{0}, \Sigma_{y_0} &\sim N(\ba_{0}, \Sigma_{y_0}), \\
\by_{\vec} \mid \ba, \Sigma_{y} &\sim N(\ba, \Sigma_{y}).
\end{align*}

Next, we consider the priors for $\ba_0$ and $\ba$. Denote by
\begin{align*}
\btheta_0 &= (\bbeta_0, \bb_0), \\
\btheta &= (\bbeta, \bb, \bphi_1, \bphi_2, \bphi_3).
\end{align*} 
Let $\bD_0$ and $\bD$ denote the exponential distance matrices for $\ba_0$ and $\ba$,
\begin{align*}
[\bD_{0}]_{ijs, i'j's'}  &= D_0(\bv_{is}, j, s ;  \bv_{i's}, j', s'), \\
[\bD]_{ijs, i'j's'}  &= D(\bv_{is}, j, s ;  \bv_{i's}, j', s').
\end{align*}
We have
\begin{align*}
\ba_0 \mid \btheta_0, \kappa_0^2 &\sim N(X_{\theta_0} \btheta_0, \kappa_0^2 \bD_0 + \tkappa_0^2 I), 
\\
\ba \mid \btheta, \kappa^2 &\sim N(X_{\theta} \btheta, \kappa^2 \bD + \tkappa^2 I),
\end{align*}
where $X_{\theta_0}$ and $X_{\theta}$ are the design matrices corresponding to Equation \eqref{eq:GP_mean}.

Denote by $\bC_0 = \kappa_0^2 \bD_0 + \tkappa_0^2 I$ and $\bC = \kappa^2 \bD + \tkappa^2 I$. Integrating out $\ba_0$ and $\ba$, the (marginal) likelihoods become
\begin{align*}
\by_{\vec 0} \mid \btheta_0, \Sigma_{y_0}, \kappa_0^2 &\sim N(X_{\theta_0} \btheta_0, \Sigma_{y_0} + \bC_0), \\
\by_{\vec} \mid \btheta, \Sigma_{y}, \kappa^2 &\sim N(X_{\theta} \btheta, \Sigma_{y} + \bC).
\end{align*}

\paragraph{Update $\ba_0$ and $\ba$.} 
It is not unusual to integrate out $\ba_0$ and $\ba$ for posterior inference on Gaussian process. However, we find that including $\ba_0$ and $\ba$ in the posterior inference would improve the mixing of the Markov chain. Therefore, we update $\ba_0$ and $\ba$ at each iteration.

\noindent 1. The likelihood and prior for $\ba_0$ are
\begin{align*}
\by_{\vec 0} \mid \ba_0, \Sigma_{y_0} &\sim N(\ba_0, \Sigma_{y_0}), \\
\ba_0 \mid \btheta_0, \kappa_0^2 &\sim N(X_{\theta_0} \btheta_0, C_0), 
\end{align*}
which lead to the posterior
\begin{align*}
\ba_0 \mid \btheta_0, \kappa_0^2, \Sigma_{y_0}, \by_{\vec 0} &\sim N(\ba_0^*, \Sigma_{a_0}^*), \; \text{where} \\
\Sigma_{a_0}^* &= [C_0^{-1} +  \Sigma_{y_0}^{-1}]^{-1}, \\
\ba_0^* &= \Sigma_{a_0}^*[C_0^{-1} X_{\theta_0} \btheta_0 + \Sigma_{y_0}^{-1} \by_{\vec 0}].
\end{align*}

\noindent 2. The likelihood and prior for $\ba$ are
\begin{align*}
\by_{\vec} \mid \ba, \Sigma_{y} &\sim N(\ba, \Sigma_{y}), \\
\ba \mid \btheta, \kappa^2 &\sim N(X_{\theta} \btheta, C), 
\end{align*}
which lead to the posterior
\begin{align*}
\ba \mid \btheta, \kappa^2, \Sigma_{y}, \by_{\vec} &\sim N(\ba^*, \Sigma_{a}^*), \; \text{where} \\
\Sigma_{a}^* &= [C^{-1} +  \Sigma_{y}^{-1}]^{-1}, \\
\ba^* &= \Sigma_{a}^*[C^{-1} X_{\theta} \btheta + \Sigma_{y}^{-1} \by_{\vec}].
\end{align*}

\paragraph{Update $\kappa_0^2$ and $\kappa^2$.} 
\noindent 1. The likelihood and prior for $\kappa_0^2$ are
\begin{align*}
\ba_0 \mid \btheta_0, \kappa_0^2 &\sim N(X_{\theta_0} \btheta_0, \kappa_0^2 \bD_0 + \tkappa_0^2 I), \\
\kappa_0^2 &\sim \IG(\lambda_1^{\kappa_0}, \lambda_2^{\kappa_0}).
\end{align*}
The posterior for $\kappa_0^2$ is 
\begin{align*}
p(\kappa_0^2 \mid \btheta_0, \ba_0) \propto p_N (\ba_0 \mid X_{\theta_0} \btheta_0, \kappa_0^2 \bD_0 + \tkappa_0^2 I) \cdot p_{\IG} (\kappa_0^2 \mid \lambda_1^{\kappa_0}, \lambda_2^{\kappa_0}),
\end{align*}
where $p_N(\bx \mid \bmu, \Sigma)$ represents (multivariate) normal density at $\bx$ with mean $\bmu$ and covariance matrix $\Sigma$, and $p_{\IG}(x \mid a, b)$ represents inverse gamma density at $x$ with shape parameter $a$ and rate parameter $b$. We use Metropolis-Hastings step to update $\kappa_0^2$.

\noindent 2. The likelihood and prior for $\kappa^2$ are
\begin{align*}
\ba \mid \btheta, \kappa^2 &\sim N(X_{\theta} \btheta, \kappa^2 \bD + \tkappa^2 I), \\
\kappa^2 &\sim \IG(\lambda_1^{\kappa}, \lambda_2^{\kappa}).
\end{align*}
The posterior for $\kappa^2$ is 
\begin{align*}
p(\kappa^2 \mid \btheta, \ba) \propto p_N (\ba \mid X_{\theta} \btheta, \kappa^2 \bD + \tkappa^2 I) \cdot p_{\IG} (\kappa^2 \mid \lambda_1^{\kappa}, \lambda_2^{\kappa}).
\end{align*}
We use Metropolis-Hastings step to update $\kappa^2$.

\paragraph{Update $\Sigma_{y_0}$ and $\Sigma_{y}$.} 
The likelihood and prior for $\sigma_{js}^2$ are
\begin{align*}
\by_{js} \mid \ba_{js}, \sigma_{js}^2 &\sim N(\ba_{js}, \sigma_{js}^2 I), \\
\sigma_{js}^2 \mid \lambda_{\sigma}, g_{\sigma} &\sim \IG(\lambda_{\sigma}, \lambda_{\sigma} g_{\sigma}).
\end{align*}
The posterior for $\sigma_{js}^2$ is 
\begin{align*}
\sigma_{js}^2 \mid \lambda_{\sigma}, g_{\sigma}, \ba_{js}, \by_{js} \sim \IG \left( \lambda_{\sigma} + \frac{N_s}{2}, \; \lambda_{\sigma} g_{\sigma} + \frac{RSS_{js}}{2} \right),
\end{align*}
where $RSS_{js} = \| \by_{js} - \ba_{js} \|_2^2$.

There are two hyperparameters related to $\sigma_{js}^2$: $\lambda_{\sigma}$ and $g_{\sigma}$. Their conditional posteriors are 
\begin{multline*}
p(\lambda_{\sigma} \mid  \{ \sigma_{js}^2 \}, g_{\sigma}) \propto 
\frac{(g_{\sigma} \lambda_{\sigma})^{(2+J)(J-1)\lambda_{\sigma}/2}} {\Gamma(\lambda_{\sigma})^{(2+J)(J-1)/2}} \prod_{j, s} 
\left( \sigma_{js}^2 \right)^{-(\lambda_{\sigma} - 1)}  \cdot \\
\exp \left( -\sum_{j,s} \frac{g_{\sigma}}{\sigma_{js}^2} \lambda_{\sigma} \right) \exp \left( -\frac{1}{\lambda_{\sigma} - 2} \right),
\end{multline*}
and
\begin{align*}
g_{\sigma} \mid \{ \sigma_{js}^2 \}, \lambda_{\sigma} \sim \text{Gamma} \left(   \frac{(2+J)(J-1)}{2} \lambda_{\sigma} + 1, \; \sum_{j, s} \frac{\lambda_{\sigma}}{\sigma_{js}^2} + 1 \right).
\end{align*}
We use Metropolis-Hastings step to update $\lambda_{\sigma}$.

\paragraph{Update $\btheta_0$ and $\btheta$.}
We integrate out $\ba_0$  and $\ba$ to update $\btheta_0$ and $\btheta$.  The likelihoods become
\begin{align*}
\by_{\vec 0} \mid \btheta_0, \Sigma_{y_0}, \kappa_0^2 &\sim N(X_{\theta_0} \btheta_0, \Sigma_{y_0} + C_0 ), \\
\by_{\vec}  \mid \btheta, \Sigma_{y}, \kappa^2 &\sim N(X_{\theta} \btheta, \Sigma_{y} + C ).
\end{align*}

\noindent 1. For $\btheta_0$, the prior is 
\begin{align*}
\btheta_0 \mid \tbbeta_0, \sigma_{\beta_0}^2, \rho_0, \tb_0, \sigma_{b_0}^2, \gamma_{b_0} \sim N(\tilde{\bm \theta}_0, \Sigma_{\theta_0}),
\end{align*}
where $\tilde{\bm \theta}_0 = (X_{\beta} \tbbeta_0, \bone \tb_0)$, and
\begin{align*}
\Sigma_{\theta} = 
\diag \big( \sigma_{\beta_0}^2 \Sigma_{\beta}(\rho_0),
 \sigma_{b_0}^2 (I - \gamma_{b_0} W_{b_0})^{-1} \mathcal{N}_{b_0} \big).
\end{align*}
Thus, the  posterior of $\btheta_0$ is
\begin{align*}
&\btheta_0 \mid \by_{\vec 0}, \ldots \sim N(\btheta_0^*, \Sigma_{\theta_0}^*), \quad \text{where} \\
&\Sigma_{\theta_0}^* = \left[ \Sigma_{\theta_0}^{-1} + X_{\theta_0}^T (\Sigma_{y_0} + C_0 )^{-1} X_{\theta_0} \right]^{-1}, \\
&\btheta_0^* = \Sigma_{\theta_0}^*  \left[ \Sigma_{\theta_0}^{-1} \tilde{\bm \theta}_0 + X_{\theta_0}^T (\Sigma_{y_0} + C_0 )^{-1} \by_{\vec 0} \right].
\end{align*}

\noindent 2. For $\btheta$, the prior is  
\begin{align*}
\btheta \mid \tbbeta, \sigma_{\beta}^2, \rho, \tb, \sigma_{b}^2, \gamma_{b},  \tphi_1, \sigma_{\phi_1}^2, \gamma_{\phi_1}, \sigma_{\phi_2}^2, \sigma_{\phi_3}^2 
\sim N(\tilde{\bm \theta}, \Sigma_{\theta}),
\end{align*}
where $\tilde{\bm \theta} = (X_{\beta} \tbbeta, \bone \tb, \bone \tphi_1, \bm 0, \bm 0)$, and
\begin{multline*}
\Sigma_{\theta} = 
\diag \big( \sigma_{\beta}^2 \Sigma_{\beta}(\rho),  \sigma_{b}^2 (I - \gamma_b W_b)^{-1} \mathcal{N}_b,  \\
\sigma_{\phi_1}^2 (I - \gamma_{\phi_1} W_{\phi_1})^{-1} \mathcal{N}_{\phi_1}, \sigma_{\phi_2}^2 I, \sigma_{\phi_3}^2 I  \big).
\end{multline*}
Thus, the  posterior of $\btheta$ is
\begin{align*}
\btheta \mid \by_{\vec}, \ldots &\sim N(\btheta^*, \Sigma_{\theta}^*), \quad \text{where} \\
\Sigma_{\theta}^* &= \left[ \Sigma_{\theta}^{-1} + X_{\theta}^T  (\Sigma_{y} + C)^{-1} X_{\theta} \right]^{-1}, \\
\btheta^* &= \Sigma_{\theta}^*  \left[ \Sigma_{\theta}^{-1} \tilde{\bm \theta} + X_{\theta}^T (\Sigma_{y} + C)^{-1} \by_{\vec} \right].
\end{align*}


\paragraph{Hyperparameters related to $\bbeta$ and $\bbeta_0$. }
There are three hyperparameters related to $\bbeta$: $\tbbeta$, $\sigma_{\beta}^2$ and $\rho$. The conditional posteriors are as follows.

\noindent 1. Conditional posterior of $\tbbeta$:
\begin{align*}
\tbbeta \mid \bbeta, \sigma_{\beta}^2, \rho &\sim N(\tbbeta^*, \Sigma_{\tbbeta}^*), \quad \text{where} \\
\Sigma_{\tbbeta}^* &= \left[ \frac{1}{\delta_{\beta}^2} I + \frac{1}{\sigma_{\beta}^2} X_{\beta}'  \Sigma_{\beta}(\rho)^{-1}  X_{\beta} \right]^{-1},\\
\tbbeta^* &= \Sigma_{\tbbeta}^* \left[ \frac{1}{\sigma_{\beta}^2} X_{\beta}'   \Sigma_{\beta}(\rho)^{-1} \bbeta \right].
\end{align*}

\noindent 2. Conditional posterior of $\sigma_{\beta}^2$:
\begin{align*}
\sigma_{\beta}^2 \mid \bbeta, \tbbeta, \rho \sim \text{IG} \bigg[ \lambda_1^{\beta} + \frac{(J-1) Q}{2}, 
\lambda_2^{\beta} + \frac{1}{2} (\bbeta - X_{\beta} \tbbeta)' \Sigma_{\beta}(\rho)^{-1} (\bbeta - X_{\beta} \tbbeta)  \bigg].
\end{align*}

\noindent 3. Conditional posterior of $\rho$:
\begin{align*}
&p(\rho \mid  \bbeta, \tbbeta, \sigma_{\beta}^2) \\
\propto {}& \text{det}[\sigma_{\beta}^{-2} \Sigma_{\beta}(\rho)^{-1}]^{1/2}   \exp \left[  - \frac{1}{2 \sigma_{\beta}^2}  (\bbeta - X_{\beta} \tbbeta)' \Sigma_{\beta}(\rho)^{-1} (\bbeta - X_{\beta} \tbbeta)   \right] \\ 
\propto {}&(1 - \rho^2)^{Q / 2} \exp \left[ - \frac{1}{2 \sigma_{\beta}^2}
\left( \rho^2 R_{\beta 1}  -   2\rho R_{\beta 2} \right) \right],
\end{align*}
where 
\begin{align*}
R_{\beta 1} = \sum_{s = 3}^{J-1} \|  \bbeta_s - \tbbeta \|_2^2, \qquad
R_{\beta 2} = \sum_{s = 3}^{J}  (\bbeta_s - \tbbeta)'(\bbeta_{s-1} - \tbbeta).
\end{align*}
We use the following properties to derive the conditional posterior of $\rho$. The inverse and determinant of $\Sigma_{\beta}(\rho)$ are
\begin{align*}
\Sigma_{\beta}(\rho)^{-1} = 
\left( \begin{array}{cccccc}
I &  - \rho I &  & &  & \\
- \rho I & (1 + \rho^2) I & - \rho I &  &  & \\
  & - \rho I & (1 + \rho^2) I & - \rho I & &  \\
  & & - \rho I  & \ddots & \ddots & \\
  & & & \ddots & (1 + \rho^2) I & - \rho I \\
  & & & & - \rho I & I
\end{array} \right),
\end{align*}
and $\det[\Sigma_{\beta}(\rho)^{-1}] = (1 - \rho^2)^Q$, respectively.
To update $\tbbeta$ and $\sigma_{\beta}^2$, we use regular Gibbs steps. To update $\rho$, given $\{\bbeta, \tbbeta, \sigma_{\beta}^2 \}$ we can easily evaluate its posterior on the $[0, 1]$ grid, and sample from it.

Similarly, there are three hyperparameters related to $\bbeta_0$: $\tbbeta_0$, $\sigma_{\beta_0}^2$ and $\rho_0$. Their conditional posteriors have exactly the same form as those for $\tbbeta$, $\sigma_{\beta}^2$ and $\rho$.

\paragraph{Hyperparameters related to $\bb$ and $\bb_0$. }
There are three hyperparameters related to $\bb$: $\tb$, $\sigma_b^2$ and $\gamma_b$. The conditional posteriors are as follows.

\noindent 1. Conditional posterior of $\tb$:
\begin{align*}
\tb \mid \bb, \sigma_b^2, \gamma_b &\sim N(\tb^*, \delta_{\tb}^{*2}), \quad \text{where} \\
\delta_{\tb}^{*2} &= \left[  \frac{1}{\delta_{b}^{2}} + \frac{1}{\sigma_b^2}\bone^T \mathcal{N}_b^{-1} (I - \gamma_b W_b) \bone \right]^{-1},   \\
\tb^* &= \delta_{\tb}^{*2}   \left[  \frac{1}{\sigma_b^2}\bone^T \mathcal{N}_b^{-1} (I - \gamma_b W_b) \bb \right].
\end{align*}

\noindent 2. Conditional posterior of $\sigma_b^2$:
\begin{align*}
\sigma_b^2 \mid \bb, \tb, \gamma_b \sim \text{IG} \bigg[ \lambda_1^b + \frac{\text{dim}(\bb)}{2}, 
\lambda_2^b + \frac{1}{2} (\bb - \bone \tb)' \mathcal{N}_b^{-1} (I - \gamma_b W_b) (\bb - \bone \tb) \bigg].
\end{align*}

\noindent 3. Conditional posterior of $\gamma_b$:
\begin{multline*}
p(\gamma_b \mid  \bb, \tb, \sigma_b^2) \propto \det(I - \gamma_b W_b)^{1/2}  \cdot \\
\exp \left[ \gamma_b \cdot \frac{1}{2 \sigma_b^2} (\bb - \bone \tb)' \mathcal{N}_b^{-1} W_b (\bb - \bone \tb) \right ].
\end{multline*}
To update $\tb$ and $\sigma_b^2$, we use regular Gibbs steps. To update $\gamma_b$, given $\{  \bb, \tb, \sigma_b^2 \}$ we can easily evaluate its posterior on the $[0, 1]$ grid, and sample from it. 
To facilitate computation, we can calculate $\det(I - \gamma_b W_b)^{1/2}$ on the $[0, 1]$ grid, save the values and use it at each iteration.

Similarly, there are three hyperparameters related to $\bb_0$: $\tb_0$, $\sigma_{b_0}^2$ and $\gamma_{b_0}$. Their conditional posteriors have exactly the same form as those for $\tb$, $\sigma_{b}^2$ and $\gamma_{b}$.

\paragraph{Hyperparameters related to $\bphi_1$. }
There are three hyperparameters related to $\bphi_1$: $\tphi_1$, $\sigma_{\phi_1}^2$ and $\gamma_{\phi_1}$. The conditional posteriors are as follows.

\noindent 1. Conditional posterior of $\tphi_1$:
\begin{align*}
\tphi_1 \mid \bphi_1, \sigma_{\phi_1}^2, \gamma_{\phi} &\sim N(\tphi^*, \delta_{\tphi}^{*2}), \quad \text{where} \\
\delta_{\tphi}^{*2} &= \left[  \frac{1}{\delta_{\phi_1}^{2}} + \frac{1}{\sigma_{\phi_1}^2}\bone' \mathcal{N}_{\phi_1}^{-1} (I - \gamma_{\phi_1} W_{\phi_1}) \bone \right]^{-1},   \\
\tphi^* &= \delta_{\tphi}^{*2}   \left[ \frac{1}{\delta_{\phi_1}^{2}} \cdot 1 +  \frac{1}{\sigma_{\phi_1}^2}\bone' \mathcal{N}_{\phi_1}^{-1} (I - \gamma_{\phi_1} W_{\phi_1}) \bphi_1 \right].
\end{align*}

\noindent 2. Conditional posterior of $\sigma_{\phi_1}^2$:
\begin{multline*}
\sigma_{\phi_1}^2 \mid  \bphi_1, \tphi_1, \gamma_{\phi_1} \sim \text{IG} \bigg[ \lambda_1^{\phi_1} + \frac{\text{dim}(\bphi_1)}{2}, \\ \lambda_2^{\phi_1} + \frac{1}{2} (\bphi_1 - \bone \tphi_1)' \mathcal{N}_{\phi_1}^{-1} (I - \gamma_{\phi_1} W_{\phi_1}) (\bphi_1 - \bone \tphi_1) \bigg].
\end{multline*}

\noindent 3. Conditional posterior of $\gamma_{\phi_1}$:
\begin{multline*}
p(\gamma_{\phi_1} \mid  \bphi_1, \tphi_1, \sigma_{\phi_1}^2) \propto \det(I - \gamma_{\phi_1} W_{\phi_1})^{1/2} \cdot \\
\exp \left[ \gamma_{\phi_1} \cdot \frac{1}{2 \sigma_{\phi_1}^2} (\bphi_1 - \bone \tphi_1)' \mathcal{N}_{\phi_1}^{-1} W_{\phi_1} (\bphi_1 - \bone \tphi_1) \right ].
\end{multline*}

\paragraph{Hyperparameters related to $\bphi_2$ and $\bphi_3$. }
There is one hyperparameter related to $\bphi_2$: $\sigma_{\phi_2}^2$. The conditional posterior is 
\begin{align*}
\sigma_{\phi_2}^2 \mid \bphi_2 \sim \IG \left[ \lambda_{1}^{\phi_2} + \frac{1}{2} \text{dim}(\bphi_2), \; \lambda_{2}^{\phi_2} + \frac{1}{2} \bphi_2^T \bphi_2 \right].
\end{align*}
Similarly, there is one hyperparameter related to $\bphi_3$: $\sigma_{\phi_3}^2$. The conditional posterior is 
\begin{align*}
\sigma_{\phi_3}^2 \mid \bphi_3 \sim \IG \left[ \lambda_{1}^{\phi_3} + \frac{1}{2} \text{dim}(\bphi_3), \; \lambda_{2}^{\phi_3} + \frac{1}{2} \bphi_3^T \bphi_3 \right].
\end{align*}

\paragraph{Update intermittent missing responses.} 
The focus of our method is dealing with monotone missing data. Sometimes there are (typically few) intermittent missing responses, and we impute it under the partial ignorability assumption \citep{harel2009partial}. Suppose $y_{ijs}$ is missing. Its conditional distribution is
\begin{align*}
p \left( y_{ijs} \mid y_{-ijs}, \bpi \right) \propto p \left( \by_{\vec 0}, \by_{\vec} \mid \bpi \right),
\end{align*}
We use a Metropolis-Hastings step to update $y_{ijs}$. 
We use a symmetric normal proposal distribution, $y_{ijs}^{\text{pro}} \sim N \left( y_{ijs}^{\text{cur}}, \; 0.5 \times \text{sd}(\by_{\vec 0}, \by_{\vec}) \right)$.

\section{G-computation Implementation Details}
\label{app-bnpmis-gcomp}
We describe in detail how to draw the pseudo responses using Gaussian process prediction rule, i.e. steps 3 and 4 in Algorithm \ref{G-computation}. We generally use a superscript $*$ to denote the pseudo subject and response. 

\paragraph{Observed response.}
To draw a vector of pseudo observed responses $\bar{\bm Y}_s^* = \bby_s^*$ from  $p(\bby_s^* \mid s^*, \bv^*, \bpi)$, we do the following.

\noindent 1. Draw $y_1^*$ from $p(y_1^* \mid s^*, \bv^*, \bpi)$.
Consider the joint distribution of $a_{1s}^* = a_{0}(\bv^*, 1, s^*)$ and the training data points $\by_{\vec 0}$,
\begin{align*}
\left( \begin{array}{c}
\by_{\vec 0}  \\
a_{1s}^*  \end{array} \right) \sim N \left[ 
\left( \begin{array}{c}
X_{\theta_0} \btheta_0 \\
\mu_{1s}^*  \end{array} \right),
\left( \begin{array}{cc}
\Sigma_{y0} + C_{0} & C_{1s*}\\
C_{1s*}^T &  C_{1s**} \end{array} \right)
\right],
\end{align*}
where
\begin{align*}
\mu_{1s}^* &= \mu_0(\bv^*, 1, s^*), \\
C_{1s*} &= C_0(V, \bm j_{\vec 0}, \bm s_{\vec 0}; \bv^*, 1, s^*), \\ 
C_{1s**} &= C_0(\bv^*, 1, s^*; \bv^*, 1, s^*),
\end{align*}
$\bm j_{\vec 0}$ and $\bm s_{\vec 0}$ are vectors of times and patterns corresponding to responses $\by_{\vec 0}$.
The predictive distribution for $a_{1s}^*$ is thus
\begin{multline*}
a_{1s}^* \mid \by_{\vec 0}, \bpi \sim N \big[ \mu_{1s}^* + C_{1s*}^T (\Sigma_{y0} + C_{0})^{-1} (\by_{\vec 0} - X_{\theta_0} \btheta_0), \\ 
C_{1s**} - C_{1s*}^T (\Sigma_{y0} + C_{0})^{-1} C_{1s*} \big],
\end{multline*}
and we can draw
\begin{align*}
y_1^* \sim N(a_{1s}^*, \sigma_{1s^*}^2).
\end{align*}

\noindent 2. Draw $y_j^*$ from $p(y_j^* \mid \bby_{j-1}^*, s^*, \bv^*, \bpi)$, ($1 < j \leq s^*$).  The joint distribution of $a_{js}^* = a(\bv^*, j, s^*) + \bby_{j-1}^{*T} \bPhi_{js^*}$ and the training data points $\by_{\vec}$ is
\begin{align*}
\left( \begin{array}{c}
\by_{\vec}  \\
a_{js}^{*}  \end{array} \right) \sim N \left[ 
\left( \begin{array}{c}
X_{\theta} \btheta \\
\mu_{js}^* + \bby_{j-1}^{*T} \bPhi_{js^*}  \end{array} \right),
\left( \begin{array}{cc}
\Sigma_{y} + C & C_{js*}\\
C_{js*}^T &  C_{js**} \end{array} \right)
\right],
\end{align*}
where
\begin{align*}
\mu_{js}^* &= \mu(\bv^*, j, s^*), \\
C_{js*} &= C(V, \bm j_{\vec}, \bm s_{\vec}; \bv^*, j, s^*), \\ 
C_{js**} &= C(\bv^*, j, s^*; \bv^*, j, s^*),
\end{align*}
$\bm j_{\vec}$ and $\bm s_{\vec}$ are vectors of times and patterns corresponding to responses $\by_{\vec}$.
The predictive distribution for $a_{js}^*$ is thus
\begin{multline*}
a_{js}^* \mid \bby_{j-1}^{*}, \by_{\vec}, \bpi \sim N \big[ \mu_{js}^* + \bby_{j-1}^{*T} \bPhi_{js^*} + C_{js*}^T (\Sigma_{y} + C)^{-1} (\by_{\vec} - X_{\theta} \btheta ), \\ 
C_{js**} - C_{js*}^T (\Sigma_{y} + C)^{-1} C_{js*} \big],
\end{multline*}
and we can draw
\begin{align*}
y_j^* \sim N(a_{js}^*, \sigma_{js^*}^2).
\end{align*}

\paragraph{Missing response.}
To draw a pseudo response $Y_j^* = y_j^*$ from the extrapolation distribution $p(y_j^* \mid \bby_{j-1}^*, s^*, \bv^*, \bomega)$ ($j > s^*$), do the following.

\noindent(I) Under MAR,
\begin{align}
p(y_j^* \mid \bby_{j-1}^*, \bv^*, S = s^*, \bomega) &= p(y_j^* \mid \bby_{j-1}^*, \bv^*, S \geq j, \bomega) \nonumber\\
&= \sum_{k = j}^J \alpha_{kj} p(y_j^* \mid \bby_{j-1}^*, \bv^*, S = k, \bomega),
\label{eq:mixture_pattern}
\end{align}
where
\begin{align*}
\alpha_{kj} &= \alpha_{kj}(\bby_{j-1}^*, \bv^*) = p(S = k \mid \bby_{j-1}^*, \bv^*, S \geq j)  \\
&= \frac{p(\bby_{j - 1}^* \mid \bv^*, S = k) \; p(S = k \mid \bv^*, S \geq j)}{\sum_{k=j}^J p(\bby_{j - 1}^* \mid \bv^*, S = k) \; p(S = k \mid \bv^*, S \geq j)}, \quad k = j, \ldots, J
\end{align*}
The above expression can be calculated by
\begin{align*}
p(\bby_{j - 1}^* \mid \bv^*, S = k) = p_k(y_1^* \mid \bv^*) \cdot \prod_{l = 2}^{j-1} p_k(y_l^* \mid \bby_{l-1}^*, \bv^*)
\end{align*}
where 
\begin{align*}
p_k (y_1^* \mid \bv^*) &= p_N \big[ y_1^* \mid \mu_{1k}^* + C_{1k*}^T (\Sigma_{y0} + C_{0})^{-1} (\by_{\vec 0} - X_{\theta_0} \btheta_0), \\
&\pushright{C_{1k**} - C_{1k*}^T (\Sigma_{y0} + C_{0})^{-1} C_{1k*}  \big]}, \\
p_k(y_l^* \mid \bby_{l-1}^*, \bv^*) &= p_N \big[ y_l^* \mid \mu_{lk}^* + \bby_{l-1}^{*T} \bPhi_{lk} + C_{lk*}^T (\Sigma_{y} + C)^{-1}  \cdot \\
&\pushright{(\by_{\vec} - X_{\theta} \btheta ), C_{lk**} - C_{lk*}^T (\Sigma_{y} + C)^{-1} C_{lk*} \big]},
\end{align*}
and
\begin{align*}
&p(S = k \mid \bv^*, S \geq j) \\
= {}& p(S = k \mid \bv^*, S \geq k) \cdot \prod_{l = j}^{k-1} p(S \geq l+1 \mid \bv^*, S \geq l) \\
= {}& p(S = k \mid \bv^*, S \geq k) \cdot \prod_{l = j}^{k-1} \left[ 1 - p(S = l \mid \bv^*, S \geq l) \right].
\end{align*}

\noindent To sample from \eqref{eq:mixture_pattern}, after calculating $(\alpha_{jj}, \ldots, \alpha_{Jj})$, we can draw $K = k$ with probability $\alpha_{kj}$, and sample $Y_j^* = y_j^*$ from $p_k(y_j^* \mid \bby_{j-1}^*, \bv^*, \bomega)$.

\noindent(II) Under NFD.
 
\noindent(II-1) For $j = s^*+1$,
\begin{align*}
[Y_{j} \mid \bm \bY_{j-1}, S = j-1, \bV, \bomega] \eqind
\left[ Y_{j} + \tau_{j} \mid \bm \bY_{j-1}, S \geq j, \bV, \bomega \right].
\end{align*}

We first sample from $p_{\geq j}(y_{j}^* \mid \bby_{j-1}^*, \bv^*, \bomega)$. Then, we apply the location shift \eqref{location_shift} with
$$
\tau_{j}(\bby_{j-1}^*, \bv^*) = \tilde{\tau} \cdot \Delta_j(\bby_{j-1}^*, \bv^*),
$$
where $\Delta_j(\bby_{j-1}^*, \bv^*)$ is chosen to be the standard deviation of $p_{j-1}(y_{j}^* \mid \bby_{j-1}^*, \bv^*, \bomega)$ under MAR, i.e. $p_{\geq j}(y_{j}^* \mid \bby_{j-1}^*, \bv^*, \bomega)$. We have 
$$
p_{\geq j}(y_{j}^* \mid \bby_{j-1}^*, \bv^*, \bomega) = \sum_{k = j}^J \alpha_{kj} N(\tilde{\mu}_{jk}, \tilde{\sigma}_{jk}^2).
$$
The standard deviation of this normal mixture is given by
$$
\Delta_j(\bby_{j-1}^*, \bv^*) = \sqrt{\sum_{k = j}^J \alpha_{kj} \tilde{\sigma}_{jk}^2 + \sum_{k = j}^J \alpha_{kj} \tilde{\mu}_{jk}^{2} - \left( \sum_{k = j}^J \alpha_{kj} \tilde{\mu}_{jk} \right)^2}.
$$

\noindent(II-2) For $j > s^*+1$,
\begin{align}
p(y_j^* \mid \bby_{j-1}^*, \bv^*, S = s^*, \bomega) &= p(y_j^* \mid \bby_{j-1}^*, \bv^*, S \geq j-1, \bomega) \nonumber\\
&= \sum_{k = j-1}^J \alpha_{k,j-1} p(y_j^* \mid \bby_{j-1}^*, \bv^*, S = k, \bomega),
\label{eq:mixture_pattern2}
\end{align}
where
\begin{align*}
\alpha_{k,j-1} &= \alpha_{k,j-1}(\bby_{j-1}^*, \bv^*) = p(S = k \mid \bby_{j-1}^*, \bv^*, S \geq j-1)  \\
&= \frac{p(\bby_{j - 1}^* \mid \bv^*, S = k) \; p(S = k \mid \bv^*, S \geq j-1)}{\sum_{k=j-1}^J p(\bby_{j - 1}^* \mid \bv^*, S = k) \; p(S = k \mid \bv^*, S \geq j-1)},
\quad k = j-1, \ldots, J.
\end{align*}
To sample from \eqref{eq:mixture_pattern2}, after calculating $(\alpha_{j-1, j-1}, \ldots, \alpha_{J, j-1})$, we can draw $K = k$ with probability $\alpha_{k, j-1}$. 

\noindent(II-2a)  If $k = j-1$, draw again $K' = k'$ with probability $\alpha_{k', j-1}/(1-\alpha_{j-1, j-1})$ for $k' = j, \ldots, J$. Then, sample $Y_j^* = y_j^*$ from $p_{k'}(y_j^* \mid \bby_{j-1}^*, \bv^*, \bomega)$, and apply the location shift \eqref{location_shift}.

\noindent(II-2b) If $k \in \{ j, \ldots, J \}$, sample $Y_j^* = y_j^*$ from $p_{k}(y_j^* \mid \bby_{j-1}^*, \bv^*, \bomega)$.

The steps for sampling the pseudo response $\bm Y^* = \by^*$ from $p(\by^* \mid s^*, \bv^*, \bomega)$ are summarized in Algorithm \ref{DrawPseudoY}, where
\begin{align*}
\tilde{\mu}_{1s} &= \mu_{1s}^* + C_{1s*}^T (\Sigma_{y0} + C_{0})^{-1} (\by_{\vec 0} - X_{\theta_0} \btheta_0), \\
\tilde{\sigma}_{js}^2 &= C_{1s**} - C_{1s*}^T (\Sigma_{y0} + C_{0})^{-1} C_{1s*} + \sigma_{1s^*}^2,
\end{align*}
and
\begin{align*}
\tilde{\mu}_{js} &= \mu_{js}^* + \bby_{j-1}^{*T} \bPhi_{js} + C_{js*}^T (\Sigma_{y} + C)^{-1} (\by_{\vec} - X_{\theta} \btheta ), \\
\tilde{\sigma}_{js}^2 &= C_{js**} - C_{js*}^T (\Sigma_{y} + C)^{-1} C_{js*} + \sigma_{js^*}^2,
\end{align*}
for $j = 2, \ldots, J$.

\begin{center}
\begin{minipage}{\textwidth}
\begin{algorithm}[H]
\caption{Draw $\bm Y^* = \by^*$ from $p(\by^* \mid s^*, \bv^*, \bomega)$}
\label{DrawPseudoY}
\begin{algorithmic}[1]
\State Draw 
$Y_1^* = y_1^* \sim N(\tilde{\mu}_{1s}, \tilde{\sigma}_{1s}^2)$
\For{$j$  in $2, \ldots, s^*$ } 
\State Draw 
$Y_j^* = y_j^* \sim N(\tilde{\mu}_{js}, \tilde{\sigma}_{js}^2)$
\EndFor
\If{MAR}
\For{$j$  in $s^* + 1, \ldots, J$ } 
\State Calculate $\bm \alpha_j (\bby_{j-1}^*, \bv^*) = (\alpha_{jj}, \ldots, \alpha_{Jj})$
\State Draw  $K = k \sim \mbox{Categorical}[(j, \ldots, J); \bm \alpha_j]$
\State Draw 
$y_j^* \sim N(\tilde{\mu}_{jk}, \tilde{\sigma}_{jk}^2)$
\EndFor
\ElsIf{NFD}
\State Set $j = s^* + 1$ 
\State Calculate $\bm \alpha_j (\bby_{j-1}^*, \bv^*) = (\alpha_{jj}, \ldots, \alpha_{Jj})$
\State Draw  $K = k \sim \mbox{Categorical}[(j, \ldots, J); \bm \alpha_j ]$
\State Calculate $\tau_{j}(\bby_{j-1}^*, \bv^*) = \tilde{\tau} \cdot \Delta_j(\bby_{j-1}^*, \bv^*)$
\State Draw $y_{j}^* \sim N(\tilde{\mu}_{jk} + \tau_{j}, \tilde{\sigma}_{jk}^2)$
\For{$j$  in $s^* + 2, \ldots, J$ } 
\State Calculate $\bm \alpha_{j-1} (\bby_{j-1}^*, \bv^*) = (\alpha_{j-1, j-1}, \ldots, \alpha_{J,j-1})$
\State Draw  $K = k \sim \mbox{Categorical}[(j-1, \ldots, J); \bm \alpha_{j-1}]$
\If{$k = j-1$}
\State Calculate $\bm \alpha_{j}'  = ( \alpha_{j, j-1}, \ldots, \alpha_{J, j-1} )  / (1 - \alpha_{j-1, j-1})$
\State Draw  $K' = k' \sim \mbox{Categorical}[(j, \ldots, J); \bm \alpha_{j}']$
\State Calculate $\tau_{j}(\bby_{j-1}^*, \bv^*) = \tilde{\tau} \cdot \Delta_j(\bby_{j-1}^*, \bv^*)$
\State Draw $y_{j}^* \sim N(\tilde{\mu}_{jk'} + \tau_{j}, \tilde{\sigma}_{jk'}^2)$
\Else 
\State Draw 
$y_j^* \sim N(\tilde{\mu}_{jk}, \tilde{\sigma}_{jk}^2)$.
\EndIf
\EndFor
\EndIf
\end{algorithmic}
\end{algorithm}
\end{minipage}
\end{center}

\section{Simulation Details}
\label{app-bnpmis-sim}
\paragraph{Prior and hyperprior parameters.} 
We set the prior and hyperprior parameters at standard noninformative choices. Table \ref{tbl:simu_hp} shows the exact values.

\begin{table}[h!]
\centering
\begin{tabular}{|cr|cr|cr|cr|}
\hline
$\delta_{\beta_0}^2$ & 30 &  $\delta_{b_0}^2$ & 30  & $\delta_{\phi_1}^2$ & 30 & $\lambda_1^{\phi_3}$ &100\\
$\lambda_1^{\beta_0}$ &1 & $\lambda_1^{b_0}$ &1 & $\lambda_1^{\phi_1}$ &1 & $\lambda_2^{\phi_3}$ &1 \\
$\lambda_2^{\beta_0}$ &1 & $\lambda_2^{b_0}$ &1 & $\lambda_2^{\phi_1}$ &1 & $d_{\eta}$                   & 0.1\\
$\delta_{\beta}^2$ & 30  & $\delta_{b}^2$ & 30      & $\lambda_1^{\phi_2}$ & 30 &&\\
$\lambda_1^{\beta}$ &1 & $\lambda_1^{b}$ &1     & $\lambda_2^{\phi_2}$ &1 &&\\
$\lambda_2^{\beta}$ &1 & $\lambda_2^{b}$ &1     &  &&&\\
\hline
\end{tabular}
\caption{Choices of prior and hyperprior parameters in the observed data model. These parameters are used for simulations and real data analysis.}
\label{tbl:simu_hp}
\end{table}

\paragraph{Scenario 1.} The covariance matrix for generating $\bV$ is
\begin{align*}
\Sigma_{vv} &= \left( \begin{array}{cccc}
1.0    &  0.52  & -0.22  &  0.07 \\
0.52   &  1.0   & -0.23  & -0.02 \\
-0.22  & -0.23  &  1.0   &  0.45 \\
0.07   & -0.02  &  0.45  &  1.0 \end{array} \right),
\end{align*}
which is the correlation matrix of the subjects' numerical auxiliary covariates from the schizophrenia clinical trial dataset.

The parameters for generating $S$ are
\begin{align*}
\bm \zeta = (-4.346, -2.193, -2.606, -2.678)^T, 
\end{align*}
where $\zeta_s$ corresponds to the $(s-1)$-th element ($s = 2, \ldots, 5$), and
\begin{align*}
\bm \xi = \left( \begin{array}{rrrr}
-1.057 &  0.328 & -0.121 &  0.273 \\
-0.826 &  0.128 &  0.525 & -0.781 \\
-0.487 &  0.479 &  0.534 & -0.480 \\
 0.642 &  0.129 &  0.448 &  0.122\end{array} \right),
\end{align*}
where $\bm \xi_s$ corresponds to the $(s-1)$-th column ($s = 2, \ldots, 5$).
These parameters come from fitting the sequential logistic regression model to the test drug arm of the schizophrenia clinical trial dataset and taking posterior mean of each parameter.

The parameters for generating $\bar{\bm Y}_S$ are 
\begin{align*}
\{ \sigma_{js}^2 \} &= \left( \begin{array}{rrrrrr}
0.232   &  0.221   &              &            & & \\
0.365   &  0.243   & 0.196    &            & & \\
0.403   &  0.222   & 0.228    &  0.941 & & \\
0.438   &  0.228  &  0.225    &  0.213 & 0.284  & \\
0.335   &  0.192  &  0.265    & 0.140  & 0.167  & 0.160 
\end{array} \right),
\end{align*}
where $\sigma_{js}^2$ corresponds to the element in the $(s -1)$-th row and $j$-th column;
\begin{align*}
(\bb_0, \bm b) &= \left( \begin{array}{rrrrrr}
0.069   & -0.191   &   &   & & \\
0.507   &  0.219   & 0.302  &  & & \\
0.393   &  0.060    & -0.022   &  0.399 & & \\
0.798   &  0.048  &  -0.051  &  0.051 & 0.362 & \\
0.384   & -0.107  &  -0.250   & -0.367 &-0.250  & -0.321 \end{array} \right),
\end{align*}
where $b_{js}$ corresponds to the element in the $(s -1)$-th row and $j$-th column;
\begin{align*}
\bbeta_0 &= \left( \begin{array}{rrrrr}
-0.046 &  0.174 & -0.005 &  0.024  &  0.230 \\
-0.200 & -0.099 & -0.124 & -0.451  & -0.163 \\
-0.315 & -0.191 & -0.104 &  0.140  &  0.032 \\
-0.053 &  0.065 &  0.003 & -0.044  & -0.092 \end{array} \right),
\end{align*}
where $\bbeta_{0s}$ corresponds to the $(s-1)$-th column;
\begin{align*}
\bbeta = \left( \begin{array}{rrrrr}
-0.080 & -0.117 & -0.118 &  0.010  &  0.066 \\
-0.044 & -0.113 &  0.023 & -0.035  & -0.030 \\
-0.109 & -0.020 & -0.014 & -0.022  &  0.056 \\
 0.170 &  0.127 &  0.166 & -0.060  &  0.002 \end{array} \right), 
\end{align*}
where $\bbeta_{s}$ corresponds to the $(s-1)$-th column;
\begin{align*}
(\bphi_1) = \left( \begin{array}{rrrrrr}
1.078   &   &   & & \\
1.088   & 0.938  &  & & \\
0.830   & 0.893   &  0.830 & & \\
0.637  &  0.877  &  0.907 & 1.065 & \\
0.881  &  0.871   & 0.842 & 0.929  & 0.943 \end{array} \right),
\end{align*}
where $\phi_{1js}$ corresponds to the element in the $(s -1)$-th row and $(j-1)$-th column;
\begin{align*}
(\bphi_2) &= \left( \begin{array}{rrrrr}
-0.045   &   &   & \\
0.040   & -0.025  &  & \\
0.021    &  0.022   &  0.035 & \\
0.089  &  0.129  &  0.019 & -0.020 
\end{array} \right),
\end{align*}
where $\phi_{2js}$ corresponds to the element in the $(s -2)$-th row and $(j-2)$-th column; and
\begin{align*}
(\bphi_3) &= \left( \begin{array}{ccc}
0.011   &   &   \\
0.037   & \left( \begin{array}{r} 0.074 \\ 0.037 \end{array} \right) &   \\
0.078    & \left( \begin{array}{r} -0.027 \\ -0.086 \end{array} \right)   & \left( \begin{array}{r} 0.021 \\ 0.010 \\ -0.009  \end{array} \right) 
\end{array} \right),
\end{align*}
where $\bphi_{3js}$ corresponds to the element in the $(s -3)$-th row and $(j-3)$-th column.
These parameters come from fitting the linear regression model to the test drug arm of the schizophrenia clinical trial dataset and taking posterior mean of each parameter.

\paragraph{Scenario 2.}
We use the same choices of $\bb_0$, $\bb$, $\bphi_1$, $\bphi_2$ and $\bphi_3$ as in Scenario 1. We set 
\begin{align*}
\Sigma_{vv} = \left( \begin{array}{ccc}
1.0    &  0.52  & -0.22   \\
0.52   &  1.0   & -0.23   \\
-0.22  & -0.23  &  1.0    
\end{array} \right),
\end{align*}
i.e. the upper left $3 \times 3$ submatrix of $\Sigma_{vv}$ in Scenario 1.
We change $\{ \sigma_{js}^2\}$, $\bm \zeta$, $\bm \xi$, $\bbeta_0$ and $\bbeta$ to
\begin{align*}
\{ \sigma_{js}^2 \} &= \left( \begin{array}{rrrrrr}
0.155   &  0.101   &              &            & & \\
0.217   &  0.133   & 0.112    &            & & \\
0.099   &  0.082   & 0.101    &  0.115 & & \\
0.141   &  0.127   &  0.169    &  0.132 & 0.107  & \\
0.106   &  0.119   &  0.095    & 0.081  & 0.266  & 0.174 
\end{array} \right),
\end{align*}
where $\sigma_{js}^2$ corresponds to the element in the $(s -1)$-th row and $j$-th column;
\begin{align*}
\bm \zeta = (-3.0,  -2.1,  -1.6,  -1.3)^T, 
\end{align*}
where $\zeta_s$ corresponds to the $(s-1)$-th element ($s = 2, \ldots, 5$), and
\begin{align*}
\bm \xi = \left( \begin{array}{rrrr}
-1.057 &  0.328 & -0.121 &  0.273 \\
-0.826 &  0.128 &  0.525 & -0.781 \\
-0.487 &  0.479 &  0.534 & -0.480 \\
-0.528 &  0.164 & -0.061 &  0.136 \\
-0.413 &  0.064 &  0.263 & -0.390 \\
-0.244 &  0.239 &  0.267 & -0.240 \\
 0.321 &  0.064 &  0.224 &  0.061 \\
-0.528 &  0.164 & -0.061 &  0.136 \\
-0.413 &  0.064 &  0.263 & -0.390
\end{array} \right),
\end{align*}
where $\bm \xi_s$ corresponds to the $(s-1)$-th column ($s = 2, \ldots, 5$).
\begin{align*}
\bbeta_0 &= \left( \begin{array}{rrrrr}
-0.530  &  -0.508  &  -0.561 &  -0.507  &  -0.525 \\
-0.366  &  -0.377  &  -0.421 &  -0.417  &  -0.386 \\
 0.351  &   0.309  &   0.323  &   0.318  &  0.346 \\
 0.283  &   0.291  &   0.282  &   0.277  &  0.275 \\
-0.316  &  -0.321  &  -0.319  &  -0.319  & -0.316 \\
 0.288  &   0.285   &   0.293  &  0.288   &  0.289 \\
 0.033  &   0.030   &   0.033  &  0.020   & 0.033 \\
-0.083  &  -0.087   &  -0.094  & -0.082   & -0.092 \\
 0.124  &   0.125   &   0.115  &   0.120   & 0.116 \end{array} \right),
\end{align*}
where $\bbeta_{0s}$ corresponds to the $(s-1)$-th column;
\begin{align*}
\bbeta = \left( \begin{array}{rrrrr}
-0.395   &   -0.387   &   -0.427   &   -0.434   &   -0.443  \\
 0.320   &    0.337   &    0.339   &    0.317   &    0.338  \\
 0.331   &    0.349   &    0.400   &    0.385   &    0.356  \\
 0.317   &    0.315   &    0.309   &    0.313   &    0.310  \\
 0.354   &    0.355   &    0.342   &    0.354   &    0.349  \\
-0.301   &   -0.299   &   -0.303   &   -0.306   &   -0.306  \\
-0.082   &   -0.082   &   -0.073   &   -0.068   &   -0.079  \\
-0.077   &   -0.088   &   -0.082   &   -0.085   &   -0.081  \\
-0.129   &   -0.126   &   -0.130   &   -0.133   &   -0.128   \\
\end{array} \right), 
\end{align*}
where $\bbeta_{s}$ corresponds to the $(s-1)$-th column.

\paragraph{Scenario 3.}
The parameter for generating $K$ is 
\begin{align*}
\bm \pi = (0.119, 0.579, 0.001, 0.115, 0.186), 
\end{align*}
which is taken from \cite{linero2015flexible} by fitting the mixture model to the active control arm of the schizophrenia clinical trial dataset.

The parameters for the joint distribution of $\bm Y$ and $\bV$ are specified and generated as follows.
Within mixture component $k$, the joint distribution of $\bm Y$ and $\bV$ is
\begin{align*}
\left( \begin{array}{c}
\bm{Y} \\
\bV  \end{array} \right) \mid K = k \sim  N\left[ \bm{\mu}^{(k)}, \Omega^{(k)}\right], 
\end{align*}
where 
\begin{align*}
\bm{\mu}^{(k)} &= \left( \begin{array}{c}
\bm{\mu}_y^{(k)} \\
\bm 0  \end{array} \right), \\
\Omega^{(k)} &\sim \mathcal{W}^{-1}\left( (\nu - J - Q - 1) \Omega_0^{(k)}, \nu \right), \\
\Omega_0^{(k)} &= \left( \begin{array}{cc}
\Sigma_{yy}^{(k)} & \Sigma_{yv}^{(k)} \\
\Sigma_{vy}^{(k)} & \Sigma_{vv} \end{array} \right).
\end{align*}
Here $\bm{\mu}_y^{(k)}$ and $\Omega_0^{(k)}$ correspond to a linear model of $(\bm Y \mid \bV)$, where
\begin{align*}
\bV \mid K = k &\sim N(\bm 0, \Sigma_{vv}), \\
Y_1 \mid \bV, K = k &\sim N \left( b_1^{(k)} + \bV^T \bbeta_0^{(k)}, \; \sigma_{1}^{2(k)} \right),  \\
Y_j \mid \bar{\bm Y}_{j-1}, \bV, K = k &\sim N \left( b_j^{(k)} + \bV^T \bbeta^{(k)} + \phi_{j}^{(k)} Y_{j-1}, \; \sigma_{j}^{2(k)} \right), \;\; j = 2, \ldots, J.
\end{align*}
Let $\bb^{(k)} = (b_1^{(k)}, \ldots, b_J^{(k)})^T$, $B^{(k)} = (\bbeta_0^{(k)}, \bbeta^{(k)}, \ldots, \bbeta^{(k)})$, $\Sigma_0^{(k)} = \diag(\sigma_1^{2(k)}, \ldots, \sigma_J^{2(k)})$,
\begin{align*}
\Phi^{(k)} = 
\left( \begin{array}{ccccc}
0           &    0   &    0   &   \cdots    & 0 \\
\phi_2^{(k)}   &    0   &    0   &    \cdots   & 0  \\
0           &  \phi_3^{(k)}    &  0     &   \cdots    &  0  \\
\vdots   &   \ddots    &   \ddots    & \ddots  & \vdots   \\
0          &    \cdots   & 0 &    \phi_J^{(k)}  &    0     \\
\end{array} \right),
\end{align*}
and $\Psi^{(k)} = \left( I - \Phi^{(k)} \right)^{-1}$.
We have 
\begin{align*}
\bm{\mu}_y^{(k)} &= \Psi^{(k)} \bb^{(k)}, \\
\Sigma_{yy}^{(k)} &= \Psi^{(k)} B^{(k)T} \Sigma_{vv} B^{(k)} \Psi^{(k)T}  + \Psi^{(k)} \Sigma_0^{(k)} \Psi^{(k)T}, \\
\Sigma_{yv}^{(k)} &=  \Psi^{(k)} B^{(k)T} \Sigma_{vv}.
\end{align*}

We use the same $\Sigma_{vv}$ as in Scenario 2. The parameters $\{ \bm \mu_y^{(k)} \}$ and $\Sigma_0^{(k)}$ are taken from \cite{linero2015flexible} (after standardization), which are generated by fitting the model to the active control arm of the schizophrenia clinical trial dataset. In particular,
\begin{align*}
\{ \bm \mu_y^{(k)} \} = \left( \begin{array}{rrrrr}
 0.715   &   0.559   &   -0.649   &   -0.085   &   0.677   \\
 0.581   &   0.406   &   -1.368   &   -0.207   &   0.799   \\
 0.329   &   0.175   &   -1.404   &   -0.851   &   0.944   \\
 0.319   &  -0.217   &   -1.650   &   -1.181   &   1.276   \\
 0.889   &  -0.473   &   -1.765   &   -1.363   &   0.483   \\
-0.664   &  -0.593   &   -3.195   &   -1.562   &   1.081   \\
\end{array} \right), 
\end{align*}
where $\bm \mu_y^{(k)}$ corresponds to the $k$-th column. Then, we add the effects of auxiliary covariates by randomly generating $B^{(k)}$ and $\Phi^{(k)}$ (values not shown). Based on $B^{(k)}$, $\Phi^{(k)}$, $\Sigma_{vv}$ and $\Sigma_0^{(k)}$ we calculate $\Omega_0^{(k)}$. 
Finally, we generate $\Omega^{(k)} \sim \mathcal{W}^{-1}\left( (\nu - J - Q - 1) \Omega_0^{(k)}, \nu \right)$ and get
\begin{align*}
\Omega^{(1)} = \left( \begin{array}{rrrrrr|rrr}
0.9 &  1.3  & 1.7  & 1.9 &  2.3  & 2.6  &-1.0  & -0.4  & 0.4   \\
1.3  & 2.2  & 2.9  & 3.4 &  4.2 &  4.9 & -1.6 & -0.4  & 0.9   \\
1.7  & 2.9 &  4.1 &  4.8  & 5.9 &  7.0  & -2.1 & -0.4  & 1.4   \\
1.9 &  3.4  & 4.8  & 5.8 &  7.1 &  8.3 & -2.4 & -0.3  & 1.7   \\
2.3 &  4.2  & 5.9 &  7.1 &  8.8 & 10.4 & -3.0 &  -0.4 &  2.2   \\
2.6 &  4.9  & 7.0  &  8.3  &10.4 & 12.2 & -3.5 & -0.4 &  2.6   \\ \hline 
-1.0  & -1.6 & -2.1&  -2.4  &-3.0 &  -3.5  & 1.7 &  0.5 & -0.2   \\
-0.4 & -0.4 & -0.4 & -0.3  &-0.4 & -0.4 &  0.5 &  0.7 & -0.1   \\
0.4  & 0.9  & 1.4 &  1.7  & 2.2  & 2.6 & -0.2 & -0.1 &  1.2   \\
\end{array} \right),
\end{align*}
\begin{align*}
\Omega^{(2)} = \left( \begin{array}{rrrrrr|rrr}
0.2 &   0.3 &   0.3 &   0.4 &   0.5 &   0.6 &  -0.2 &  -0.3 &   0.3 \\
0.3 &   0.6 &   0.8 &   1.0  &   1.3 &   1.6 &  -0.2 &  -0.3 &   0.7 \\
0.3 &   0.8 &   1.2 &   1.5 &   1.9 &   2.4 &  -0.3 &  -0.2 &   1.0  \\ 
0.4 &   1.0  &   1.5 &   2.0  &   2.5 &   3.1 &  -0.4 &  -0.1 &   1.2 \\
0.5 &   1.3 &   1.9 &   2.5 &   3.2 &   4.0  &  -0.4 &  -0.2 &   1.6 \\
0.6 &   1.6 &   2.4 &   3.1 &   4.0  &   5.0  &  -0.4 &  -0.2 &   2.1 \\  \hline 
-0.2 &  -0.2 &  -0.3 &  -0.4 &  -0.4 &  -0.4 &   0.5 &   0.1 &   0.1 \\
-0.3 &  -0.3 &  -0.2 &  -0.1 &  -0.2 &  -0.2 &   0.1 &   0.9 &  -0.4 \\
 0.3 &   0.7 &   1.0  &   1.2 &   1.6 &   2.1 &   0.1 &  -0.4 &   1.2 \\
 \end{array} \right),
\end{align*}
\begin{align*}
\Omega^{(3)} = \left( \begin{array}{rrrrrr|rrr}
1.2 &   1.3 &   1.3 &   1.3 &   1.5 &   1.6 &  -0.8 &  -0.8 &   0.4 \\
1.3 &   1.5 &   1.6 &   1.7 &   1.9 &   2.1 &  -0.9 &  -0.7 &   0.6 \\
1.3 &   1.6 &   1.7 &   1.9 &   2.2 &   2.4 &  -0.9 &  -0.5 &   0.7 \\
1.3 &   1.7 &   1.9 &   2.1 &   2.4 &   2.7 &  -0.9 &  -0.4 &   0.8 \\
1.5 &   1.9 &   2.2 &   2.4 &   2.9 &   3.3 &  -1.1 &  -0.4 &   0.9 \\
1.6 &   2.1 &   2.4 &   2.7 &   3.3 &   3.7 &  -1.2 &  -0.3 &   1.1 \\  \hline
-0.8 &  -0.9 &  -0.9 &  -0.9 &  -1.1 &  -1.2 &   0.8 &   0.5 &  -0.1 \\
-0.8 &  -0.7 &  -0.5 &  -0.4 &  -0.4 &  -0.3 &   0.5 &   0.9 &  -0.1 \\
0.4 &   0.6 &   0.7 &   0.8 &   0.9 &   1.1 &  -0.1 &  -0.1 &   0.6 \\
 \end{array} \right),
\end{align*}
\begin{align*}
\Omega^{(4)} = \left( \begin{array}{rrrrrr|rrr}
1.0  &   1.3 &   1.5 &   1.7 &   2.0  &   2.2 &  -0.9 &  -0.7 &   0.5 \\
1.3 &   2.0  &   2.4 &   2.7 &   3.2 &   3.6 &  -1.4 &  -0.7 &   0.6 \\
1.5 &   2.4 &   2.9 &   3.4 &   4.0  &   4.5 &  -1.7 &  -0.7 &   0.8 \\
1.7 &   2.7 &   3.4 &   4.0  &   4.7 &   5.4 &  -2.0  &  -0.7 &   0.9 \\
2.0  &   3.2 &   4.0  &   4.7 &   5.6 &   6.3 &  -2.3 &  -0.7 &   1.0  \\
2.2 &   3.6 &   4.5 &   5.4 &   6.3 &   7.3 &  -2.6 &  -0.7 &   1.2 \\  \hline
-0.9 &  -1.4 &  -1.7 &  -2.0  &  -2.3 &  -2.6 &   1.3 &   0.5 &  -0.1 \\
-0.7 &  -0.7 &  -0.7 &  -0.7 &  -0.7 &  -0.7 &   0.5 &   0.9 &  -0.2 \\
 0.5 &   0.6 &   0.8 &   0.9 &   1.0  &   1.2 &  -0.1 &  -0.2 &   0.7 \\
 \end{array} \right), 
\end{align*}
\begin{align*}
\Omega^{(5)} = \left( \begin{array}{rrrrrr|rrr}
0.8 &   1.0  &   1.3 &   1.5 &   1.7 &   2.0  &  -0.8 &  -0.4 &   0.5 \\
1.0  &   1.7 &   2.2 &   2.7 &   3.2 &   3.8 &  -1.4 &  -0.3 &   0.7 \\
1.3 &   2.2 &   2.9 &   3.7 &   4.3 &   5.1 &  -1.8 &  -0.2 &   1.0  \\
1.5 &   2.7 &   3.7 &   4.8 &   5.7 &   6.7 &  -2.2 &  -0.1 &   1.2 \\
1.7 &   3.2 &   4.3 &   5.7 &   6.8 &   8.0  &  -2.6 &  -0.1 &   1.4 \\
2.0  &   3.8 &   5.1 &   6.7 &   8.0  &   9.5 &  -3.1 &  -0.0  &   1.7 \\  \hline
-0.8 &  -1.4 &  -1.8 &  -2.2 &  -2.6 &  -3.1 &   1.4 &   0.3 &  -0.3 \\
-0.4 &  -0.3 &  -0.2 &  -0.1 &  -0.1 &  -0.0  &   0.3 &   0.5 &  -0.1 \\
0.5 &   0.7 &   1.0  &   1.2 &   1.4 &   1.7 &  -0.3 &  -0.1 &   0.7 \\
 \end{array} \right),
\end{align*}

The parameters for generating $S$ are 
\begin{align*}
\bm \zeta = (-2.61, -2.75, -2.08, -1.52)^T, 
\end{align*}
where $\zeta_s$ corresponds to the $(s-1)$-th element ($s = 2, \ldots, 5$), 
\begin{align*}
\bm \ell = (-0.96, 0.66, 0.78, 0.54)^T, 
\end{align*}
where $\ell_s$ corresponds to the $(s-1)$-th element ($s = 2, \ldots, 5$), and
\begin{align*}
\bm \xi = \left( \begin{array}{rrrr}
-1.057 &  0.328 & -0.121 &  0.273 \\
-0.826 &  0.128 &  0.525 & -0.781 \\
-0.487 &  0.479 &  0.534 & -0.480 
\end{array} \right),
\end{align*}
where $\bm \xi_s$ corresponds to the $(s-1)$-th column ($s = 2, \ldots, 5$).
The parameters are chosen to mimic the dropout rate of the real data.


\addcontentsline{toc}{chapter}{Bibliography} 
\bibliographystyle{Chicago}  
\bibliography{ref-diss}        
\index{Bibliography@\emph{Bibliography}}

\printindex

\end{document}